\documentclass[12pt]{article}
\usepackage[dvips]{graphicx}
\usepackage{amssymb}
\usepackage{amsmath}

\def\beq{\begin{equation}}\def\eeq{\end{equation}}
\def\bea{\begin{eqnarray}}\def\eea{\end{eqnarray}}

\begin{document}
 
\title{Can quantum lattice be generated through several classical ones superimposed in spacetime continuum?}
 
\author{Roman Sverdlov
\\CURRENT PLACE: Institute of Mathematical Sciences ,
\\IV Cross Road, CIT Campus, Taramani, Chennai, 600 113, Tamil Nadu, India
\\ PLACE WHERE WORK STARTED: Raman Research Institute 
\\C.V. Raman Avenue, Sadashiva Nagar, Bangalore}

\date{March 13, 2012}
\maketitle
 
\begin{abstract}
\noindent 
This paper has few different, but interrelated, goals. At first, we will propose a version of discretization of quantum field theory (Chapter 3). We will write down Lagrangians for sample bosonic fields (Section 3.1) and also attempt to generalize them to fermionic QFT (Section 3.2). At the same time, we will insist that the elements of our discrete space are embedded into a continuum. This will allow us to embed several different "lattices" into the same continuum and view them as separate "quantum" field configurations. Classical parameters will be used in order to specify "which lattice" each given element "belongs to". Furthermore, another set of "classical" parameters will be proposed in order to define so-called "probability amplitude" of each "field configuration", embodied by a corresponding lattice, "taking place" (Chapter 2). Apart from that, we will propose a set of "classical" signals that propagate throughout continuum, and define their dynamics in such a way that they produce the mathematical information consistent with the desired "quantum" effects within the lattices we are concerned about (Chapter 4).  Finally, we will take advantage of the lack of "true" quantum mechanics, and "add" gravity in such a way that avoids the issue of its quantization altogether (Chapter 5). In the process of doing so, we will propose a gravity-based collapse model of a wave function. In particular, we will claim that the collapse of a wave function is merely a result of states that violate Einstein's equation being "thrown away". The mathematical structure of this model (in particular, the appeal to gambler's ruin) will be similar to GRW collapse models. 
\end{abstract}

\subsection*{1. Introduction}

One of the main problems of interpretation of quantum mechanics is the fact that the "probability amplitude" $\psi$ does not have ontological meaning. In case of the single particle quantum mechanics, we can simply view $\psi$ as a "classical field" in par with electromagnetic one. In this case, the problem with interference is far less mysterious. In many body case, this can not be done. After all, we can not "touch" a wave function on configuration space in the same way as we can "touch" the one in ordinary one. We propose to address this issue by discretizing the configuration space and, thus, making it "smaller" than ordinary space and, therefore, "embeddable" inside the latter. In particular, if we have a "computer" inside the "classical universe", that "computer" can, in principle be "programmed" to reproduce the mathematical information of quantum mechanics. At the same time, behind the scenes, this "computer" is being "ran" by classical signals. This means that classical physics \emph{can} in principle reproduce quantum mechanics. One of the goals of this paper is to come up with a model of such a computer.

Now, the key reason why the "computer" would be able to reproduce quantum mechanical information is that its screen is \emph{not} "all there is there". There is some room "behind the scenes" where the information on the screen is being generated. In case of our model, we will view a set of discrete "lattice points" (or "particles" as we will sometimes call them) as "screen", while viewing the \emph{continuum} in which they are "embedded into" as "information behind the scenes". Now, in order to understand computer program, we don't need to understand the way in which the computer is actually built. We can explain the way program works \emph{strictly} in terms of what we see on the screen. In case of quantum field theory, the "program" is Lagrangian. Thus, Lagrangian is described in terms of discrete points, alone (Chapter 3) without reference to signals. At the same time, in reality "the program" \emph{is} the result of what is going on behind the scenes mechanically. Similarly, we will propose a mechanism that will produce the mathematical information consistent with the Lagrangians (Chapter 4).

 Nevertheless, the "program" itself is written in an "ontologically meaningful" terms. After all, even if we don't see anything that is behind the screen of the computer, we still can "touch" the "letters" we see \emph{on} the screen. Thus, if the computer will write down a solution to a certain quantum mechanical program then, still, the variation of color of the screen that "defines" the "letters" in which output has been produced is "classical". Similarly, the "output" regarding the "probability amplitudes" of various "quantum" sates is, likewise "classical". Since our "screen" is the set of discretized points, the "program output" (that is, a set of probability amplitudes) will be identified with "classical" parameters \emph{attached to} each particular "lattice point" (Chapter 3). On the other hand, the mechanism that produces the "program output" is \emph{not} on the screen; thus, it will be modeled in terms of signals propagating \emph{between} lattice points. The detailed differential equations that would cause them to produce the desired effects are given in Chapter 4.

This outlook allows us to view "parallel universes" as \emph{classical} entities co-existing within one \emph{single} space. In particular, the "lattice points" have different "charges". A given "universe" consists of all of the points of a given value of charge. Thus, if the charge takes the values from $1$ to $M$ we then have $M$ "parallel universes". Equivalently, they can be viewed as different lattices "superimposed" upon each other within the same space. This makes sense: a computer typically consist of several "parallel programs" written in it, even though the "particles" that make up computer are living within the same space.

Nevertheless, \emph{despite} the "mathematical compatibility" between our model and "parallel universes", we chose to instead side with "collapse-based" point of view, as far as Chapter 5 is concerned (although the Chapters 2, 3, and 4 are perfectly compatible with parallel universe point of view). The main reason we made this particular choice is gravity. We would \emph{like} gravity to be produced as an outcomes of "classical" phenomena as opposed to quantum mechanical since this would allow us to avoid the issues of its quantization. We propose that gravity and measurement process are intrinsically related. On the one hand, the measurement process attempts to "select" the states that violate Einstein's equation "the least". On the other hand, gravity is produced \emph{by} the states that have been affected through above-mentioned "collapse". 

While we believe that the "gravity" idea is original, the approach bears a lot of similarities with GRW model. In both cases, there is a parallel between the "collapse mechanism" and the game "gambler's ruin". As was described by Pearle (\cite{Pearle1} and \cite{Pearle2}), the game is based on players exchanging very small amounts of money at random. The game stops when one of the players has lost all of their money. It turns out that the probability of winning is proportional to the amount of money a given player had "at the start". The key to GRW model is a mechanism that would "similate" Gambler's ruin; by making quantum states "exchange" small amounts of $\vert \psi \vert^2$ "at random". 

According to "usual" GRW models, this mechanism is based on the wave function being multiplied by "very wide" Gaussians at "random" points in time and centered around "random" locations in space. In our proposal, we decided to "replace" multiplications by Gaussians "in favor of" a gravity-based mechanism. In particular, we claim that the quantum state that violates Einstein's equation "slightly more" than other quantum states will be the state "losing money".In light of the fact that the quantum state is defined in a non-local way, we don't know which state violates Einstein's equation the most since our observations are confined to our laboratory. Therefore, the violations of Einstein's equation in regions unaccessible to our observations is used as a "random coin" in "gambler's ruin" and, therefore, "produces" Born's rule. 

In the same way as with everything else, we then go on to propose a "mechanism" through which these "collapses" occur. In other words, part of "computer program" is to "lower" the value of $\vert \psi \vert^2$ of the states that violate Einstein's equation the most. But, the mechanism needs to be specified in which the computer uses a set of signals to "find out" the mathematical information consistent with "state violating Einstein's equation the most" and, likewise, we need a mechanism that \emph{results} in $\vert \psi \vert^2$ being lowered. The proposal of one possible such mechanism is one of the main purposes of Chapter 5. 

In light of the fact that the probability amplitude of configuration is defined in a non-local way, our mechanism has to be based on superluminal signals. However, we make it a point that the superluminal signals we introduce are \emph{still} moving with the finite speed, even though this speed is much larger than the speed of light. Furthermore, we assume that our universe is compact. This allows the signals to circle the universe within a very short (but finite) time. Thus, if we don't have a devices sensitive enough to detect the consequence of that very small time delay, it appears to us that physics is consistent with "true non-locality". Needless to say, the fact that our signals move faster than the speed of light means that we do \emph{not} believe in relativity. We feel that relativity might appear in the lab as a result of "coincidence in Lagrangians". At the same time, we \emph{do} believe in locality, mainly because non-locality would contradict our intuition. 

Let us now summarize the way in which this paper is put together. The Chapter 2 will attempt to outline the specific way in which the "computer screen" is meant to be "read". Neither the "program" (that is, specific Lagrangians), nor the mechanism behind them (that is, the superluminal signals), is introduced in Chapter 2. However, the information there is crucial in terms of understanding how to "read the letters" that appear "on the screen" of the computer. For example, it explains how "probability amplitudes" are being "encoded". As far as Chapter 3 is concerned, it explains the "program" itself. In other words, it explains Lagrangians for bosonic and fermionic fields. At the same time, however, it does not attempt to tackle the mechanism behind the program; only the program. Then, in Chapter 4, we are attempting to describe the mechanism that implements the "program". Finally, in Chapter 5 we "add" gravity as well as "collapse mechanism" to the framework developed.

Now, since computer is \emph{first} built and \emph{then} the program is written, Chapter 4 is logically independent from Chapter 3.While Chapter 4 refers to an abstract Lagrangian, it never specifies what it is, and for a good reason. We would like to be able to "plug in" any Lagrangian we like "into" Chapter 4; thus, we don't want to "limit ourselves" by specifying a Lagrangian. This is similar to the fact that we are able to write any program we like, once the computer is properly built. The main reason we decided to \emph{first} write Chapter 3 and \emph{then} Chapter 4 is mainly pedagogical. We believe that the "Lagrangian generator", which is defined as a three point function, is a bit too abstract to understand. Thus, we believe it might help the reader if they first see some examples as how it ends up producing ordinary quantum field theory Lagrangians that we are used to. This would allow one to gain intuitive sense regarding the "ultimate goal" of Chapter 4, before they plunge into quite a detailed material that this chapter involves.  However, since Chapter 3 quickly gets very inolved as well (particularly, Section 3.2 on fermions), it might be advisable for a reader that they just read Section 3.1 and then immediately read Chapter 4. 

At the same time, Chapter 3.2 has a lot of material that might be interesting on its own right as it attempts to generalize 3.1 to fermions. Along the way, it accomplishes several goals, including the definition of Grassmann numbers as literal mathematical objects well defined outside of path integral and shows how the integral can be "produced" as a literal "sum" (3.2.1-3.2.4) and then proceeds to produce a "bosonic" Lagrangian that "produces" the fermionic process of section 3.2.4.

The main reason we chose to put Chapter 3 and Chapter 4 into the same paper is that, from "logic" point of view, one has to "be convinced" that "all fields" (including fermions) can be modeled as Lagrangian generators in order for one to "believe" that Chapter 4 is not pointless. After all, if a certain "program" (namely, Fermionic amplitudes) is "not logically possible", then doing a good job trying to build a computer (Chapter 4) won't help us.  Nevertheless, even if one is "not convinced" they might still be interested in how things might work in "boson only toy model". In this sense Chapter 4 can be read right after Chapter 3.1. Chapter 4 is, in fact, written in terms of bosonic fields alone. It \emph{assumes} that we "agree" that fermions can be modeled as bosons per Chapter 3 and, which is why it does not have fermionic part in it. Nevertheless, this is something that pertains to motivation only. What is actually being done does not have any reference to Chapter 3. 

To make long story short, a reader interested in mathematical aspects of quantum mechanics without being interested in turning it "classical" is advised to focus on Chapter 3. A reader interested in "encoding" quantum mechanics into a classical framework is advised to focus on Chapter 4. A reader interested in questions of quantum measurement as well as gravity is advised to focus on Chapter 5. All three chapters have Chapter 2 as a prerequisite and, therefore, Chapter 2 should be understood by all readers (although Chapter 3 relies on Chapter 2 a lot less than the other two chapters). Chapter 4 is logically independent from Chapter 3. Reading Chapter 3.1 might give some intuition for Chapter 4, but nothing more is necessary. Chapter 5 has Chapter 4 as a prerequisite. But, at the same time, it is substantially simpler than Chapter 4; therefore, superficial reading of Chapter 4 would probably suffice. As far as Chapter 5 is concerned, the Chapter 3 is not needed at all. 

\subsection*{2. More elementary concepts of the proposal}

\subsection*{2.1 Quantum states as a superluminal-local scenario in $\mathbb{R}^4$}

We would like to "embed" the mathematical information of quantum field theory, presented in path integral formulation, into ordinary three dimensional space. The key reason why such embedding is possible is that we will first discretize quantum field theory. Thus, the \emph{discretized} configuration space will be "smaller" than the ordinary \emph{continuum} we are trying to embed it into. Our aimed for discretized theory consists of fixed time slices, separated by the time $\delta t$, and large, but finite, number of "lattice particles", which are distributed chaotically in space. 

From the point of view of "classical" field theory, a "lattice point" number $k$ has internal degree of freedom $\phi_k (t)$ and $A_k^{\mu} (t)$, which tells us the value of $\phi (\vec{x}_k, t)$ and $A^{\mu} (\vec{x}_k, t)$, respectively. In order to make notation more intuitive, we can also denote them as $\phi (k, t)$ and $A^{\mu} (k, t)$:
\beq \phi_k (t) = \phi (k, t) \; ; \; A^{\mu}_k (t) = A^{\mu} (k, t) \eeq
We will freely go between these two choices of notation, depending on our convenience in any given situation. In order to "quantize" the above, we introduce dynamics (see section 2.3) that "forces" points to "group into clusters" in such a way that the size of each cluster is several magnitudes smaller than the distances between clusters. This allows us to replace a "point number $k$" with a "cluster of points $C_k$".  Since \emph{each} of the points within the same cluster has its \emph{own} values of $A_k^{\mu} = A^{\mu} (k)$ and $\phi_k = \phi (k)$, we can "change" these fields by "going" from one point to another, \emph{while staying within the same cluster}. Since the distance between the points within the same cluster is very small, we "don't notice" how we "move" within the cluster; instead, the only thing we "see" is alteration in the field. This is a very vague idea of how we "embed" the quantized fields into regular spacetime. 

We will identify a "point" by a number it represents, such as "point particle number $i$". Likewise, we will also number clusters. We will use small indexes to number points and capital indexes to number clusters. We will identify cluster number $K$ with the set of point particles that it contains, 
\beq C_k = {p_{k1}, \cdots, p_{k \sharp C_k}}, \eeq
where $p_i$ is point number $i$ and $\sharp C_k$ denotes the number of points in the cluster. In addition to this, the "charges" of the point particles will be introduced. These charges are not to be confused with the charges seen in electromagnetism. For one thing, they took values between $1$ and $M$, where $M$ is some very large number. Furthermore, there are two "charge" parameters: $q$ and $q'$. We will denote their respective values for particle number $k$ by $q_k$ and $q'_k$.

As we mentioned earlier, in "classical" lattice field theory, we define fields in terms of internal degrees of freedom of each lattice point, $\phi_i (t)$ and $A_i^{\mu} (t)$. Quantum mechanically, we would like to \emph{superimpose} different "possibilities" of $\phi_i (t) = \phi (i, t)$ and $A_i^{\mu} (t) = A^{\mu} (i,t)$ upon each other. We will do so by replacing "point-based" $\phi_i (t)= \phi (i, t)$ and $A_i^{\mu} (t) = A^{\mu} (i,t)$ with "cluster-based" $\phi_{C_I} (t) = \phi (C_I, t)$ and $A_{C_I}^{\mu} (t) = A^{\mu} (C_I, t)$. We will then "generate" $M$ different "possibilities" of these fields. A "possibility" number $q$ is given by 
\beq \phi_{C_I q} = \phi_q (C_I) = \frac{1}{\sharp{C_{Iq}}} \sum_{i \in C_{iq}} \phi_i \; ; \; A^{\mu}_{C_I q} = A_q^{\mu} (C_I) = \frac{1}{\sharp{C_{Iq}}} \sum_{i \in C_{iq}} A^{\mu}_i \eeq
where $C_{Iq}$ is a set of particles that belong to cluster $C_I$ and have "charge" $q$:
\beq C_{Iq} = \{i \vert i \in C_i \; ; \; q_i = q \} \eeq
At the same time, these \emph{so-called} "possibilities" are \emph{physically present} in the form of internal (classical) degrees of freedom of corresponding particles. Thus, they are not really possibilities but rather \emph{classical quantities}. The only reason we call them "possibilities" is to draw a link to a "standard" quantum field theory. Furthermore, due to the dynamics in Chapter 4, a "consistency condition" is satisfied, namely
\beq ({\rm Consistency \; of \;} \phi_q) \; \Rightarrow ((C_i = C_j) \wedge (q_i = q_j) \Rightarrow (\phi_i \approx \phi_j))  \eeq
\beq ({\rm Consistency \; of \;} A^{\mu}_q) \; \Rightarrow ((C_i = C_j) \wedge (q_i = q_j) \Rightarrow (A^{\mu}_i \approx A^{\mu}_j))  \eeq
Apart from that, we need to have "probability amplitude" of any given field configuration just described, and a "transition amplitude" from one configuration to the other. In order to encode the "probability amplitudes", we introduce internal degree of freedom $\psi_i (t)$, where $i$ is a "number" of the particle this internal degree of freedom is attached to.  In light of the above constraint, we can identify the spacelike-average of $\psi_i$ with a probability amplitude of $(\phi, A^{\mu})= (\phi_q, A^{\mu}_q)$:
\beq \psi [(\phi, A^{\mu}) = (\phi_q, A_q^{\mu}); t] \; =  \; \psi_{q=q_i} (t)= \frac{1}{\sharp D_q} \sum_{i \in D_q} \psi_i (t), \eeq
where we have assumed a "preferred time" $t$, and the set $D_q$ consists of all particles with charge $q$: 
\beq D_q = \{i \vert q_i =q \} \eeq
 In order for the probability amplitude to be "tangible" locally, we subject $\psi_i (t)$ to a constraint 
\beq {\rm Consistency \; of \;} \psi \Rightarrow [(q_i = q_j) \Rightarrow (\psi_i (t) = \psi_j (t))], \eeq
where again we assume the absolute time. This allows us to "rewrite" the above condition as 
\beq {\rm Consistency \; of \;} \psi \Rightarrow \psi [(\phi, A^{\mu}) = (\phi_{q_i}, A_{q_i}^{\mu}); t] \; =  \psi_{q=q_i} (t) = \psi_i (t) \eeq
Now, the transition amplitudes from $(\phi, A^{\mu}) = (\phi_{q_1}, A_{q_1}^{\mu})$ to $(\phi, A^{\mu}) = (\phi_{q_2}, A_{q_2}^{\mu})$ is given by 
\beq e^{-iS ((\phi_{q_1}, A_{q_1}^{\mu}) \rightarrow (\phi_{q_2}, A_{q_2}^{\mu}))} = e^{-iS_{q_1 q_2}} = exp \Big(- \frac{i}{\sharp D_{q_1 q_2}} \sum_{j \in D_{q_1 q_2}} S_j \Big) \eeq
where $D_{q_1 q_2}$ is defined as 
\beq D_{q_1 q_2} = \{i \vert q_i = q_1 ; q'_i = q_2 \} \eeq
The use of $q'_i$ in the above definition of $D_{q_1 q_2}$ is the sole purpose of why the "additional charge" $q'$ is introduced. A particle number $i$ "computes" a probability amplitude of transition from $(\phi_{q_i}, A_{q_i})$ to $(\phi_{q'_i}, A_{q'_i})$. Similarly to what we did with $\psi_i$, we will impose a constraint on $S_i$:
\beq (q_i, q'_i) = (q_j, q'_j) \Rightarrow S_i = S_j \eeq
This will allow us to rewrite the probability amplitude as 
\beq e^{-iS \big(\big(\phi_{q_i}, A_{q_i}^{\mu} \big) \rightarrow \big(\phi_{q'_i}, A_{q'_i}^{\mu} \big) \big)} = e^{-iS_i} \eeq
The evolution of $\psi_i$ under $S_i$ is, likewise, written in terms of the internal degrees of freedom of lattice particles; namely, 
\beq \psi_i (t) = \sum_{ \{j \vert q'_j = q_i \}} e^{-iS_j} \psi_j (t- \delta t) \eeq
Now, since there is no correlation between $q_j$ and $q'_j$, the sum on the right hand side goes over \emph{all} possible values of $q_j$. Thus, it can be broken down as 
\beq \psi_i (t) = \sum_a \sum_{ \{j \vert q_j = a; q'_j = q_i \}} e^{-iS_j} \psi_j (t- \delta t) \eeq
Now, based on the global correlations we have specified, we know that 
\beq (q_j = a; q'_j = q_i) \Rightarrow (\psi_j (t)= \psi_{q=a} (t); S_j = S_{a,q_i}) \eeq
Upon substitution of this into the above sum, we obtain 
\beq \psi_i (t) = \sum_a \sum_{ \{j \vert q_j = a; q'_j = q_i \}} e^{-iS_{a,q_i}} \psi_{q=a} (t- \delta t) \eeq
Finally, if we assume that the number of particles of $q=a$ and $q'=b$ is approximately the same, regardless of the values of $a$ and $b$. In this case, we can replace the second "sum" with a "very large" coefficient $K$, and obtain 
\beq \psi_i (t) = K \sum_a  e^{-iS_{a,q_i}} \psi_{q=a} (t- \delta t) \eeq
Now, a \emph{path} is simply a sequence of integers $\cdots a_{-1}, a_0, a_1, \cdots$ that tells us that "at a time $t=n \delta t$ the system 'picked' a state corresponding to $q=a_n$". The action corresponding to that path is given by 
\beq  S(\cdots, a_{-1}, a_0, a_1, \cdots) = \epsilon \; \sum_{n= - \infty}^{n= \infty} S (q=q_{a_n} \; ; \; q'=q_{a_{n+1}})\eeq
where $\epsilon$ is a very small constant. The action corresponding to that path is given by 
\beq S(\cdots, a_{-1}, a_0, a_1, \cdots) = \epsilon \; \sum_{n= - \infty}^{n= \infty} S_{k_n} \eeq
where $k_n$ is the "index" of a particle which is selected in such a way that 
\beq q_{k_n} = a_n \; ; \; q'_{k_n} = a_{n+1} \eeq

It is now easy to see that iteration of the above equation produces Feynmann path integral, as desired. This implies that "time derivative" can now be "replaced" with comparison between $q$-related quantity and $q'$-related one taken "at the same time":
\beq \frac{\partial \phi}{\partial t} \rightarrow \frac{\phi_{q'} (C_q, t) - \phi_{q} (C_q, t)}{\epsilon} \eeq
where $\epsilon$ is a small, but finite, constant. In order to remind ourselves of the intended physical meaning, we will replace $\epsilon$ with $\delta t$. However, it is important to remember that we are dealing with $t= \rm const$ time slice; the parameter $\delta t$ is merely a "constant" that \emph{happens} to be equal to time increment in discrete settings. 

\subsection*{2.2 Lagrangian generators}

Let us now discuss how we can write down the probability amplitude of transition more explicitly. Furthermore, we will also show more explicitly that we will, indeed, reproduce the expected expression for Feynman path integral. In case of single-particle, non-relativistic, quantum mechanics, we have the following intuitive picture. A particle has a probability $1+ i \epsilon S(\vec{x}, \vec{x}+ \delta \vec{x})$ of going from the point $\vec{x}$ to the point $\vec{x} + \delta \vec{x}$ within time interval $\delta t$. Thus, the "probability amplitude" of traveling along some fixed path is a "product" of "probability amplitudes" of travelling along each of the above mentioned infinitesimal segments of that path. After that, we are "summing over" all possible paths. 

The naive picture of "taking product", however, is not completely accurate. If we explicitly write down Lagrangian as a difference between kinetic and potential term, ${\cal L}=K-V$, then its exponent will be the "product" of these terms, $e^{-i \cal L} = e^{-iK} e^{iV}$. Now, $e^{-iK}$ represents the probability amplitude of "going" from $\vec{x}$ to $\vec{x} + \delta \vec{x}$. On the other hand, $e^{iV}$ represents the probability amplitude of "being" at a point $\vec{x}$. Thus, on a segment from $\vec{x}$ to $\vec{x} + \delta \vec{x}$ we take the product of \emph{three} probability amplitudes: starting at $\vec{x}$, jumping from $\vec{x}$ to $\vec{x} + \delta \vec{x}$ and then arriving at $\vec{x} + \delta \vec{x}$. This is done \emph{despite} the fact that the first two events "logically imply" the third one.

This feature has been noticed by Johnston (\cite{Johnston1} and \cite{Johnston2}) in the context of the formulation of Green's function in discretized \emph{relativistic} spacetime. He termed this as "hops" and "stops": a particle first "stops" at point $a_1$, then it "hops" from point $a_1$ to point $a_2$, then it "stops" at $a_2$, then "hops" from $a_2$ to $a_3$, then "stops" at $a_3$, and so forth. In his terms, we have to take the product of the probabilities of both "hops" \emph{and} "stops", \emph{despite} the fact that any particle that makes a "hop" is guaranteed to make a "stop". Our approach is different from his in several ways. For one thing, his work (being the one in causal set theory) assumes absence of any kind of geometry apart from causal relations; apart from that, his work is in particle-based representation while this paper is field-based. Nevertheless, the element of "overlap" between "hops" and "stops" is one of the few things that is being shared by both approaches. 

When we describe path integral over a field distributions, things become even more interesting. For one thing, the spacelike information goes into Lagrangian \emph{along with} timelike. For example, in case of scalar field, one of the "multiples" of $e^{iS}$ is $1+ \epsilon (\partial \phi/ \partial x)^2$. This can be interpreted in several ways. We can \emph{either} think of it as a "probability" of "co-occurrence" $\phi (x_0) = \phi_0$ and $\phi (x_0 + \delta x) = \phi_0 + \delta \phi$ \emph{but} we can \emph{also} think of it as a probabiity of "co-occurrence" $\phi (x_0) = \phi_0$ and $\phi (x_0 - \delta x) = \phi_0 - \delta \phi$.  Even if we look at something "timelike", we get similar dilemma. The event $\phi (\vec{x}_0, t_0) = \phi_0$ can \emph{either} be viewed as "co-occurring" with $\phi (\vec{x}_0 + (\delta \vec{x})_1, t_0 + \delta t) = \phi_0 + (\delta \phi)_1$, \emph{or} it can be seen as "co-occurring" with $\phi (\vec{x}_0 + (\delta \vec{x})_2, t_0 + \delta t) = \phi_0 + (\delta \phi)_2$ (here, we assumed that the value of $\delta t$ is the same in both cases in order to be consistent with the Section 2.1 that implies that $q'$-charge helps us make predictions regarding "uniform future" $t+ \delta t$).
In this paper, we propose to view a behavior of a field as a set of co-occurrences of \emph{triples} of events. In case of the scalar field, the "co-occurrences" that are under consideration are the following:
\beq {\rm Option \; 1:} \; \phi (\vec{x}_1, t) = \phi_1 \; ; \; \phi (\vec{x}_2, t) = \phi_2 \; ; \; \phi (\vec{x}_3, t) = \phi_3 \eeq
\beq {\rm Option \; 2:} \; \phi (\vec{x}_1, t) = \phi_1 \; ; \; \phi (\vec{x}_2, t) = \phi_2 \; ; \; \phi (\vec{x}_3, t + \delta t) = \phi_3 \eeq
\beq {\rm Option \; 3:} \; \phi (\vec{x}_1, t) = \phi_1 \; ; \; \phi (\vec{x}_2, t+ \delta t) = \phi_2 \; ; \; \phi (\vec{x}_3, t + \delta t) = \phi_3 \eeq
This implies even more gross overlap between different correlations. For example, all three options listed above "overlap" regarding their claim that $\phi (\vec{x}_1, t) = \phi_1$. Yet, their probabilities should be \emph{separately} computed and \emph{then} multiplied by each other. Thus, we have to \emph{first} take into account $\phi (\vec{x}_1) = \phi_1$ in the process of computing the probability of Option 1, and \emph{then} take \emph{the same} information into account while computing the probability of Option 2. The respective "probability amplitudes" of these three options are defined as exponentials of \emph{Lagrangian generators} $\cal K$:
\beq {\rm Amp (Option \; 1)} = \exp \; {\cal K} (\vec{x}_1, \phi_1, 0 \; ; \; \vec{x}_2, \phi_2, 0 \; ; \; \vec{x}_3, \phi_3, 0) \nonumber \eeq
\beq {\rm Amp (Option \; 1)} = \exp \; {\cal K} (\vec{x}_1, \phi_1, 0 \; ; \; \vec{x}_2, \phi_2, 0 \; ; \; \vec{x}_3, \phi_3, \delta t) \eeq
\beq {\rm Amp (Option \; 1)} = \exp \; {\cal K} (\vec{x}_1, \phi_1, 0 \; ; \; \vec{x}_2, \phi_2, \delta t \; ; \; \vec{x}_3, \phi_3, \delta t) \nonumber \eeq
where we have used the translational symmetry in time in order to replace $t$ and $t+ \delta t$ with $0$ and $t$, respectively. Thus, the product of amplitudes turn into exponent of the sum. In general, if we are attempting to compute "collective probability" of several options, then we use
\beq {\rm Amp} (({\rm Option \; 1}, \cdots, {\rm Option \; n}) = \prod_j {\rm Amp} (j) = \nonumber \eeq
\beq = \prod_j \exp {\cal K} ({\rm Option \; j}) = \exp \Big(\sum_j {\cal K} ({\rm Option \; j}) \Big) \eeq
Now, since we know that the probability amplitude is given as $e^{iS}$ we can immediately "read off" that $S$ is a sum of the Lagrangian generators:
\beq S = \sum_{i, j, k} {\cal K} (i, j, k) \label {SK}\eeq
where $i$, $j$ and $k$ are the indexes of the three points we are selecting. Thus, the product rule works for probability amplitudes:
\beq e^{iS} = \prod e^{i {\cal K} (i, j, k)} \eeq
Now, we can break down \ref{SK} in the following way: 
\beq S = \sum_k \sum_{i, j} {\cal K} (i, j, k) \eeq
This will allow us to further rewrite it as
\beq S ({\rm field})= \sum_k {\cal L}_k \; , \; {\rm where} \; {\cal L}_k = \sum_{i, j} {\cal K} ({\rm field} ; i, j, k) \eeq
The expression ${\cal L}_k$ is referred to as \emph{Lagrangian density} at $k$. In section 3.1 we will show that it is possible to define $\cal K$ in such a way that ${\cal L}_k$ will approximate the Lagrangian density at $k$ of the bosonic field we are interested in, provided that the latter is smooth and slowly varying with respect to our discreteness scale.

One should also notice that, as far as time is concerned, we only have two options: $t$ or $t+ \delta t$, where $\delta t$ in all cases is taken to be the same. This grossly contradicts relativity. But, it is in line with the philosophy of this paper: we believe contradiction of relativity is necessary, anyway, in order to "explain" quantum mechanics in a classical way, which is our ultimate goal. This, of course, should not be confused with \cite{Johnston1} and \cite{Johnston2} where entire framework is intrinsically relativistic. The only similarity between what we are doing and \cite{Johnston1} is the fact that in both cases we have to "count" the same event "more than once" while "taking the product" of probabilities. 

\subsection*{2.3 Mechanism of formation of clusters}

In light of the fact that we have "mechanistic" view in this paper, if we are to have "clusters of points", there has to be a mechanism through which these clusters are formed. Thus we would like to claim that the formation of clusters is due to the interaction the particles have, via certain field. We will postulate that the field obeys the following dynamical equation: 
\beq \nabla_s^{\alpha} \nabla_{s; \alpha} U + m^2_U U + \zeta \partial_0 U = \sum_{j=1}^N Q_j (t) \delta^3 (\vec{x} - \vec{x}_j), \eeq
where by $\nabla_s$ we mean that it is based on "superluminal" metric; that is, the metric that generates the superluminal signals discussed in the previous section. The $\zeta$-term notably violates Lorentz covariance and its purpose is the attinuation of fields so that they are only short-acting; violation of Lorentz covariance is ''okay'' since it has to be violated regardless, due to the combination of $c_o$ and $c_s$. Furthermore, we will assume that for the \emph{most} particles $j$, $Q_j$ is zero; it only has non-zero value for the \emph{very few} partiles. The mass $m_U$ of these fields is very small, so that they can circle universe several times. But, at the same time, its finite nature prevents them from circling the universe "for too long"; in particular, their lifetime is "too short" to create a resonance anywhere and the only region where the field is "strong" is a vicinity of a given point. 

Now, we would like to say that our particles are \emph{classical} and they are subjected to some potential $V (x)$. Their trajectory is give by 
\beq m_p \frac{d^2 \vec{x}_i}{dt^2} = - \vec{\nabla} V - \lambda \frac{d \vec{x}_i}{dt}, \eeq
where $m_p$ is a particle mass and is \emph{not to be confused} with $m_U$. We were "allowed" to introduce a "friction" $\lambda$ due to the fact that we \emph{are} allowed to violate relativity, as was stated earlier. Now, the potential $V(x)$ is a function of $U(x)$ given by 
\beq V (x) = a_1 V^{b_1} (x) - a_2 V^{b_2} (x) + a_3 V^{b_3} (x) \eeq
where $a_1 < a_2 < a_3$, and $b_1$, $b_2$ and $b_3$ are some other appropriately chosen coefficients. This means that $V$ has local maximum at $U_1$, local minimum at $U_2$, and diverges at $U \rightarrow \infty$, where $U_1 < U_2$.  Now, we can reasonably conjecture that the dynamics of $U$, as described above, will imply that $U$ is singular at location of the particle and then steadily decreases around it. If $U$ is symmetric, then $V$ will have "local minimum" at a distance $r_1$ from the particle and the local maximum at the distance $r_2$, where $r_1 < r_2$. 

 Now, the \emph{only} place where $Q$ is present is the dynamics of $U$. At the same time, $Q$ is \emph{absent} from the definition of $V$ based on $U$ and it is also \emph{absent} in the equation for acceleration. This implies that particle with $Q \neq 0$ influences \emph{all} other particles, \emph{including} the ones with $Q=0$; but at the same time, particles with $Q=0$ do not influence \emph{any} other particles (regardless of their charge). In other words, if we have one particle with $Q \neq 0$ and the other with $Q =0$, then the former influences the latter, but the latter does \emph{not} influence the former. 

Now, as we stated earlier, we assume that our universe is compact. Let us also assume that the number of particles in the universe is so large, that in order for them to be able to "fit" there, each particle has to has \emph{a lot} of other particles within a ball of radius $r_1$ surrounding that particle. Furthermore, we will assume that the number of particles with $Q \neq 0$ is many times smaller than the number of particles with $Q=0$. At the same time, we have enough particles with $Q \neq 0$ so that any given ball of radius $r_1$ has at least one such particle; but, at the same time, the number of such particles within that ball can not be much greater than $1$, either. 

In this case, the particles with $Q=0$, being influenced by the particle with $Q=1$ will gather around its local minimum. Thus, they will be somewhere in the vicinity of sphere of radius $r_1$ surrounding $Q=0$ particle. At the same time, the particles with $Q=0$ will \emph{not} interact with each other. Thus, we don't have to be bothered by the fact that their density will be very hight. Since the number of particles "stuck" in that minimum region is statistically the same, we can compute the average expected distance between them. 

Finally, we will say that the particles with $Q \neq 0$ have $q=q'=0$, while the particles with $Q=0$ have $q \neq 0 \neq q'$. This means that the particle "at the center" is \emph{not} participating in physics. The only particles that participate are the ones concentrated around the spheres surrounding the latter. This allows us to say that the particles within each cluster that "participate in physics" have on average the same expected distance from each other (after all, we are "ignoring" article at the center, which is considerably further away).

\subsection*{2.4 The necessity of superluminal signals}

As we have seen in section 2.1, we define the probability amplitudes non-locally. Despite that one might think that path integral formulation is inherently relativistic, in order to define the outcomes of path integral at any point other than infinity, we do need to violate relativity. Furthermore, we need the probability amplitude to change "instantaneousy" whenever we "update" any information. This implies that we need superluminal signals. However, our goal is to make everything agree with our intuition. Therefore, we insist that superluminal signals still move with finite speed. In order for them to produce effects consistent with infinite speed of propagation, we will postulate that our univers is compact. Thus, these signals sweep the universe within a very small period of time. In Chaper 4 we will, in fact, write down a detailed dynamics that these equations are obeyed. This, in particular, means that we will view superluminal speed as a "constant" and we simply don't know its value yet. We will denote that speed by $c_s$, where letter "s" stands for "superluminal". On the other hand, the speed of light will be denoted from now on by $c_o$, where "o" stands for "ordinary". Thus, if the size of the universe is $L$, then the phenomenology consistent with quantum field theory is produced if 
\beq \frac{L}{c_s} \ll \delta t \eeq
where $\delta t$ is the smallest time scale that produces detectable effects. We fully acknowledge that making the speed of signal finite does not save relativity. We believe that relativity is merely an illusion that is due to the "coincidence" in Lagrangian. Likewise, the constant $c_o$ is merely a number that happened to be part of Lagrangian. Apart from Lagrangian, $c_s$ is the only fundamental constant. Nevertheless, $c_s$-based relativity is not obeyed either. For one thing, we don't have any experimental reasons to assume that it does. So we simply believe in "preferred" time direction on all scales, including the ones compatible with $c_s$. Nevertheless, we insist that there is locality since that is what our intuition demands. 

\subsection*{3.Lagrangian generators for specific fields}

\subsection*{3.1 Bosonic fields}

\subsection*{3.1.1. Derivatives in integral form}

We would now like to come up with discretized expression for Lagrangians. In case of cubic lattice, we can simply replace derivatives with appropriate "differences". In our case we do not have a cubic structure. However, we can still come up with expression for derivatives, provided we do an important modification. In case of cubic lattice, a point had exactly six "neighbors"; the derivative along $x$ axis was "defined" based on exactly \emph{two} of them. In our case, several different points can be viewed as "neighbors", and several of them can be "useful" for the definition of derivative with respect to $x$. At the same time, none of them are "perfect" for that purpose, either. After all, each given neighbor "happened" to also be shifted in $y$ and $z$ direction as well. Thus, instead of looking at any one "perfect neighbor" we have to "add up" the contributions of several "non-perfect" ones in such a way that their contributions to $y$ and $z$ derivatives cancel out, while $x$ derivative survives. 

We will identify the linear combination in the following way. First of all, we will assume we are on a continuum. We will further assume that the field $\phi$ is linear. We will then single out integrals that would be equal to space and time derivatives of $\phi$. We will write down the integrals in such a way that only the points that are "very close" to the point of our interest play any role. Thus, we will be able to drop the assumption that $\phi$ is linear since any well behaved function is \emph{approximately} linear in a small (but finite) neighborhood of a point we are looking at. Later, we will discretize everything by simply replacing the integrals with the sums. 

In light of translational symmetry, we can move the origin towards the point we want to compute derivatives at. Thus, we are interested in $(\partial_k \phi) (\vec{x}=\vec{0})$ and $(\partial_0 \phi) (\vec{x}=\vec{0})$, where $t = x^0$ and $\vec{x} = (x_1, \cdots, x_d)$ is a projection of $\vec{x}$ onto $t= \rm const$ hyperplane for the "preferred" coordinate $t$ (which corresponds to the "preferred frame" demanded by our interpretation of quantum mechanics). In order to single out the "neighbors" of $\vec{0}$, we have to but a "weight factor" $e^{- \frac{\alpha}{2}}$ in our integrand. For the linear case, $\alpha$ can be anythign we like; but, as soon as $\phi$ is non-linear we have to make sure $\alpha$ is "very large"; this would limit us to a region in which the linearity of $\phi$ is preserved up to a very good approximation. 

Now, $\partial_i \phi$ changes sign upon the "flip" of the $x^i$ coordinate and it stays unchanged upon the "flip" of any other coordinate. Thus, if we are to have any hope that the integral gives $\partial_i \phi$, we should make sure that the integrand has similar properties. This can be easily done by sticking $x^i$ inside our integrand. Thus, we are lead to try the expression
\beq \int d^d x e^{- \frac{\alpha}{2} \vert \vec{x} \vert^2} x^i (\phi (\vec{x}) - \phi (\vec{0}))  \nonumber \eeq
Now, since we are assuming that $\phi$ is linear, we can replace $\phi (\vec{x}) - \phi (\vec{0})$ with $x^j \partial_j \phi$, where we keep in mind that $\partial_j \phi$ is constant. When we will generalize it to non-linear case, we will say that we are using $x^j \partial_j \phi$ as a first order \emph{approximation} for $\phi (\vec{x}) - \phi (\vec{0})$. In order for that "approximation" to be easily computed, we will replace $\partial_i \phi$ with $(\partial_i \phi) (\vec{x} = \vec{0})$ and, therefore, still treat it as constant. Thus, our integral is 
\beq \int d^d x e^{- \frac{\alpha}{2} \vert \vec{x} \vert^2} x^i (\phi (\vec{x}) - \phi (\vec{0})) \approx \int d^d x d^d y e^{- \frac{\alpha}{2} \vert \vec{x} \vert^2} x^i x^j (\partial_j \phi) \vert_{0} \eeq
Now, if $i \neq j$ then the above integral will be zero by antisymmetry. On the other hand, if $i=j$ then we can replace $x^i x^j$ with $(x^i)^2$. These two statements can be summarized as 
\beq \int d^d x x^i x^j e^{- \frac{\alpha}{2} \vert \vec{x} \vert^2} = \delta_i^j \int d^d x (x^i)^2 e^{- \frac{\alpha}{2} \vert \vec{x} \vert^2} \eeq
Now, we can split the integral on the right hand side as a product of separate integrals over each $x_a$. In case of $a \neq i$, the integrand will be simply the exponential weight factor $e^{- \frac{a}{2} (x^a)^2}$. For the $a=i$ case, there will also be an additional multiple of $(x^i)^2$: 
\beq \int d^d x x^i x^j e^{- \frac{\alpha}{2} \vert \vec{x} \vert^2} = \delta_i^j \Big( \int dx^i (x^i)^2 e^{- \frac{\alpha}{2} (x^i)^2} \Big) \Big(\prod_{a \neq i} \int dx^a e^{- \frac{\alpha}{2} (x^a)^2} \Big) \eeq
Now, by symmetry, we can rename indexes and replace $x^i$ with $x^i$. Furthermore, we notice that each of the integrals over $a \neq i$ are identical. Thus, we can replace $d-1$ "copies" of them with a single integral over $x^2$, taken to the power of $d-1$. Thus, we obtain
\beq \int d^d x x^i x^j e^{- \frac{\alpha}{2} \vert \vec{x} \vert^2} = \delta_i^j \Big( \int dx^1 (x^1)^2 e^{- \frac{\alpha}{2} (x^1)^2} \Big) \Big(\int dx^2 e^{- \frac{\alpha}{2} (x^2)^2} \Big)^{d-1}. \eeq
Now, it is easy to show that these two key integrals are given by 
\beq \int e^{- \frac{\alpha}{2} x^2} dx = \Big(\frac{2 \pi}{\alpha} \Big)^{1/2} \; ; \; \int x^2 e^{- \frac{\alpha}{2} x^2} dx = \frac{(2 \pi)^{1/2}}{\alpha^{3/2}}. \eeq
By substituting these expressions, we obtain 
\beq \int d^d x x^i x^j e^{- \frac{\alpha}{2} \vert \vec{x} \vert^2} = \delta_i^j \frac{(2 \pi)^{1/2}}{\alpha^{3/2}}\Big(\frac{2 \pi}{\alpha} \Big)^{(d-1)/2}, \eeq
which, after some simple algebra, becomes 
\beq \int d^d x x^i x^j e^{- \frac{\alpha}{2} \vert \vec{x} \vert^2} = \delta_i^j \frac{(2 \pi)^{d/2}}{\alpha^{(d+2)/2}} \eeq
Now, we have earlier stated that our original integral was given by 
\beq \int d^d x e^{- \frac{\alpha}{2} \vert \vec{x} \vert^2} x^i (\phi (\vec{x}) - \phi (\vec{0}))  \approx \int d^d x e^{- \frac{\alpha}{2} \vert \vec{x} \vert^2} x^i x^j (\partial_j \phi)\vert_{\vec{x} = \vec{0}} \eeq
The right hand side coincides with the integral we have just computed. By substituting our answer into the right hand side, we obtain 
\beq \int d^d x d^d y  e^{- \frac{\alpha}{2} \vert \vec{x} \vert^2} x^i (\phi (\vec{x}) - \phi (\vec{0}))  \approx \delta_i^j \frac{(2 \pi)^{d/2}}{\alpha^{(d+2)/2}} \partial_j \phi = \frac{(2 \pi)^{d/2}}{\alpha^{(d+2)/2}} (\partial_i \phi)\vert_{\vec{x} = \vec{0}} \eeq
We can now move the coefficient of the right hand side to the left which gives us the expression for $\partial_i \phi$: 
\beq (\partial_i \phi)\vert_{\vec{x} = \vec{0}} \approx \frac{\alpha^{(d+2)/2}}{(2 \pi)^{d/2}} \int d^d x e^{- \frac{\alpha}{2} \vert \vec{x} \vert^2} x^i (\phi (\vec{x}) - \phi (\vec{0})) \label{eqn:33}\eeq
It is easy to see that in our discussion we have \emph{not} made any use of the fact that $\phi$ is spin $0$. Thus, the same argument can be repeated if we replace $\phi$ with $A^{\mu}$; the index $\mu$ will simply "come for the ride". At the same time, we \emph{do} have to worry about the index of $\partial_{\mu}$; after all, as we will soon see, the expression for $\partial_0$ will be substantially different from what we have for $\partial_k$. In case we forget that we can be "a lot sloppier" when it comes to $A_0$ verses $A_k$, let us formally rewrite our result by replacing $\phi$ with $A_0$ and $A_k$:
\beq \partial_i A_0 = \frac{\alpha^{(d+2)/2}}{(2 \pi)^{d/2}} \int d^d x e^{- \frac{\alpha}{2} \vert \vec{x} \vert^2} x^i (A_0 (\vec{x}) - A_0 (\vec{0})) \eeq
\beq \partial_i A_j = \frac{\alpha^{(d+2)/2}}{(2 \pi)^{d/2}} \int d^d x e^{- \frac{\alpha}{2} \vert \vec{x} \vert^2} x^i (A_j (\vec{x}) - A_j (\vec{0})) \eeq
Let us now move on to writing an expression for time derivatives. On the first glance it might appear that the presence of "preferred" time coordinate frees us from the need of integrating. We can simply go "vertically" in time. Or, from a discrete point of view, every "point" in $\mathbb{R}^3$ is a line in $\mathbb{R}^4$. In order to do space derivative, we have to look at the "neighboring lines". Since there is no "preferred" way of selecting a "neighboring line", we have to "add up" different "choices" of neighboring lines, which amounts to space integral. On the other hand, in order to perform time derivative we don't have to look at  \emph{any} other line. Rather, we have to differentiate along the world line of the specific "point" we are interested in. Thus, no integration is necessary.

This view, however, does not harmonize with the framework of our paper. We recall from the discussion of Section 2.1 that, for any given particle $i$, the value of $\phi_i$ does \emph{not} change in time. What "changes" is the fact that we "stop looking" at particle $i$ and decide to "look" at the particle $j$, instead. If particle $j$ happens to be within the same "cluster" as the particle $i$, then due to such a small distance between them we "wrongly" believe that "the same particle" have "changed" its value of $\phi$ from $\phi_i$ to $\phi_j$. Now, we have at least $M$ "options" for selection of such particle $j$. After all, within any given cluster, and for each given value of $q$, we have a \emph{selection} of particles with all possible values of $q'$; and there are $M$ values of the latter. Therefore, we again are forced to perform integration. 

In principle, it would have been possible to "limit" our integration to a very small radius so that only the particles within the given cluster fall inside the region of integration. However, since from the "formal" point of view we still have to "take an integral", this would not imply any significant simplification to the theory. On the other hand, if we are to take integral over all space, we will \emph{still} be able to restore time derivative as long as we make sure that integrand is even with respect to spacelike coordinates and odd with respect to timelike one. One good candidate that appears to give time derivative based on symmetry argument is 
\beq \int d^d x e^{- \frac{\alpha}{2} \vert \vec{x} \vert^2} (\phi (\vec{x}, t+ \delta t) - \phi (\vec{0}, t)) \nonumber \eeq
This choice is also preferable from the point of view of "flexibility" of the theory. As discussed in more details in Conclusion, we are considering the possibility of "dropping" the assumption of "clusters" in our future work. If we are to do that, we will \emph{still} have to look at "other particles" in order to "switch" from $\phi (\vec{x}, t)$ to $\phi (\vec{x}, t+ \delta t)$. \emph{But} the "other particles" will no longer be within the same cluster. Thus, they will not contribute to the "sum" if we were to limit our um to the radius of a cluster.  On the other hand, if we are to integrate over all space, we will be able to equally accommodate the cluster-based and cluster-free scenarios, without modifications to our formulae. This is what makes "unbounded integration" preferable. 

Let us now verify that the integral we are proposing does, in fact, produce time derivative. Again, we expand the field $\phi$ up to first order around $\vec{0}$. We notice that we have different reasons for linear approximations in space and in time. The linear approximation in space is due to the fact that $\alpha$ is "large", while the linear approximation in time is due to the fact that $\delta t$ is "small". Thus, just to be on a safe side, we are justified in including the "mixed derivative" $\partial_0 \partial_k \phi$ as a part of the expansion for $\phi$. Thus, our integral is
\beq \int d^d x e^{- \frac{\alpha}{2} \vert \vec{x} \vert^2} (\phi (\vec{x}, t+ \delta t) - \phi (\vec{0}, t)) = \int d^d x e^{- \frac{\alpha}{2} \vert \vec{x} \vert^2} (x^k \partial_k \phi + (\delta t) (\partial_0 \phi)\vert_0 + x^k (\delta t) (\partial_0 \partial_k \phi)\vert_0) \eeq
At the same time, however, the "mixed derivative" $\partial_0 \partial_k \phi$ is dropped out anyway, due to the fact that it is odd in $x^k$. For the same reason, the \emph{first-order} term $x^k \partial_k$ drops out as well. Thus, on the right hand side, we are left with only \emph{one} term, the one containing $\partial_0 \phi$:
\beq \int d^d x e^{- \frac{\alpha}{2} \vert \vec{x} \vert^2} (\phi (\vec{x}, t+ \delta t) - \phi (\vec{0}, t)) = \int d^d x e^{- \frac{\alpha}{2} \vert \vec{x} \vert^2} (\delta t) (\partial_0 \phi)\vert_0 \eeq
Now, we recall from the discussion in Section 2.1, $\delta t$ is a small, but finite, constant. Likewise $(\partial_0 \phi)\vert_0$ is constant as well since it is taken \emph{at} $(\vec{x}=0; t)$. Thus, we can pull both of them out of integration as constants. Now the integrand, which is $e^{-\frac{\alpha}{2} \vert \vec{x}^2 \vert}$ is simply a product of $e^{-\frac{\alpha}{2} (x^k)^2}$ taken over $d$ possible values of $k \in \{1, \cdots, d \}$. Each of these is integrated to $(2 \pi/ \alpha)^{1/2}$. Thus, their product is $(2 \pi/ \alpha)^{d/2}$. This means that our integral becomes
\beq \int d^d x e^{- \frac{\alpha}{2} \vert \vec{x} \vert^2} (\phi (\vec{x}, t+ \delta t) - \phi (\vec{0}, t)) = (\delta t) (\partial_0 \phi)\vert_0 \Big(\frac{2 \pi}{\alpha} \Big)^{d/2} \eeq
Now, by moving both $\delta t$ and $(2 \pi/a)^{d/2}$ to the left hand side, we obtain an expression for $(\partial_0 \phi)\vert_0$; namely, 
\beq (\partial_0 \phi) \vert_{\vec{x}=\vec{0}} = \frac{\alpha^{d/2}}{(2 \pi)^{d/2} \delta t}  \int d^d x e^{- \frac{\alpha}{2} \vert \vec{x} \vert^2} (\phi (\vec{x}, t+ \delta t) - \phi (\vec{0}, t)) \label{eqn:39}\eeq
Now, just like in case of the space derivatives, we can freely replace $\phi$ with either $A_0$ or $A_i$. The index the "vector field" carries simply "comes for the ride". While the switch from $\partial_i$ to $\partial_0$ implies new algebra, the switch from $A_0$ to $A_i$ leaves algebra unchanged. Thus, we can get the expressions for $\partial_0 A_0$ and $\partial_0 A_k$ by "blindly copying" the expression for $\partial_0 \phi$ with appropriate substitutions:
\beq (\partial_0 A_0) \vert_{\vec{x} = \vec{0}} = \frac{\alpha^{d/2}}{(2 \pi)^{d/2} \delta t}  \int d^d x e^{- \frac{\alpha}{2} \vert \vec{x} \vert^2} (A_0 (\vec{x}, t+ \delta t) - A_0 (\vec{0}, t)) \eeq
\beq (\partial_0 A_i) \vert_{\vec{x} = \vec{0}} = \frac{\alpha^{d/2}}{(2 \pi)^{d/2} \delta t}  \int d^d x e^{- \frac{\alpha}{2} \vert \vec{x} \vert^2} (A_i (\vec{x}, t+ \delta t) - A_i (\vec{0}, t)) \eeq
This means that the corresponding discretized expressions take the form
\beq (\partial_0 A_0) \vert_{\vec{x} = \vec{0}} = \frac{v_0 \alpha^{d/2}}{(2 \pi)^{d/2} \delta t}  \sum_i e^{- \frac{\alpha}{2} \vert \vec{x}_i - \vec{x}_k \vert^2} (A_0 (\vec{x}_i, t+ \delta t) - A_0 (\vec{x}_k, t)) \eeq
\beq (\partial_0 A_i) \vert_{\vec{x} = \vec{0}} = \frac{v_0 \alpha^{d/2}}{(2 \pi)^{d/2} \delta t}  \sum_i e^{- \frac{\alpha}{2} \vert \vec{x}_i - \vec{x}_k \vert^2} (A_i (\vec{x}_i, t+ \delta t) - A_i (\vec{x}_k, t)) \eeq 
where $v_0$ is a constant volume element which will be extensively discussed towards the end of section 3.1.2. While turning the integral into the sum might seem trivial, this is, in fact, our ultimate goal. After all, in order to be able to "fit" the configuration space inside the ordinary space, we have to make the former "smaller" by discretizing it. This means that we can \emph{neither} use continuum derivative \emph{nor} continuum integral in a definition of Lagrangian. The discretization of derivative only has two terms, while discretization of the integral has several. In the discrete context, the more terms we have, the closer we are to symmetric outcome; this is the reason we picked the discretization of integral over the discretization of the derivative. If we didn't have to discretize things to begin with, neither derivative nor the integral would have violated symmetries and, therefore, there would be no need in converting the former into the latter. 

\subsection*{3.1.2 Kinetic term for charged scalar field}

So far we have written down the space and time "derivatives" in the integral form. Now, the Lagrangian densities can be obtained by simple contraction of these derivatives. In this section we will work out the kinetic term for charged scalar field. In the section 3.1.3 we will incorporate its interaction with the electromagnetic field and, finally in the section 3.1.4 we will work out the kinetic term for the electromagnetic field, itself. For the purposes of Chapter 5, we will assume a background curvature in all three cases. At the same time, we will \emph{not} do the Lagrangian for gravity because, according to Chapter 5 (as motivated in Section 5.1-5.3, and then extensively used in Sections 5.6 and 5.10) gravity is on a completely different status. It "evolves" according to \emph{classical} Einstein's equation with appropriate modifications, and is \emph{not} subject to the "quantum mechanics" that we are subjecting other fields to.

Now, the kinetic term of charged scalar field in the presence of gravity is given by 
\beq \partial^{\mu} \phi^* \partial_{\mu} \phi = g^{\mu \nu} \partial_{\mu} \phi^* \partial_{\nu} \phi \eeq
From what we have seen in the previous section, however, the expressions for space and time derivatives are different from each other. This means that we have to hold them "on a different footing"; this is consistent with our theory since, for the "quantum mechanical" purposes we need a "preferred frame" anyway. Thus, we rewrite the above expression as 
\beq \partial^{\mu} \phi^* \partial_{\mu} \phi = g^{00} \partial_0 \phi^* \partial_0 \phi + g^{0i} \partial_0 \phi^* \partial_i \phi + g^{i0} \partial_i \phi^* \partial_0 \phi + g^{ij} \partial_i \phi^* \partial_j \phi \eeq
Let us now evaluate the above expression term by term. We will start with the "time derivative" term, $g^{00} \partial_0 \phi^* \partial_0 \phi$. In the previous section we have found that the integral expression for $\partial_0 \phi$ is given by
\beq \partial_0 \phi = \frac{\alpha^{d/2}}{(2 \pi)^{d/2} \delta t}  \int d^d x e^{- \frac{\alpha}{2} \vert \vec{x} \vert^2} (\phi (\vec{x}, t+ \delta t) - \phi (\vec{0}, t)) \eeq
We can now "take a complex conjugate" of the above expression to obtain similar expression for $\partial_0 \phi^*$. For the notational convenience, we will use the integration over $\vec{x}$ for $\partial_0 \phi$ and the integration over $\vec{y}$ for $\partial_0 \phi^*$. Thus, we have
\beq \partial_0 \phi^* = \frac{\alpha^{d/2}}{(2 \pi)^{d/2} \delta t}  \int d^d y e^{- \frac{\alpha}{2} \vert \vec{y} \vert^2} (\phi^* (\vec{y}, t+ \delta t) - \phi^* (\vec{0}, t)) \eeq
We can now "multiply" the above two expressions by each other. For the purposes of our future convenience we will combine both integrals into a signle double integral:
\beq \partial_0 \phi^* \partial_0 \phi = \frac{\alpha^d}{(2 \pi)^d (\delta t)^2} \int d^d x d^d y e^{-\frac{\alpha}{2} (\vert \vec{x} \vert^2 + \vert \vec{y} \vert^2)} (\phi^* (\vec{y}, t+ \delta t) - \phi^* (\vec{0}, t)) (\phi (\vec{x}, t+ \delta t) - \phi (\vec{0}, t))\eeq
Now, in light of the fact that the metric can be off-diagonal, we also have to compute $g^{0i} \partial_0 \phi^* \partial_i \phi$. We recall from the previous section that the expression for $\partial_i \phi$ is given by 
\beq \partial_i \phi = \frac{\alpha^{(d+2)/2}}{(2 \pi)^{d/2}} \int d^d x e^{- \frac{\alpha}{2} \vert \vec{x} \vert^2} x^i (\phi (\vec{x}) - \phi (\vec{0})) \eeq
By multiplying it by the expression for $\partial_0 \phi^*$ that we have just written and, again, combining the two integrals into a single double integral, we obtain
\beq \partial_0 \phi^* \partial_i \phi = \frac{\alpha^{d+1}}{(2 \pi)^d \delta t} \int d^d x d^d y e^{- \frac{\alpha}{2} (\vert \vec{x} \vert^2 + \vert \vec{y} \vert^2)} x^i (\phi^* (\vec{y}, t+ \delta t) - \phi^* (\vec{0}, t)) (\phi (\vec{x}) - \phi (\vec{0})) \eeq
The expression for $\partial_i \phi^* \partial_0 \phi$ is a carbon copy of the one for $\partial_0 \phi^* \partial_i  \phi$ with "rearranged" $\phi^*$ and $\phi$. Thus, 
\beq \partial_i \phi^* \partial_0 \phi = \frac{\alpha^{d+1}}{(2 \pi)^d \delta t} \int d^d x d^d y e^{- \frac{\alpha}{2} (\vert \vec{x} \vert^2 + \vert \vec{y} \vert^2)} x^i (\phi (\vec{y}, t+ \delta t) - \phi (\vec{0}, t)) (\phi^* (\vec{x}) - \phi^* (\vec{0})) \eeq
Finally, let us write down an expression for $\partial_i \phi^* \partial_j \phi$. The expression for $\partial_i \phi^*$ can be obtained as a carbon copy of the one for $\partial_i \phi$ where $\phi$ is replaced with $\phi^*$:
\beq \partial_i \phi^* = \frac{\alpha^{(d+2)/2}}{(2 \pi)^{d/2}} \int d^d x e^{- \frac{\alpha}{2} \vert \vec{x} \vert^2} x^i (\phi^* (\vec{x}) - \phi^* (\vec{0})) \eeq
On the other hand, the expression for $\partial_j \phi$ can be obtained as a carbon copy of the one for $\partial_i \phi$ where $i$ is replaced by $j$ and $\vec{x}$-integration is replaced by $\vec{y}$-integration:
\beq \partial_j \phi = \frac{\alpha^{(d+2)/2}}{(2 \pi)^{d/2}} \int d^d y e^{- \frac{\alpha}{2} \vert \vec{y} \vert^2} y^j (\phi (\vec{y}) - \phi (\vec{0})) \eeq
If we now multiply these two expressions by each other and combine the two integrals into a double-integral, we obtain
\beq \partial_i \phi^* \partial_j \phi = \frac{\alpha^{d+2}}{(2 \pi)^d} \int d^d x d^d y e^{- \frac{\alpha}{2} (\vert \vec{x} \vert^2 + \vert \vec{y} \vert^2)} x^i y^j (\phi^* (\vec{x}) - \phi^* (\vec{0}))(\phi (\vec{y}) - \phi (\vec{0})) \eeq
We are now ready to finally write down our original sought-for expression, $\partial^{\mu} \phi^* \partial_{\mu} \phi$. As we recall, we represent this expression as a sum of time derivative, space derivative and mixed derivative terms:
\beq \partial^{\mu} \phi^* \partial_{\mu} \phi = g^{00} \partial_0 \phi^* \partial_0 \phi + g^{0i} \partial_0 \phi^* \partial_i \phi + g^{i0} \partial_i \phi^* \partial_0 \phi + g^{ij} \partial_i \phi^* \partial_j \phi \eeq
Now, we have just found that each of these terms can be represented as a double integral. We can now replace $g^{00}$, $g^{0i}$ and $g^{ij}$ with "constants" $g^{00} (\vec{0}, t)$, $g^{0i}(\vec{0}, t)$ and $g^{ij} (\vec{0}, t)$, and move them \emph{under} the integration. Once we have done that, we can replace the "sum of double integrals" with "double integral of the sum". Thus, we get
\beq \partial^{\mu} \phi^* \partial_{\mu} \phi = \int d^d x d^d y e^{-\frac{\alpha}{2} (\vert \vec{x} \vert^2 + \vert \vec{y} \vert^2)} \times \nonumber \eeq
\beq \times \Big( g^{00} (\vec{0}, t) \frac{\alpha^d}{(2 \pi)^d (\delta t)^2} (\phi^* (\vec{y}, t+ \delta t) - \phi^* (\vec{0}, t)) (\phi (\vec{x}, t+ \delta t) - \phi (\vec{0}, t)) \nonumber \eeq
\beq + 2 g^{0i} (\vec{0}, t) \frac{\alpha^{d+1}}{(2 \pi)^d \delta t} x^i (\phi^* (\vec{y}, t+ \delta t) - \phi^* (\vec{0}, t)) (\phi (\vec{x}, t) - \phi (\vec{0}, t)) + \eeq
\beq + g^{ij} (\vec{0}, t) \frac{\alpha^{d+2}}{(2 \pi)^d}  x^i y^j (\phi^* (\vec{x}, t) - \phi^* (\vec{0}, t))(\phi (\vec{y}, t) - \phi (\vec{0}, t)) \Big) \nonumber \eeq
The reason we insist on combining everything under \emph{one} double-integral is that we will be able to identify the expression \emph{under the integral} with "Lagrangian generator", thus accommodating it into the framework outlined in Section 2.2. 

Now, a "path" is represented by a sequence of integers 
\beq \cdots , a_{-1}, a_0, a_1, \cdots \eeq
The action associated with each path is given by 
\beq S (\cdots, a_{-1}, a_0, a_1, \cdots) = \sum_{n=-\infty}^{\infty} S(q=a_n; q'=a_{n+1}) \eeq
At the same time, we also know that the action can be interpreted as 
\beq S (\cdots, a_{-1}, a_0, a_1, \cdots) = \eeq
\beq =  S [\cdots, \psi (t= - \epsilon) = \psi_{q=a_{-1}}, \psi (t=0) = \psi_{q=a_0}, \psi (t = \delta t) = \psi_{q= \delta t}, \cdots] \nonumber \eeq
This means that we can replace $\psi (t_0+ \delta t)$ with $\psi_{q=q'_i} (t_0)$ as long as we are careful to put $q=q'_i$ instead of $q=q_i$, while using $\psi_{q=q_i} (t_0)$  in a reference to what we "secretly" think of as $t_0$.  We can, therefore, rewrite the above integral as 
\beq {\cal L} (\vec{x} =\vec{0};  q=a, q'=b) = \int d^d x d^d y e^{-\frac{\alpha}{2} (\vert \vec{x} \vert^2 + \vert \vec{y} \vert^2)} \times \nonumber \eeq
\beq \times \Big( g^{00} (\vec{0}, t) \frac{\alpha^d}{(2 \pi)^d (\delta t)^2} (\phi_{q=b}^*(\vec{y}, t) - \phi^*_{q=a} (\vec{0}, t)) (\phi_{q=b} (\vec{x}, t) - \phi_{q=a} (\vec{0}, t)) \nonumber \eeq
\beq +2  g^{0i} (\vec{0}, t) \frac{\alpha^{d+1}}{(2 \pi)^d \delta t} x^i (\phi^*_{q=b}(\vec{y}, t) - \phi^*_{q=a} (\vec{0}, t)) (\phi_{q=a} (\vec{x}, t) - \phi_{q=a} (\vec{0}, t)) + \eeq
\beq + g^{ij} (\vec{0}, t) \frac{\alpha^{d+2}}{(2 \pi)^d}  x^i y^j (\phi^*_{q=a} (\vec{x}, t) - \phi^*_{q=a} (\vec{0}, t))(\phi_{q=a} (\vec{y}, t) - \phi_{q=a} (\vec{0}, t)) \Big) \nonumber \eeq
As one notices, we changed all of the $t+ \delta t$ back into $t$. After all, in order to "decide" how likely are we to "transition" into a given state, we have to "look" at the information available regarding that state. That information has to be available at $t=t_0$, \emph{not} at $t= t_0 + \delta t$. In other words, we have a "tour guide" in front of us \emph{at $t=t_0$} that tells us what will happen at $t=t_0 + \delta t$ \emph{if} we decide to go somewhere. Thus, from this point of view, letter "$t$" could be dropped altogether since Hamiltonian is invariant under time translation. The only reason we keep letter $t$ is the time-dependence of $g_{\mu \nu}$ which, for the reasons explained in Section 5.1, is viewed as a "classical background" of the so-called "quantum" theory.

Now, whenever we say $\phi_{q=a} (\vec{x}, t)$ or $\phi_{q=b} (\vec{x}, t)$, we mean, of course, the points with charges $q=a$ and $q=b$ located \emph{near} $\vec{x}$. Obviously, saying that there is a particle of charge $q$ and \emph{also} a particle of charge $q'$ \emph{exactly at} $\vec{x}$ would be a contradiction. That is why we say "near $\vec{x}$" rather than "at $\vec{x}$, and we assume we have a good understanding of what we mean by "near". This means that both $\phi_{q=a}$ and $\phi_{q=b}$ are taken over \emph{discrete sets}; namely, sets occupied by corresponding type of particles. This, in turn, implies that integrals should be converted into discrete sums. We would like, therefore, to find the coefficient that we have to insert into these sums. 

As we said in Section 2.4, in order for the superluminal signals to "sweep" the universe within "very small" (but finite) time, we propose that the universe is compact. Therefore, if the number of points is $N$, then the "average volume per point" is $N/V$. However, we \emph{only} want to count points with "matching charges". Whenever we are integrating over hyperplane $t=t_0$, we are "summing over" $\{j \vert q_j = q_i \}$ and whenever we are integrating over the hyperplace $t=t_0 + \delta t$, we are "counting" points $\{j \vert q_j = q'_i \}$. If there are $M$ possible values of charge, this means that in each case we count $N/M$ points. On the other hand, if we impose restrictions both on $q$ and $q'$, then the number of points we count would be $N/M^2$. 

Let us consider an example to show what role counting (or not counting) $q'$ might play. Suppose we have specified the path we are interested in. Thus, we have specified that $q (t=0) = a_0$, $q (t= \delta t) = a_1$ and $q (t = 2 \delta t) = a_2$. Now suppose there are two points, $i$ and $j$. As far as the point $i$ is concerned, everything is fine: $q_i = a$ and $q'_i =b$; thus, $i$ "fits right into" the $t=0$ section of the path. However, as far as the point $j$ is concerned, things are a bit "complicated". On the one hand, $q_j =a_1$, which implies that $j$ "belongs to" $t= \delta t$ section of the path; but, on the other hand, $q'_j \neq a_2$, which means that the point $j$ is "departing from" the path. At the same time, the charges $q$ and $q'$ don't change in time (just like the definition of a specific quantum state does not). Thus, point $j$ is "permanently on the path, and permanently departing". So, should we count the contribution of $(i, j)$-pair towards the Lagrangian or not? 

The good news is that, as long as the path is differentiable, the answer to the above question will only affect the normalization factor. After all, even if $j$ is "departing from the path", the fact that it is still "one foot there" implies that it probably contributes approximately the same thing as the points that are "fully there" do. It is, however, important to avoid counting the points that are "not there at all". Let us make a choice to "disregard" as few points as necessary. Thus, we \emph{will} count the points that are "half foot in the path". In this case, we have only \emph{one} restriction for every point. Since there are $N$ points and $M$ charges, one restriction implies that there are $N/M$ points we are looking at. Since we are adding pairs of points, this means that the number of pairs we are looking at is $N^2/M^2$. 

There is also another issue that needs to be addressed. Namely, as we recall from Chapter 2, the points are grouped into "clusters". Furthermore, we have stated that a "point" in lattice field theory is equivalent to "clusters of points" in our case. Since our "integral" sums over "standard QFT" version of points, do we really sum over points or clusters of points in our case? It turns out that there are two opposite effects that cancel each other out. If we count over clusters of points, there are "fewer" of them, so, given the same volume of the universe, $\delta V$ has to be larger. On the other hand, if we count over points, and \emph{wrongly} pretend that they are "evenly distributed" then we will use smaller $\delta V$ but, at the same time, we will have a lot more terms. 

Of course, if even versus not even distribution of points has an impact then it is important to sum over clusters. This, too, will be taken into account once we go from integral to the sum. After all, when we write down the sum, we are not claiming that distribution is even. Rather, we summing over whatever values the coordinates of the points \emph{actually have}, whether even or otherwise. If they are distributed in clusters, what will go into the sum is the coordinates of these clusters. Thus, the coordinate of each cluster will be multiplied by small $\delta V$, but will be counted many times, which will effectively be the same thing as multiplying it by large $\delta V$. 

There is another reason why we made this choice. In particular, part of the agenda of this paper is to come up with "classical" universe that "simulates" the "quantum mechanics" we are describing. While in this chapter we are simply postulating Lagrangians (which is, of course, something the "true believer" in quantum mechanics would do, as well), in the Chapter 4 we "switch" to "classical physics" and attempt to "build a machine" that would "generate" our Lagrangians through \emph{classical} means (and, of course, we identify that machine with the "true" description of our universe). Now, since part of the agenda is to make sure that physics is local, our machine utilizes the signals emitted by \emph{points} rather than clusters. Therefore, if we are to count "clusters", we can't blindly keep receiving signals from points with matching charge. We have to "write down" which cluster that point belongs to; and also we have to "look at our list" to see if we have already received a signal from that cluster, in which case we would disregard a signal at hand. In principle, this can be done (as one can see from Chapter 4, we were able to "teach" our machine to "remember" a lot of things). But, from a very human need of saving some time, we would rather not do extra work unless we have to. So we will take easier rout and just define our Lagrangian as a sum over individual points. 

From what we said so far, we have finally reached an agreement of just how many points we are counting. First of all, we \emph{do} count the points that are "in the process of departing from the path". Thus, our restriction is on $q$ and \emph{not} on $q'$. Secondly, we count individual points rather than clusters of points; thus, we do \emph{not} have to perform any multiplication or devision by the number of points per cluster. This identifies the number of terms we are counting with $N/M$ (where $N$ is the number of points and $M$ is the number of available charges). This, in term, specifies our volume element to be 
\beq \delta V = \frac{VM}{N} \eeq
where $V$ is the volume of the universe. In light of "preferred frame", by "volume of the universe" we mean three dimensional volume of a hyperplane $t=t_0$. Of course, in case of double integral, we square the volume element:
\beq \int d^d x d^d y (\cdots) = \Big(\frac{VM}{N} \Big)^2 \sum (\cdots) \eeq
Now, if we are to write the above integral in the "sum" form, it no longer makes sense to write $\phi_{q=a}$ or $\phi_{q=b}$. After all, if we are looking at a particle we are evaluating $\phi$ \emph{at}, we already \emph{know} what value of "charge" that particle has; and there is only \emph{one} field $\phi$ to be evaluated; not two fields.Thus, we make the following replacements:
\beq \phi_{q=a} (i) \rightarrow \phi_i \delta^{q_i}_a \; ; \; \phi_{q=b} (i) \rightarrow \phi_i \delta^{q_i}_b \eeq
Furthermore, we already \emph{know} what $a$ and $b$ is. Namely, if the particle at the origin is particle number $k$, then 
\beq a \rightarrow q_k \; ; \; b \rightarrow q'_k \eeq
Thus, the above substitutions become
\beq \phi_{q=a} (i) \rightarrow \phi_i \delta^{q_i}_{q_k} \; ; \; \phi_{q=b} (i) \rightarrow \phi_i \delta^{q_i}_{q'_k}\eeq
Furthermore, the fact that we identify the origin with the point $k$ means that 
\beq \phi (\vec{0}) \rightarrow \phi_k \eeq
and we skipped $\delta$ part because we know that $\delta_k^k=1$. We have written $\phi_i$ instead of $\phi (i)$ in order to emphasize that $\phi_i$ does not change in time; it is a "constant" attached to point $i$ and the only thing that changes is whether or not we are "looking at $i$", which, of course, determined by whether or not we are looking at \emph{all} points of charge $q_i$. However, we still retain $\phi(k)$ as one of the "valid" notations we can use, and throughout the paper we use $\phi_i$ and $\phi (i)$ interchangeably:
\beq \phi (i) = \phi_i \eeq
 We change the "integrals" over $\vec{x}$ and $\vec{y}$ with respective "sums" over points $i$ and $j$; thus, $\vec{x}$ and $\vec{y}$ will be changed into $\vec{x}_i$ and $\vec{x}_j$, respectively. Furthermore, we have identified what we have previously viewed as origin, with the location of point $k$. Dropping our preferred choice of origin, we have
\beq \vec{x} \rightarrow \vec{x}_i - \vec{x}_k \; ; \; \vec{y} \rightarrow \vec{x}_j - \vec{x}_k \; ; \; \vec{0} \rightarrow \vec{x}_k \eeq
which, together with the discretization volume scale defined earlier, implies that 
\beq \int d^d x d^d y f(\vec{x}, \vec{y}) \rightarrow \Big(\frac{VM}{N} \Big)^2 \sum_{i, j} f(\vec{x}_i - \vec{x}_k, \vec{x}_j - \vec{x}_k) \eeq
The values of $\phi (\vec{x})$, $\phi (\vec{y})$ and $\phi (\vec{0})$, evaluated at $t=0$ and $t= \delta t$, now become
The Lagrangian density is now taken \emph{at a point} $k$, and is defined in terms of quasi-local expression that is given by "turning" the integral into the sum in the way just described. In particular, 
\beq {\cal L}_k = \Big(\frac{VM}{N} \Big)^2 \sum_{i,j} e^{-\frac{\alpha}{2} (\vert \vec{x}_i - \vec{x}_k \vert^2 + \vert \vec{x}_j - \vec{x}_k \vert^2)}  \Big( g^{00} (\vec{x}_k, t) \frac{\alpha^d}{(2 \pi)^d (\delta t)^2} (\phi_j^*\delta^{q_j}_{q'_k}  - \phi^*_k) (\phi_i \delta^{q_i}_{q'_k}  - \phi_k ) + \nonumber \eeq
\beq +2 g^{0i} (\vec{x}_k, t) \frac{\alpha^{d+1}}{(2 \pi)^d \delta t} x_k^i (\phi^*_j \delta^{q_j}_{q'_k}  - \phi^*_k  ) (\phi_i \delta^{q_i}_{q_k} - \phi_k)  + g^{ij} (\vec{0}, t) \frac{\alpha^{d+2}}{(2 \pi)^d}  x_k^i y_k^j (\phi^*_i \delta^{q_i}_{q_k} - \phi^*_k) (\phi^{q_i}_{q_k} - \phi_k) \Big)  \eeq
Finally, we define \emph{Lagrangian generator} $\cal K$ to be a three-point function that we need to be "summing over" in order to obtain Lagrangian density. Thus, we simply rewrite the above equation while dropping sum term and treating what used to be "dummy indexes" in the sum as the "input parameters" into the Lagrangian generator:
\beq {\cal K}_{\phi; kin} (i, j, k) = \Big(\frac{VM}{N} \Big)^2 e^{-\frac{\alpha}{2} (\vert \vec{x}_i - \vec{x}_k \vert^2 + \vert \vec{x}_j - \vec{x}_k \vert^2)}  \Big( g^{00} (\vec{x}_k, t) \frac{\alpha^d}{(2 \pi)^d (\delta t)^2} (\phi_j^*\delta^{q_j}_{q'_k}  - \phi^*_k) (\phi_i \delta^{q_i}_{q'_k}  - \phi_k ) + \nonumber \eeq
\beq +2 g^{0i} (\vec{x}_k, t) \frac{\alpha^{d+1}}{(2 \pi)^d \delta t} x_k^i (\phi^*_j \delta^{q_j}_{q'_k}  - \phi^*_k  ) (\phi_i \delta^{q_i}_{q_k} - \phi_k)  + g^{ij} (\vec{0}, t) \frac{\alpha^{d+2}}{(2 \pi)^d}  x_k^i y_k^j (\phi^*_i \delta^{q_i}_{q_k} - \phi^*_k) (\phi^{q_i}_{q_k} - \phi_k) \Big)  \eeq
While the last step might seem trivial, this step is crucial for the purposes of Chapter 4. Part of our agenda of turning everything into "classical physics" is making things "local". In particular, we would like to describe the universe as a system that would "compute" the above sum by means of various signals the particles exchange. Since each "ingredient" of the system (that is, a particle) will have limited "memory", it will record "intermediate information" in a form of $\cal K$; and then, through some step by step procedure, $\cal L$ will \emph{eventually} be produced in such a way that it approximates $\cal K$. It turns out that the "intermediate information" will be the three-index quantity ${\cal K}_{ijk}$ we are summing. Since our "mashine" will be explicitly utilizing $\cal K$, we have no choice but introduce it.

\subsection*{3.1.3 Interaction of charged scalar field with electromagnetic field}

So far we have described the Lagrangian for non-interacting charged scalar field. Let us now include its interaction with Gauge field (while we leave the Lagrangian of Gauge field, itself, for the next section). We recall that interaction with Gauge field is equivalent to replacement the "derivative" $\partial^{\mu}$ with a "covariant derivative" $D^{\mu} = \partial^{\mu} + ieA^{\mu}$. Thus, we can simply rewrite the non-interacting Lagrangian with this replacement. This gives us
\beq D^{\mu} \phi^* D_{\mu} \phi = (\partial^{\mu} \phi + ieA^{\mu} \phi)^*(\partial_{\mu} \phi + ieA_{\mu} \phi) \eeq
Now, the evaluation of complex conjugate amounts to replacement of $\phi$ by $\phi^*$ and $+i$ with $-i$. Thus, we obtain
\beq D^{\mu} \phi^* D_{\mu} \phi = (\partial^{\mu} \phi^* - ieA^{\mu} \phi^*)(\partial_{\mu} \phi + ieA_{\mu} \phi) \eeq
We can now perform a straightforward term by term multiplication and recombine the above expression in the following form:
\beq D^{\mu} \phi^* D_{\mu} \phi = \partial^{\mu} \phi^* \partial_{\mu} \phi + e^2 A^{\mu} A_{\mu} \phi^* \phi + ieA^{\mu} (\phi \partial_{\mu} \phi^* - \phi^* \partial_{\mu} \phi) \eeq
This is not too surprising. After all, we know that the "current" associated with charged Klein Gordon field is given by 
\beq j_{\mu} = i (\phi \partial_{\mu} \phi^* - \phi^* \partial_{\mu} \phi) \eeq
Thus the above expression can be rewritten as 
\beq D^{\mu} \phi^* D_{\mu} \phi = \partial^{\mu} \phi^* \partial_{\mu} \phi + e^2 A^{\mu} A_{\mu} \phi^* \phi + eA^{\mu} j_{\mu}\eeq
The first term is a kinetic term, the last term is a coupling of electromagnetic field to a current. The only "surprising" term is $A^{\mu} A_{\mu} \phi^* \phi$. However, after some simple reasoning it is evident that the "four vortex" is to be expected if $\phi$ is spin 0: after all, both $\phi$ and $A^{\mu}$ have dimension $+1$, so the four-vortex has a dimension $+4$ and, therefore, it is renormalizable. The only reason this appears unusual is that we are used to dealing with spin $1/2$ field $\psi$ which is of dimension $+3/2$ in which case the $\psi^{\dagger} \psi A^{\mu} A_{\mu}$ vortex is of dimension $+5$ and, therefore, non-renomalizable. This vortex does not appear anyway in such cases since the Dirac Lagrangian has only \emph{one} copy of $\partial^{\mu}$ which can only "generate" one copy of $A^{\mu}$. 

Let us now come up with the integral expression for the above. We will evaluate it term by term. Let us start from the "single-derivative" term, 
\beq A^{\mu} \phi^* \partial_{\mu} \phi = A^0 \phi^* \partial_0 \phi + A^k \phi^* \partial_k \phi \eeq
Now, we have found in the Section 3.1.1 that the time and space derivatives of $\phi$ are given by
\beq \partial_0 \phi = \frac{\alpha^{d/2}}{(2 \pi)^{d/2} \delta t}  \int d^d x e^{- \frac{\alpha}{2} \vert \vec{x} \vert^2} (\phi (\vec{x}, t+ \delta t) - \phi (\vec{0}, t)) \eeq
\beq \partial_k \phi = \frac{\alpha^{(d+2)/2}}{(2 \pi)^{d/2}} \int d^d x e^{- \frac{\alpha}{2} \vert \vec{x} \vert^2} x^k (\phi (\vec{x}) - \phi (\vec{0})) \eeq
By substituting these into the above expression for $A^{\mu} \phi^* \partial_{\mu} \phi$, we obtain
\beq A^{\mu} \phi^* \partial_{\mu} \phi = \frac{\alpha^{d/2}}{(2 \pi)^{d/2} \delta t} A^0 (\vec{0}) \phi^* (\vec{0}) \int d^d x e^{- \frac{\alpha}{2} \vert \vec{x} \vert^2} (\phi (\vec{x}, t+ \delta t) - \phi (\vec{0}, t)) + \nonumber \eeq
\beq + \frac{\alpha^{(d+2)/2}}{(2 \pi)^{d/2}} A^k (\vec{0}) \phi^* (\vec{0})\int d^d x e^{- \frac{\alpha}{2} \vert \vec{x} \vert^2} x^k (\phi (\vec{x}) - \phi (\vec{0})) \eeq
In principle, we could have said that we are done with this specific term. However, in light of the framework outlined in Section 2.2, we want to express it as an integral \emph{of} a Lagrangian generator, as opposed to a linear combination of integrals. Therefore, we pull the constant coefficients inside the respective integrals and rewrite the sum of integrals in a form of an integral of a sum. This gives us
\beq A^{\mu} \phi^* \partial_{\mu} \phi = \int d^d x e^{- \frac{\alpha}{2} \vert \vec{x} \vert^2} \Big( \frac{\alpha^{d/2}}{(2 \pi)^{d/2} \delta t} A^0 (\vec{0}) \phi^* (\vec{0})  (\phi (\vec{x}, t+ \delta t) - \phi (\vec{0}, t)) + \nonumber \eeq
\beq + \frac{\alpha^{(d+2)/2}}{(2 \pi)^{d/2}} A^k (\vec{0}) \phi^* (\vec{0}) x^k (\phi (\vec{x}) - \phi (\vec{0})) \Big) \eeq
Now, the other singe-derivative term, $A^{\mu} \phi \partial_{\mu} \phi^*$, can be obtained by simply taking a "complex conjugate" of the expression we have just written for $A^{\mu} \phi^* \partial_{\mu} \phi$:
\beq A^{\mu} \phi \partial_{\mu} \phi^* = \int d^d x e^{- \frac{\alpha}{2} \vert \vec{x} \vert^2} \Big( \frac{\alpha^{d/2}}{(2 \pi)^{d/2} \delta t} A^0 (\vec{0}) \phi (\vec{0})  (\phi^* (\vec{x}, t+ \delta t) - \phi^* (\vec{0}, t)) + \nonumber \eeq
\beq + \frac{\alpha^{(d+2)/2}}{(2 \pi)^{d/2}} A^k (\vec{0}) \phi (\vec{0}) x^k (\phi^* (\vec{x}) - \phi^* (\vec{0})) \Big) \eeq
Finally, we can subtract one of these two terms from the other in order to obtain the sought-after $eA^{\mu} j_{\mu}$. Again, for the purposes of agreement with the setup of Section 2.2, we represent the difference of integrals in a form of an integral of difference. Thus, we obtain
\beq ieA^{\mu} (\phi \partial_{\mu} \phi^* - \phi^* \partial_{\mu} \phi) = \int d^d x e^{- \frac{\alpha}{2} \vert \vec{x} \vert^2} \Big( \frac{\alpha^{d/2}}{(2 \pi)^{d/2} \delta t} A^0 (\vec{0}) \phi (\vec{0})  (\phi^* (\vec{x}, t+ \delta t) - \phi^* (\vec{0}, t)) + \nonumber \eeq
\beq + \frac{\alpha^{(d+2)/2}}{(2 \pi)^{d/2}} A^k (\vec{0}) \phi (\vec{0}) x^k (\phi^* (\vec{x}) - \phi^* (\vec{0})) - \frac{\alpha^{d/2}}{(2 \pi)^{d/2} \delta t} A^0 (\vec{0}) \phi^* (\vec{0})  (\phi (\vec{x}, t+ \delta t) - \phi (\vec{0}, t)) - \nonumber \eeq
\beq -  \frac{\alpha^{(d+2)/2}}{(2 \pi)^{d/2}} A^k (\vec{0}) \phi^* (\vec{0}) x^k (\phi (\vec{x}) - \phi (\vec{0})) \Big) \eeq
Now, according to the formalism of Section 2.2, we would like to have a \emph{double} integral over \emph{two} variables $\vec{x}$ and $\vec{y}$. We can easily accomplish this by multiplying the right hand side by the "unit coefficient" given by
\beq 1 = \Big(\frac{\alpha}{2 \pi} \Big)^{d/2} \int d^d y e^{- \frac{\alpha}{2} \vert \vec{y} \vert^2} \label{unit1} \eeq
Here, the power of $d/2$ instead of $1/2$ is due to the presence of "$d$ copies" of the integral over the variables $x_k$ where ranges through $\{1, \cdots d \}$. With this multiplication by a unit, we obtain a double-integral expression: 
\beq ieA^{\mu} (\phi \partial_{\mu} \phi^* - \phi^* \partial_{\mu} \phi) = \int d^d x d^d y e^{- \frac{\alpha}{2} (\vert \vec{x} \vert^2 + \vert \vec{y} \vert^2)} \Big( \frac{\alpha^d}{(2 \pi)^d \delta t} A^0 (\vec{0}) \phi (\vec{0})  (\phi^* (\vec{x}, t+ \delta t) - \phi^* (\vec{0}, t)) + \nonumber \eeq
\beq + \frac{\alpha^{d+1}}{(2 \pi)^d} A^k (\vec{0}) \phi (\vec{0}) x^k (\phi^* (\vec{x}) - \phi^* (\vec{0})) - \frac{\alpha^d}{(2 \pi)^d \delta t} A^0 (\vec{0}) \phi^* (\vec{0})  (\phi (\vec{x}, t+ \delta t) - \phi (\vec{0}, t)) - \nonumber \eeq
\beq -  \frac{\alpha^{d+1}}{(2 \pi)^d} A^k (\vec{0}) \phi^* (\vec{0}) x^k (\phi (\vec{x}) - \phi (\vec{0})) \Big) \label{coupling} \eeq
Finally, let us describe the four-vortex $A^{\mu} A_{\mu} \phi^* \phi$. In light of the lack of "derivatives" we \emph{could} have left it as is by simply "evaluating" everything at a point $\vec{x} = \vec{0}$. However, again, because of the formalism of Section 2.2, we would like to have double integral. We will use the same trick: namely, we will multiply our expression by the "unit coefficient" that has integrals in it. Previously, we had \emph{one} integral coming from the evaluation of \emph{single} $\partial^{\mu}$; thus we needed \emph{one} additional integral coming from unit coefficient. This time, we have \emph{no} integrals on our own; thus, we need \emph{two} additional integrals from the unit coefficient. Thus, we write down the "unit coefficient" as a double integral:
\beq 1 = \Big(\frac{\alpha}{2 \pi} \Big)^d \int d^d x d^d y e^{- \frac{\alpha}{2} (\vert \vec{x} \vert^2 + \vert \vec{y} \vert^2)} \eeq
This time the coefficient has power of $d$ instead of $d/2$ because of the product of the powers of $d/2$ in from of the $\vec{x}$- and $\vec{y}$- integrals. Now, to keep things as simple as possible we will \emph{still} evaluate each of the fields \emph{at zero}; thus, we will replace $A^{\mu} A_{\mu} \phi^* \phi$ with $A^{\mu} (\vec{0}) A_{\mu}  (\vec{0}) \phi^*  (\vec{0}) \phi  (\vec{0})$. But, at the same time, for the "formal" purposes we will pull this  "constant" factor \emph{under} the integration. Thus, we have
\beq e^2 A^{\mu} A_{\mu} \phi^* \phi  = \int d^d x d^d y e^{- \frac{\alpha}{2} (\vert \vec{x} \vert^2 + \vert \vec{y} \vert^2)} \Big(\frac{\alpha}{2 \pi} \Big)^d e^2 A^{\mu} (\vec{0}) A_{\mu} (\vec{0}) \phi^* (\vec{0}) \phi (\vec{0}) \label{eqn:DiscreteBoson}\eeq
Finally, the kinetic term, $\partial^{\mu} \phi^* \partial_{\mu} \phi$ is merely a copy of our answer from the previous section and, therefore, does not need to be repeated here. Let us now turn both of the new terms from integrals into sums. We have previously established that 
\beq \int d^d x d^d y f (\vec{x}, \vec{y}, \vec{0}) \rightarrow  \frac{V^2 M^2}{N^2} \sum_{i, j} f(\vec{x}_i - \vec{x}_k, \vec{x}_j - \vec{x}_i, \vec{x}_k ) \eeq
which is motivated by the change in notation
\beq ieA^{\mu} (\phi \partial_{\mu} \phi^* - \phi^* \partial_{\mu} \phi) = \frac{V^2 M^2}{N^2} \sum_{i, j} e^{- \frac{\alpha}{2} (\vert \vec{x}_i - \vec{x}_k \vert^2 + \vert \vec{x}_j - \vec{x}_k \vert^2)} \times \nonumber \eeq
\beq \times \Big( \frac{\alpha^d}{(2 \pi)^d \delta t} A^0 (\vec{x}_k, t) \phi (\vec{x}_k, t)  (\phi^* (\vec{x}_i, t+ \delta t) - \phi^* (\vec{x_k}, t)) + \nonumber \eeq
\beq + \frac{\alpha^{d+1}}{(2 \pi)^d} A^l (\vec{x}_k, t) \phi (\vec{x}_k, t) (x_i^l - x_k^l) (\phi^* (\vec{x}_i , t) - \phi^* (\vec{x}_k, t)) - \label{coupling}  \eeq
\beq - \frac{\alpha^d}{(2 \pi)^d \delta t} A^0 (\vec{x}_k, t) \phi^* (\vec{x}_k, t)  (\phi (\vec{x}_i, t+ \delta t) - \phi (\vec{x}_k, t)) - \nonumber \eeq
\beq -  \frac{\alpha^{d+1}}{(2 \pi)^d} A^k (\vec{x}_k, t) \phi^* (\vec{x}_k, t) (x_i^l - x_k^l) (\phi (\vec{x}_i, t) - \phi (\vec{x}_k, t)) \Big)  \nonumber \eeq
Furthermore, we perform another replacement (which is, again, identical to what we did before),
\beq \phi (\vec{x}_i, t) \rightarrow \delta_{q_k}^{q_i} \phi (\vec{x}_k, t) \; ; \; \phi (\vec{x}_i, t+ \delta t) \rightarrow \delta_{q'_k}^{q_i} \phi (\vec{x}_k, t), \eeq
\beq A^{\mu} (\vec{x}_i, t) \rightarrow \delta_{q_k}^{q_i} A^{\mu} (\vec{x}_k, t) \; ; \; A^{\mu} (\vec{x}_i, t+ \delta t) \rightarrow \delta_{q'_k}^{q_i} A^{\mu} (\vec{x}_k, t), \eeq
where the use of $q'$ allows us to turn $t+ \delta t$ into $t$ for the same reasoning as presented before. Thus, the above expression becomes 
\beq ieA^{\mu} (\phi \partial_{\mu} \phi^* - \phi^* \partial_{\mu} \phi) = \frac{V^2 M^2}{N^2} \sum_{i, j} e^{- \frac{\alpha}{2} (\vert \vec{x}_i - \vec{x}_k \vert^2 + \vert \vec{x}_j - \vec{x}_k \vert^2)} \times \nonumber \eeq
\beq \times \Big( \frac{\alpha^d}{(2 \pi)^d \delta t} A^0 (\vec{x}_k, t) \phi (\vec{x}_k, t)  (\delta_{q'_k}^{q_i} \phi^* (\vec{x}_i, t) - \phi^* (\vec{x_k}, t)) + \nonumber \eeq
\beq + \frac{\alpha^{d+1}}{(2 \pi)^d} A^l (\vec{x}_k, t) \phi (\vec{x}_k, t) (x_i^l - x_k^l) (\delta_{q_k}^{q_i} \phi^* (\vec{x}_i , t) - \phi^* (\vec{x}_k, t)) - \label{coupling}  \eeq
\beq - \frac{\alpha^d}{(2 \pi)^d \delta t} A^0 (\vec{x}_k, t) \phi^* (\vec{x}_k, t)  (\delta_{q'_k}^{q_i}\phi (\vec{x}_i, t) - \phi (\vec{x}_k, t)) - \nonumber \eeq
\beq -  \frac{\alpha^{d+1}}{(2 \pi)^d} A^k (\vec{x}_k, t) \phi^* (\vec{x}_k, t) (x_i^l - x_k^l) (\delta_{q_k}^{q_i} \phi (\vec{x}_i, t) - \phi (\vec{x}_k, t)) \Big)  \nonumber \eeq
This means that the Lagrangian generator for $j^{\mu} A_{\mu}$ is given by 
\beq {\cal K}_{jA} (i, j, k)  = \frac{V^2 M^2}{N^2}  e^{- \frac{\alpha}{2} (\vert \vec{x}_i - \vec{x}_k \vert^2 + \vert \vec{x}_j - \vec{x}_k \vert^2)} \times \nonumber \eeq
\beq \times \Big( \frac{\alpha^d}{(2 \pi)^d \delta t} A^0 (\vec{x}_k, t) \phi (\vec{x}_k, t)  (\delta_{q'_k}^{q_i} \phi^* (\vec{x}_i, t) - \phi^* (\vec{x_k}, t)) + \nonumber \eeq
\beq + \frac{\alpha^{d+1}}{(2 \pi)^d} A^l (\vec{x}_k, t) \phi (\vec{x}_k, t) (x_i^l - x_k^l) (\delta_{q_k}^{q_i} \phi^* (\vec{x}_i , t) - \phi^* (\vec{x}_k, t)) - \label{coupling}  \eeq
\beq - \frac{\alpha^d}{(2 \pi)^d \delta t} A^0 (\vec{x}_k, t) \phi^* (\vec{x}_k, t)  (\delta_{q'_k}^{q_i}\phi (\vec{x}_i, t) - \phi (\vec{x}_k, t)) - \nonumber \eeq
\beq -  \frac{\alpha^{d+1}}{(2 \pi)^d} A^k (\vec{x}_k, t) \phi^* (\vec{x}_k, t) (x_i^l - x_k^l) (\delta_{q_k}^{q_i} \phi (\vec{x}_i, t) - \phi (\vec{x}_k, t)) \Big)  \nonumber \eeq
Finally, let us do the four-vertex term (\ref{eqn:DiscreteBoson})
\beq e^2 A^{\mu} A_{\mu} \phi^* \phi  = \int d^d x d^d y e^{- \frac{\alpha}{2} (\vert \vec{x} \vert^2 + \vert \vec{y} \vert^2)} \Big(\frac{\alpha}{2 \pi} \Big)^d e^2 A^{\mu} (\vec{0}) A_{\mu} (\vec{0}) \phi^* (\vec{0}) \phi (\vec{0}) \eeq
This time, substitution is far more straightforward since everything is evaluated at $\vec{0}$. This means that everything will be evaluated at $\vec{x}_k$ \emph{as opposed to} either $\vec{x}_i$ or $\vec{x}_j$. Furthermore, we don't need any $\delta$-s, since they all evaluate to $\delta_{q_k}^{q_k}=1$. Thus, we obtain
\beq e^2 A^{\mu} A_{\mu} \phi^* \phi  = \frac{V^2M^2}{N^2} \sum_{i, j} e^{- \frac{\alpha}{2} (\vert \vec{x}_i - \vec{x}_k \vert^2 + \vert \vec{x}_j - \vec{x}_k \vert^2)} \Big(\frac{\alpha}{2 \pi} \Big)^d e^2 A^{\mu} (\vec{x}_k) A_{\mu} (\vec{x}_k) \phi^* (\vec{x}_k) \phi (\vec{x}_k) \eeq
Here, the expression under the sum is independent of $i$ and $j$. Thus, the summation over $i$ and $j$ amounts to the multiplication by the constant. The reason why the summation is there is in order to make our equation formally match the results of section 2.2. This means that the Lagrangian generator for a four-vortex $A^{\mu} A_{\mu} \phi^* \phi$ is given by
\beq {\cal K}_{AA\phi \phi}(i, j, k) = \frac{V^2M^2}{N^2}  e^{- \frac{\alpha}{2} (\vert \vec{x}_i - \vec{x}_k \vert^2 + \vert \vec{x}_j - \vec{x}_k \vert^2)} \Big(\frac{\alpha}{2 \pi} \Big)^d e^2 A^{\mu} (\vec{x}_k) A_{\mu} (\vec{x}_k) \phi^* (\vec{x}_k) \phi (\vec{x}_k) \eeq

\subsection*{3.1.4 Kinetic term for electromagnetic field}

We have now finished discussing the scalar field. Our discussion of gauge field, however, was limited to its interaction with the scalar field. Let us now turn to the question of kinetic term of gauge field, 
\beq F^{\mu \nu} F_{\mu \nu} = g^{\mu \rho} g^{\nu \sigma} F_{\mu \nu} F_{\rho \sigma} \eeq
Similarly to what we have done in other cases, we split the above into the spacelike terms and mixed terms (timelike terms are absent due to the fact that $F_{00}=0$). Upon some simple algebra it is easy to show that the above expression can be rewritten as
\beq F^{\mu \nu} F_{\mu \nu} = F_{ab} F_{cd} g^{ac} g^{bd} + 2F_{a0} F_{c0} (g^{00} g^{ac} - g^{a0} g^{c0}) + 4 F_{ab} F_{c0} g^{ac} g^{b0} \eeq
Let us first write down the expressions for $F_{0a}$ and $F_{ab}$ and then we will write their products afterwords. In order to write down the expression for $F_{ab}$ we need to know $\partial_a A_b$ and $\partial_b A_a$. We can obtain the expression for $\partial_a A_b$ by writing down the expression for $\partial_a \phi$ and replacing $\phi$ with $A_b$ (it is easy to see that we can replace $\phi$ with $A_b$ in the derivation for $\partial_a \phi$ and the index $b$ will simply "come for the ride"). Thus, we have 
\beq \partial_a A_b= \frac{\alpha^{(d+2)/2}}{(2 \pi)^{d/2}} \int d^d x e^{- \frac{\alpha}{2} \vert \vec{x} \vert^2} x^a (A_b (\vec{x}) - A_b (\vec{0})) \eeq
Now, we can write the expression for $\partial_b A_a$ by "copying" the expression for $\partial_a A_b$ while "interchanging" $a$ and $b$. Thus, we have 
\beq \partial_b A_a= \frac{\alpha^{(d+2)/2}}{(2 \pi)^{d/2}} \int d^d x e^{- \frac{\alpha}{2} \vert \vec{x} \vert^2} x^b (A_a (\vec{x}) - A_a (\vec{0})) \eeq
In order to obtain $F_{ab}$ we can simply subtract one of these expressions from the other. Just like in the previous section, we want to combine everything under the same integral in order to be as close as possible to the formalism of Section 2.2. Thus, we write $F_{ab}$ as
\beq F_{ab} = \frac{\alpha^{(d+2)/2}}{(2 \pi)^{d/2}} \int d^d x e^{- \frac{\alpha}{2} \vert \vec{x} \vert^2} (x^a (A_b (\vec{x}) - A_b (\vec{0})) - x^b (A_a (\vec{x}) - A_a (\vec{0})) ) \eeq
Let us now write down the expression for $F_{0a}$. In order to do this, we need to compute $\partial_0 A_a$ and $\partial_i A_0$. This time this will be slightly less trivial. In case of $\partial_0 A_a$ we need to borrow the expression from $\partial_0 \phi$ while index $i$ "comes for the ride"; on the other hand, in case of $\partial_a A_0$ we need to borrow the expression from $\partial_a \phi$, while index $0$ "comes for the ride". Thus, the two expressions have different forms, which correspond with the differences between the expressions for $\partial_a \phi$ and $\partial_0 \phi$: 
\beq \partial_a A_0 = \frac{\alpha^{(d+2)/2}}{(2 \pi)^{d/2}} \int d^d x e^{- \frac{\alpha}{2} \vert \vec{x} \vert^2} x^a (A_0 (\vec{x}) - A_0 (\vec{0})) \eeq
\beq \partial_0 A_a = \frac{\alpha^{d/2}}{(2 \pi)^{d/2} \delta t}  \int d^d x e^{- \frac{\alpha}{2} \vert \vec{x} \vert^2} (A_a (\vec{x}, t+ \delta t) - A_a (\vec{0}, t)) \eeq
Notably, $x^a$ is present in $\partial_a A_0$ and is absent in $\partial_0 A_a$. In both cases there is one index $a$ which implies the odd symmetry with respect to the spacelike coordinates. The same odd symmetry is preserved on the right hand side. In case of the right hand side of $\partial_a A_0$, this symmetry is due to $x^a$, while in case of the right hand side of $\partial_0 A_a$ it is due to $A_a$. At the same time, if we are to replace $A_a$ with $\phi$, the odd symmetry due to $x^a$ will be preserved while the one due to $A_a$ will disappear. This corresponds to the fact that $\partial_a \phi$ has odd symmetry (just like $\partial_a A_0$ does), while $\partial_0 \phi$ does \emph{not} have odd symmetry (\emph{despite} the fact that $\partial_0 A_a$ does).  Finally, we can subtract the expression for $\partial_a A_0$ from the expression for $\partial_0 A_a$ in order to obtain the expression for $F_{0a}$. As before, we combine these two expressions into one integral, and simply subtract the expressions \emph{under} the integral sign: 
\beq F_{0a} = \Big(\frac{\alpha}{2 \pi} \Big)^{d/2} \int d^d x e^{- \frac{\alpha}{2} \vert \vec{x} \vert^2} \Big( \frac{A_a (\vec{x}, t+ \delta t) - A_a (\vec{0}, t)}{\delta t} - \alpha x^a (A_0 (\vec{x}) - A_0 (\vec{0})) \Big)\eeq
We are now ready to calculate the contraction $F^{\mu \nu} F_{\mu \nu}$. We will separately evaluate the spacelike and mixed terms (timelike terms are absent since $F_{00}=0$) and then combine them at the end. Let us start from the spacelike term. In light of the fact that the metric $g^{\mu \nu}$ can have off-diagonal components, we can \emph{not} assume that any of the indexes match. Thus, we are interested in computing a general expression of the form $F_{ab} F_{cd}$. We will define $F_{ab}$ based on the corresponding integral we have just written. On the other hand, the expression for $F_{cd}$ will be given by a "copy" of that integral where $a$ is replaced with $c$, $b$ is replaced with $d$ and $\vec{x}$ is replaced with $\vec{y}$. Thus, we have
\beq F_{ab} F_{cd} = \frac{\alpha^{(d+2)/2}}{(2 \pi)^{d/2}} \int d^d x e^{- \frac{\alpha}{2} \vert \vec{x} \vert^2} (x^a (A_b (\vec{x}) - A_b (\vec{0})) - x^b (A_a (\vec{x}) - A_a (\vec{0})) ) \times \nonumber \eeq
\beq \times \frac{\alpha^{(d+2)/2}}{(2 \pi)^{d/2}} \int d^d y e^{- \frac{\alpha}{2} \vert \vec{y} \vert^2} (y^c (A_d (\vec{x}) - A_d (\vec{0})) - y^l (A_c (\vec{x}) - A_c (\vec{0})) ) \nonumber \eeq
Again, in order to accomodate the framework layed out in Section 2.2, we combine these two integrals into a double integral. We furthermore combine the two coefficients into one. Thus, we have
\beq F_{ab} F_{cd} = \frac{\alpha^{d+2}}{(2 \pi)^d} \int d^d x d^d y e^{- \frac{\alpha}{2} (\vert \vec{x} \vert^2 + \vert \vec{y} \vert^2)} (x^a (A_b (\vec{x}) - A_b (\vec{0})) - x^b (A_i (\vec{x}) - A_a (\vec{0})) )  \times \nonumber \eeq
\beq \times  (y^c (A_d (\vec{x}) - A_d (\vec{0})) - y^d (A_c (\vec{x}) - A_c (\vec{0})) ) \eeq
Our next task is compute "mixed terms". Again, because of the off-diagonal terms of $g^{\mu \nu}$, we can not assume that indexes match. Thus, there are two kinds of "mixed" terms: $F_{a0} F_{c0}$ and $F_{ab} F_{c0}$. Let us start with $F_{a0} F_{c0}$. For the $F_{a0}$ we simply copy the "integral" expression we already have. For the $F_{c0}$, we replace $a$ with $c$ and $\vec{x}$ with $\vec{y}$ in that integral. Thus, we have 
\beq F_{a0} F_{c0} = \Big(\frac{\alpha}{2 \pi} \Big)^{d/2} \int d^d x e^{- \frac{\alpha}{2} \vert \vec{x} \vert^2} \Big( \frac{A_a (\vec{x}, t+ \delta t) - A_a (\vec{0}, t)}{\delta t} - \alpha x^a (A_0 (\vec{x}) - A_0 (\vec{0})) \Big) \times \nonumber \eeq
\beq \times \Big(\frac{\alpha}{2 \pi} \Big)^{d/2} \int d^d y e^{- \frac{\alpha}{2} \vert \vec{y} \vert^2} \Big( \frac{A_a (\vec{y}, t+ \delta t) - A_a (\vec{0}, t)}{\delta t} - \alpha y^a (A_0 (\vec{y}) - A_0 (\vec{0})) \Big) \nonumber \eeq
Again, we rewrite the product of single-variable integral as a double variable integral, and we combine the coefficients in front. Thus, we obtain 
\beq F_{a0} F_{c0} = \Big( \frac{\alpha}{2 \pi} \Big)^d \int d^d x d^d y e^{- \frac{\alpha}{2} ( \vert \vec{x} \vert^2 + \vert \vec{y} \vert^2)} \Big( \frac{A_a (\vec{x}, t+ \delta t) - A_a (\vec{0}, t)}{\delta t} - \alpha x^a (A_0 (\vec{x}) - A_0 (\vec{0})) \Big) \times \nonumber \eeq
\beq \times \Big( \frac{A_a (\vec{y}, t+ \delta t) - A_a (\vec{0}, t)}{\delta t} - \alpha y^a (A_0 (\vec{y}) - A_0 (\vec{0})) \Big) \eeq
Again, due to the off-diagonal nature of $g^{\mu \nu}$ we have one more "mixed" term to take care of; namely, $F_{ab} F_{c0}$. In this case the computation is slighty less trivial since the expressions for $F_{ab}$ and $F_{c0}$ have different form (the former being "spacelike" and the latter being "mixed"). But, since we already have separate expressions for both, the substitution is still quite trivial:
\beq F_{ab} F_{c0} = \frac{\alpha^{(d+2)/2}}{(2 \pi)^{d/2}} \int d^d x e^{- \frac{\alpha}{2} \vert \vec{x} \vert^2} (x^a (A_b (\vec{x}) - A_b (\vec{0})) - x^b (A_a (\vec{x}) - A_a (\vec{0})) ) \times \nonumber \eeq
\beq \times \Big(\frac{\alpha}{2 \pi} \Big)^{d/2} \int d^d y e^{- \frac{\alpha}{2} \vert \vec{y} \vert^2} \Big( \frac{A_c (\vec{y}, t+ \delta t) - A_c (\vec{0}, t)}{\delta t} - \alpha y^c (A_0 (\vec{y}) - A_0 (\vec{0})) \Big)\eeq
Again, we rewrite a product of integrals in terms of one double-integral, and we combine coefficients in front (which now happened to be different from each other making the algebra slightly more complicated), and obtain
\beq F_{ab} F_{c0} = \frac{\alpha^{d+1}}{(2 \pi)^d} \int d^d x d^d y e^{- \frac{\alpha}{2} (\vert \vec{x} \vert^2 + \vert \vec{y} \vert^2)} (x^a (A_b (\vec{x}) - A_b (\vec{0})) - x^b (A_a (\vec{x}) - A_a (\vec{0})) ) \times \nonumber \eeq
\beq \times \Big( \frac{A_c (\vec{y}, t+ \delta t) - A_c (\vec{0}, t)}{\delta t} - \alpha y^c (A_0 (\vec{y}) - A_0 (\vec{0})) \Big)\eeq
Thus, we have done the "spacelike term" ($F_{ab} F_{cd}$) as well as the two possible kinds of "mixed terms" ($F_{a0} F_{c0}$ and $F_{ab} F_{c0}$). In light of the fact that $F_{00}=0$ \emph{regardless of the metric}, we do \emph{not} need to evaluate "timelike term" (which is identically zero). Thus, we are now ready to combine our results into $F^{\mu \nu} F_{\mu \nu}$. As we said earlier, the latter is given by 
\beq F^{\mu \nu} F_{\mu \nu} = F_{ab} F_{cd} g^{ac} g^{bd} + 2F_{a0} F_{c0} (g^{00} g^{ac} - g^{a0} g^{c0}) + 4 F_{ab} F_{c0} g^{ac} g^{b0} \eeq
We now take the values of metric \emph{at} a point $\vec{x} = \vec{0}$, while substituting the corresponding integrals for the products of $F$-s. Thus, we have 
\beq  F^{\mu \nu} F_{\mu \nu} = g^{ac} (\vec{0}) g^{bd} (\vec{0}) \frac{\alpha^{d+2}}{(2 \pi)^d} \int d^d x d^d y e^{- \frac{\alpha}{2} (\vert \vec{x} \vert^2 + \vert \vec{y} \vert^2)} \times \nonumber \eeq
\beq \times (x^a (A_b (\vec{x}) - A_b (\vec{0})) - x^b (A_a (\vec{x}) - A_a (\vec{0})) )  \times \nonumber \eeq
\beq \times  (y^c (A_d (\vec{x}) - A_d (\vec{0})) - y^d (A_c (\vec{x}) - A_c (\vec{0})) ) + \nonumber \eeq
\beq + 2 (g^{00} (\vec{0})g^{ac}(\vec{0}) - g^{a0}(\vec{0}) g^{c0}(\vec{0})) \Big( \frac{\alpha}{2 \pi} \Big)^d \int d^d x d^d y e^{- \frac{\alpha}{2} ( \vert \vec{x} \vert^2 + \vert \vec{y} \vert^2)} \times \nonumber \eeq
\beq \times \Big( \frac{A_a (\vec{x}, t+ \delta t) - A_a (\vec{0}, t)}{\delta t} - \alpha x^a (A_0 (\vec{x}) - A_0 (\vec{0})) \Big) \times  \eeq
\beq \times \Big( \frac{A_a (\vec{y}, t+ \delta t) - A_a (\vec{0}, t)}{\delta t} - \alpha y^a (A_0 (\vec{y}) - A_0 (\vec{0})) \Big) +  \nonumber \eeq
\beq + 4 g^{ac} g^{b0} \frac{\alpha^{d+1}}{(2 \pi)^d} \int d^d x d^d y e^{- \frac{\alpha}{2} (\vert \vec{x} \vert^2 + \vert \vec{y} \vert^2)} \times \nonumber \eeq
\beq \times (x^a (A_b (\vec{x}) - A_b (\vec{0})) - x^b (A_a (\vec{x}) - A_a (\vec{0})) ) \times \nonumber \eeq
\beq \times \Big( \frac{A_c (\vec{y}, t+ \delta t) - A_c (\vec{0}, t)}{\delta t} - \alpha y^c (A_0 (\vec{y}) - A_0 (\vec{0})) \Big) \nonumber \eeq
In order to accommodate the formalism of Section 2.2, we prefer to have a single integral instead of a linear combination of integrals. Thus, we pull the constants (including $g^{\mu \nu} (\vec{0})$) \emph{under} the integrals, and rewrite the sum of integrals in a form of integral of a sum. Thus, we get 
\beq  F^{\mu \nu} F_{\mu \nu} = \int d^d x d^d y e^{- \frac{\alpha}{2} (\vert \vec{x} \vert^2 + \vert \vec{y} \vert^2)}\times \nonumber \eeq
\beq \times  \Big(g^{ac} g^{bd} \frac{\alpha^{d+2}}{(2 \pi)^d}  (x^a (A_b (\vec{x},t) - A_b (\vec{0},t)) - x^b (A_a (\vec{x},t) - A_a (\vec{0},t)) )  \times \nonumber \eeq
\beq \times  (y^c (A_d (\vec{x},t) - A_d (\vec{0},t)) - y^d (A_c (\vec{x},t) - A_c (\vec{0},t)) ) + \nonumber \eeq
\beq + 2 (g^{00} g^{ac} - g^{a0} g^{c0}) \Big( \frac{\alpha}{2 \pi} \Big)^d \times \nonumber \eeq
\beq \times \Big( \frac{A_a (\vec{x}, t+ \delta t) - A_a (\vec{0}, t)}{\delta t} - \alpha x^a (A_0 (\vec{x},t) - A_0 (\vec{0},t)) \Big) \times  \eeq
\beq \times \Big( \frac{A_a (\vec{y}, t+ \delta t) - A_a (\vec{0}, t)}{\delta t} - \alpha y^a (A_0 (\vec{y},t) - A_0 (\vec{0},t)) \Big) +  \nonumber \eeq
\beq + 4 g^{ac} g^{b0} \frac{\alpha^{d+1}}{(2 \pi)^d}  (x^i (A_b (\vec{x},t) - A_b (\vec{0},t)) - x^b (A_a (\vec{x},t) - A_a (\vec{0},t)) ) \times \nonumber \eeq
\beq \times \Big( \frac{A_c (\vec{y}, t+ \delta t) - A_c (\vec{0}, t)}{\delta t} - \alpha y^c (A_0 (\vec{y},t) - A_0 (\vec{0},t)) \Big) \Big) \nonumber \eeq
Seeing that we \emph{already} have a double integral, as desired, we do \emph{not} need to increase the number of integrals by any extra unit factors. Thus, the above is our final integral-based expression for $F^{\mu \nu} F_{\mu \nu}$. Now we will again try to turn integral into the sum by the usual prescription. Now, in order to turn this integral into the sum we again, as usual, use the following substitutions:
\beq \int d^d x d^d y (\cdots) \rightarrow \frac{M^2 V^2}{N^2} \sum_{i, j} (\cdots) \nonumber \eeq
\beq x^a \rightarrow x_i^a - x_k^i \; ; \; y^a \rightarrow x_j^a - x_k^i \; ; \; \vec{0} \rightarrow \vec{x}_k \eeq
\beq \phi (\vec{x}_i, t) \rightarrow \delta_{q_k}^{q_i} \phi (\vec{x}_i, t) \; ; \; \phi (\vec{x}_i, t+ \delta t) \rightarrow \delta_{q'_k}^{q_i} \phi (\vec{x}, t)  \nonumber  \eeq
\beq A^{\mu} (\vec{x}_i, t) \rightarrow \delta_{q_k}^{q_i} A^{\mu} (\vec{x}_i, t) \; ; \; A^{\mu} (\vec{x}_i, t+ \delta t) \rightarrow \delta_{q'_k}^{q_i} A^{\mu} (\vec{x}, t) \nonumber \eeq
By making the above substitutions, our discretized expression becomes 
\beq  F^{\mu \nu} F_{\mu \nu} = \frac{M^2V^2}{N^2} \sum_{i,j} e^{- \frac{\alpha}{2} (\vert \vec{x}_i - \vec{x}_k \vert^2 + \vert \vec{x}_j - \vec{x}_k \vert^2)}\times \nonumber \eeq
\beq \times  \Big(g^{ac} g^{bd} \frac{\alpha^{d+2}}{(2 \pi)^d}  (x_i^a (\delta_{e_k}^{e_i} A_b (\vec{x}_i, t) - A_b (\vec{x}_k, t)) - x_i^b (\delta_{e_k}^{e_i} A_a (\vec{x}_i, t) - A_a (\vec{x}_k, t)) )  \times \nonumber \eeq
\beq \times  (x_j^c (\delta_{e_k}^{e_i} A_d (\vec{x}_i, t) - A_d (\vec{x}_k, t)) - x_j^d (\delta_{e_k}^{e_i} A_c (\vec{x}_i,t) - A_c (\vec{x}_3, t)) ) + \nonumber \eeq
\beq + 2 (g^{00} g^{ac} - g^{a0} g^{c0}) \Big( \frac{\alpha}{2 \pi} \Big)^d \times \nonumber \eeq
\beq \times \Big( \frac{\delta_{e'_k}^{e_i} A_a (\vec{x}_i, t) - A_a (\vec{x}_k, t)}{\delta t} - \alpha x_i^a (\delta_{e_i}^{e_k} A_0 (\vec{x}_i) - A_0 (\vec{x}_k, t)) \Big) \times  \eeq
\beq \times \Big( \frac{\delta_{e_k'}^{e_j} A_a (\vec{x}_j, t) - A_a (\vec{x}_k, t)}{\delta t} - \alpha x_j^a (\delta_{e_k}^{e_j} A_0 (\vec{x}_j) - A_0 (\vec{x}_k, t)) \Big) +  \nonumber \eeq
\beq + 4 g^{ac} g^{b0} \frac{\alpha^{d+1}}{(2 \pi)^d}  (x_i^a (A_b (\vec{x}_i) - A_b (\vec{x}_k, t)) - x_i^b (A_a (\vec{x}_i, t) - A_a (\vec{x}_k, t)) ) \times \nonumber \eeq
\beq \times \Big( \frac{\delta_{e'_k}^{e_j}A_c  (\vec{x}_j) - A_c (\vec{0}, t)}{\delta t} - \alpha x_j^c (\delta_{e _k}^{e_j}A_0 (\vec{x}_j) - A_0 (\vec{x}_k, t)) \Big) \Big) \nonumber \eeq
From this we read off the Lagrangian generator to be
\beq {\cal K}_{EM; kin} (i, j, k) = \frac{M^2V^2}{N^2} e^{- \frac{\alpha}{2} (\vert \vec{x}_i - \vec{x}_k \vert^2 + \vert \vec{x}_j - \vec{x}_k \vert^2)}\times \nonumber \eeq
\beq \times  \Big(g^{ac} g^{bd} \frac{\alpha^{d+2}}{(2 \pi)^d}  (x_i^a (\delta_{e_k}^{e_i} A_b (\vec{x}_i, t) - A_b (\vec{x}_k, t)) - x_i^b (\delta_{e_k}^{e_i} A_a (\vec{x}_i, t) - A_a (\vec{x}_k, t)) )  \times \nonumber \eeq
\beq \times  (x_j^c (\delta_{e_k}^{e_i} A_d (\vec{x}_i, t) - A_d (\vec{x}_k, t)) - x_j^d (\delta_{e_k}^{e_i} A_c (\vec{x}_i,t) - A_c (\vec{x}_3, t)) ) + \nonumber \eeq
\beq + 2 (g^{00} g^{ac} - g^{a0} g^{c0}) \Big( \frac{\alpha}{2 \pi} \Big)^d \times \nonumber \eeq
\beq \times \Big( \frac{\delta_{e'_k}^{e_i} A_a (\vec{x}_i, t) - A_a (\vec{x}_k, t)}{\delta t} - \alpha x_i^a (\delta_{e_i}^{e_k} A_0 (\vec{x}_i) - A_0 (\vec{x}_k, t)) \Big) \times  \eeq
\beq \times \Big( \frac{\delta_{e_k'}^{e_j} A_a (\vec{x}_j, t) - A_a (\vec{x}_k, t)}{\delta t} - \alpha x_j^a (\delta_{e_k}^{e_j} A_0 (\vec{x}_j) - A_0 (\vec{x}_k, t)) \Big) +  \nonumber \eeq
\beq + 4 g^{ac} g^{b0} \frac{\alpha^{d+1}}{(2 \pi)^d}  (x_i^a (A_b (\vec{x}_i) - A_b (\vec{x}_k, t)) - x_i^b (A_a (\vec{x}_i, t) - A_a (\vec{x}_k, t)) ) \times \nonumber \eeq
\beq \times \Big( \frac{\delta_{e'_k}^{e_j}A_c  (\vec{x}_j) - A_c (\vec{0}, t)}{\delta t} - \alpha x_j^c (\delta_{e _k}^{e_j}A_0 (\vec{x}_j) - A_0 (\vec{x}_k, t)) \Big) \Big) \nonumber \eeq

\subsection*{3.2 Fermions}

\subsection*{3.2.1 Can Grassmann integration be viewed as literal "sum"?}

We would now like to turn to the issue of fermions. In order to incorporate fermionic field into our theory, we would like to define fermions as ontologically meaningful quantity. In particular, we would like Grassmann numbers to be well defined outside path integral, and to view path integral as literally a sum. We will view Grassmann numbers to be elements of some \emph{countable} set $G \subset P$. The set $P$ is equipped with anticommuting wedge product ($\wedge$) \emph{along with} a dot product ($\cdot$) which neither commutes, nor anti-commutes:
\beq p_1 \wedge p_2 = - p_2 \wedge p_1 \; ; \; p_2 \cdot p_1 \neq p_1 \cdot p_2 \neq - p_2 \cdot p_1 \eeq
Apart from this, the set $P$ is equipped with addition and scalar multiplication, both of which are commuting and transitive: 
\beq p_1 + p_2 = p_2 + p_1 \; ; \; p_1 + (p_2 + p_3) = (p_1 + p_2) + p_3 \eeq
\beq c \in \mathbb{C} \Rightarrow cp = pc \; ; \; c_1 \in \mathbb{C} \wedge c_2 \in \mathbb{C} \Rightarrow c_1 (c_2 p) = (c_1 c_2) p \nonumber \eeq
The set $G$ "generates" the set $p$ in a sense that any element $p \in P$ can be expressed as 
\beq p = \sum_{n=1}^N c_n \theta_{n1} \wedge \cdots \wedge \theta_{na_n} \eeq
where $\theta_{na} \in G$. We are going to "break down" $G$ into the "smaller" sets $G_k$, \emph{each of which contains only one element}:  
\beq G = G_1 \cup G_2 \cup \cdots \; ; \; G_k = \{ \theta_k \} \eeq
The set $G_k$ represents the domain of integration over a variable $\theta_k$. Thus, integration is \emph{literally} over a one-point set. We can "turn" the integration over "countable set" $G$ into the integration over one-point set $G_k$ by defining the set of differential elements 
\beq (d_i \theta) \vert_{\theta = \theta_j} = \delta_i^j \theta_j \eeq
The generic expression for the integral takes the form 
\beq \int (d_1 \eta_1 \wedge \cdots \wedge d_n \eta_n) \cdot f(\eta_1, \cdots, \eta_n) = \sum_{\eta_k \in G} (d_1 \eta_1 \wedge \cdots \wedge d_n \eta_n) \cdot f(\eta_1, \cdots, \eta_n), \label{eqn:91}\eeq
where the "limit" was skipped because $G$ is countable and, for the same reason, $d_i \eta_i$ are treated as finite. In light of the fact that it is "literally" a sum, $f(\eta_1, \cdots, \eta_n)$ can be any function we like; it does \emph{not} have to take any algebraic form. Nevertheless, in the usual "physics" situation it takes the form of a polynomials of \emph{wedge} products. This is, in particular, true because both the Dirac Lagrangian, as well as the "exponentiation" are defined in terms of wedge products:
\beq {\cal L} = \overline{\eta} \wedge (\gamma^{\mu}  (\partial_{\mu} +ieA_{\mu}) \eta) = \gamma^k_{ac} \gamma^0_{cb} (\eta_{ar}-i \eta_{ai}) \wedge (\partial_{\mu} \eta_{ar} + i \partial_{\mu} \eta_{ai} + ieA_{\mu} \eta_{ar} - eA_{\mu} \eta_{ai}) \nonumber \eeq
\beq e^{-iS} = \sum_{n=0}^{\infty} \frac{(-iS)^n}{n!} \; , \; (iS)^{n+1} = iS \wedge (iS)^n \eeq
where $a$, $b$, and $c$ are elements of $\{1, 2, 3, 4 \}$, "$r$" and "$i$" stands for real and imaginary part,
\beq \eta_a = \eta_{ar} + i \eta_{ai} \; ; \; \eta_a^{\dagger} = \eta_a^* = \eta_{ar} - i \eta_{ai} \eeq
and the "derivative" is a discretized expression which will be explicitly defined in Section 3.2.6. The only relevent thing to this section is that in the definition of Dirac Lagrangian we \emph{only} used wedge product and \emph{not} the dot-product; the same is true for the definition of exponential. Therefore, the generic integral involves wedge product between "differentials" \emph{dotted} with \emph{another} wedge product between "finite parts".  Now, if we substitute the expression 
\beq (d_i \eta_k) \vert_{\eta_k = \theta_j} = \delta_i^j \theta_j \label{eqn:94}\eeq
into the definition of the integral in (\ref{eqn:91}), we get 
\beq \int (d_1 \eta_1 \wedge \cdots \wedge d_n \eta_n) \cdot f(\eta_1, \cdots, \eta_n) = (\theta_1 \wedge \cdots \wedge \theta_n) \cdot f(\theta_1, \cdots, \theta_n). \eeq
In light of $\delta_i^j$ present in the definition of $d_i \eta_k$, the "sum" on the left hand side consists of exactly one term, which is why the summation was skipped altogether on the right hand side. The key in reproducing the desired properties of Grassmann integral is a definition of a \emph{dot} product. For example, we would "like" to have 
\beq \int d \eta \eta = 1 \eeq
In our notation this means 
\beq \int d_1 \eta \cdot \eta = 1 \eeq
But, if we know dot product, then the left hand side evaluates to 
\beq \int d_1 \eta \cdot \eta = \theta_1 \cdot \theta_1 \eeq
Thus, we would like to demand that 
\beq \theta_1 \cdot \theta_1 = 1 \eeq
On the other hand, we would like the integral of constant to be zero:
\beq \int d \theta = 0 \eeq
In our notation, this means 
\beq \int d_1 \eta \cdot 1 =0 \eeq
But, if we evaluate this integral formally, we have
\beq \int d_1 \eta \cdot 1 = \theta_1 \cdot 1 \eeq
Thus, we have to demand that 
\beq \theta_1 \cdot 1 = 0 \eeq
Now, in case of the two-dimensional integral we would like to have (contrary to the sign convention that is usually being used)
\beq \int d \eta d \xi \eta \xi = \int d \eta \Big( \int d \xi \eta \xi \Big) = - \int d \eta \Big(\eta \int d \xi \xi \Big) = - \int d \eta \eta = -1 \eeq
This means that we should be able to write it down as 
\beq \int (d_1 \eta \wedge d_2 \xi) \cdot (\eta \wedge \xi) = -1 \eeq
\emph{But} we already know that the integral on the left hand side is given by 
\beq \int (d_1 \eta \wedge d_2 \xi) \cdot (\eta \wedge \xi) = (\theta_1 \wedge \theta_2) \cdot (\theta_1 \wedge \theta_2) \eeq
This means that we would like to define dot-product in such a way that 
\beq (\theta_1 \wedge \theta_2) \cdot (\theta_1 \wedge \theta_2) = -1 \eeq
We would like to \emph{also} be able to write the above integral as 
\beq \int \Big( d_1 \eta \cdot \Big( \int d_2 \xi \cdot (\eta \wedge \xi) \Big) \Big) = -1 \eeq
But, again, if we know the wedge product, we can "read off" the left hand side of the above expression to be
\beq \int \Big( d_1 \eta \cdot \Big( \int d_2 \xi \cdot (\eta \wedge \xi) \Big) \Big) = \theta_1 \cdot (\theta_2 \cdot (\theta_1 \wedge \theta_2)) \eeq
This means that we would like to have 
\beq \theta_1 \cdot (\theta_2 \cdot (\theta_1 \wedge \theta_2)) = -1 \eeq
Since we already know that $\theta_1 \cdot \theta_1 =1$. This means that the above equation will be produced if we insist that
\beq \theta_2 \cdot (\theta_1 \wedge \theta_2) = - 1 \eeq
Now, since we would like integration to anticommute, we would like to have
\beq  \int \Big(d_2 \xi \cdot \Big( \int d_1 \eta \cdot (\eta \wedge \xi) \Big) \Big) = - \int \Big(d_1 \eta \cdot \Big( \int d_2 \xi \cdot (\eta \wedge \xi) \Big) \Big) = +1\eeq
This implies that 
\beq \theta_2 \cdot (\theta_1 \cdot (\theta_1 \wedge \theta_2)) = -  \theta_1 \cdot (\theta_2 \cdot (\theta_1 \wedge \theta_2)) = +1 \eeq
But, just like we have shown that $\theta_1 \cdot \theta_1 =1$, we can also show that $\theta_2 \cdot \theta_2 = 1$. Therefore, the above equation can be satisfied if we demand that 
\beq \theta_1 \cdot (\theta_1 \wedge \theta_2) = + \theta_2 \eeq
As we have just seen, we identify the integrals with "too few" differentials with non-zero quantities. At the same time, however, we would like the integrals with "too many" differentials to be set to zero. For example, 
\beq \int (d_1 \xi \wedge d_2 \eta) \cdot \xi =  \int (d_1 \xi \wedge d_2 \eta) \cdot \eta = 0 \eeq
This means that we would like to have 
\beq (\theta_1 \wedge \theta_2) \cdot \theta_1 = (\theta_1 \wedge \theta_2) \cdot \theta_2 = 0 \eeq
At the same time, we also stick to what we said easier that 
\beq \theta_1 \cdot (\theta_1 \wedge  \theta_2) = - \theta_2 \cdot (\theta_1 \wedge \theta_2) = +1 \eeq
This shows that the dot product \emph{neither commutes nor anticommutes}. We can now guess a general definition of a dot-product of the form $ (\theta_{a_1} \wedge \cdots \wedge \theta_{a_m}) \cdot (\theta_{b_1} \wedge \cdots \wedge \theta_{b_n})$:
\beq (\theta_{a_1} \wedge \cdots \wedge \theta_{a_m}) \cdot (\theta_{b_1} \wedge \cdots \wedge \theta_{b_n}) = 
\left\{
	\begin{array}{ll}
		0 \; , \; \exists i (\forall j (b_j \neq a_i) \\
		(-1)^{\sharp \{i, j \vert a_i >b_j \}} \theta_{c_1} \wedge \cdots \wedge \theta_{c_p} \; , \; \forall i (\exists j (b_j = a_i)
	\end{array}
\right.
\eeq
where 
\beq \{c_1, \cdots, c_p \} = \{a_1, \cdots, a_m \} \setminus \{b_1, \cdots, b_n \} \; ; \; c_1 < \cdots < c_p \eeq
The first line on the right hand side tells us that if we are "integrating over" at least one of the variables that is absent in the integrand, itself, we will get zero. The simplest example is the integral of the constant being zero. The second line tells us that if everything we are integrating over \emph{is} present in the integrand, we are guaranteed to have non-zero result. Furthermore, even if integrand includes some variables we are \emph{not} integrating over, we will still get non-zero; these variables will simply "factor out" and produce Grassmann-valued result, as was seen in the few examples we presented. This will allow us to "go back" and integrate over them "later on" to produce unity (with either plus or minus sign). The fact that having something in differential part and not in the integrand will give zero while the reverse situation will not, implies that the dot product neither commutes nor anticommutes. 

\subsection*{3.2.2. Rigorous definition of Grassmann polynomials and their products}

In the previous section, we got an intuitive sense of what Grassmann numbers, their "polynomials" as well as dot and wedge products to look like. Let us now attempt to define them in a mathematically rigorous way. A general finite polynomial of Grassmann numbers with complex-valued coefficients takes the form
\beq c_1 \theta_{a_{11}} \wedge \cdots \wedge \theta_{a_{1n_1}}  + \cdots +  c_m \theta_{a_{m1}}  \wedge \cdots \wedge \theta_{a_{mn_m}} \eeq
We will borrow a well known common trick from abstract algebra and identify polynomical with a function, whose values are identified with coefficients of a polynomial. In our case, our "function" is of the type $p \colon {\cal P} \mathbb{N} \rightarrow \mathbb{C}$, where ${\cal P} \mathbb{N}$ is a "power set of $\mathbb{N}$"; or, in other words, a set of subsets of $\mathbb{N}$. We explicitly specify that function as follows: 
\beq p = c_1 \theta_{a_{11}} \wedge \cdots \wedge \theta_{a_{1n_1}}  + \cdots +  c_m \theta_{a_{m1}}  \wedge \cdots \wedge \theta_{a_{mn_m}}  \Rightarrow \nonumber \eeq
\beq \Rightarrow 
\left\{
	\begin{array}{ll}
		p (\{a_{k1}, \cdots, a_{kn_k} \}) = c_k \; , \; a_{k1} < \cdots < a_{kn_k}\\
		(\forall k (A \neq \{a_{k1}, \cdots, a_{kn_k} \})) \Rightarrow (p (A) = 0)
	\end{array}
\right.
\eeq
Thus, we can formally define the Grassmann polynomial in the following way:

{\bf Definition 1} Let $p \colon {\cal P} \mathbb{N} \rightarrow \mathbb{C}$ be a complex valued function. The function $p$ is said to be \emph{Grassmann polynomial} if $p(A) = 0$ for any infinite set $A$. 

We can likewise define the individual Grassmann numbers:

{\bf Definition 2} Let $n$ be an integer. Then a \emph{Grassmann number} $\theta_n$ is a Grassmann polynomial defined in the following way:
\beq \theta_n (A) = \left\{
	\begin{array}{ll}
		1 \; , \; A = \{n \}\\
		0 \; , \; A \neq \{n \}
	\end{array}
\right.
\eeq
 
Finally, we can define a "scalar polynomial", which is distinct mathematical object from the corresponding complex number, even though there is an obvious bijection between one and the other:

{\bf Definition 3} Let $c$ be a complex number. Then the "scalar" $\underline{c}$ is a Grassmann polynomial defined as 

\beq \underline{c} (A) = \left\{
	\begin{array}{ll}
		c \; , \; A = \emptyset \\
		0 \; , \; A \neq \emptyset
	\end{array}
\right.
\eeq

Now, from basic notational convention of set theory, $\{1, 2 \} = \{2, 1 \}$; thus, a simple notational taughtology tells us that  
\beq p(\{1, 2 \}) = + p (\{2, 1 \}) \eeq
But, at the same time, we would like to demand that 
\beq \theta_1 \wedge \theta_2 = - \theta_2 \wedge \theta_1 \eeq
Therefore, our "definition" can not continue to hold if we rearrange $\theta_1$ and $\theta_2$. We choose to demand that it \emph{only} holds for $\theta_1 \wedge \theta_2$ \emph{as opposed to} $\theta_2 \wedge \theta_1$. Thus, 
\beq (\theta_1 \wedge \theta_2)(\{1, 2\}) =  (\theta_1 \wedge \theta_2)(\{2, 1\}) = 1 \; ; \; (\theta_2 \wedge \theta_1)(\{1, 2\}) =  (\theta_2 \wedge \theta_1)(\{2, 1\}) = -1 \eeq
It is easy to see from above expression that $\theta_1$ and $\theta_2$ anticommute and, \emph{at the same time}, $1$ and $2$ in $\{1, 2\}$ commute, as desired.  In order to be able to write above in the form
\beq \theta_1 \wedge \theta_2 = - \theta_2 \wedge \theta_1 \eeq
without "acting" on $\{1, 2\}$, we have to define "scalar multiplication by $-1$". This is merely a special case of a general definition of scalar multiplication by a complex number $a$, which is given as 
\beq (ap)(A) = a(p(A)) \eeq
It is possible to check by induction that the expected properties of wedge product (namely, anticommutativity and linearity) are satisfied if we define it in the following way:
\beq (p_1 \wedge p_2)(T) = \sum_{U \cup V = T ; U \cap V = \emptyset} (-1)^{ \sharp \{ ((\vec{x}_1, \alpha_1, \beta_1) \in U , (\vec{x}_2, \alpha_2, \beta_2) \in V) \vert (\vec{x}_1, \alpha_1, \beta_1) \succ (\vec{x}_2, \alpha_2, \beta_2) \} } p_1 (U) p_2 (V) \eeq
Interestingly enough, wedge product happens to coincide with scalar product: 
\beq c \in \mathbb{C} \Rightarrow \underline{c} \wedge \theta_1 = + \theta_1 \wedge \underline{c} = + \theta_1 \eeq
The plus-sign will no longer be surprising if we remind ourselves of plus-signs appearing in other situations, such as
\beq (\theta_1 \wedge \theta_2) \wedge (\theta_3 \wedge \theta_4) = + (\theta_3 \wedge \theta_4) \wedge (\theta_1 \wedge \theta_2) \eeq
The general principle is that 
\beq (\theta_{a_1} \wedge \cdots \wedge \theta_{a_m}) \wedge (\theta_{b_1} \wedge \cdots \wedge \theta_{b_n}) = (-1)^{mn} (\theta_{b_1} \wedge \cdots \wedge \theta_{b_n}) \wedge  (\theta_{a_1} \wedge \cdots \wedge \theta_{a_m}) \eeq
Thus, we get plus-sign whenever either $m$ or $n$ is even. This of course includes the possibility of one of them being zero, which is what we have when we do scalar product. In light of this, wedging a scalar with itself does \emph{not} return $0$. For example,
\beq \underline{1} \wedge \underline{1} = \underline{1} \eeq
At the same time, in all other cases we \emph{do} get zero, even if we have a string of even number of $\theta$-s:
\beq 0 = \theta_1 \wedge \theta_1 = (\theta_1 \wedge \theta_2) \wedge  (\theta_1 \wedge \theta_2) = \cdots \eeq
We are now ready to write down the formal definition of wedge and dot products:

{\bf Definition 4} Let $p_1$ and $p_2$ be two arbitrary Grassmann polynomials. In this case, there are Grassmann polynomials $p_1 \wedge p_2$ and $p_1 \cdot p_2$ which, respectively, are referred to as \emph{wedge product} and \emph{dot product} of $p_1$ and $p_2$. If $C$ is an arbitrary subset of $\mathbb{N}$, then $(p_1 \wedge p_2)(C)$ and $(p_1 \cdot p_2)(C)$ are defined as

\beq (p_1 \wedge p_2)(C) = \sum_{A \cup B = C \; ; \; A \cap B = \emptyset} (-1)^{\sharp \{a>b \vert a \in A, b \in B \}} p_1 (A) p_2 (B) \eeq
and
\beq (p_1 \cdot p_2)(C) = \sum_{A \cup C = B \; ; \; A \cap C = \emptyset}  (-1)^{\sharp \{a>b \vert a \in A, b \in B \}} p_1 (A) p_2 (B) \eeq 

\subsection*{3.2.3 Key properties of wedge and dot products}

In the previous  section we formulated mathematically rigorous definitions of Grassmann numbers. Let us now use these definitions in order to write down a mathematically rigorous proof that the key properties that we assume while performing Grassmann integration are, in fact, true. The most obvious property that we would like to hold is associativity of wedge product:
\beq p_1 \wedge (p_2 \wedge p_3) = (p_1 \wedge p_2) \wedge p_3 \eeq
Apart from that, we would like to be able to "split" our integrals in the following way:
\beq \int (d_1 \eta_1 \wedge \cdots \wedge d_{n_1} \eta_{n_1} \wedge \cdots \wedge d_{n_2} \eta_{n_2}) \cdot f(\eta_1, \cdots, \eta_{n_2}) = \eeq
\beq = \int \Big( (d_1 \eta_1 \wedge \cdots \wedge d_{n_1} \eta_{n_1}) \cdot \int (d_{n_1+1} \eta_{n_1+1} \wedge \cdots \wedge d_{n_2} \eta_{n_2}) \cdot f(\eta_1, \cdots, \eta_{n_2}) \Big) \nonumber \eeq
In order to be able to do that, we have to assume that 
\beq (p_1 \wedge p_2) \cdot p_3 = p_1 \cdot (p_2 \cdot p_3) \eeq
Finally, we would also be able to pull "constants" under the integral:
\beq \int \Big( (d_1 \eta_1 \wedge \cdots \wedge d_{n_1} \eta_{n_1}) \cdot \int (d_{n_1 + 1} \eta_{n_1 +1} \wedge \cdots \wedge d_{n_2} \eta_{n_2}) \cdot (f(\eta_1, \cdots, \eta_{n_1}) \wedge g(\eta_{n_1 +1}, \cdots, \eta_{n_2})) \Big) = \nonumber \eeq
\beq =  \int \Big( (d_1 \eta_1 \wedge \cdots \wedge d_{n_1} \eta_{n_2}) \cdot \Big( f(\eta_1, \cdots, \eta_{n_1}) \wedge \int (d_{n_1 + 1} \eta_{n_1 +1} \wedge \cdots \wedge d_{n_2} \eta_{n_2}) \cdot g(\eta_{n_1 +1}, \cdots, \eta_{n_2}) \Big) \Big) \nonumber \eeq
In the spirit of what we have been saying, we should be able to "get rid of" the integral over $d_1 \eta_1 \wedge \cdots \wedge d_{n_1} \eta_{n_1}$ and retain the equality between Grassmann-valued outcomes of the integrals inside:
\beq \int (d_{n_1 + 1} \eta_{n_1 +1} \wedge \cdots \wedge d_{n_2} \eta_{n_2}) \cdot (f(\eta_1, \cdots, \eta_{n_1}) \wedge g(\eta_{n_1 +1}, \cdots, \eta_{n_2})) = \nonumber \eeq
\beq =  f(\eta_1, \cdots, \eta_{n_1}) \wedge \int (d_{n_1 + 1} \eta_{n_1 +1} \wedge \cdots \wedge d_{n_2} \eta_{n_2}) \cdot  g(\eta_{n_1 +1}, \cdots, \eta_{n_2})   \eeq
If we identify the product of differentials with $p_1$, $g$ with $p_2$ and $f$ with $p_3$, we see that the above requires an assumption that $p_3 \wedge (p_1 \cdot p_2) = p_1 \cdot (p_3 \wedge p_2)$. It turns out, however, that this does not work for $p_1 = \theta_1 \wedge \theta_2$, $p_2 = \theta_1$ and $p_3 = \theta_2$. After all, in this case $p_1 \cdot (p_3 \wedge p_2) = 1$ while $p_3 \wedge (p_1 \cdot p_2) =0$. This can be understood based on 
\beq \int (d_1 \eta \wedge d_2 \xi) \cdot (\xi \wedge \eta) = 1 \eeq
\beq \xi \wedge \int (d_1 \eta  \wedge d_2 \xi) \cdot \eta = \xi \wedge 0 = 0 \nonumber \eeq
That equation likewise does not hold when $p_1 = \theta_1$, $p_2 = \theta_1 \wedge \theta_2$ and $p_3 = \theta_3$. In this case, $p_1 \cdot (p_3 \wedge p_2) = + \theta_2 \wedge \theta_3$ but $p_3 \wedge (p_1 \cdot p_2) = - \theta_2 \wedge \theta_3$. This can be understood from
\beq \int d_1 \eta \cdot (\chi \wedge \eta \wedge \xi) = - \int d_1 \eta \cdot (\eta \wedge \chi \wedge \xi) = - \chi \wedge \xi = + \xi \wedge \chi \nonumber \eeq
\beq \chi \wedge \int d \eta \cdot (\eta \wedge \xi) = + \chi \wedge \xi = - \xi \wedge \chi \eeq
Nevertheless, $p_3 \wedge (p_1 \cdot p_2) = p_1 \cdot (p_3 \wedge p_2)$ \emph{does} hold if the following two conditions are met. First of all, $p_1$ and $p_3$ need to be \emph{disjoint}, which means that they do not share the same Grassmann factors. For example, $\eta \wedge \xi$ and $\xi \wedge \chi$ are \emph{not} disjoint because of common factor $\xi$; on the other hand, $\eta \wedge \xi$ and $\chi \wedge \lambda$ \emph{are} disjoint. The linear combination of "disjoint" and "not disjoint" is \emph{not} disjoint. For example, $\eta \xi$ and $\xi \wedge \chi + \chi \wedge \lambda$ are \emph{not} disjoint. Secondly, \emph{either} $p_1$ \emph{or} $p_3$ (or both) have to be "even". By even we mean that they contain even number of Grassmann factors. For example, $\eta \wedge \xi$ is even, $\eta \wedge \xi \wedge \chi$ is odd, while $\eta \wedge \xi + \eta \wedge \xi \wedge \chi$ is neither even nor odd.

Let us now give rigorous definition of these two properties. Let us start from "disjoint" property:

{\bf Definition 5:} Let $p_1$ and $p_2$ be two Grassmann polynomials. These two Grassmann polynomials are said to be \emph{disjoint} if for any sets $A$ and $B$ \emph{either} $p_1 (A) p_2 (B) =0$ holds, \emph{or} $A \cap B = \emptyset$ holds. 

Thus $p_1 = \theta_1 \wedge \theta_2$ and $p_3 = \theta_2$ are \emph{not} disjoint. After all, if we set $A= \{1, 2 \}$ and $B = \{2 \}$, then $p_1 (A) p_3 (B) = 1 \times 1 = 1 \neq 0$; at the same time $A \cap B = \{1 \} \neq \emptyset$. On the other hand, $p_2 = \theta_1$ and $p_3 = \theta_2$ \emph{are} disjoint. After all, the only $A$ and $B$ that satisfy $p_1 (A) p_2 (B) \neq 0$ are $A = \{1 \}$ and $B = \{2 \}$. For these specific $A$ and $B$, we have $A \cap B = \emptyset$. On the other hand, replacing them with any other $A'$ and $B'$ will lead to $p_1 (A') p_2 (B') =0$. Thus, one of the two conditions always hold, which means that $p_2= \theta_1$ and $p_3 = \theta_2$ are disjoint. 

{\bf Definition 6} Polynomial $p$ is considered \emph{even} if any set $A$ which satisfies $p(A) \neq 0$ has even number of elements. Likewise, it is considered \emph{odd} if any set $A$ which satisfies $p(A) \neq 0$ has odd number of elements.

Thus $p_3 = \theta_2$ is odd. After all, the only set $A$ that satisfies $p_3 (A) \neq 0$ is $A = \{2 \}$, which has $1$ element, and $1$ is odd number. On the other hand, $p_2 = \theta_1 \wedge \theta_2$ is even. After all, the only set $B$ that satisfies $p_2 (B) \neq 0$ is $B = \{1, 2 \}$. Thus, $B$ has exactly $2$ elements, and $2$ is an even number. Finally, a polynomial $p_4 = \theta_2 + \theta_1 \wedge \theta_2$ is neither even nor odd. After all, $p_4 (A) \neq 0$ and $p_4 (B) \neq 0$, where $A = \{2 \}$ and $B = \{1, 2 \}$. Since $A$ has odd number of elements, $p_4$ is \emph{not} even. On the other hand, since $B$ has even number of elements, $p_4$ is \emph{not} odd. Thus, $p_4$ is \emph{neither even nor odd}. 

To summarize our discussion so far, we would like to show that the following three properties hold:

{\bf Property 1:} $p_1 \wedge (p_2 \wedge p_3) = (p_1 \wedge p_2) \wedge p_3$

{\bf Property 2:} $(p_1 \wedge p_2) \cdot p_3 = p_1 \cdot (p_2 \cdot p_3)$

{\bf Property 3:} \emph{If} $p_1$ and $p_3$ are disjoint, and \emph{if} either $p_1$ is even, or $p_3$ is even, or both, \emph{then} $p_3 \wedge (p_1 \cdot p_2) = p_1 \cdot (p_3 \wedge p_2)$.

These three properties will be the subjects of Theorems 1, 2, and 3, respectively. We will devote the rest of this section to formally proving these theorems. 

{\bf Theorem 1:} Let $p_1$, $p_2$, and $p_3$ be Grassmann polynomials. Then $p_1 \wedge (p_2 \wedge p_3) = (p_1 \wedge p_2) \wedge p_3$

{\bf Proof:} Let $E$ be some arbitrary subset of $\mathbb{N}$. We will separately compute $(p_1 \wedge (p_2 \wedge p_3))(E)$ and $((p_1 \wedge p_2) \wedge p_3)(E)$. We will then compare the results and show that they are the same. Let us start from $(p_1 \wedge (p_2 \wedge p_3))(E)$. The expression for $(p_1 \wedge (p_2 \wedge p_3))(E)$ can be obtained by "copying" the expression for $(p_1 \wedge p_2)(E)$, while substitutting $p_2 \wedge p_3$ in place for $p_2$. This gives us 
\beq (p_1 \wedge (p_2 \wedge p_3))(E) = \sum_{C \cup D = E \; ; \; C \cap D = \emptyset} (-1)^{\sharp \{c>d \vert c \in C, d \in D \}} p_1 (C) (p_2 \wedge p_3) (D) \eeq
Now, if we replace $p_1$ with $p_2$ and $p_2$ with $p_3$, we can obtain the expression for $(p_2 \wedge p_3)(D)$. By substituting that expression into the above equation, we obtain
\beq (p_1 \wedge (p_2 \wedge p_3))(E) = \sum_{C \cup D = E \; ; \; C \cap D = \emptyset} \Big( (-1)^{\sharp \{c>d \vert c \in C, d \in D \}} p_1 (C) \times \nonumber \eeq
\beq \times \sum_{A \cup B = D \; ; \; A \cap B = \emptyset} (-1)^{\sharp \{a>b \vert a \in A, b \in B \}} p_2 (A) p_3 (B) \Big) \eeq
By moving all of the summation signs towards the left, we can rewrite it as 
\beq (p_1 \wedge (p_2 \wedge p_3))(E) = \sum_{A \cap B = C \cap D = \emptyset \; ; \; A \cup B = D \; ; \; C \cup D = E} p_1 (C) p_2 (A) p_3 (B) \times \nonumber \eeq
\beq \times (-1)^{\sharp \{c>d \vert c \in C, d \in D \} + \sharp \{a>b \vert a \in A, b \in B \}} \eeq
It will be convenient for us to rename the sets in such a way that the expression has the form $p_1 (A) p_2 (B) p_3 (C)$. Thus, we will replace $C$ and $c$ with $A$ and $a$, respectively; $A$ and $a$ with $B$ and $b$, respectively; and $B$ and $b$ with $C$ and $c$, respectively. Thus, we obtain
 \beq (p_1 \wedge (p_2 \wedge p_3))(E) = \sum_{B \cap C = A \cap D = \emptyset \; ; \; B \cup C = D \; ; \; A \cup D = E} p_1 (A) p_2 (B) p_3 (C) \times \nonumber \eeq
\beq \times (-1)^{\sharp \{a>d \vert a \in A, d \in D \} + \sharp \{b>c \vert b \in B, c \in C \}} \eeq
Let us now inspect the conditions under the sum to see if we can rewrite them in a more straightforward way. By substitutting $D= B \cup C$ into $A \cap D = \emptyset$, we find that $A \cap (B \cup C) = \emptyset$. This implies that $A \cup B = A \cup C = \emptyset$. By combining it with the condition under the sum $B \cap C = \emptyset$, we see that 
\beq A \cap B = A \cap C = B \cap C = \emptyset \eeq
Let us now express $D$ and $E$ in terms of $A$, $B$, and $C$. From the condition under the sum we read off 
\beq D  = B \cup C \eeq
Another condition under the sum tells us that $E = A \cup D$. By substitutting $B \cup C$ in place of $D$, we obtain
\beq E = A \cup B \cup C \eeq
It should be noted that $E$ is our original set, as our goal is to evaluate $(p_1 \wedge (p_2 \wedge p_3))(E)$. Thus, we can \emph{not} "get rid" of $E$. The sets $A$, $B$, and $C$ are "dummy indexes" so to speak (or "dummy sets" if you will) that are selected in such a way that $A \cup B \cup C = E$ always holds, for \emph{fixed}, afore-given, set $E$. On the other hand, the set $D$ \emph{can} be  freely replaced with $B \cup C$. Thus, we obtain
\beq (p_1 \wedge (p_2 \wedge p_3))(E) = \sum_{A \cap B = A \cap C = B \cap C = \emptyset; E=A \cup B \cup C}  p_1 (A) p_2 (B) p_3 (C) \times \nonumber \eeq
\beq \times (-1)^{\sharp \{a>d \vert a \in A; d \in B \cup C \} + \sharp \{b>c \vert b \in B; c \in C \}} \eeq
 By using the fact that $B \cap C = \emptyset$, we see that 
\beq \sharp \{a>d \vert a \in A ; d \in B \cup C \}= \sharp \{a>b \vert a \in A ; b \in B \} + \sharp \{a>c \vert a \in A ; c \in C \} \eeq
By substitutting this into the power of $-1$, and also by replacing the conditions under sum with $A \cap B = A \cap C = B \cap C = \emptyset$, our expression becomes
\beq (p_1 \wedge (p_2 \wedge p_3))(E) = \sum_{A \cap B = A \cap C = B \cap C = \emptyset; E=A \cup B \cup C}  p_1 (A) p_2 (B) p_3 (C) \times \nonumber \eeq
\beq \times (-1)^{\sharp \{a>b \vert a \in A; b \in B \} +  \sharp \{a>c \vert a \in A; c \in C \} + \sharp \{b>c \vert b \in B; c \in C \}} \eeq
Let us now evaluate the expression $((p_1 \wedge p_2) \wedge p_3) (E)$, and then compare the two expressions at the end. Again, we copy the definition of $(p_1 \wedge p_2)(E)$. This time, however, we put $p_1 \wedge p_2$ in place of $p_1$, and we put $p_3$ in place of $p_2$. Thus, we obtain 
\beq ((p_1 \wedge p_2) \wedge p_3)(E) = \sum_{C \cup D = E \; ; \; C \cap D = \emptyset} (-1)^{\sharp \{c>d \vert c \in C, d \in D \}} (p_1 \wedge p_2) (C) p_3 (D) \eeq
Now, by replacing $(p_1 \wedge p_2)(C)$ in favor of the corresponding explicit expression, we obtain
\beq ((p_1 \wedge p_2) \wedge p_3)(E) = \sum_{C \cup D = E \; ; \; C \cap D = \emptyset} \Big( (-1)^{\sharp \{c>d \vert c \in C, d \in D \}} p_3 (D) \times \nonumber \eeq
\beq \times \sum_{A \cup B = C ; A \cap B = \emptyset} (-1)^{\sharp \{a >b \vert a \in A, b \in B \} } p_1 (A) p_2 (B) \Big) \eeq
By moving all of the summation signs to the left, we obtain
\beq ((p_1 \wedge p_2) \wedge p_3)(E) = \sum_{C \cup D = E  ;  C \cap D = \emptyset; A \cup B = C ; A \cap B = \emptyset} \Big(   p_1 (A) p_2 (B) p_3 (D) \times \nonumber \eeq
\beq \times  (-1)^{\sharp \{a >b \vert a \in A, b \in B \} + \sharp \{c>d \vert c \in C, d \in D \}} \Big) \eeq
In order to put it in the form $p_1 (A) p_2 (B) p_3 (D)$ we change notation slightly by putting $C$ and $c$ in place of $D$ and $d$ respectively and likewise putting $D$ and $d$ in place of $C$ and $c$ respectively. Thus, we obtain
\beq ((p_1 \wedge p_2) \wedge p_3)(E) = \sum_{D \cup C = E  ;  D \cap C = \emptyset; A \cup B = D ; A \cap B = \emptyset} \Big(   p_1 (A) p_2 (B) p_3 (C) \times \nonumber \eeq
\beq \times  (-1)^{\sharp \{a >b \vert a \in A, b \in B \} + \sharp \{d>c \vert d \in D, c \in C \}} \Big) \eeq
Let us now simplify the conditions under the sum. By substitutting $A \cup B = D$ into the condition $D \cap C = \emptyset$, we obtain $(A \cup B) \cap C = \emptyset$. This implies that $A \cap C = B \cap C = \emptyset$. By combining it with the condition $A \cap B = \emptyset$, we obtain
\beq A \cap B = A \cap C = B \cap C = \emptyset \eeq
Furthermore, by substitutting $D = A \cup B$ into $E = D \cup C$, we obtain
\beq E = A \cup B \cup C \eeq
As before, we remind ourselves that $E$ is an apriori-given set, while $A$, $B$, and $C$ are so-called "dummy sets" we are "summing over", which are "constrained" to satisfy the above condition. Thus, replacing $E$ with $A \cup B \cup C$ is \emph{not} appropriate. However, we \emph{can} replace $D$ with $A \cup B$ since $D$ is also a "dummy set". Thus, we obtain
\beq ((p_1 \wedge p_2) \wedge p_3)(E) = \sum_{A \cap B = A \cap C = B \cap C = \emptyset; A \cup B \cup C =E} \Big( p_1 (A) p_2 (B) p_3 (C) \times \nonumber \eeq
\beq \times  (-1)^{\sharp \{a >b \vert a \in A, b \in B \} + \sharp \{d>c \vert d \in A \cup B, c \in C \}} \Big) \eeq
By using the fact that $A \cap B = \emptyset$, we have
\beq \sharp \{d>c \vert d \in A \cup B, c \in C \}= \sharp \{a>c \vert a A, c \in C \}+\sharp \{b>c \vert b \in B, c \in C \} \eeq
By substitutting this into the above expression we have 
\beq ((p_1 \wedge p_2) \wedge p_3)(E) = \sum_{A \cap B = A \cap C = B \cap C = \emptyset; A \cup B \cup C =E} \Big( p_1 (A) p_2 (B) p_3 (C) \times \nonumber \eeq
\beq \times  (-1)^{\sharp \{a >b \vert a \in A, b \in B \} + \sharp \{a>c \vert a \in A, c \in C \}+\sharp \{b>c \vert b \in B, c \in C \}} \Big) \eeq
This expression exactly coincides with the one for $(p_1 \wedge (p_2 \wedge p_3))(E)$. Since this argument works for the arbitrary $E$, we have shown that 
\beq p_1 \wedge (p_2 \wedge p_3) = (p_1 \wedge p_2) \wedge p_3 \eeq
This proves the statement of the theorem.  {\bf QED.}

{\bf Theorem 2:} Let $p_1$, $p_2$ and $p_3$ be arbitrary Grassmann polynomials. Then $(p_1 \wedge p_2) \cdot p_3 = p_1 \cdot (p_2 \cdot p_3)$. 

{\bf Proof:} We will do the same as we did in the previous proof. Namely, we will pick arbitrary finite set $E \subset \mathbb{N}$, separately evaluate $((p_1 \wedge p_2) \cdot p_3)(E)$ and $(p_1 \cdot (p_2 \cdot p_3))(E)$, and then show that the two expressions are, indeed, the same. Let us start from $((p_1 \wedge p_2) \cdot p_3)(E)$. By "copying" the expression for $(p_1 \cdot p_2)(E)$, and replacing $p_1$ with $p_1 \wedge p_2$ and $p_2$ with $p_3$, we obtain
\beq ((p_1 \wedge p_2) \cdot p_3)(E) = \sum_{C \cup E = D, C \cap E = \emptyset} (-1)^{\sharp \{c>d \vert c \in C ; d \in D \}} (p_1 \wedge p_2)(C) p_3 (D) \eeq
Now, by replacing $(p_1 \wedge p_2)(C)$ with a corresponding explicit expression, we obtain
\beq ((p_1 \wedge p_2) \cdot p_3)(E) = \sum_{C \cup E = D, C \cap E = \emptyset} \Big( (-1)^{\sharp \{c>d \vert c \in C ; d \in D \}} p_3 (D) \times \nonumber \eeq
\beq \times \sum_{A \cup B = C; A \cap B = \emptyset}  (-1)^{ \sharp \{a>b \vert a \in A ; b \in B \}} p_1 (A) p_2 (B) \Big) \eeq
By moving all of the summation signs towards the left, and combining everything the expressions under the sums, we obtain
\beq ((p_1 \wedge p_2) \cdot p_3)(E) = \sum_{A \cup B = C; C \cup E = D; A \cap B = C \cap E = \emptyset}\Big(  p_1 (A) p_2 (B) p_3 (D) \times \nonumber \eeq
\beq \times (-1)^{\sharp \{a>b \vert a \in A ; b \in B \} +\sharp \{c>d \vert c \in C ; d \in D \} } \Big) \eeq
For the purposes of the notational convenience, we would like to have expression in the form $p_1 (A) p_2 (B) p_3 (C)$. Thus, we "rearrange" $C$ with $D$ and, likewise, $c$ with $d$, without not touching $A$ or $B$. Thus, we obtain
\beq ((p_1 \wedge p_2) \cdot p_3)(E) = \sum_{A \cup B = C; D \cup E = C; A \cap B = D \cap E = \emptyset}\Big(  p_1 (A) p_2 (B) p_3 (C) \times \nonumber \eeq
\beq \times (-1)^{\sharp \{a>b \vert a \in A ; b \in B \} +\sharp \{d>c \vert d \in D ; c \in C \} } \Big) \eeq
Let us now try to simiplify the condition under the sum. If we substitute $D = A \cup B$ into $C = D \cup E$, we obtain
\beq A \cup B \cup E = C \eeq
If we substitute $D = A \cup B$ into $D \cap E = \emptyset$, we obtain $(A \cup B) \cap E = \emptyset$. This implies that $A \cap E = B \cap E = \emptyset$. If we now combine it with the condition $A \cap B = \emptyset$, we obtain
\beq A \cap B = A \cap E = B \cap E = \emptyset \eeq
Let us now evaluate the expression in the power of $-1$. By substitutting $C = A \cup B \cup C$ and $D = A \cup B$, this expression becomes 
\beq \sharp \{a>b \vert a \in A; b \in B \} + \sharp \{d>c \vert d \in D; c \in C \} = \eeq
\beq = \sharp \{a>b \vert a \in A; b \in B \} + \sharp \{d>c \vert d \in A \cup B; c \in A \cup B \cup E \}  \nonumber \eeq
By using the fact that $A$, $B$ and $E$ do not intersect, we can "split" $d \in A \cup B$ into $a \in A$ and $b \cup B$. Likewise, we can also "split" $c \in A \cup B \cup C$ into $a' \in A$, $b' \in B$ and $e' \in E$. Thus, $1+1=2$ terms will "split" into $1+3 \times 2=7$ terms. The expression becomes
\beq \sharp \{a>b \vert a \in A; b \in B \} + \sharp \{d>c \vert d \in D; c \in C \}  = \sharp \{a>b \vert a \in A; b \in B \} + \nonumber \eeq
\beq + \sharp \{a>a' \vert a \in A, a' \in A \}  + \sharp \{a>b' \vert a \in A, b' \in B \} + \sharp \{a>e' \vert a \in A, e' \in E \}   \eeq
\beq + \sharp \{b>a' \vert b \in B, a' \in A \} + \sharp \{b>b' \vert b \in B, b' \in B \} + \sharp \{b>e' \vert b \in B, e' \in E \} + \nonumber \eeq
Now, two of the terms above consist of pairs of elements from the same set. These two terms simplify according to 
\beq \sharp \{a>a' \vert a \in A, a' \in A \} = \frac{(\sharp A)(\sharp A - 1)}{2} \; ; \; \sharp \{b>b' \vert b \in B, b' \in B \} = \frac{(\sharp B)(\sharp B - 1)}{2} \eeq
Furthermore, if we add the term involving $a>b'$ together with the term involving $b>a'$, this becomes a sum over \emph{all} $a \in A$ and $b \in B$, which simplifiese according to 
\beq \sharp \{a>b' \vert a \in A, b' \in B \} + \sharp \{b>a' \vert b \in B, a' \in A \} = \sharp \{ a \in A, b \in B \} = (\sharp A)(\sharp B) \eeq
As a result, the power of $-1$ simplifies according to 
\beq \sharp \{a>b \vert a \in A; b \in B \} + \sharp \{d>c \vert d \in D; c \in C \}  = \frac{(\sharp A)(\sharp A - 1)}{2} +  \frac{(\sharp B)(\sharp B - 1)}{2} + (\sharp A)(\sharp B) + \nonumber \eeq
\beq + \sharp \{ a> b \vert a \in A, b \in B \} + \sharp \{a>e' \vert a \in A, e' \in E \} + \sharp \{b>e' \vert b \in B, e' \in E \} \eeq
We can now go back to the expression for $((p_1 \wedge p_2) \cdot p_3)(E)$. We now replace the conditions under sum with the "simplified" ones; namely, $A \cap B = A \cap E = B \cap E = \emptyset$, and $A \cup B \cup E = C$. Furthermore, we substitute the power of $-1$ that we have just derived. Thus, we have 
\beq ((p_1 \wedge p_2) \cdot p_3)(E) = \sum_{A \cap B = A \cap E = B \cap E = \emptyset; A \cup B \cup E = C }\Big(  p_1 (A) p_2 (B) p_3 (C) \times \nonumber \eeq
\beq \times (-1)^{ \frac{(\sharp A)(\sharp A - 1)}{2} +  \frac{(\sharp B)(\sharp B - 1)}{2} + (\sharp A)(\sharp B) + \sharp \{ a> b \vert a \in A, b \in B \} + \sharp \{a>e' \vert a \in A, e' \in E \} + \sharp \{b>e' \vert b \in B, e' \in E \} } \Big) \eeq
Finally, we can "get rid of $C$" by replacing it with $A \cup B \cup E$, which renders the condition $A \cup B \cup E = C$ under the sum unnecessary. Thus, we obtain 
\beq ((p_1 \wedge p_2) \cdot p_3)(E) = \sum_{A \cap B = A \cap E = B \cap E = \emptyset}\Big(  p_1 (A) p_2 (B) p_3 (A \cup B \cup E) \times \nonumber \eeq
\beq \times (-1)^{ \frac{(\sharp A)(\sharp A - 1)}{2} +  \frac{(\sharp B)(\sharp B - 1)}{2} + (\sharp A)(\sharp B) + \sharp \{ a> b \vert a \in A, b \in B \} + \sharp \{a>e' \vert a \in A, e' \in E \} + \sharp \{b>e' \vert b \in B, e' \in E \} } \Big) \eeq
Let us now evaluate $(p_1 \cdot (p_2 \cdot p_3))(E)$, and then compare results at the end. By taking the definition of $(p_1 \cdot p_2)$ and substitutting $p_2 \cdot p_3$ on place of $p_2$, we obtain
\beq (p_1 \cdot (p_2 \cdot p_3))(E)  = \sum_{C \cup E = D; C \cap E = \emptyset} (-1)^{\sharp \{c>d \vert c \in C, d \in D \}} p_1 (C) (p_2 \cdot p_3)(D) \eeq
Now, we replace $(p_2 \cdot p_3)(D)$ with corresponding explicit expression, and obtain
\beq (p_1 \cdot (p_2 \cdot p_3))(E)  = \sum_{C \cup E = D; C \cap E = \emptyset} \Big((-1)^{\sharp \{c>d \vert c \in C, d \in D \}} p_1 (C) \times \nonumber \eeq
\beq \times \sum_{A \cup D = B, A \cap D = \emptyset} (-1)^{\sharp \{ a>b \vert  a \in A, b \in B \}} p_2 (A) p_3 (B) \Big) \eeq
By pulling the sum signs towards the left, and combining the expressions under the sums, we obtain
\beq (p_1 \cdot (p_2 \cdot p_3))(E)  = \sum_{A \cup D = B; C \cup E = D; A \cap D= C \cap E = \emptyset} \Big( p_1 (C) p_2 (A) p_3 (B) \times \nonumber \eeq
\beq \times (-1)^{ \sharp \{ c>d \vert c \in C, d \in D \} + \sharp \{a>b \vert a \in A, b \in B \} } \Big) \eeq
For the notational convenience, we would like to rename the sets in such a way that we get $p_1 (A) p_2 (B) p_3 (C)$. Thus, we will replace $C$ with $A$, replace $A$ with $B$, and replace $B$ with $C$. Accordingly, we will also replace $c$ with $a$, $a$ with $b$, and $b$ with $c$. Thus, we get 
\beq (p_1 \cdot (p_2 \cdot p_3))(E)  = \sum_{B \cup D = C; A \cup E = D; B \cap D= A \cap E = \emptyset} \Big( p_1 (A) p_2 (B) p_3 (C) \times \nonumber \eeq
\beq \times (-1)^{ \sharp \{ a>d \vert a \in A, d \in D \} + \sharp \{b>c \vert b \in B, c \in C \} } \Big) \eeq
Let us now try to simplify the conditions under the sum. If we substitute $D = A \cup E$ into $C=B \cup D$, we obtain
\beq C= A \cup B \cup E \eeq
Also, by substitutting $D = A \cup E$ into $B \cap D = \emptyset$, we get $B \cap (A \cup E) = \emptyset$, which implies $B \cap A = B \cap E = \emptyset$. By combining it with another condition under the sum, $A \cap E = \emptyset$, we obtain
\beq A \cap B = A \cap E = B \cap E = \emptyset \eeq
Let us now evaluate the power of the $-1$. We will use the fact that $A \cup E = D$ and $A \cap E = \emptyset$ and "split" $\{a>d \vert a \in A, d \in D \}$ into $\{ a>a' \vert a \in A, a' \in A \}$ and $\{a>e \vert a \in A, e \in E \}$. We will also use the fact that $C=A \cup B \cup E$, together with $A \cap B = A \cap E = B \cap E = \emptyset$, and "split" $\{ b>c \vert b \in B, c \in C \}$ into $\{ b>c \vert b \in B, a \in A \}$, $\{ b>b' \vert b \in B, b' \in B \}$ and $\{ b>e \vert b \in B, e \in E \}$. Thus, instead of having $1 \times 1+1 \times 1 = 2$ terms, we have $1 \times 2 + 1 \times 3 = 5$ terms. Namely, 
\beq \sharp \{ a>d \vert a \in A, d \in D \} + \sharp \{b>c \vert b \in B, c \in C \} = \nonumber \eeq
\beq \sharp \{ a>a' \vert a \in A, a' \in A \} + \sharp \{ a>e \vert a \in A, e \in E \} + \eeq
\beq + \sharp \{b>a \vert b \in B, a \in A \} + \sharp \{b>b' \vert b \in B, b' \in B \} + \sharp \{b>e \vert b \in B, e \in E \} \nonumber \eeq
Now, two of the above terms involve pairs of elements from the same set. These can be easilly evaluated accordign to 
\beq \sharp \{ a>a' \vert a \in A, a' \in A \}  = \frac{(\sharp A)(\sharp A -1)}{2} \; ; \; \sharp \{ b>b' \vert b \in B, b' \in B \}  = \frac{(\sharp B)(\sharp B -1)}{2} \eeq
The term involving $b>a$ is a bit unusual since $b$ goes after $a$ in the alphabet. So, in order to be able to compare our expressions more easilly, we will rewrite the "$b>a$" sum in terms of "$a>b$" sum by using 
\beq \sharp \{b>a \vert b \in B, a \in A \} = (\sharp A)(\sharp B) -  \sharp \{a>b \vert a \in A, b \in B \} \eeq
By making these two substitutions, the expression becomes 
\beq \sharp \{ a>d \vert a \in A, d \in D \} + \sharp \{b>c \vert b \in B, c \in C \} = \nonumber \eeq
\beq \frac{(\sharp A)(\sharp A -1)}{2} + \frac{(\sharp B)(\sharp B -1)}{2} +  (\sharp A)(\sharp B) +  \eeq
\beq + \sharp \{ a>e \vert a \in A, e \in E \}  - \sharp \{a>b \vert a \in A, b \in B \}   + \sharp \{b>e \vert b \in B, e \in E \} \nonumber \eeq
We are now ready to go back to the expression for $(p_1 \cdot (p_2 \cdot p_3))(E)$. We have found out that the conditions under the sum simplify to $A \cap B = A \cap E = B \cap E = \emptyset$, $A \cup E = D$, and $A \cup B \cup E = C$. Furthermore, we can use the last two conditions to eliminate $C$ and $D$ in favor of $A \cup B \cup E$ and $A \cup E$, respectively. If we do that, we no longer need to write down these last two conditions since $C$ and $D$ is no longer present. Thus, the only condition we have to write down is $A \cap B = A \cap E = B \cap E = \emptyset$. Apart from that, we also substitute the expression we have just computted for the power of $-1$. This gives us
\beq (p_1 \cdot (p_2 \cdot p_3))(E)  = \sum_{A \cap B = A \cap E = B \cap E = \emptyset} \Big( p_1 (A) p_2 (B) p_3 (A \cup B \cup E) \times \nonumber \eeq
\beq \times (-1)^{\frac{(\sharp A)(\sharp A -1)}{2} + \frac{(\sharp B)(\sharp B -1)}{2}+ (\sharp A)(\sharp B) + \sharp \{ a>e \vert a \in A, e \in E \}  - \sharp \{a>b \vert a \in A, b \in B \}   + \sharp \{b>e \vert b \in B, e \in E \}} \Big) \eeq
This expression matches the expression for $((p_1 \wedge p_2) \cdot p_3)(E)$ almost exactly, except for one difference. The term $\sharp \{a>b \vert a \in A, b \in B \}$ in the power of $-1$ comes with the plus sign in case of $((p_1 \wedge p_2) \cdot p_3)(E)$, and it comes with the minus sign in case of $(p_1 \cdot (p_2 \cdot p_3))(E)$. However, since $\sharp \{a>b \vert a \in A, b \in B \}$ is integer, and it only appears in the power of $-1$, the result is independent of its sign:
\beq (-1)^{\sharp \{a>b \vert a \in A, b \in B \}} = (-1)^{-\sharp \{a>b \vert a \in A, b \in B \}} \eeq
This eliminates the one and only difference between the two expressions, thus proving that $((p_1 \wedge p_2) \cdot p_3)(E)= (p_1 \cdot (p_2 \cdot p_3))(E)$. Since this argument works for arbitrary set $E$, this implies that 
\beq (p_1 \wedge p_2) \cdot p_3 = p_1 \cdot (p_2 \cdot p_3) \eeq
This proves the statement of the theorem. {\bf QED.}

{\bf Theorem 3:} Suppose $p_1$, $p_2$ and $p_3$ are Grassmann polynomials. Assume that $p_1$ and $p_3$ are disjoint. Appart from that, assume that \emph{either} $p_1$ \emph{or} $p_3$ is even (or both). In this case, $p_3 \wedge (p_1 \cdot p_2) = p_1 \cdot (p_3 \wedge p_2)$. 

{\bf Proof:} Let us separately evaluate $(p_3 \wedge (p_1 \cdot p_2))(E)$ and $(p_1 \cdot (p_3 \wedge p_2))(E)$, and then show that the answers match, provided that $p_1$, $p_2$ and $p_3$ satisfy afore-stated conditions. Let us start with $(p_3 \wedge (p_1 \cdot p_2))(E)$. By writing down the expression for $p_1 \wedge p_2$ and then replacing $p_1$ with $p_3$ and $p_2$ with $p_1 \cdot p_2$, we obtain
\beq (p_3 \wedge (p_1 \cdot p_2))(E) = \sum_{C \cup D = E; C \cap D = \emptyset} (-1)^{\sharp \{c>d \vert c \in C, d \in D \}} p_3 (C) (p_1 \cdot p_2)(D) \eeq
By substitutting into $(p_1 \cdot p_2)(D)$ a corresponding explicit expression, we obtain
\beq (p_3 \wedge (p_1 \cdot p_2))(E) = \sum_{C \cup D = E; C \cap D = \emptyset} \Big( (-1)^{\sharp \{c>d \vert c \in C, d \in D \}} p_3 (C) \times \nonumber \eeq
\beq \times \sum_{A \cup D = B, A \cap D = \emptyset} (-1)^{\sharp \{a>b \vert a \in A, b \in B \}} p_1 (A) p_2 (B) \Big)\eeq
Moving the sumation signs to the left, and combining everything under the sumation, we obtain
\beq  (p_3 \wedge (p_1 \cdot p_2))(E) = \sum_{A \cup D = B, C \cup D = E; A \cap D = C \cap D = \emptyset} p_1 (A) p_2 (B) p_3 (C) \times \nonumber \eeq
\beq \times (-1)^{ \sharp \{c>d \vert c \in C, d \in D \} + \sharp \{a>b \vert a \in A, b \in B \}} \eeq
Let us now see the relationships between the sets that we have and attempt to rewrite it in a way that is as transparent as possible. Since $p_1$ and $p_3$ are disjoint, we know that the only terms that have non-zero contributions are the ones where $A \cap C = \emptyset$. Furthermore, one of the conditions of the sum tells us that $A \cap D = C \cap D = \emptyset$. We can now write them together as 
\beq A \cap C = A \cap D = C \cap D = \emptyset \eeq
We also have a conditions $A \cup D = B$ and $E= C \cup D$ under the sum. We can now replace $B$ with $A \cup D$ wherever $B$ arizes, rendering this condition unnecessary. However, we can \emph{not} get rid of $E = C \cup D$ in the similar way; after all, $E$ is a \emph{fixed} afore-given set, while all other sets are "dummy indexes" we are "summing over". Thus, $E = C \cup D$ is a specific restriction of our sum, \emph{as opposed to} merely a "notational issue" such as $A \cup D = B$. Thus, we arrive at the expression 
\beq  (p_3 \wedge (p_1 \cdot p_2))(E) = \sum_{A \cap C = A \cap D = C \cap D = \emptyset; E = C \cup D} p_1 (A) p_2 (A \cup D) p_3 (C) \times \nonumber \eeq
\beq \times (-1)^{ \sharp \{c>d \vert c \in C, d \in D \} + \sharp \{a>b \vert a \in A, b \in A \cup D \}} \eeq
We can now use the fact that $A \cap D = \emptyset$ and write 
\beq \sharp \{a>b \vert a \in A, b \in A \cup D \} = \sharp \{a>a' \vert a \in A, a' \in \cup A \} + \sharp \{a>d \vert a \in A, d \in D \} \eeq
By further realizing that the number of choices of  $a>a'$ is given by 
\beq \sharp \{a>a' \vert a \in A, a' \in \cup A \} = \frac{(\sharp A)(\sharp A - 1)}{2} \eeq
the above expression becomes 
\beq \sharp \{a>b \vert a \in A, b \in A \cup D \} = \frac{(\sharp A)(\sharp A - 1)}{2} + \sharp \{a>d \vert a \in A, d \in D \} \eeq
By substitutting it into the power of $-1$, we arrive at 
\beq  (p_3 \wedge (p_1 \cdot p_2))(E) = \sum_{A \cap C = A \cap D = C \cap D = \emptyset; E = C \cup D} p_1 (A) p_2 (A \cup D) p_3 (C) \times \nonumber \eeq
\beq \times (-1)^{ \sharp \{c>d \vert c \in C, d \in D \} + \sharp \{a>d \vert a \in A, d \in D \} +  \frac{(\sharp A)(\sharp A - 1)}{2} } \eeq
Let us now evaluate $(p_1 \cdot (p_3 \wedge p_2))(E)$. By "copying" the expression for $(p_1 \cdot p_2)(E)$ while replacing $p_2$ with $p_3 \wedge p_2$, we obtain
\beq (p_1 \cdot (p_3 \wedge p_2))(E) = \sum_{C \cup E = D; C \cap E = \emptyset} (-1)^{ \sharp \{c>d \vert c \in C, d \in D \} } p_1 (C) (p_3 \wedge p_2)(D) \eeq
By replacing $(p_3 \wedge p_2)(D)$ with corresponding explicit expression, we obtain
\beq (p_1 \cdot (p_3 \wedge p_2))(E) = \sum_{C \cup E = D; C \cap E = \emptyset} \Big( (-1)^{ \sharp \{c>d \vert c \in C, d \in D \} } p_1 (C) \times \nonumber \eeq
\beq \times \sum_{A \cup B = D; A \cap B = \emptyset} (-1)^{\sharp \{ a>b \vert a \in A, b \in B \} } p_3 (A) p_2 (B) \Big) \eeq
By moving all of the summation signs to the left, and combining the expressions under the sums, we obtain
\beq (p_1 \cdot (p_3 \wedge p_2))(E) = \sum_{A \cup B = C \cup E = D; A \cap B =C \cap E = \emptyset} p_1 (C) p_3 (A) p_2 (B) \times \nonumber \eeq
\beq \times (-1)^{\sharp \{ c>d \vert c \in C, d \in D \} + \sharp \{a>b \vert a \in A, b \in B \}} \eeq
Now, for the purposes of notational convenience, we would like to re-name sets in such a way that we get an expression of the form $p_1 (A) p_2 (B) p_3 (C)$. This can be accomplished by interchanging $A$ and $C$, while leaving $B$ untouched. Likewise, we also interchange $a$ and $c$, without moving $b$. Thus, we have 
\beq (p_1 \cdot (p_3 \wedge p_2))(E) = \sum_{C \cup B = A \cup E = D; C \cap B =A \cap E = \emptyset} p_1 (A)  p_2 (B) p_3 (C) \times \nonumber \eeq
\beq \times (-1)^{\sharp \{ a>d \vert a \in A, d \in D \} + \sharp \{c>b \vert c \in C, b \in B \}} \eeq
Let us now see the relationship between the sets involved. Since $A \cup E = D$ and also $C \cup B = D$, we immediately know that $A \cup E = C \cup B$; in fact we can eliminate $D$ in favor of one of these two quantities. Now, one of the conditions of the sum is $A \cap E = \emptyset$. If we combine it with $A \cup E = C \cup B$, we get that $E = (C \cup B) \setminus A$. Now, since $p_1$ and $p_3$ are disjoint, in order for a given term in the sum to be non-zero we need to have $A \cap C = \emptyset$. This means that we can rewrite $E = (C \cup B) \setminus A$ as $E= C \cup (B \setminus A)$. Now, let us identify $B \setminus A$ with $F$:
\beq F=^{\rm def} B \setminus A \eeq
In this case, the expression $E = C \cup (B \setminus A)$ can be rewritten as 
\beq E = C \cup F \eeq
Now, one of the conditions of the sum is that $C \cap B = \emptyset$. But, since $p_1$ and $p_3$ are disjoint, we can also assume that $A \cap C = \emptyset$. Thus, $C \cap B = \emptyset$ is equivalent to $C \cap (B \setminus A) = \emptyset$. But, by definition, we have denoted $B \setminus A$ by $F$. Thus, we can rewrite it as $C \cap F = \emptyset$. Finally, the fact that $F= B \setminus A$ implies that $A \cap F = \emptyset$. We can now combine our findings as
\beq A \cap C = A \cap F = C \cap F = \emptyset \eeq
Now, if we again look at the conditions $A \cup E = D$ and $C \cup B = D$ under sum, we get $A \cup E = C \cup B$. Furthermore, we have recently found out that $E = C \cup F$. Substitutting the latter into the former, we obtain $A \cup (C \cup F) = C \cup B$. We can rewrite it as $C \cup (A \cup F) = C \cup B$. But, from the fact that $C \cap A = C \cap F = \emptyset$, we also know that $C \cap (A \cup F) = \emptyset$. By combining it with $C \cup (A \cup F) = C \cup B$, we get $A \cup F = (C \cup B) \setminus C$. This further evaluastes to $A \cup F = B \setminus C$. Now, one of the conditions under the sum is $C \cap B = \emptyset$. Thus $A \cup F = B \setminus C$ becomes 
\beq B = A \cup F \eeq
By substitutting $B = A \cup F$ into $D= C \cup B$, we obtain
\beq D = A \cup C \cup F \eeq
We  can, therefore, replace $B$ and $C$ under the sum in favor of $A \cup F$ and $A \cup C \cup F$, respectively. Thus, we no longer have to explicitly write these conditions under the sum, since the letters $B$ and $C$ no longer exist. At the same time, we \emph{do} still have to write $E = C \cup F$. After all, $E$ is a \emph{fixed} set we are supposed to have \emph{apriori} knowledge about; this means we can \emph{not} replace it in favor of any other set. Thus, we can rewrite the sum as 
\beq (p_1 \cdot (p_3 \wedge p_2))(E) = \sum_{A \cap C = A \cap F = C \cap F = \emptyset; C \cup F = E} \Big( p_1 (A)  p_2 (A \cup F) p_3 (C) \times \nonumber \eeq
\beq \times (-1)^{\sharp \{ a>d \vert a \in A, d \in A \cup C \cup F \} + \sharp \{c>b \vert c \in C, b \in A \cup F \}} \Big) \label{eqn:207} \eeq
Let us now evaluate the power of $-1$. By using $A \cap C = A \cap F = C \cap F = \emptyset$, we can split the first term into the sum of $1 \times 3$ terms, and second term into the sum of $1 \times 2$ terms, thus having a total of $1 \times 3 + 1 \times 2 = 5$ terms:
\beq \sharp \{ a>d \vert a \in A, d \in A \cup C \cup F \} + \sharp \{c>b \vert c \in C, b \in A \cup F \} = \nonumber \eeq
\beq = \sharp \{ a>a' \vert a \in A, a' \in A \} + \sharp \{ a>c \vert a \in A, c \in  C  \} + \sharp \{ a>f \vert a \in A, f \in F \} + \eeq
\beq + \sharp \{c>a \vert c \in C, a \in A  \} + \sharp \{c>f \vert c \in C, f \in  F \}  \nonumber \eeq
The first term in the right hand side of the above expression evaluates to 
\beq \sharp \{ a>a' \vert a \in A, a' \in A \} =  \frac{(\sharp A)(\sharp A -1)}{2} \eeq
Appart from that, we can combine $a>c$ and $ca$ terms into
\beq \sharp \{a>c \vert a \in A, c \in C \} + \sharp \{c>a \vert a \in A, c \in C \} = \sharp \{a \in A, c \in C \} = (\sharp A)(\sharp C) \eeq
Thus, the expression in the power of $-1$ becomes 
 $1 \times 3$ terms, and second term into the sum of $1 \times 2$ terms, thus having a total of $1 \times 3 + 1 \times 2 = 5$ terms:
\beq \sharp \{ a>d \vert a \in A, d \in A \cup C \cup F \} + \sharp \{c>b \vert c \in C, b \in A \cup F \} = \eeq
\beq = \frac{(\sharp A)(\sharp A -1)}{2}  + (\sharp A)(\sharp C) + \sharp \{ a>f \vert a \in A, f \in F \}  + \sharp \{c>f \vert c \in C, f \in  F \}  \nonumber \eeq
By substituting this into the the equation (\ref{eqn:207}), we get 
\beq (p_1 \cdot (p_3 \wedge p_2))(E) = \sum_{A \cap C = A \cap F = C \cap F = \emptyset; C \cup F = E} \Big( p_1 (A)  p_2 (A \cup F) p_3 (C) \times \nonumber \eeq
\beq \times (-1)^{\frac{(\sharp A)(\sharp A -1)}{2}  + (\sharp A)(\sharp C) + \sharp \{ a>f \vert a \in A, f \in F \}  + \sharp \{c>f \vert c \in C, f \in  F \} } \Big) \eeq
Let us now compare the expression we just derived for $(p_1 \cdot (p_3 \wedge p_2))(E)$ with the expression for $(p_3 \wedge (p_1 \cdot p_2))(E)$ derived earlier. It is easy to see that $F$ in the final expression for $(p_1 \cdot (p_3 \wedge p_2))(E)$ plays the same role as $D$ in the final expression for $(p_3 \wedge (p_1 \cdot p_2))(E)$. Once we change the notation accordingly, we see that the only difference between these two expressions is the factor $(-1)^{(\sharp A)(\sharp C)}$. But, if \emph{either} $p_1$ \emph{or} $p_3$ is even, then \emph{either} $\sharp A$ \emph{or} $\sharp C$ will be even, accordingly. In either case, we will get
\beq (-1)^{(\sharp A)(\sharp C)} = 1 \eeq
which will imply that $(p_1 \cdot (p_3 \wedge p_2))(E)$ and $(p_3 \wedge (p_1 \cdot p_2))(E)$ will match, as claimed. {\bf QED.}

\subsection*{3.2.4 Sequential process that generates fermionic path integral}

In the previous section we have developed a way of "splitting" the integral into several parts. This is crucial for our purposes. After all, the fermionic propagator is normally written down as an integral over all space \emph{and} all time. On the other hand, our agenda is to postulate a process that "develops in time". Thus, we simply don't have our "future" available to us in order to do what the "official definition" prescribes. At the same time, we can't be taking "wedge products" while putting "dot product" on hold, either; if the nature were to put dot product on hold, we would have never learned about its consequences since we would still be "waiting". The only logical thing to to is alternate between dot products and wedge products. Namely, "take the integral" over each slice as we go along. The fact that we would still get results consistent with the "official" definition of path integral is a consequence of the theorems proven in the previous section. 

 We will now "split" this integral in the following way. First, we exponentiate the "part" of action relevent to time slices $t=0$ and $t= \delta t$. That part of the action has some terms that couple one slice to the other, particularly from the "time derivative" part of Dirac equation. Then we integrate it over the slice $t=0$ (and it is understood that "integrating" is synonymous with "taking dot product"). The outcome of the integration might be Grassmannian. Since we integrated over \emph{all} fermionic degrees of freedom at $t=0$, we can't possibly have any of the Grassmann variables from $t=0$ left. However, we \emph{can} have \emph{some} Grassmann variables comming from $t = \delta t$.

 After that, we again do a wedge product. In particular, we take the outcome of the above integration and then multiply it \emph{on the outside} by the \emph{new} wedge product \emph{which we have not integrated yet}. That \emph{new} product is an exponent of the part of the action comming from slices $t= \delta t$ and $t = 2 \delta t$.  The integral itself, as we stated earlier, has Grassmann parameters comming from $\delta t$ alone. The exponent we multiply it by from outside has a coupling beteen $t= \delta t$ and $t= 2 \delta t$ due to the derivative term in Dirac equation. Thus, the product between these two terms will also be a coupling between $t= \delta t$ and $t = 2 \delta t$. Then we integrate over $t= \delta t$ (that is, we again take "dot product"). Thus, we no longer have any of the Grassmann parameters comming from $t = \delta t$, and the \emph{only} thing we have is $t= 2 \delta t$. Then we again "wedge" it from outside and get a coupling between $t= 2 \delta t$ and $t = 3 \delta t$; and we keep going in similar fashion. 

From what we have just said, \emph{right after} the integration, we only have to face the Grassmann variables comming from \emph{one} slice. In particular, if we have integrated over a slice $t = t_0 - \delta t$, we will be "left" with Grassmann variables comming from the slice $t_0 $. On the other hand, during the \emph{process} of exponentiating, we are dealing with Grassmann variables from \emph{two} slices. For example, \emph{after} we integrated over slice $t_0 - \delta t$, the subsequent exponentiation will involve slices $t_0$ and $t_0 + \delta t$. Thus, we will have Grassmann variables from \emph{both} of them, \emph{until} we "clear them up" during the \emph{subsequent} integration over the slice $t=t_0$, \emph{which we have not performed yet} while we are still "in the process" of "taking exponent". The exact picture behind these vague statements will be seen in Theorem 4 and its proof.

Be it as it may, the fact remains that \emph{two} time slices is by far "simpler" than arbitrary many time slices present in the definition of fermionic path integral. We would like, therefore, to attempt to model a two-slice process that can be shown to represent the mathematical information produced by multiple-slice one. Our trick will be the following. After we have performed the integration over the slice $t=0$, we are left with Grassmann variables on the slice $t = \delta t$. Since the slice $t=0$ is "clear" (and, therefore, "useless"), we might as well "upload" the information from the slice $t= \delta t$ onto the slice $t=0$. After we done that uploading, we no longer need to retain the information on slice $t= \delta t$. So we get rid of it. Since the slice $t= \delta t$ is now empty, we can use \emph{this} slice to "play a role" of a slice $t= 2 \delta t$. Then after the slice $t = \delta t$ has generated the information that \emph{should have been} on a slice $t= 2 \delta t$, we again "upload it" onto the slice $t=0$ and use the slice $t= \delta t$ to generate information that "should have been" generated by slice $t = 3 \delta t$, and so forth. This means that we don't need the slices $t= 2 \delta t$, $t= 3 \delta t$, and so forth, to be present to begin with! Our entire scenario is based \emph{only} on slices $t=0$ and $t= \delta t$.

Even though this picture looks intuitive, one can be reasonably skeptical about it. After all, using the same variables several times might create some unforeseen errors; and this problem only magnifies in case of Grassmann variables. Therefore, it is necessary to work this out in a mathematically rigorous way. In fact, this was one of the main motivations in devoting the previous two sections to mathematically rigorous definitions and proofs.  We are now ready to apply that machinery in order to state and prove the theorem that the above algorithm works. This will be the subject of Theorem 4. Before we turn to that theorem, let us first write down some definitions. The "uploading" that we have just described can be defined through the following two definitions: 

{\bf Definition 7:} Let $A \subset \mathbb{C}$ be a set and $c \in \mathbb{C}$ be an arbitrary complex number. Then $A+c$ is another subset of $\mathbb{C}$ which is defined as 
\beq A + c = \{a \vert a-c \in A \} \eeq

{\bf Definition 8:} Let $p$ be a Grassmann polynomial, and let $n$ be an integer, which is either positive or negative. Then \emph{$n$-th upload} of $p$ is a polynomial $u_n (p)$, defined by 
\beq (u_n (p))(A) = p (A +n) \eeq
where $A$ is an arbitrary subset of $\mathbb{N}$.

{\bf Note:} In the above definition, we allow $n$ to be both positive \emph{and} negative; likewise, we can have $\theta_{-1}$, $\theta_{-2}$ and so forth \emph{along with} $\theta_1$, $\theta_2$, etc. This is needed in order to make sure that "uploading" is \emph{always} well defined. We do not need to use negative indexes for our purpose; but it is important to make sure that all definitions are correct in order to prevent unforeseable errors. 

The other necessary part of the upcomming theorem is exponentiation (which is not surprising, since that is what we claim to "reproduce" through our step-by-step process). In light of the fact that we have two different products, it is important to officially state that exponent is based on the wedge product \emph{as opposed to} dot product. Since the definition of exponent is based on power series, our goal can be accomplished by officially stating that a \emph{power} of a Grassmann polynomial is defined strictly in terms of wedge product: 

{\bf Definition 9:} Let $p$ be Grassmann polynomial. Then the positive integer powers of $p$ are defined based by induction, according to 
\beq p^0 = \underline{1} \; ; \; p^{n+1} = p \wedge p^n \eeq

Once we have defined what we mean by $p^n$, one would expect that the definition of exponent will "go through". This, however, might be a little suddle, in light of the fact that we haven't defined limits in Grassmann space. In principle it won't be too much work to do it. But, in order not to digress, let us choose simpler way of dealing with it. Namely, we will \emph{first} act with $p^n$ on a set $A$ and then "take the sum" \emph{afterwords}. This means that we will only have to take sum of scalars, which we know how to do. Thus, we define exponent as follows:

{\bf Definition 10:} Let $p$ be Grassmann polynomial. Then the \emph{exponent} of $p$ is another Grassmann polynomial defined according to 
\beq (\exp p)(A) = \sum_{n=0}^{\infty} \frac{(p^n)(A)}{n!} \eeq
where $A$ is an arbitrary subset of $\mathbb{Z}$. 

Finally, we have to officially define what me mean by a statement "we are only looking at two slices". In order to make the statement to this effect, we will say "a polynomial is living within a set $A$", and then define set $A$ in such a way that it corresponds to the "two slices" we are talking about. Vaguely speaking, a polynomial "living within $A$" contains only $\theta$-s whose indexes are elements of $A$. The machinery developed in previous two sections allows us to say it in a more rigorous way, as follows: 

{\bf Definition 11:} Let $p$ be a Grassmann polynomial and let $A \subset \mathbb{Z}$ be some subset of integers. The polynomial $p$ is said to be \emph{living inside $A$} if $p (B) = 0$ whenever $B \setminus A \neq \emptyset$. 

We are now ready to proceed to the goal of this section. In particular, we will outline the proposed process and then prove that this will, in fact, reproduce the path integral. As we have seen in the last two sections, "integral" is synonymous with "taking a dot product". Therefore, we are interested in producing the expression in the form
\beq (\theta_1 \wedge \cdots ) \cdot \exp (\cdots) \nonumber \eeq
The $\theta$-s in the above expression come from \emph{many different} slices, and our challenge is to reproduce it through two-slice process. For the purposes of notational convenience, we will state that the "small part" of action $K_k$ we are "wedging" at a particular time might come from a slices "far away" from $t=0$. \emph{But} in our "process" we will use the \emph{upload} of $K_k$, instead of $K_k$ itself.  Now, we "enumerate" $K_k$ in the order in which we "look" at them. Thus, we first look at $K_1$, then $K_2$, and so forth. We will further assume that, at each "slice" we look at $n$ different $K$-s. Thus, $K_k$ belongs to the \emph{pair} of slices $c+1$ and $c+2$ if $cn \leq k < c(n+1)$. 

Now, since $K_k$ is wedge product of \emph{more than one} $\theta$, the number of $K$-s we "go through" in each slice is \emph{much greater} than the number of $\theta_s$ each slice contains. For example, one of the $K$-s can be $1+ a\theta_1 \wedge \theta_2$ while the other might be $1+ b\theta_1 \wedge \theta_3$. Going through both will \emph{not} give zero simply because in each case we "take exponent". Thus, we might be getting 
\beq (1+ a \theta_1 \wedge \theta_2) \wedge (1 + b \theta_1 \wedge \theta_3) = 1 + a \theta_1 \wedge \theta_2 + b \theta_1 \wedge \theta_3 \eeq
In light of this, if the number of $K$-s is $n$, then the relevent number of $\theta$-s is $m \ll n$. Now, if we want to "upload" slices $c+1$ and $c+2$ onto the slices $1$ and $2$, we need to go "$c$ slices back". Since each slice has $m$ elements, we need to go "$cm$ elements back". This means tha we have to use $u_{-cm} (K_k)$ whenever $cn \leq k < c(n+1)-1$. We were using $m$ in the former expression and $n$ in the latter because we were "counting" $\theta$-s and their products, respectively. This means that when we are taking wedge product with corresponding $K_k$, we are explicitly doing
\beq p_{k+1} = \exp (-iu_{-cm}(K_k))] \wedge p_k \eeq
This means that, if $K_k$ lives inside $A \subset \mathbb{N}$, then the things we are "wedging" live inside $A-cm$. 

Now, since we are coupling \emph{two} slices, according to our definition $K_k$ "lives inside" $\{cn+1, \cdots, c (n+2) \}$. This can be seen by substitutting $n=0$. In this case, $cn \leq K < c(n+1)$ becomes $0 \leq K < c-1$, which implies that we are "enumerating" $K$-s comming from the first slice. But we know that we couple two consequetive slices. Since we don't have slice number $0$, we \emph{know} that slice number $1$ is coupled to slice number $2$. In other words, we \emph{know} that $K$ lives inside $\{0, \cdots, 2c-1 \}$. Now, if we set $m=0$, then $1=cm+1$ and $2c=c(n+2)$. Thus, we can generalize it by saying that $K$ lives within $\{cm+1, \cdots, c (m+2) \}$. Now, this implies that $u_{-cm} (K_k)$ lives inside $\{cm+1, \cdots, c (m+2) \} - cm = \{1, \cdots, 2c \}$. This means that we are only taking wedge product with the elements of \emph{first two} slices, which is what we want.

Finally, during the time we are "integrating", we take the integral \emph{over $t=0$ slice} each time. The reason for this is that wedge product, which involves $u_{-cm} (K_k)$, is \emph{already} living inside $t=0$ and $t= \delta t$ slices, so we don't need to make any further uploadings to "bring it" there. From what we have seen in the previous sections, the integration amounts to a "dot product". Thus, we take the "dot product" with slice $t=0$. That slice is represented by $\theta_1 \wedge \cdots \wedge \theta_m$, and our dot product is represented by 
\beq p_k \rightarrow (\theta_1 \wedge \cdots \wedge \theta_m) \cdot p_k \eeq
However, \emph{after} we have taken the integral, we have to made \emph{one more} uploading, and upload the slice $t= \delta t$ onto the slice $t=0$. After all, we need to "clear" slice $t= \delta t$ for our next set of wedge-ing. We do it through
\beq p_k \rightarrow u_{-c} (p_k) \eeq
The two steps above, taken in sequence, really take place during a \emph{single step} of going from $p_{c(m+1)-1}$ to $p_{c(m+1)}$. Thus, it can be summarized as summarized as 
\beq u_{-c}((\theta_1 \wedge \cdots \wedge \theta_m) \cdot p_k) \eeq
Now, the "dot product" takes place when we "go" from $k=cn-1$ to $k=cn$. On the other hand, the "wedge product", 
\beq p_{k+1} = \exp (-iu_{-cm}(K_k))] \wedge p_k \eeq
takes place for all \emph{other} values of $k$. This means that the two very different processes can be combined by a "case by case" inductive statement as follows: 
\beq p_{k+1} = 
\left\{
	\begin{array}{ll}
		[\exp (-iu_{-cm}(K_k))] \wedge p_k \; , \; cn \leq k < c (n+1)-1 \\
		u_{-c}((\theta_1 \wedge \cdots \wedge \theta_m) \cdot p_k) \; , \; k=cn-1
	\end{array}
\right.
\eeq
This implies that if we indeed succeed in producing exponent of the "sum" of $K_k$, that "sum" would "skip" the values $K_k$ corresponding to $K=cn-1$. In order to make it a sum over \emph{all} $K_k$, we would like to set the "skipped" values to zero:  
\beq K_{cn-1} = 0 \eeq
Let us now state the theorem. We will outline the exact process we have in mind in the statement of the theorem, and then prove why it works:

{\bf Theorem 4:} Suppose $\{K_1, K_2, \cdots \}$ is a sequence of Grassmann polynomials. Suppose that $m$ and $n$ are positive integers.  Furthermore, suppose that, for any $k$ satisfying $an \leq k < c(n+1)$, $K_k$ lives in $\{cm+1, \cdots, c(m+2) \}$. Let $\{p_1, p_2, \cdots \}$ be a sequence of  Grassmann polynomials defined by induction: $p_1 = \underline{1}$ and 
\beq p_{k+1} = 
\left\{
	\begin{array}{ll}
		[\exp (-iu_{-cm}(K_k))] \wedge p_k \; , \; cn \leq k < c (n+1)-1 \\
		u_{-c}((\theta_1 \wedge \cdots \wedge \theta_m) \cdot p_k) \; , \; k=cn-1
	\end{array}
\right.
\label{eqn:226} \eeq
Suppose that $k= an+b$ for some $a \in \mathbb{Z}$ and $b \in \mathbb{N}$ satisfying $0 \leq b <n$. Furthermore, suppose that all of $K_k$ are even. Finally, suppose that $K_{cn-1}= \underline{0}$ for all integer $c$. Then 
\beq p_k = u_{-cm} \Big( (\theta_1 \wedge \cdots \wedge \theta_{cm}) \cdot \exp \Big(-i \sum_{j=0}^{k-1} K_j \Big)\Big) \label{eqn:227}\eeq

{\bf Proof:} The above condition is clearly satisfied for $k=1$. Now suppose it is satisfied for some $k$. We would like to see whether it is also satisfied for $k+1$. We will split it into two cases: in caes $1$ we will assume that $k+1$ is \emph{not} divisible by $n$, and in case number $2$ we will assume that it is. Let us start from case $1$. If $k+1$ is not divisible by $n$, then we can find $c$ such that 
\beq cn \leq k < c(n+1) - 1 \eeq
Based on the inductive definition of $p_k$ postulated in the statement of the theorem, this implies that 
\beq p_{k+1} =  [\exp (-iu_{-cm}(K_k))] \wedge p_k \eeq
Now, since we assume that the statement in question is true for $p_k$, we know that 
\beq p_k = u_{-cm} \Big( (\theta_1 \wedge \cdots \wedge \theta_{cm}) \cdot \exp \Big(-i \sum_{j=0}^{k-1} K_j \Big)\Big)\eeq
This immediately tells us that 
\beq p_{k+1} =  [\exp (-iu_{-cm}(K_k))] \wedge u_{-cm} \Big( (\theta_1 \wedge \cdots \wedge \theta_{cm}) \cdot \exp \Big(-i \sum_{j=0}^{k-1} K_j \Big)\Big) \eeq
It can easilly shown that for any two polynomials $q_1$ and $q_2$, and any integer $a$, 
\beq u_a (q_1) \wedge u_a (q_2) = u_a (q_1 \wedge q_2) \eeq
It can furthermore be shown that, for any integer $a$, and any polynomial $q$, 
\beq \exp(-iu_a (q)) = u_a (\exp(-iu_a (q))) \eeq
Based on these two facts, we can rewrite $p_{k+1}$ as 
\beq p_{k+1} =  u_{-cm} \Big[ [\exp (-i K_k)] \wedge \Big( (\theta_1 \wedge \cdots \wedge \theta_{cm}) \cdot \exp \Big(-i \sum_{j=0}^{k-1} K_j \Big) \Big) \Big] \label{eqn:234}\eeq
Now, in the statement of the theorem it was assumed that $K_k$ is living in $\{cm+1, \cdots, c(m+1) \}$. It can easy to show that this implies that $\exp (-iK_k)$ is living in $\{cm+1, \cdots, c(m+2) \}$ as well. This, in turn, implies that $\exp (-iK_k)$ and $\theta_1 \wedge \cdots \wedge \theta_{cm}$ are disjoint. Furthermore, in the statement of the theorem it was also assumed that all of $K_j$ are even. This  implies that $exp (\cdots)$ is even. Thus, both of the "conditions" of Theorem 3 are met. Therefore, Theorem 3 implies that 
\beq [\exp (-i K_k)] \wedge \Big( (\theta_1 \wedge \cdots \wedge \theta_{cm}) \cdot \exp \Big(-i \sum_{j=0}^{k-1} K_j \Big) \Big) = \eeq
\beq = (\theta_1 \wedge \cdots \wedge \theta_{cm}) \cdot \Big[[\exp (-i K_k)] \wedge \exp \Big(-i \sum_{j=0}^{k-1} K_j \Big) \Big] \nonumber \eeq 
Again, by using the assumption that all of $K_j$ are even, we know that they commute. The fact that they commute allows us to say that the product of exponents is equal to the exponent of the sum: 
\beq [{\rm all \; K_i \; are \; even}] \Longrightarrow \Big[ [\exp (-i K_k)] \wedge \exp \Big(-i \sum_{j=0}^{j-1} K_k \Big) = \exp \Big(-i \sum_{j=0}^k K_j \Big) \Big] \eeq
This, upon substitution into the previous equation, produces 
\beq [\exp (-i K_k)] \wedge \Big( (\theta_1 \wedge \cdots \wedge \theta_{cm}) \cdot \exp \Big(-i \sum_{j=0}^{k-1} K_j \Big) \Big) = (\theta_1 \wedge \cdots \wedge \theta_{cm}) \cdot \exp \Big(-i \sum_{j=0}^k K_j \Big) \eeq 
Finally, if we the above into the equation (\ref{eqn:234}), we obtain 
\beq p_{k+1} =  u_{-cm} \Big[ (\theta_1 \wedge \cdots \wedge \theta_{cm}) \cdot \exp \Big(-i \sum_{j=0}^k K_j \Big) \Big] \eeq
Thus, we have just shown that the induction step goes through when $k+1$ is \emph{not} divisible by $n$. Let us now consider the other case, when it is. Thus, we will now assume that $k+1 =n(c+1)$ (we used $c+1$ instead of $c$ simply to make the notation a little bit cleaner in the upcomming argument). The inductive definition of $p_k$ given in the statement of the theorem, therefore, implies that 
\beq p_{k+1} =u_{-c}( (\theta_1 \wedge \cdots \wedge \theta_m) \cdot  p_k) \eeq
Now, since we are assumming that the induction hypothesis is true for $k$, we know that 
\beq p_k = u_{-cm} \Big( (\theta_1 \wedge \cdots \wedge \theta_{cm}) \cdot \exp \Big(-i \sum_{j=0}^{k-1} K_j \Big)\Big)\eeq
By substitutting this into the previous equation, we obtain 
\beq p_{k+1} =  u_{-c} \Big[ (\theta_1 \wedge \cdots \wedge \theta_m)  \cdot u_{-cm} \Big( (\theta_1 \wedge \cdots \wedge \theta_{cm}) \cdot \exp \Big(-i \sum_{j=0}^{k-1} K_j \Big)\Big) \Big] \eeq
Now, it can be easilly shown that, for any Grassmann polynomials $q_1$ and $q_2$, 
\beq q_1  \cdot u_{-cm} (q_2) = u_{-cm} ((u_{cm}(q_1)) \cdot q_2) \eeq
Furthermore, it can be shown that 
\beq u_{-c} (u_{-cm} (q)) = u_{-c(m+1)} (q) \eeq
Therefore 
\beq u_{-c} (q_1 \cdot u_{-cm} (q_2))= u_{-c} (u_{-cm} ((u_{cm} (q_1)) \cdot q_2))= u_{-c(m+1)}((u_{cm} (q_1)) \cdot q_2) \eeq
Based on this, we can rewrite $p_{k+1}$ as 
\beq p_{k+1} =  u_{-c(m+1)} \Big[ (u_{cm} (\theta_1 \wedge \cdots \wedge \theta_m))  \cdot \Big( (\theta_1 \wedge \cdots \wedge \theta_{cm}) \cdot \exp \Big(-i \sum_{j=0}^{k-1} K_j \Big)\Big) \Big] \eeq
By using
\beq u_{cm} (\theta_1, \cdots, \theta_m) = (\theta_{cm+1}, \cdots, \theta_{c(m+1)}) \eeq
we can further rewrite it as 
\beq p_{k+1} =  u_{-c(m+1)} \Big[ (\theta_{cm+1}, \cdots, \theta_{c(m+1)})  \cdot  \Big( (\theta_1 \wedge \cdots \wedge \theta_{cm}) \cdot \exp \Big(-i \sum_{j=0}^{k-1} K_j \Big)\Big) \Big] \eeq
By Theorem 2, we can rewrite it as 
\beq p_{k+1} = u_{-c(m+1)} \Big[ ((\theta_{cm+1} \wedge \cdots \wedge \theta_{c(m+1)}) \wedge  (\theta_1 \wedge \cdots \wedge \theta_{cm})) \cdot \exp \Big(-i \sum_{j=0}^{k-1} K_j \Big) \Big] \eeq
Now, by using the assumption that $m$ is even, together with Theorem 1, we obtain 
\beq (\theta_{cm+1} \wedge \cdots \wedge \theta_{c(m+1)}) \wedge  (\theta_1 \wedge \cdots \wedge \theta_{cm}) = \theta_1 \wedge \cdots \wedge \theta_{c(m+1)} \eeq
Thus, the expression for $p_{k+1}$ becomes 
\beq p_{k+1} = u_{-c(m+1)} \Big[ (\theta_1 \wedge \cdots \wedge \theta_{c(m+1)}) \cdot \exp \Big(-i \sum_{j=0}^{k-1} K_j \Big) \Big] \eeq
Finally, we recall that one of the hypotheses of the theorem is that $K_k= \underline{0}$ whenever $k+1$ is divisible by $n$. This is, indeed, the case right now. Therefore, $K_k = \underline{0}$, which means that we can freely add it to our sum in the exponent. In other words, we can change the upper limit of the sum from $k-1$ to $k$: 
\beq p_{k+1} = u_{-c(m+1)} \Big[ (\theta_1 \wedge \cdots \wedge \theta_{c(m+1)}) \cdot \exp \Big(-i \sum_{j=0}^k K_j \Big) \Big] \eeq
This proves that the induction step goes through whenever $k+1$ \emph{is} divisible by $n$.  The situation when $k+1$ is \emph{not} divisible by $n$ was "covered" in the earlier part of the proof. We have shown that in that case, too, induction step goes through. Thus, induction step goes through in \emph{all} cases, which proves the theorem. {\bf QED.}

Let us now explicitly see how the "uploading" in the theorem we just proved "gets rid" of Grassmann degrees of freedom. By from translational symmetry in time, we expect the prescription for taking wedge product to be the same when we come across different slices. The only reason $u_{-cm} (K_{k+ cm})$ is distinct from $K_k$ is that the bosonic fields, $A^{\mu}$ and $\phi$, change between the time slices, and they interact with the Fermionic fields we are concerned about. In light of this, we can write
\beq u_{-cm} K_{a+cm} = K_a (\phi (c \delta t), A^{\mu} (c \delta t)) \; , \; 1 \leq a < c \eeq
This means that we can remove $u_{-cm}$, and change $cn \leq k   < c(n+1) -1$ into $0 \leq n < n-1$, as long as we \emph{insert} $A^{\mu}$-dependance and $\phi$-dependance since the time-dependance of the latter "replaces" the time dependance we have removed:  
\beq p_{k+1} = 
\left\{
	\begin{array}{ll}
		[\exp (-i K_k(A^{\mu} (t), \phi (t)))] \wedge p_k \; , \; 0 \leq k < c-1 \\
		(\theta_1 \wedge \cdots \wedge \theta_m) \cdot p_{n-1} \; , \; k=n-1
	\end{array}
\right.
\label{eqn:253}\eeq
The above equation is our ultimate "definition" of the Fermionic dynamics, and we will devote the next section to defining the action that produces this dynamics 

\subsection*{3.2.5 "Bosonic" action leading to Grassmann-based fermionic one}

In the previous section, we have described a step by step process that produces mathematical information consistent with fermionic path integral. From the "non local" perspective, we can simply postulate that process, and leave it at that. However, as we stated earlier, our goal is to "model" everything "non local" in terms of superluminal signals, equipped with differential equations describing how they get emitted, absorbed, how the "absorbed" information is being processed, retained, and so forth.  Thus, the more we "postulate" in this chapter, the more work we have to do in Chapter 4. As one can see by reading Chapter 4, coming up with a "construction" for a bosonic field \emph{alone} is \emph{already} very complicated. We would hope, therefore, to be able to use that for fermions as well, so that we won't have to redo it all over again. 

It turns out that this is possible. After all, we have defined Grassmann numbers in terms of functions over sets. Now, a given set $A$ is in one-to-one correspondence with an "evaluation" function $ev_A$ given by
\beq ev_A (a) = 
\left\{
	\begin{array}{ll}
		1\; , \; a \in A \\
		0 \; , \; A \not\in A 
	\end{array}
\right.
\eeq
Now, the fact that $0$ and $1$ are commutting scalars, the function $ev_A$ is "bosonic". At the same time, the sets $A$, $B$ and $C$ were used in defining Grassmann integration. This means that the Grassmann integration can be described in terms of \emph{bosonic} functions $ev_A$, $ev_B$ and $ev_C$! There will, however, be a noticeable difference from "truly bosonic" case in a sense that the process of interest can not be viewed as smooth. After all, we "jump" between $0$ and $1$ whenever we do any of the "wedge" or "dot" product. Nevertheless, in light of the discrete nature of the theory, the "time" between different "products" is finite, even if very small. Thus, the discrete jumps can be accounted for by "very large" (but still finite) "bosonic" Lagrangian.

Instead of speculating what the "bosonic" action might be, let us instead analyze the results we have so far and "follow our noses" until we can "read off" the action.  Our ultimate goal is to attach "probability amplitudes" to field distribution. Now, based on our definitions in Section 3.2.1 (see eq \ref{eqn:94}), Grassmann variables are subject to a measure
\beq d_i \eta \vert_{\eta = \theta_j} = \delta_i^j \label{eqn:255}\eeq
This implies that they are restricted to afore-given values and there is no room for "fluctuation". This has been reflected in the fact that "path integral" ended up being a "sum" over one single element, which is why the integral (or sum) sign dropped out. In light of this, we attach probability amplitudes \emph{only} to bosonic fields. The fermionic fields are simply "dummy indexes" we are integrating over in order to produce the result.  On the other hand, we also have an opposite surprise. Normally we assume that the probability amplitude we attach is a scalar. But, in Section 3.2.4 we have seen that, contrary to "standard" point of view, path integral can have Grassmannian values. Thus, we have a \emph{complex-valued} wave function with \emph{Grassmann-valued} probability amplitude. In other words, the probability amplitude takes the form 
\beq \psi (A^{\mu}, \phi) = p \eeq
where $p$ is a Grassmann polynomial. We know that $p$ is an actual polynomial as opposed to just one element because we expect $p$ to reproduce the "step by step" defined in equation (\ref{eqn:253}). Now, based on Section 3.2.2, the Grassmann polynomial is defined to be a complex-valued function over $\mathbb{N}$. Thus, 
\beq E \subset \mathbb{N} \Rightarrow p (E) \in \mathbb{C} \eeq
But, if we substitute the left hand side, $\psi (A^{\mu}, \phi)$, in place of $p$, we obtain 
\beq E \subset \mathbb{N} \Rightarrow (\psi (A^{\mu}, \phi))(E) \in \mathbb{C} \eeq
Now, a generic "step" in the "scheme" outlined in Equation (\ref{eqn:253}) can be summarized as "going" from one Grassmann polynomial to another:
\beq (\psi (A^{\mu}, \phi) = p_1 ) \longrightarrow (\psi (A^{\mu}, \phi) = p_2) \eeq
Again, when we say $\psi (A^{\mu}, \phi) = p_1$, we \emph{really} mean that $(\psi (A^{\mu}, \phi))(E) = p_1 (E)$ for all $E$. Thus, the above transformation takes the form 
\beq ((\psi (A^{\mu}, \phi))(E) = p_1 (E)) \longrightarrow ((\psi (A^{\mu}, \phi))(E) = p_2 (E)) \eeq
Now, we replace $E$ with the zero-or-one field $ev_E$, described earlier. Furthermore, we replace $(\psi (a, b))(c)$ with $\psi (a, b, c)$ in the following way: 
\beq \psi (A^{\mu}, \phi, ev_E) = (\psi (A^{\mu}, \phi))(E) \eeq
Thus, our transformation becomes 
\beq (\psi (A^{\mu}, \phi, ev_E) = p_1 (E) ) \longrightarrow (\psi (A^{\mu}, \phi, ev_E) = p_2 (E)) \label{eqn:262}\eeq
Now, \emph{if} we were to succeed to find bosonic-like action, it would take the form 
\beq S(A_1^{\mu}, \phi_1, ev_{E_1}; A_2^{\mu}, \phi_2, ev_{E_2}), \eeq
and our transformation equation would take the form
\beq (\psi (A^{\mu}, \phi, ev_E) = p_1 (E)) \longrightarrow \label{eqn:264}\eeq
\beq \longrightarrow \Big( \psi (A^{\mu}, \phi, ev_E) = \sum_{(A^{\prime \mu}, \phi^{\prime}, ev_{E'})} e^{-iS(A_1^{\mu}, \phi_1, ev_{E_1}; A_2^{\mu}, \phi_2, ev_{E_2})} \psi (A^{\prime \mu}, \phi^{\prime}, ev_{E'}) \Big) \nonumber \eeq
Now, we can apply the right hand side of the equation (\ref{eqn:262}) to the "sum" in the present one. This will amount to replacing $\psi (A^{\prime \mu}, \phi', ev_{E'})$ with $p_1 (E')$: 
\beq (\psi (A^{\mu}, \phi, ev_E) = p_1 (E)) \longrightarrow \eeq
\beq \longrightarrow \Big( \psi (A^{\mu}, \phi, ev_E) = \sum_{(A^{\prime \mu}, \phi^{\prime}, ev_{E'})} e^{-iS(A_1^{\mu}, \phi_1, ev_{E_1}; A_2^{\mu}, \phi_2, ev_{E_2})} p_2 (E') \Big) \nonumber \eeq
By viewing $\psi (A^{\mu}, \phi, ev_E)$ as "synonymous" with $p_1 (E)$, and replacing the former with the latter, we can rewrite the above as 
\beq p_1 (E) \longrightarrow \sum_{(A^{\prime \mu}, \phi^{\prime}, ev_{E'})} e^{-iS(A_1^{\mu}, \phi_1, ev_{E_1}; A_2^{\mu}, \phi_2, ev_{E_2})} p_2 (E')  \eeq
But we know from Equation (\ref{eqn:253}) that, in order to reproduce the fermionic path integral, we would like to "update" $p$ according to 
\beq p_{k+1} = 
\left\{
	\begin{array}{ll}
		[\exp (-i K_k(A^{\mu} (t), \phi (t), A^{\prime \mu} (t), \phi^{\prime} (t)))] \wedge p_k \; , \; 0 \leq k < c-1 \\
		(\theta_1 \wedge \cdots \wedge \theta_m) \cdot p_{n-1} \; , \; n-1
	\end{array}
\right.
\eeq
Now we can combine these two results. In order not to confuse notation, we will replace $p_1$ with $p_k$ and $p_2$ with $p_k'$, and make similar replacements for $A$, $\phi$ and $E$. Thus, we get 
\beq \sum_{(A^{\prime \mu}, \phi^{\prime}, ev_{E'})} e^{-iS(A^{\mu}, \phi, ev_{E}; A^{\prime\mu}, \phi^{\prime}, ev_{E^{\prime}})} p_k (E')  = \eeq
\beq =  \left\{
	\begin{array}{ll}
		[[\exp (-i K_k(A^{\mu} (t), \phi (t), A^{\prime \mu} (t), \phi^{\prime} ( t)))] \wedge p_k] (E) \; , \; 0 \leq k < c-1 \\
		((\theta_1 \wedge \cdots \wedge \theta_m) \cdot p_{n-1}) (E) \; , \; k=n-1
	\end{array}
\right.
\nonumber \eeq
Now, if we "take seriously" the fact that $S$ is, indeed, an action, we want to be able to view $k$ as a physical quantity. Therefore, we will assume that 
\beq k = I (t), \label{eqn:269}\eeq
where $I(t)$ is an integer-valued \emph{dynamical} quantity. Yet, we want it to evolve throughout the space in the same fashion. Therefore, we will assume that there is \emph{one} special particle in the universe (we will call it $I$-partile, where $I$ stands for the word "index") that emits some signals that eventually get converted into $k$l; and no other particle does that. Now, that $K$-particle has internal degrees of freedom, $e_{I1}$ and $\beta_{I2}$ that evolve according to 
\beq \frac{de_{I1}}{dt} = \frac{\omega e_{I2}}{\sqrt{e_{I1}^2 + e_{I2}^2}} \; ; \; \frac{de_{I_2}}{dt} = -\frac{\omega e_{I1}}{\sqrt{e_{I1}^2 + e_{I2}^2}} \label{eqn:270}\eeq
Now, these "rotating" internal degrees of freedom of particle $I$ emit corresponding "messenger" fields, $\mu_{I_1}$ and $\mu_{I_2}$ according to 
\beq \nabla_s^{\alpha} \nabla_{s \alpha} \mu_{I_1} + m_I^2 \mu_{I_1} = e_{I_1} \delta^3 (\vec{x} - \vec{x}_I) e_{I_1} \nonumber \eeq
\beq \nabla_s^{\alpha} \nabla_{s \alpha} \mu_{I_2} + m_I^2 \mu_{I_2} = e_{I_1} \delta^3 (\vec{x} - \vec{x}_I) e_{I_2} \label{eqn:271}\eeq
where the letter $s$ in $\nabla_s$ stands for "superluminal". In other words, $\nabla$ is defined based on the metric that produces superluminal sigals; this implies that the fields $\mu_{K_1}$ and $\mu_{K_2}$ are superluminal, themselves. As a result of compactness of the universe, this means that they circle the universe within a very short period of time. Now, the "integer" $I (\mu_{I_1}, \mu_{I_2})$ we are seeking is merely the angle of "rotation" of the "messenger fields" $\mu_{I_1}$ and $\mu_{I_2}$. Now, angle, of course, doesn't have to be integer. However, if it is multiplied by a large number, it \emph{passes by} the "integers" as it goes along. Thus, whenever the angle is "very close" to a certain integer, we will set $I (\mu_{l_1}, \mu_{l_2})$ to be that integer; otherwise, we will set it to $-1$, thus preventing it from being "relevent" to any of our "truth statements". Since we want the time period during which $I$ is non-zero to be finite, we will introduce a small number $\epsilon$; the value of $I (\mu_{l_1}, \mu_{l_2})$ is non-zero integer if and only if the angle \emph{approximates} that integer \emph{up to} $\epsilon$; otherise, we set $I=-1$. This can be accomplished through
\beq I (\mu_{I_1}, \mu_{I_2}) = -1 + \sum l T \Big( l - \epsilon < C \tan^{-1} \frac{\mu_{I_2}}{\mu_{I_1}} < l + \epsilon \Big) \label{eqn:272}\eeq
Now from this point till the end of this section we will continue to use notation $k$ instead of $I (\mu_{I_1}, \mu_{I_2})$ just to keep notation brief. Then, at the end, we will go back and write it out in terms of $I (\mu_{I_1}, \mu_{I_2})$.

 Now, in order to be able to compare both sides more easilly, we would like to have a \emph{single} sum on the right hand side, instead of "case by case" sum. We accomplish this by rewriting right hand side in terms of "truth values" $T(k=n-1)$ and $T (k \neq n-1) = 1 - T(k = n-1)$ (where $T({\rm true}) = 1$ and $T ({\rm false}) =0$)
in the following way:
\beq \sum_{(A^{\prime \mu}, \phi^{\prime}, ev_{E'})} e^{-iS(A^{\mu}, \phi, ev_{E}; A^{\prime\mu}, \phi^{\prime}, ev_{E^{\prime}})} p_k (E')  = \nonumber \eeq
\beq = [[\exp (-i K_k(A^{\mu} (t), \phi (t), A^{\prime \mu} (t), \phi^{\prime} ( t)))] \wedge p_k] (E)  T(0 \leq k < n-1) + \eeq
\beq +  ((\theta_1 \wedge \cdots \wedge \theta_m) \cdot p_{n-1}) (E) T(k=n-1) \nonumber \eeq
We will also keep on the back of our mind that action can be both real and imaginary. Thus, if we "want" our right hand side to be zero we can simply stick very large (but finite) imaginary part into $S$. 

Now, we know from fermionic field theory that the Lagrangian is a product of only \emph{two} Grassmann variables. Furthermore, we will assume that the choice of Grassmann variables for $K_k$ is independant of bosonic fields, such as $\phi$ or $A^{\mu}$. The latter only deterine coefficients. Thus,  
\beq K_k (A^{\mu}, \phi, A^{\prime \mu}, \phi^{\prime}) = (\theta_{a_k} \wedge \theta_{b_k}) J_k (A^{\mu}, \phi, A^{\prime \mu}, \phi^{\prime}) \label{J} \eeq
where $J_k (A^{\mu}, \phi, A^{\prime \mu}, \phi^{\prime})$ is a complex valued quantity. This implies that 
\beq \exp (-i  K_k (A^{\mu}, \phi, A^{\prime \mu}, \phi^{\prime})) = 1 - i (\theta_{a_k} \wedge \theta_{b_k}) J_k (A^{\mu}, \phi, A^{\prime \mu}, \phi^{\prime}) \eeq
and, therefore, 
\beq \sum_{(A^{\prime \mu}, \phi^{\prime}, ev_{E'})} e^{-iS(A^{\mu}, \phi, ev_{E}; A^{\prime\mu}, \phi^{\prime}, ev_{E^{\prime}})} p_k (E')  = \nonumber \eeq
\beq =  (p_k (E) - i (\theta_{a_k} \wedge \theta_{b_k} \wedge p_k) (E) J_k (A^{\mu}, \phi, A^{\prime \mu}, \phi^{\prime})) T(0 \leq k < n-1) + \label{eqn:276}\eeq
\beq +  ((\theta_1 \wedge \cdots \wedge \theta_m) \cdot p_{n-1}) (E) T(k=n-1)  \nonumber \eeq
Now we recall that our definitions of wedge and dot product are given by 
\beq (p_1 \wedge p_2)(C) = \sum_{A \cup B = C \; ; \; A \cap B = \emptyset} (-1)^{\sharp \{j_1>j_2 \vert j_1 \in A, j_2 \in B \}} p_1 (A) p_2 (B) \eeq
and
\beq (p_1 \cdot p_2)(C) = \sum_{A \cup C = B \; ; \; A \cap C = \emptyset} (-1)^{\sharp \{j_1>j_2 \vert j_1 \in A, j_2 \in B \}} p_1 (A) p_2 (B) \eeq 
By combining it with the fact that
\beq (\theta_{a_k} \wedge \theta_{b_k})(\{a_k, b_k \}) = (-1)^{T(a_k > b_k)} \eeq
We can, therefore, rewrite (\ref{eqn:276}) as 
\beq \sum_{(A^{\prime \mu}, \phi^{\prime}, ev_{E'})} e^{-iS(A^{\mu}, \phi, ev_{E}; A^{\prime\mu}, \phi^{\prime}, ev_{E^{\prime}})} p_k (E')  = \nonumber \eeq
\beq =  p_k (E) T(0 \leq k < n-1) - i T(0 \leq k < n-1) J_k (A^{\mu}, \phi, A^{\prime \mu}, \phi^{\prime})   \times \nonumber \eeq
\beq \times \sum_{\{a_k, b_k \} \cup E' = E \; ; \; \{a_k, b_k \} \cap E' = \emptyset} (-1)^{T(a_k > b_k)}  (-1)^{\sharp \{j_1>j_2 \vert j_1 \in \{a_k, b_k \} , j_2 \in E' \}}  p_k (E')   + \label{eqn:280} \eeq
\beq +  \sum_{\{1, \cdots, m \} \cup E= E' \; ; \; \{1, \cdots, m \} \cap E = \emptyset}  T(k=n-1)  (-1)^{\sharp \{j_1>j_2 \vert j_1 \in \{1, \cdots, m \}, j_2 \in E'} p_k (E') \nonumber \eeq 
In the above expression, we have deliberately refrained from evaluating $E'$ even though we eacilly could have (namely, $E' = E \setminus \{a_k, b_k \}$ for the first sum and $E' = \emptyset$ in case of the second sum). The reason we didn't do these substitutions is that we would like $S(\cdots, ev_{E'})$ to be defined for \emph{all} $E'$, not just the specific value of $E'$ we found out. In case of the "wrong" $E'$ we simply add a large imaginary part to the action, thus setting $e^{-iS}$ to zero:
\beq {\rm wrong \; E'} \Rightarrow S (\cdots, ev_{E'}) = ({\rm real \; part}) - iN \Rightarrow e^{-iS} = e^{-i \times ({\rm real part })} \times e^{-N} \approx 0 \eeq
Motivated by what we have just said, we would like to rewrite the "restricted sums" in the right hand side in a form of non-restricted ones. We will do that by inserting "truth statements":
\beq \sum_{\rm restriction} (\cdots) = \sum (\cdots) T({\rm restriction}) \Rightarrow S = S_0 +i \; \ln  T({\rm restriction }) \eeq
where it is understood that the "logarithm" is defined in such a way that logarithm of zero is still finite; for example, we can define logarithm in the following way:
\beq \exp (z_1) = z_2 \Rightarrow^{def} \ln z_2 = \frac{z_1}{\vert z_1 \vert} \min (\vert z_1 \vert, N_{\ln}) \label{eqn:ln0}\eeq
where $N_{ln}$ is some very small number; thus, $N_{\ln}$ "cuts off" the absolute value of logarithm. In order to gain the definition of action in above-described form, we will now rewrite the equation (\ref{eqn:280}) in terms of a sum over \emph{all} $E^{\prime}$, where all of the restrictions are replaced by corresponding "truth values":
\beq \sum_{(A^{\prime \mu}, \phi^{\prime}, ev_{E'})} e^{-iS(A^{\mu}, \phi, ev_{E}; A^{\prime\mu}, \phi^{\prime}, ev_{E^{\prime}})} p_k (E')   = \nonumber \eeq
\beq =  p_k (E) T(0 \leq k < n-1)  - i J_k (A^{\mu}, \phi, A^{\prime \mu}, \phi^{\prime}) T(0 \leq k < n-1)  \times  \label{eqn:284}\eeq
\beq \times \sum p_k (E') (-1)^{T(a_k > b_k) +\sharp \{j_1>j_2 \vert j_1 \in \{a_k, b_k \} , j_2 \in E' \}}  T(\{a_k, b_k \} \cup E' = E) T( \{a_k, b_k \} \cap E' = \emptyset) + \nonumber \eeq
\beq +  \sum p_k (E')  (-1)^{ \sharp \{j_1>j_2 \vert j_1 \in \{1, \cdots, m \}, j_2 \in E'} T(k=n-1)  T (\{1, \cdots, m \} \cup E= E') T(\{1, \cdots, m \} \cap E = \emptyset) \nonumber \eeq 
We now rewrite $p_k (E)$ as
\beq p_k (E) = \sum_{E'} p_k (E') T(E'=E) \eeq
and we also notice that, due to the fact that $E \subset \{1, \cdots m \}$ and $E' \subset \{1, \cdots m \}$, the following holds:
\beq T(\{1, \cdots, m \} \cup E = E') T(\{1, \cdots, m \} \cap E = \emptyset) = T(E' = \{1, \cdots, m \}) T (E = \emptyset ) \eeq
Using these last two equations, we can now combine equation (\ref{eqn:284}) into a single sum and obtain 
\beq \sum_{(A^{\prime \mu}, \phi^{\prime}, ev_{E'})} e^{-iS(A^{\mu}, \phi, ev_{E}; A^{\prime\mu}, \phi^{\prime}, ev_{E^{\prime}})} p_k (E')   =  \nonumber \eeq
\beq = \sum p_k (E') [T(0 \leq k < n-1) T(E'=E) - i J_k (A^{\mu}, \phi, A^{\prime \mu}, \phi^{\prime}) T(0 \leq k < n-1)  \times \nonumber \eeq
\beq \times  (-1)^{T(a_k > b_k)+ \sharp \{j_1>j_2 \vert j_1 \in \{a_k, b_k \} , j_2 \in E' \}}  T(\{a_k, b_k \} \cup E' = E) T( \{a_k, b_k \} \cap E' = \emptyset) + \eeq
\beq +   (-1)^{\sharp \{j_1>j_2 \vert j_1 \in \{1, \cdots, m \}, j_2 \in E'}  T(k=n-1) T(E' = \{1, \cdots, m \}) T (E = \emptyset )] \nonumber \eeq 
From this we immediately read off the definition of $S$:
\beq S (A^{\mu}, \phi, ev_E; A^{\prime\mu}, \phi^{\prime}, ev_{E^{\prime}}) = i \; \ln [T(0 \leq k < n-1) T(E'=E) - \nonumber \eeq
\beq - i J_k (A^{\mu}, \phi, A^{\prime \mu}, \phi^{\prime})   (-1)^{T(a_k > b_k) + \sharp \{j_1>j_2 \vert j_1 \in \{a_k, b_k \} , j_2 \in E' \}}  \times \nonumber \eeq
\beq \times  T(0 \leq k < n-1) T(\{a_k, b_k \} \cup E' = E) T( \{a_k, b_k \} \cap E' = \emptyset) + \eeq
\beq +   (-1)^{\sharp \{j_1>j_2 \vert j_1 \in \{1, \cdots, m \}, j_2 \in E'}  T(k=n-1)  T(E' = \{1, \cdots, m \}) T (E = \emptyset )] \nonumber \eeq 
Now, on the first glance, it might seem that the logarithm of the sum is "bad news", since it makes it impossible to decouple these terms.  \emph{What comes to our rescue}, however, is the fact that, at any given time, only \emph{one} of the three terms under logarithm is non-zero. First of all, the third term has $T(k=n-1)$ while the first two terms contain $T(0 \leq k < n-1)$. Thus, if third term is non-zero, then both first and second term are zero. Thus, the only thing left to show is that first two terms can't be simulteneously non-zero. Now, the statement $\{a_k, b_k \} \cup E' =E$ on the second term implies that $\{a_k, b_k \} \cap E = \{a_k, b_k \}$. At the same time, the second term contains $T(\{a_k, b_k \} \cap E' \neq \emptyset)$. Together, they imply that, if the second term is non-zero, then $E \neq E'$. This will, in term, imply that first term is zero, since it contains $T(E'=E)$. Now, it is easy to check that, for any three numbers $a$, $b$, $c$ the following is true:
\beq ab=ac=bc=0 \Rightarrow \ln (a+b+c) = \ln a + \ln b + \ln c \eeq
Since we have just shown that the three terms we have under the logarithm satisfy the above condition, we can go ahead and decouple them: 
\beq S (A^{\mu}, \phi, ev_E; A^{\prime\mu}, \phi^{\prime}, ev_{E^{\prime}}) = i \; \ln (T(0 \leq k < n-1) T(E'=E)) + \nonumber \eeq
\beq + i \ln \Big(- i J_k (A^{\mu}, \phi, A^{\prime \mu}, \phi^{\prime})   (-1)^{T(a_k > b_k) + \sharp \{j_1>j_2 \vert j_1 \in \{a_k, b_k \} , j_2 \in E' \}}  \times \nonumber \eeq
\beq \times  T(0 \leq k < n-1) T(\{a_k, b_k \} \cup E' = E) T( \{a_k, b_k \} \cap E' = \emptyset) \Big) + \eeq
\beq +  i \ln \Big( (-1)^{\sharp \{j_1>j_2 \vert j_1 \in \{1, \cdots, m \}, j_2 \in E'}  T(k=n-1)  T(E' = \{1, \cdots, m \}) T (E = \emptyset )] \Big) \nonumber \eeq 
Thus, we are only left with logarithms of the products. Now, we can split them, too, into the sums of logarithms and obtain 
\beq S (A^{\mu}, \phi, ev_E; A^{\prime\mu}, \phi^{\prime}, ev_{E^{\prime}}) = i \; \ln T(0 \leq k < n-1) + i \; \ln T(E'=E) + \nonumber \eeq
\beq + i \ln (-i) + i \ln J_k (A^{\mu}, \phi, A^{\prime \mu}, \phi^{\prime})  + i \ln (-1)^{T(a_k > b_k) + \sharp \{j_1>j_2 \vert j_1 \in \{a_k, b_k \} , j_2 \in E' \}}  + \nonumber \eeq
\beq + i \ln  T(0 \leq k < n-1) + i \ln T(\{a_k, b_k \} \cup E' = E) + i \ln T( \{a_k, b_k \} \cap E' = \emptyset) ) + \eeq
\beq +  i \ln  (-1)^{\sharp \{j_1>j_2 \vert j_1 \in \{1, \cdots, m \}, j_2 \in E'}  + i \ln T(k=n-1)  + i \ln T(E' = \{1, \cdots, m \}) + i \ln T (E = \emptyset )  \nonumber \eeq 
We would now like to be able to express the above in terms of the sums of three-element functions. This will allow us to identify these functions with \emph{Lagrangian generators}. Let us, furthermore, attempt to express everything in terms of "evaluation functions" $ev_E$, $ev_{E'}$ and $ev_{\{a_k, b_k \}}$. The first two will be treated as "bosonic fields" which are subjected to our Lagrangian. The last is strictly classical, and it obeys equations (\ref{eqn:269}-\ref{eqn:272}). It is easy to verify that the following conditions are true (we will list them in the same order as the terms in the above equation, in order to make it easy to track down how we did the substitution)
\beq T(0 \leq k < n-1) = \delta^0_{n-1} \eeq
\beq T(E'=E) =  \prod_j  (1- (ev_{E'} (j) - ev_E (j))^2) \eeq
\beq \ln i = i \pi \; ; \; \ln (-i) = -i \pi \; ; \; \ln (-1) = i \pi \eeq
\beq \ln (-1)^{T(a_k > b_k) + \sharp \{j_1 > j_2 \vert j_1 \in \{a_k, b_k \}, j_2 \in E' \} }= i \pi \delta_{\vert a_k - b_k \vert}^{a_k - b_k} + i \pi \sum_{j_1,j_2} \delta^{j_1-j_2}_{\vert j_1-j_2 \vert} ev_{\{a_k, b_k \}} (a) ev_{E'} (j_2) \eeq
\beq T(\{a_k, b_k \} \cup E' = E) = \prod_j (1 - (ev_{ \{a_k, b_k \}} (x) ev_{E'} (j) - ev_E (j))^2) \eeq 
\beq T (\{a_k, b_k \} \cap E' = \emptyset) = \prod_j (1 - ev_{\{a_k, b_k \}}(j) ev_{E'}(j)) \eeq
\beq \ln (-1)^{\sharp \{j_1>j_2 \vert j_1 \in \{1, \cdots, m \}, j_2 \in E' \} } = i \pi \sum \delta^{j_1-j_2}_{\vert j_1-j_2 \vert} ev_{E'} (j_2) \eeq
\beq T(E' = \{1, \cdots, m \}) = \prod_j ev_{E'} (j) \eeq
\beq T (E = \emptyset) = \prod_j (1- ev_E (j)) \eeq
By substitutting these into the equation for $S$, and using the fact that logarithm of the product is equal to sum of logarithms, we obtain 
\beq S (A^{\mu}, \phi, ev_E; A^{\prime\mu}, \phi^{\prime}, ev_{E^{\prime}}) =2 i \; \ln \delta^k_{n-1} + i \sum \ln (1- (ev_{E'} (j) - ev_E (j))^2) + \nonumber \eeq
\beq + \pi + i \ln J_k (A^{\mu}, \phi, A^{\prime \mu}, \phi^{\prime})  +  i \pi \delta_{\vert a_k - b_k \vert}^{a_k - b_k} - \pi \sum_{j_1,j_2} \delta^{j_1-j_2}_{\vert j_1-j_2 \vert} ev_{\{a_k, b_k \}} (j_1) ev_{E'} (j_2) + \nonumber \eeq
\beq  + i \sum \ln (1 - (ev_{ \{a_k, b_k \}} (x) ev_{E'} (j) - ev_E (j))^2)  +  \eeq
\beq + i \sum \ln (1 - ev_{\{a_k, b_k \}}(j) ev_{E'}(j)) - \pi \sum \delta^{j_1-j_2}_{\vert j_1-j_2 \vert} ev_{E'} (j_2)   + \nonumber \eeq
\beq + i \sum \ln ev_{E'} (j)  + i \sum \ln (1- ev_E(j)) , \nonumber \eeq 
where the factor of $2$ on the first term comes from adding it with the sixth one. Finally, in order to make it in the format of Lagrangian generators, we have to sum over triplets. Therefore, we will rewrite the above sums according to the following prescription:
\beq 1 = \sum_{j_1,j_2, j_3} N_p^{-3} \; ; \; \sum_{j_1} C_{j_1} = \sum_{j_1, j_2, j_3} N_p^{-2} k_{j_1} \; ; \; \sum_{j_1, j_2} C_{j_1,j_2} = \sum_{j_1, j_2, j_3} N_p^{-1} C_{j_1, j_2, j_3} \eeq
where $N_p$ is the number of lattice particles. Thus, we rewrite the action as 
\beq S (A^{\mu}, \phi, ev_E; A^{\prime\mu}, \phi^{\prime}, ev_{E^{\prime}}) = \sum_{j_1, j_2, j_3} \big( - \pi N^{-1} \delta^{j_1-j_2}_{\vert j_1-j_2 \vert} ev_{\{a_k, b_k \}} (j_1) ev_{E'} (j_2) \nonumber \eeq
\beq - \pi N^{-1} \delta^{j_1-j_2}_{\vert j_1-j_2 \vert} ev_{E'} (j_2)   + i \pi N^{-1} \delta_{\vert j_1 - j_2 \vert}^{j_1 - j_2} \delta^{j_1}_{a_k} \delta^{j_2}_{b_k}+ i N_p^{-2} \ln (1- (ev_{E'} (j_1) - ev_E (j_1))^2)    + \nonumber \eeq
\beq +  i N^{-2} \ln (1 - (ev_{ \{a_k, b_k \}} (j_1) ev_{E'} (j_1) - ev_E (j_1))^2)  +  \eeq
\beq + i N^{-2} \ln (1 - ev_{\{a_k, b_k \}}(j_1) ev_{E'}(j_1))  + i N^{-2} \ln ev_{E'} (j_1)  + \nonumber \eeq
\beq +  i N^{-2}\ln T (1- ev_E(j_1)) + N^{-3} \pi + i N^{-3} \ln J_k (A^{\mu}, \phi, A^{\prime \mu}, \phi^{\prime}) + 2 i N_p^{-3} \; \ln \delta^k_{n-1} ) \nonumber \eeq
where we have regrouped the terms in such a way that we go from $N^{-1}$ to $N^{-2}$ to $N^{-3}$ terms. Finally, in order to keep it consistent with the framework of Chapter 2 as well as Chapter 4, we have to "attach" the "bosonic" fields to the lattice points themselves. We choose to attach $A^{\mu}$ and $\phi$ to the point $a_k$ and we attach $A^{\prime \mu}$ and $\phi^{\prime}$ to the point $b_k$. Thus, we get rid of primes:
\beq (A^{\mu}, \phi) \rightarrow (A^{\mu} (j_1), \phi (j_1)) \; ; \; (A^{\prime \mu}, \phi^{\prime}) \rightarrow (A^{\mu} (j_2), \phi (j_2)) \eeq
Therefore, we can rewrite corresponding sum as 
\beq \sum iN^{-3} \ln J_k (A^{\mu}, \phi, A^{\prime \mu}, \phi^{\prime} = \sum iN^{-1} \delta_{a_k}^{j_1} \delta_{b_k}^{j_2} \ln J_k (A^{\mu}(j_1) , \phi (j_1) , A^{ \mu} (j_2) , \phi (j_2) ) \eeq
where we have replaced $N^{-3}$ with $N^{-1}$ due to the fact that we introduced $\delta_{a_k}^{j_1} \delta_{b_k}^{j_2}$ which replaced $N^2$ copies of the same term with a single term. Thus, we can rewrite action as 
\beq S (A^{\mu}, \phi, ev_E; A^{\prime\mu}, \phi^{\prime}, ev_{E^{\prime}}) = \sum_{j_1, j_2, j_3} \big( - \pi N^{-1} \delta^{j_1-j_2}_{\vert j_1-j_2 \vert} ev_{\{a_k, b_k \}} (a) ev_{E'} (b) -  \eeq
\beq - \pi N^{-1} \delta^{j_1-j_2}_{\vert j_1-j_2 \vert} ev_{E'} (j_2)   + i \pi N^{-1} \delta_{\vert a - b \vert}^{a - b} \delta^a_{a_k} \delta^b_{b_k}+iN^{-1} \delta_{a_k}^{j_1} \delta_{b_k}^{j_2} \ln J_k (A^{\mu}(j_1) , \phi (j_1) , A^{ \mu} (j_2) , \phi (j_2) ) \nonumber \eeq
\beq +  i N_p^{-2} \ln (1- (ev_{E'} (j_1) - ev_E (j_1))^2)   +  i N^{-2} \ln (1 - (ev_{ \{a_k, b_k \}} (j_1) ev_{E'} (a) - ev_E (j_1))^2)  +  \nonumber \eeq
\beq + i N^{-2} \ln (1 - ev_{\{a_k, b_k \}}(j_1) ev_{E'}(j_1))  + i N^{-2} \ln ev_{E'} (j_1)  +  i N^{-2}\ln (1- ev_E(j_1)) +  \nonumber \eeq
\beq  + N^{-3} \pi + 2 i N_p^{-3} \; \ln \delta^k_{n-1} ) \nonumber \eeq
Finally, in order to be able to write down the Lagrangian generator, we have to remove all of the indexes, other than the two indexes we are explicitly summing over. As we recall from equation (\ref{eqn:269}), $k$ is not just an index but rather a \emph{classical field} whose behavior is described by 
\beq k = I (\mu_{I_1} (t), \mu_{I_2} (t)) \eeq
where the dynamics of $I(a, b)$ has been specified by (\ref{eqn:270}-\ref{eqn:272}).  Furthermore, we will replace $ev_E$ and $ev_{E'}$ with another pair of "classical" fields $\cal E$ and ${\cal E}'$, whose values are restricted to $0$ or $1$:
\beq ev_{E} (p) = {\cal E} (a) \; ; \; ev_{E'} (p) = {\cal E}' (a) \eeq
In fact, we will view ${\cal E}$ and ${\cal E}'$ as "more fundamental" than $E$ and $E'$. Thus, we \emph{first} introduce $\cal E$ and ${\cal E}'$, \emph{then} we declare that they are either $0$ or $1$, and \emph{then} we will define the sets $E$ and $E'$ in terms of them:
\beq {\cal E} \in \{0, 1 \} \; ; \; {\cal E}' \in \{0, 1 \} \; ; \; E = \{j \vert {\cal E } (j) = 1 \} \; ; \;  E' = \{j \vert {\cal E}' (j) = 1 \} \eeq
As far as $ev_{a_k, b_k}$ is concerned, we can replace it with 
\beq ev_{a_k, b_k} (i) = \delta_{a_{I (\mu_{I_1}, \mu_{I_2})}}^i + \delta_{b_{I (\mu_{I_1}, \mu_{I_2})}}^i \eeq
where we assume some afore-given definition of $a_i$ and $b_i$ for any given integer $i$. Thus, we can rewrite our action as  
\beq S (A^{\mu} , \phi , {\cal E}, {\cal E}' ) = \sum_{j_1, j_2, j_3} \big( - \pi N^{-1} \delta^{j_1-j_2}_{\vert j_1-j_2 \vert} (\delta_{a_{I (\mu_{I_1}, \mu_{I_2})}}^{j_1}+ \delta_{b_{I (\mu_{I_1}, \mu_{I_2})}}^{j_1}) {\cal E}' (j_2) -  \nonumber \eeq
\beq - \pi N^{-1} \delta^{j_1-j_2}_{\vert j_1-j_2 \vert} {\cal E}' (j_2)   + i \pi N^{-1} \delta_{\vert j_1 - j_2 \vert}^{j_1 - j_2} \delta^a_{a_k} \delta^b_{b_k}+iN^{-1} \delta_{a_k}^{j_1} \delta_{b_k}^{j_2} \ln J_k (A^{\mu}(j_1) , \phi (j_1) , A^{ \mu} (j_2) , \phi (j_2) ) \nonumber \eeq
\beq +  i N_p^{-2} \ln (1- ({\cal E}' (j_1) - {\cal E} (j_1))^2)   +  i N^{-2} \ln (1 - ((\delta_{a_{I (\mu_{I_1}, \mu_{I_2})}}^{j_1} + \delta_{b_{I (\mu_{I_1}, \mu_{I_2})}}^{j_1}) {\cal E}' (j_1) - {\cal E} (j_1))^2)  +  \nonumber \eeq
\beq + i N^{-2} \ln (1 - (\delta_{a_{I (\mu_{I_1}, \mu_{I_2})}}^{j_1} + \delta_{b_{I (\mu_{I_1}, \mu_{I_2})}}^{j_1}) {\cal E}'(j_1))  + i N^{-2} \ln {\cal E}' (j_1)  +  i N^{-2}\ln (1- {\cal E} (j_1)) + \nonumber \eeq
\beq + N^{-3} \pi + 2 i N_p^{-3} \; \ln \delta^k_{n-1} )  \label{FermionLagrangian} \eeq
where, just to repeat the things stated earlier, 
\beq  \ln 0 = N_{ln} \in \mathbb{R}\; ; \; \exp (z_1) = z_2 \Rightarrow^{def} \ln z_2 = \frac{z_1}{\vert z_1 \vert} \min (\vert z_1 \vert, N_{\ln}) \nonumber \eeq
\beq {\cal E} (a) \in \{0, 1 \} \; ; \; {\cal E}' (a) \in \{0, 1 \} \nonumber \eeq
\beq \frac{de_{I1}}{dt} = \frac{\omega e_{I2}}{\sqrt{e_{I1}^2 + e_{I2}^2}} \; ; \; \frac{de_{I_2}}{dt} = -\frac{\omega e_{I1}}{\sqrt{e_{I1}^2 + e_{I2}^2}}  \nonumber \eeq
\beq \nabla_s^{\alpha} \nabla_{s \alpha} \mu_{I_1} + m_I^2 \mu_{I_1} = e_{I_1} \delta^3 (\vec{x} - \vec{x}_I) e_{I_1}  \label{Conditions} \eeq
\beq \nabla_s^{\alpha} \nabla_{s \alpha} \mu_{I_2} + m_I^2 \mu_{I_2} = e_{I_1} \delta^3 (\vec{x} - \vec{x}_I) e_{I_2} \nonumber \eeq
\beq k (\mu_{I_1}, \mu_{I_2}) = -1 + \sum l T \Big( l - \epsilon < C \tan^{-1} \frac{\mu_{I_2}}{\mu_{I_1}} < l + \epsilon \Big) \nonumber \eeq
\beq T({\rm true})= 1 \; ; \; T({\rm false}) = 0 \nonumber \eeq
From this, we read off the Lagrangian generator to be 
\beq {\cal K} (A^{\mu} (j_1), \phi (j_1), {\cal E} (j_1), {\cal E}' (j_1); A^{\mu} (j_2), \phi (j_2), {\cal E}^{\prime}; A^{\mu} (j_2), \phi (j_2), {\cal E} (j_2), {\cal E}' (j_2); A^{\mu} (j_3), \phi (j_3), {\cal E} (j_3), {\cal E}' (j_3)) =  \nonumber \eeq
\beq = - \pi N^{-1} \delta^{j_1-j_2}_{\vert j_1-j_2 \vert} (\delta_{a_{I (\mu_{I_1}, \mu_{I_2})}}^{j_1}+ \delta_{b_{I (\mu_{I_1}, \mu_{I_2})}}^{j_1}) {\cal E}' (j_2)  - \pi N^{-1} \delta^{j_1-j_2}_{\vert j_1-j_2 \vert} {\cal E}' (j_2)  + i \pi N^{-1} \delta_{\vert j_1 - j_2 \vert}^{j_1 - j_2} \delta^a_{a_k} \delta^b_{b_k}+ \nonumber \eeq
\beq + iN^{-1} \delta_{a_k}^{j_1} \delta_{b_k}^{j_2} \ln J_k (A^{\mu}(j_1) , \phi (j_1) , A^{ \mu} (j_2) , \phi (j_2) ) +  i N_p^{-2} \ln (1- ({\cal E}' (j_1) - {\cal E} (j_1))^2)   + \label{eqn:FermionGenerator}\eeq
\beq + i N^{-2} \ln (1 - ((\delta_{a_{I (\mu_{I_1}, \mu_{I_2})}}^{j_1} + \delta_{b_{I (\mu_{I_1}, \mu_{I_2})}}^{j_1}) {\cal E}' (j_1) - {\cal E} (j_1))^2)  +  + i N^{-2} \ln {\cal E}' (j_1)  + \nonumber \eeq
\beq + i N^{-2} \ln (1 - (\delta_{a_{I (\mu_{I_1}, \mu_{I_2})}}^{j_1} + \delta_{b_{I (\mu_{I_1}, \mu_{I_2})}}^{j_1}) {\cal E}'(j_1))  + i N^{-2}\ln (1- {\cal E} (j_1)) + N^{-3} \pi + 2 i N_p^{-3} \; \ln \delta^k_{n-1}  \nonumber \eeq
The only thing that has not been specified in the above Lagrangian is $J_k$. This term is actually quite important since this is the \emph{only} term in which we actually take into account the specifics of the fermionic Lagrangian; all the other terms are there only for the purposes of path integration. This, for example, is evident from the fact that the above equation does not contain any $\gamma$-matrices; all $\gamma$-matrices (in coefficient form) are hidden in $J_k$. Thus, in some sense, $J_k$ for fermions serves the same purpose as $\cal K$ for bosons: in both cases it fills the one and only "gap" in the theory; the way in which this "gap" is filled defines the fields we are dealing with, much like Lagrangian in standard physics does. Therefore. we will devote next section in trying to find $J_k$ for weak interactions, just to show the reader how it "works". 

\subsection*{3.2.6. More than one fermionic field; weak interaction}

As one can easily notice from the previous section, there doesn't seem to be a connection between the Lagrangian generator that we have written down and the actual fermionic fields we are dealing with. First of all, we have never utilized $\gamma$-matrixes. Secondly, we used only one fermionic parameter, whereas in reality there might be several different fermionic fields. In fact, apart from us "limitting ourselves" to only one fermion, it seems like we also limit ourselves to only one spin index. Thus it seems impossible to introuce $\gamma$ matrixes even if we wanted to. 

Part of the "$\gamma$-matrix objection" can be answered by pointing at the unspecified function $J_k (A^{\mu} (j_1), \phi (j_1), A^{\mu} (j_2), \phi (j_2))$ inside of \ref{eqn:FermionGenerator}. This function is supposed to be selected according to the specific fermionic Lagrangians we want to discretize. Thus, it will include $\gamma$-matrixes once we write it out explicitly. The only part of $\gamma$-matrix objection that is not adressed by this argument is the issue of us having only one fermionic field, which prevents us from having more than one spinor component that $\gamma$-matrix is supposed to "rotate". Thus, it is critical for us to learn how to introduce several fermionic fields. \emph{But} if we do learn how to do it, we will \emph{also} be able to have more than one kind of particle (such as electron plus neutrino) as a bonus. 

Our trick of adressing this issue is to first "write down" different "fields" side by side and then "merge" these lists together into one single one. For example, if we have only one particle, and our only concern was its spin indexes, we can first "write them out" as 
\beq \eta_k =  \left( \begin{array}{ccc}
\eta_{k1} \\
\eta_{k2} \\
\eta_{k3} \\
\eta_{k4}  \end{array} \right) \eeq
After that, we also write out each of these four indexes in a complex form:
\beq \eta_{k1} =\eta_{k11} + i \eta_{k12} \; ; \; \eta_{k2} = \eta_{k21} + i \eta_{k22} \; ; \;  \eta_{k3} = \eta_{k31} + i \eta_{k32} \; ; \; \eta_{k4} = \eta_{k41} + i \eta_{k42} \label{eqn:conversion} \eeq
Finally, we list \emph{all of them} under the same index $\theta$:
\beq \eta_{kc1} = \theta_{l(c-1) +k} \; ; \; \eta_{kc2} = \theta_{l(c+3) +k} \; ; \; \eta'_{jc1} = \theta_{l(c+7)+j} \; ; \; \eta'_{jc2} = \theta_{l(c+11)+j} \label{eqn:FermionSplit}\eeq
Here, $\eta'$ refers to the slice $t + \delta t$, while $\eta$ refers to slice $t$; and we remember from what we discussed in section 3.2.4, we only need two slices to be present; the two couple to each other in time derivative terms. The above indexes range from 

1) $4$ values of $c$ = spinor indexes

2) $2$ time slices = $\eta$ vs $\eta'$ = $-1 \cdots 7$ vs $8 \cdots 15$

3) $2$ options: real ($-1, \cdots, 3$, $8, \cdots 11$) or imaginary ($4, \cdots, 7$, $12, \cdots, 15$)

This means that we have $4 \times 2 \times 2 = 16$ indexes. On the other hand, according to section 3.2.4, we \emph{still} have two slices, but nothing else; in other words $2=2$. This means we now have additional factor of $8$. This additional factor comes from the fact that we no longer have $8$ fields (four real spinor components and four imaginary spinor components) but only one field (which neither have spinor components, nor real or imaginary parts). Thus, the total number of indexes is preserved. 

This, however, creates a problem. From what we have seen in the previous section, we identify $\theta_i$ with a function that sets $\{i \}$ to $1$ and sets everything else to zero. Furthermore, we have stated that we identify the elements of these sets with the lattice points. Thus, we have one-to-one correspondence between $\theta_i$ and lattice point number $i$. But then how can we define spin up and spin down field \emph{at the same point}? The answer is that we can't! Instead, these two fields arrive at two \emph{different} points that share the same cluster, and thus "look like" the same point.  

This part of the argument is very similar to what we said in bosonic case. As we recall, we have argued that the value of $\phi$ \emph{attached to the specific lattice point} never changes. What changes is our "choice" of the lattice point we are "looking at". \emph{But} the fact that the new lattice point we "look at" happened to be "within the same cluster", makes it \emph{seem like} the value of field changed, when in reality it hasn't. Thus, if we apply the same argument to fermios, we would, likewise, expect the "fermionic field" (such as spin up) to be attached to the point, as well. True, we don't have a linear combination of spin up or spin down, nor can our coefficients vary. But this is in line with the "discrete" nature of fermions. 

Things get more interesting, however, if we ask the above question in the context of electron versus neutrino (as opposed to spin up versus spin down). In this case, \emph{again} we reserve some indexes \emph{only} for electron and some \emph{only} for neutrio, which implies that at any give point we have only \emph{one} of these two fields. \emph{On the other hand} when we have two bosonic fields, say, $\phi_1$ and $\phi_2$, then \emph{both of them} are attached to each given point, not just one. True, a given point is "stuck" with $\phi_1 = 5$ and $\phi_2 =7$, but it is \emph{not} stuck with a "choice" that of $\phi_1$ \emph{instead of} $\phi_2$. On the other hand, in case of fermions, a point is "stuck" with the fact that it \emph{only} has $\phi_1=5$, while $\phi_2$ is absent altogether! 

The only way to make sense of this is to say that fermionic field is really just "one" field, while "electron" and "neutrino" are simply its different "values". This has nothing to do with SU(2) symmetry. After all, the same will apply to proton, or any other fermion we might have. \emph{All} "fermions" are defined as different values of \emph{the same} "fermionic field" $\theta$ \emph{regardless} of any symmetries they do or don't have. On the other hand, the symmetries that happened to be there are not respected. For instance, while some points encode spin-up and others spin-down, none encode a linear combination of the two. This seems to imply the "preferred" definition of "up" direction. The author of this paper sticks to the view that this "paradox" can be answered by viewing vierbies as literal physical fields, which simply happened not to play any role in observable pheomena; but this doesn't make them any less real. 

Let us now proceed to the definition of derivatives. As we recall from bosonic case (in sections 3.1.1 -- 3.1.4), we had to replace the derivatives with discrete expressions. Let us repeat the same for fermions. In bosonic case, we recall that the derivatives were given by 
\beq (\partial_0 \phi) \vert_{\vec{x}=\vec{0}} = \frac{\alpha^{d/2}}{(2 \pi)^{d/2} \delta t}  \int d^d x e^{- \frac{\alpha}{2} \vert \vec{x} \vert^2} (\phi (\vec{x}, t+ \delta t) - \phi (\vec{0}, t)) \eeq
\beq (\partial_i \phi)\vert_{\vec{x} = \vec{0}} \approx \frac{\alpha^{(d+2)/2}}{(2 \pi)^{d/2}} \int d^d x e^{- \frac{\alpha}{2} \vert \vec{x} \vert^2} x^i (\phi (\vec{x}) - \phi (\vec{0})) \eeq
Repeating this for fermions might be non-trivial. After all, the above expressions were derived based on the assumption of continuity of $\phi$. Fermions, on the other hand, are expressly discotinuous. After all, $\theta_i$ and $\theta_j$ are viewed as "separate coordinates" as opposed to slightly different values of the same coordinate. Apart from this issue, there is also a fermion doubling problem that needs to be adressed.  We will admit that we have not yet researched these issues. So, for the purposes of this paper, we will simply take "leap of faith" and "postulate" the fermion version of the above expressions as a "definition" of fermionic Lagrangian, while leave the above conceptual questions for later.  Nevertheless, we want to make sure that what we are doing is mathematically precise. In other words, we want to be able to say something along these lines:

a) We defined our derivative to be the specific expression. Therefore, \emph{by definition} it has to be true, and all the math we are doing is, likewise, correct \emph{and rigorous}

b) We don't know whether that given definition produces physics we would see "in the lab".

Now, if we stick with point $a$, we can't write an expression such as $\partial_x \theta_1$. After all, $\theta_1$ is only defined \emph{at exactly one point}. Or, if we attempt "not to refer to" one to one correspondence in the context of definition of derivative, then the other option is to view $\theta_1$ as "constant" (after all, that is how we viewed it when we "went" from "integral" to "single term"), and the derivative of a constant is zero. Neither of these two things is what we want. 

We will avoid this dillema by the following trick: we have a function $\theta \colon \mathbb{N} \rightarrow G$, where $G$ is a set of Grassamnn numbers. As we know, that function is defined to be $\theta(i) = \theta_i$ Now, while $\theta_i$ is a constant, $\theta$ without index is \emph{not}. After all, if we go from $1$ to $2$, $\theta_1$ constinues to stay $\theta_1$, while $\theta$ changes from $\theta_1$ to $\theta_2$. Therefore, we are \emph{not} differentiating $\theta_i$. Instead, we are differentiating $\theta$, and thus obtain a "new function" $\partial \theta$. Now, since the derivative of a function is also a function, $\partial \theta$ has to be a function, too. If so, then, by definition, we should be able to "apply" the "function" $\partial \theta$ to anything we want. In particular, we can apply it to element $i$. The result we can denote by $(\partial \theta)_i$, which is to be contrasted from $\partial (\theta_i)$. 

Now, even though we are \emph{first} defining a function and \emph{then} applying it to $i$, it is often the case that whenever functions are defined, the definitions are given in terms of evaluating them at some general element. We will do the same now. Thus, we will define $\partial \theta$ \emph{in terms of} the expressions for $(\partial \theta)_i$. We will formally define these expressions as follows:
\beq (\partial_0 \eta)_{lc}= \frac{\alpha^{d/2}}{(2 \pi)^{d/2} \delta t}  \sum e^{- \frac{\alpha}{2} \vert \vec{x}_j - \vec{x}_l \vert^2} (\eta'_{jc} - \eta_{lc}) \eeq
\beq (\partial_i \eta)_{ld}  = \frac{\alpha^{(d+2)/2}}{(2 \pi)^{d/2}} \sum e^{- \frac{\alpha}{2} \vert \vec{x}_j - \vec{x}_l \vert^2} (x^i_j - x^i_l) (\eta_{jd} - \eta_{ld}), \label{eqn:FermiDer}\eeq
In the above expression, $l$ designates the number of a lattice point we are looking at (thus, it can be very large), while $c$ and $d$ designate spinor index; thus, they range from $1$ to $4$. Thus, strictly speaking, the above should be rewritten in terms of single-indexed $\theta$-s by the prescription in \ref{eqn:FermionSplit}.  But, for now, we will write them in multi-index form for our convenience. This does not compromise mathematical rigour since there is one to one correspondence between these two notations. In this notation, the Dirac Lagrangian can be written as
\beq {\cal L} = i \overline{\eta}_k  \wedge (\gamma^{\mu}(\partial_{\mu} + gA_{\mu}^a T_{ij}^a) \eta_k ) \label{eqn:DiscreteDirac}\eeq
Again, this is well defined from the mathematical point of view. The only question is "physics": will we get the expected solutions to Dirac equation without doubling problem or other issues? And this is the part we are not adressing in this paper. Now we have stated previously that differet components are viewed as "separate fields" despite $SU(2)$ symmetry. Therefore, in order to obtain explicit definition, we have to "isolate" each of these "separate fields" and substitute appropriate definition on their place. Thus, the obvious first step is to write everything in component form: 
\beq {\cal L} =\sum_{c=1}^4 \sum_{d=1}^4 ( i (\gamma^0 \gamma^{\mu})_{cd} \eta_{kc}^* \wedge (\partial_{\mu} \eta)_{kd} + igA^a_{\mu}T^a_{cd}  (\gamma^0 \gamma^{\mu})_{cd} \eta^*_{kc} \wedge \eta_{kd}) \eeq
Now, in case of $\mu=0$, $\gamma^0 \gamma^{\mu}$ becomes identity matrix. Thus, the above expression can be rewritten as
\beq {\cal L} = \sum_{c=1}^4 \sum_{d=1}^4 \Big( i \delta_c^d  \eta_{kc}^* \wedge (\partial_0 \eta)_{kd} +  i (\gamma^0 \gamma^k)_{cd} \eta_{kc}^* \wedge (\partial_k \eta)_{kd} + \nonumber \eeq
\beq + igA^a_{\mu}T^a_{cd}  (\gamma^0 \gamma^0)_{cd} \eta^*_{kc} \wedge \eta_{kd} +  igA^a_{\mu}T^a_{cd}  (\gamma^0 \gamma^k)_{cd} \eta^*_{kc} \wedge \eta_{kd} \Big) \eeq
Now, the derivatives, as defined in \ref{eqn:FermiDer}, are sums over the index of a lattice point. Thus, if we substitute the products of these derivatives, we obtain the double sum. Thus, we will separately sum over $j$ going from $1$ to $l$ and over $k$ going from $1$ to $l$, where $l$ is the total nubmer of points:
\beq {\cal L} =  \sum_{c=1}^4 \sum_{d=1}^4 \Big( \frac{i\alpha^{d/2}}{(2 \pi)^{d/2} \delta t}  \sum_{j=1}^l \sum_{k=1}^l e^{- \frac{\alpha}{2} \vert \vec{x}_j - \vec{x}_k \vert^2} \eta_{kd}^* \wedge  (\eta'_{jc} - \eta_{kc})  + \nonumber \eeq
\beq +   \frac{i(\gamma^0 \gamma^k)_{cd} \alpha^{(d+2)/2}}{(2 \pi)^{d/2}} \sum_{j=1}^l \sum_{k=1}^l  e^{- \frac{\alpha}{2} \vert \vec{x}_j - \vec{x}_k \vert^2} (x^i_j - x^i_k) \eta_{kc}^* \wedge (\eta_{jd} - \eta_{kd})+ \nonumber \eeq
\beq + igA^a_{\mu}T^a_{ij}  (\gamma^0 \gamma^0)_{ij} \sum_{k=1}^l \eta^*_{kc} \wedge \eta_{kd} +  igA^a_{\mu}T^a_{cd}  (\gamma^0 \gamma^k)_{cd} \sum_{k=1}^l \eta^*_{kc} \wedge \eta_{kd} \Big) \eeq
Now, if we substitute 
\beq \eta_{kc} = \eta_{kc1} + i \eta_{kc2} \eeq
then we notice that the above expression, up to the index permutation, splits into the three key terms: $\eta^*_{kd} \wedge \eta'_{jc}$, $\eta^*_{kd} \wedge \eta_{jd}$ and $\eta^*_{kd} \wedge \eta_{kc}$. Here $\eta'$ is different from $\eta$ in a sense that $\eta'$ "secretely refers to" the "transition to" $t + \delta t$, while $eta$ does not; and $\eta^*_{kd} \wedge \eta_{jd}$ is different from $\eta^*_{kd} \wedge \eta_{kc}$ because index $k$ does not match index $j$ in the former case, while index $j$ is replaced with "matching" index $k$ in the latter case. The explicit forms for the above three fermionic expressions are given by 
\beq \eta^*_{kc} \wedge \eta'_{jc}= \eta_{kc1} \wedge \eta'_{jc1} + \eta_{kc2} \wedge \eta'_{jc2} + i \eta_{kc1} \wedge \eta'_{jc2} - i \eta_{kc2} \wedge \eta'_{jc1} \label{eqn:wedge1}\eeq
\beq \eta^*_{kc} \wedge \eta_{jd}= \eta_{kc1} \wedge \eta_{jd1} + \eta_{kc2} \wedge \eta_{jd2} + i \eta_{kc1} \wedge \eta_{jd2} - i \eta_{kc2} \wedge \eta_{jd1} \label{eqn:wedge2} \eeq
\beq \eta^*_{kc} \wedge \eta_{kd}= \eta_{kc1} \wedge \eta_{kd1} + \eta_{kc2} \wedge \eta_{kd2} + i \eta_{kc1} \wedge \eta_{kd2} - i \eta_{kc2} \wedge \eta_{kd1} \label{eqn:wedge3} \eeq
Once again, we repeat, that we interpret the indexed components listed above in the following way:

1) $4$ values of $c$ = spinor indexes

2) $2$ time slices = $\eta$ vs $\eta'$ = $-1 \cdots 7$ vs $8 \cdots 15$

3) $2$ options: real ($-1, \cdots, 3$, $8, \cdots 11$) or imaginary ($4, \cdots, 7$, $12, \cdots, 15$)

In order to be able to write down the expression for the Lagrangian generator, we have to "convert" them into a single field $\theta$. For our convenience, we will re-state the conversion equation (\ref{eqn:conversion}) that we have written earlier:
\beq \eta_{kc1} = \theta_{l(c-1) +k} \; ; \; \eta_{kc2} = \theta_{l(c+3) +k} \; ; \; \eta'_{jc1} = \theta_{l(c+7)+j} \; ; \; \eta'_{jc2} = \theta_{l(c+11)+j}\eeq
We now rewrite the equations \ref{eqn:wedge1}, \ref{eqn:wedge2} and \ref{eqn:wedge3} with the above substitutions. For our convenience, we will leave the left hand sides in the original, $\eta$-based, form, while rewriting the right hand sides in a $\theta$-based one: 
\beq \eta^*_{kc} \wedge \eta'_{jc}= \theta_{l(c-1) +k} \wedge \theta_{l(c+7)+j} + \theta_{l(c+3) +k} \wedge \theta_{l(c+7)+j} + i \theta_{l(c-1) +k} \wedge \theta_{l(c+11)+j} - i \theta_{l(c+3) +k} \wedge  \theta_{l(c+7)+j} \nonumber \eeq
\beq \eta^*_{kc} \wedge \eta_{jd}= \theta_{l(c-1) +k} \wedge \theta_{l(d-1) +j} + \theta_{l(c+3) +k} \wedge \theta_{l(d+3) +j} + i \theta_{l(c-1) +k}\wedge \theta_{l(d+3) +j}- i \theta_{l(c+3) +k} \wedge \theta_{l(d-1) +j} \nonumber \eeq
\beq \eta^*_{kc} \wedge \eta_{kd}= \theta_{l(c-1) +k} \wedge \theta_{l(d-1) +k} + \theta_{l(c+3) +k} \wedge \theta_{l(d+3) +k} + i \theta_{l(c-1) +k}\wedge \theta_{l(d+3) +k}- i \theta_{l(c+3) +k} \wedge \theta_{l(d-1) +k} \nonumber \eeq
\beq \eta^*_{kc} \wedge \eta_{kc}= 2 i \theta_{l(c-1) +k}\wedge \theta_{l(c+3) +k}  \eeq
We can now substitute the above into our discretized Dirac equation, \ref{eqn:DiscreteDirac} and obtain
\beq {\cal L} =  \sum_{c=1}^4  \Big( \frac{i\alpha^{d/2}}{(2 \pi)^{d/2} \delta t}  \sum_{j=1}^l \sum_{k=1}^l e^{- \frac{\alpha}{2} \vert \vec{x}_j - \vec{x}_k \vert^2} (\theta_{l(c-1) +k} \wedge \theta_{l(c+7)+j} + \theta_{l(c+3) +k} \wedge \theta_{l(c+7)+j} + \nonumber \eeq
\beq +  i \theta_{l(c-1) +k} \wedge \theta_{l(c+11)+j} - i \theta_{l(c+3) +k} \wedge  \theta_{l(c+7)+j} - 2 i \theta_{l(c-1) +k}\wedge \theta_{l(c+3) +k}) \nonumber \eeq
\beq +   \frac{i(\gamma^0 \gamma^k)_{cd} \alpha^{(d+2)/2}}{(2 \pi)^{d/2}} \sum_{j=1}^l \sum_{k=1}^l e^{- \frac{\alpha}{2} \vert \vec{x}_j - \vec{x}_k \vert^2}(x_j^i-x_k^i) (   \theta_{l(c-1) +k} \wedge \theta_{l(d-1) +j} + \theta_{l(c+3) +k} \wedge \theta_{l(d+3) +j} + \nonumber \eeq
\beq +  i \theta_{l(c-1) +k}\wedge \theta_{l(d+3) +j}- i \theta_{l(c+3) +k} \wedge \theta_{l(d-1) +j} -  \theta_{l(c-1) +k} \wedge \theta_{l(d-1) +k} -  \label{ThetaBased1} \eeq
\beq - \theta_{l(c+3) +k} \wedge \theta_{l(d+3) +k} -  i \theta_{l(c-1) +k}\wedge \theta_{l(d+3) +k}+ i \theta_{l(c+3) +k} \wedge \theta_{l(d-1) +k} )+ \nonumber \eeq
\beq +( igA^a_{\mu}T^a_{ij}  (\gamma^0 \gamma^0)_{ij} +  igA^a_{\mu}T^a_{cd}  (\gamma^0 \gamma^k)_{cd} )( \theta_{l(c-1) +k} \wedge \theta_{l(d-1) +k} + \nonumber \eeq
\beq + \theta_{l(c+3) +k} \wedge \theta_{l(d+3) +k} + i \theta_{l(c-1) +k}\wedge \theta_{l(d+3) +k}- i \theta_{l(c+3) +k} \wedge \theta_{l(d-1) +k} ) \nonumber \eeq
Now, the quick inspection of \ref{ThetaBased1} tells us that each of our sums have only $l$ terms, despite the fact that we end up summing over $16 \, l$ indexes of $\theta$. This is because we have expressions of the form
\beq \sum_{j=0}^l \theta_{j+kl} = \sum_{j=kl}^{(k+1)l} \theta_j \label{Renumber1}\eeq
Now, since $k$ ranges from $0$ to $15$, the sum on the right hand side can either range from $0$ to $l$, or from $15 \, l$ to $16 \, l$, and anywhere in between. Now, in order to be able to come up with an expression that we can substitute into the results of sections 3.2.4 and 3.2.5, we would like to re-number indexes in such a way that we are only summing over $\theta_k \wedge \theta_j$ (as opposed to $\theta_{k+a} \wedge \theta_{j+b}$). Thus, we will rewrite everything in terms of the right hand side of the above expression. Furthermore, we would like to be able to combine all of the above sums into a single one. For this purpose, we would like to rewrite the above expression in terms of the sum from $0$ to $16 \, l$ in the following way:
\beq \sum_{j=kl}^{(k+1)l} \theta_j = \sum_{j=0}^{16 \, l} \theta_j \delta_{\vert j-kl \vert}^{j-kl} \delta_{\vert (k+1)l -j \vert}^{(k+1)-j} \label{Renumber2}\eeq
The coefficient $\delta_{\vert j-kl \vert}^{j-kl}$ will set to zero all coefficients corresponding to $j<kl$, whereas the coefficient $\delta_{\vert (k+1)l -j \vert}^{(k+1)-j}$ will set to zero all coefficients corresponding to $j>(k+1)l$. Thus, together, these two coefficients will limit our sum to the range $kl \leq j \leq (k+1)l$, despite the fact that "officially" we are summing over $0 \leq j \leq 16 \, l$. Now, the right hand side of \ref{Renumber1} is the same as the left hand side of \ref{Renumber2}. Now, equating the \emph{left} hand side of the former with \emph{right} hand side of the latter, we obtain 
\beq \sum_{j=0}^l \theta_{j+kl} = \sum_{j=0}^{16 \, l} \theta_j \delta_{\vert j-kl \vert}^{j-kl} \delta_{\vert (k+1)l -j \vert}^{(k+1)-j} \label{Renumber3}\eeq
Things will get a little more interesting when we include the position of the particles, $\vec{x}_k$, since in this case we will have to change the position index accordingly. According to \ref{Renumber1}, whatever we used to label as $j+kl$, we now label $j$. Therefore, whatever we used to label as $j$ will now have to be labeled as $j-kl$. Thus, $\vec{x}_j$ will be labeled as $\vec{x}_{j-kl}$:
\beq \sum_{j=0}^l \theta_{j+kl} f(\vec{x}_j)= \sum_{j=0}^{16 \, l} \theta_j f(\vec{x}_{j-kl}) \delta_{\vert j-kl \vert}^{j-kl} \delta_{\vert (k+1)l -j \vert}^{(k+1)-j} \label{Renumber3}\eeq
 \emph{But} if we are summing from $j=0$, it means that we will need to define $\vec{x}_{-kl}$ which, of course, makes no sense. Part of good news is that, due to the $\delta$-s, the coefficients in front of $j<kl$ will be set to zero. Therefore, we "might as well" choose $\vec{x}_{-kl}$ at random, since it won't contribute anything to the sum. There is, however, a better answer: As we have said previously, different fields (such as electron and neutrino) as well as different spinor components of the same field (such as spin up and spin down electron), are evaluated at different points from each other. The reason it "seems like" these points are the same is that they are grouped into "clusters". We can now formally state this by writing down that their $\vec{x}$-coordinates match up to certain approximation:
\beq \vec{x}_{j+kl} \approx \vec{x}_j \label{PositionIdentification}\eeq
From the above identification it is clear that we "first" go through "all spin-up electrons" throughout the "whole space" point by point (and there are $l$ of them); \emph{after that} we \emph{go back} to the point where we have started, and go through all spin-down electrons point by point \emph{in the same order} in which we did in case of spin up electrons, and so forth. This list likewise covers neutrino and other fermions we might have. Now, we can \emph{use} the equation \label{PositionIdentification} in order to \emph{change} the right hand side of \label{Renumber3} from $\vec{x}_{j-kl}$ \emph{back} to $\vec{x}_j$: 
\beq \vec{x}_{j+kl} \approx \vec{x}_j \Rightarrow \sum_{j=0}^l \theta_{j+kl} f(\vec{x}_j) \approx \sum_{j=0}^{16 \, l} \theta_j f(\vec{x}_{j-kl}) \delta_{\vert j-kl \vert}^{j-kl}  \delta_{\vert (k+1)l -j \vert}^{(k+1)-j} \label{Renumber4}\eeq
Now the generic element of \ref{ThetaBased1} changes under \ref{Renumber4} in the following way: 
\beq \sum_{k=0}^l \sum_{j=0}^l  \sum_{a=0}^{15} \sum_{b=0}^{15} e^{- \frac{\alpha}{2} \vert \vec{x}_j - \vec{x}_k \vert} \theta_{al+k} \wedge \theta_{bl+j} = \eeq
\beq = \sum_{k=1}^{16\, l} \Big((\theta_k \wedge \theta_j) e^{- \frac{\alpha}{2} \vert \vec{x}_j - \vec{x}_k \vert}  \delta_{\vert k - al \vert}^{k-al} \delta_{\vert (a+1)l-k \vert}^{(a+1)l-k} \delta_{\vert k - bl \vert}^{k-bl} \delta_{\vert b(l+1)-k \vert}^{b(l+1)-k} \Big) \nonumber \eeq
By going through the equation \ref{ThetaBased1} and making the above change term by term, we arrive at 
\beq {\cal L} =  \frac{i\alpha^{d/2}}{(2 \pi)^{d/2} \delta t}  \sum_{k=1}^{16\, l} e^{- \frac{\alpha}{2} \vert \vec{x}_j - \vec{x}_k \vert^2} \theta_k \wedge \theta_j \Big(\delta_{\vert k - (c-1)l \vert}^{k-(c-1)l} \delta_{\vert cl -k \vert}^{cl-k} \delta_{\vert (c+8)l-j \vert}^{(c+8)l-j} \delta_{\vert j- (c+7)l \vert}^{j-(c+7)l}+ \nonumber \eeq
\beq + \delta_{\vert k - (c+3)l \vert}^{k-(c+3)l} \delta_{\vert  (c+4)l -k \vert}^{(c+4)l-k}\delta_{\vert j - (c+7)l\vert}^{j- (c+7)l} \delta_{\vert (c+8)l-j \vert}^{(c+8)l-j}+  i\delta_{\vert k - (c-1)l \vert}^{k-(c-1)l} \delta_{\vert cl-k \vert}^{cl-k}\delta_{\vert (c+12)l-j \vert}^{(c+12)l-j} \delta_{\vert j- (c+11)l\vert}^{j-(c+11)l} - \nonumber \eeq
\beq - i \delta_{\vert k - (c+3)l \vert}^{k-(c+3)l} \delta_{\vert  (c+4)l -k \vert}^{(c+4)l-k}\delta_{\vert (c+8)l-j \vert}^{(c+8)l-j} \delta_{\vert j- (c+7)l\vert}^{j- (c+7)l}- 2 i \delta_{\vert k - (c-1)l \vert}^{k-(c-1)l} \delta_{\vert cl-k \vert}^{cl - k}\delta_{\vert k- (c+3)l \vert}^{k-(c+3)l} \delta_{\vert (c+4)l-k \vert}^{(c+4)l-k} \Big) \nonumber \eeq
\beq +   \frac{i(\gamma^0 \gamma^k)_{cd} \alpha^{(d+2)/2}}{(2 \pi)^{d/2}} \sum_{k=1}^{16l} \theta_k \wedge \theta_l  e^{- \frac{\alpha}{2} \vert \vec{x}_j - \vec{x}_k \vert^2}(x_j^i-x_k^i)  \Big(   \delta_{\vert k - (c-1)l \vert}^{k-(c-1)l} \delta_{\vert cl-  k \vert}^{cl- k}\delta_{\vert dl-j \vert}^{dl-j}  \delta_{\vert j- (d-1)l \vert}^{j-(d-1)l} +  \label{TwoPointDirac}\eeq
\beq +  \delta_{\vert k - (c+3)l \vert}^{k-(c+3)l} \delta_{\vert  (c+4)l -k \vert}^{(c+4)l-k}\delta_{\vert (d+4)l-j \vert}^{(d+4)l-j} \delta_{\vert j- (d+3)l \vert}^{j-(d+3)l}  +  i \delta_{\vert k - (c-1)l \vert}^{k-(c-1)l} \delta_{\vert cl-k\vert}^{cl-k}\delta_{\vert (d+4)l-j \vert}^{(d+4)l-j} \delta_{\vert j-(d+3)l \vert}^{j- (d+3)l}-  \nonumber \eeq
\beq - i \delta_{\vert k - (c+3)l \vert}^{k-(c+3)l} \delta_{\vert  (c+4)l - k\vert}^{(c+4)l-k}\delta_{\vert dl-j \vert}^{dl-j} \delta_{\vert j- (d-1)l\vert}^{j- (d-1)l} -  \delta_{\vert k - (c-1)l \vert}^{k-(c-1)l} \delta_{\vert cl-k \vert}^{cl-k}\delta_{\vert k- (d-1)l\vert}^{k- (d-1)l} \delta_{\vert dl-k \vert}^{dl-k}- \nonumber \eeq
\beq - (\delta_{\vert k - (c+3)l \vert}^{k-(c+3)l} (\delta_{\vert  (c+4)l -k\vert}^{(c+4)l-k}\delta_{\vert (d+4)l-k \vert}^{(d+4)l-k}\delta_{\vert k- (d+3)l\vert}^{k-(d+3)l}  -  i\delta_{\vert k - (c-1)l \vert}^{k-(c-1)l} \delta_{\vert cl-k \vert}^{cl-k}\delta_{\vert (d+4)l-k \vert}^{(d+4)l-k} \delta_{\vert k- (d+3)l \vert}^{k-(d+3)l} + \nonumber \eeq
\beq +  i \delta_{\vert k - (c+3)l \vert}^{k-(c+3)l} \delta_{\vert (c+4)l  -k\vert}^{(c+4)l-k}\delta_{\vert dl-k \vert}^{dl-k} \delta_{\vert k-(d-1)l \vert}^{k-(d-1)l} ) \Big)+ \nonumber \eeq
\beq +( igA^a_{\mu}T^a_{ij}  (\gamma^0 \gamma^0)_{ij} +  igA^a_{\mu}T^a_{cd}  (\gamma^0 \gamma^k)_{cd} )( \theta_{l(c-1) +k} \wedge \theta_{l(d-1) +k} + \nonumber \eeq
\beq + \theta_{l(c+3) +k} \wedge \theta_{l(d+3) +k} + i \theta_{l(c-1) +k}\wedge \theta_{l(d+3) +k}- i \theta_{l(c+3) +k} \wedge \theta_{l(d-1) +k} ) \nonumber \eeq
Now, most of the terms in the above equation take a form of a double sum, just like we want them to have. But, the lat few terms at the end don't involve the sumation. In order to be able to formally identify a "Lagrangian generator", we would like \emph{everything} to be expressed in terms of the sum. This situation is similar to what we have encountered in bosonic case, when we had different number of integrals for different terms and we wanted the number of integrals to be the same. We will, therefore, use the same trick to adress the issue as we did back then. Namely, we multiply our expression by a "unit" defined as 
\beq f (k, k ) \rightarrow \frac{VMA^{d/2}}{(2 \pi)^{d/2}N}\sum_j e^{- \frac{A}{2} \vert \vec{x}_k - \vec{x}_j \vert^2} f(k, j) \; , \; A \gg \alpha \eeq
Here, $V$ is the volume in the universe, $N$ is the number of particles and $M$ is the number of available "charges", which means that $VM/N$ is a "volume element" of our "integral", as was explained in detail for bosonic case. By performing the multiplication by the "unit" above, we can rewrite a generic wedge product as
\beq \theta_{al+k} \wedge \theta_{bl+k} =  \frac{A^{d/2}}{(2 \pi)^{d/2}} \sum_{k=1}^{16l} \sum_{j=1}^{16l} (\theta_k \wedge \theta_j) e^{- \frac{A}{2} \vert \vec{x}_j - \vec{x}_k \vert^2} \delta_{\vert k - al \vert}^{k-al} \delta_{\vert l(a+1)-k \vert}^{l(a+1)-k}\delta_{\vert l(b+1)-j \vert}^{l(b+1)-j}\delta_{\vert j-lb\vert}^{j-lb} \label{ThreePointWedge} \eeq
We can now take the terms in \ref{TwoPointDirac} that we "don't like", and multiply them by the "unit" given \ref{ThreePointWedge}. This produces the expression 
\beq {\cal L} = \frac{iA^{d/2}}{(2 \pi)^{d/2} \delta t} \sum_{c=1}^4   \sum_{k=1}^{16l} \sum_{j=1}^{16l} e^{- \frac{A}{2} \vert \vec{x}_j - \vec{x}_k \vert^2} \Big( - 2 i \delta_{\vert k - (c-1)l \vert}^{k-(c-1)l} \delta_{\vert cl-k \vert}^{cl - k}\delta_{\vert k- (c+3)l \vert}^{k-(c+3)l} \delta_{\vert (c+4)l-k \vert}^{(c+4)l-k} \Big) +\nonumber \eeq
\beq + \frac{i\alpha^{d/2}}{(2 \pi)^{d/2} \delta t} \sum_{c=1}^4   \sum_{k=1}^{16l} \sum_{j=1}^{16l}  e^{- \frac{\alpha}{2} \vert \vec{x}_j - \vec{x}_k \vert^2} \theta_k \wedge \theta_j \Big(\delta_{\vert k - (c-1)l \vert}^{k-(c-1)l} \delta_{\vert cl -k \vert}^{cl-k} \delta_{\vert (c+8)l-j \vert}^{(c+8)l-j} \delta_{\vert j- (c+7)l \vert}^{j-(c+7)l}+ \nonumber \eeq
\beq + \delta_{\vert k - (c+3)l \vert}^{k-(c+3)l} \delta_{\vert  (c+4)l -k \vert}^{(c+4)l-k}\delta_{\vert j - (c+7)l\vert}^{j- (c+7)l} \delta_{\vert (c+8)l-j \vert}^{(c+8)l-j}+  i\delta_{\vert k - (c-1)l \vert}^{k-(c-1)l} \delta_{\vert cl-k \vert}^{cl-k}\delta_{\vert (c+12)l-j \vert}^{(c+12)l-j} \delta_{\vert j- (c+11)l\vert}^{j-(c+11)l} - \nonumber \eeq
\beq - i \delta_{\vert k - (c+3)l \vert}^{k-(c+3)l} \delta_{\vert  (c+4)l -k \vert}^{(c+4)l-k}\delta_{\vert (c+8)l-j \vert}^{(c+8)l-j} \delta_{\vert j- (c+7)l\vert}^{j- (c+7)l}\Big) + \eeq
\beq +  \sum_{c=1}^4 \sum_{d=1}^4 \Big( \frac{i(\gamma^0 \gamma^k)_{cd} \alpha^{(d+2)/2}}{(2 \pi)^{d/2}} \sum_{k=1}^{16l} \sum_{j=1}^{16l} \theta_k \wedge \theta_j e^{- \frac{\alpha}{2} \vert \vec{x}_j - \vec{x}_k \vert^2}(x_j^i-x_k^i)  \Big(   \delta_{\vert k - (c-1)l \vert}^{k-(c-1)l} \delta_{\vert cl-  k \vert}^{cl- k}\delta_{\vert dl-j \vert}^{dl-j}  \delta_{\vert j- (d-1)l \vert}^{j-(d-1)l} + \nonumber \eeq
\beq +  \delta_{\vert k - (c+3)l \vert}^{k-(c+3)l} \delta_{\vert  (c+4)l -k \vert}^{(c+4)l-k}\delta_{\vert (d+4)l-j \vert}^{(d+4)l-j} \delta_{\vert j- (d+3)l \vert}^{j-(d+3)l}  +  i \delta_{\vert k - (c-1)l \vert}^{k-(c-1)l} \delta_{\vert cl-k\vert}^{cl-k}\delta_{\vert (d+4)l-j \vert}^{(d+4)l-j} \delta_{\vert j-(d+3)l \vert}^{j- (d+3)l}-  \nonumber \eeq
\beq - i \delta_{\vert k - (c+3)l \vert}^{k-(c+3)l} \delta_{\vert  (c+4)l - k\vert}^{(c+4)l-k}\delta_{\vert dl-j \vert}^{dl-j} \delta_{\vert j- (d-1)l\vert}^{j- (d-1)l} + \nonumber \eeq
 \beq +   \frac{i(\gamma^0 \gamma^k)_{cd} A^{(d+2)/2}}{(2 \pi)^{d/2}} \sum_{k=1}^{16l} \sum_{j=1}^{16l} \theta_k \wedge \theta_j e^{- \frac{A}{2} \vert \vec{x}_j - \vec{x}_k \vert^2}(x_j^i-x_k^i)  \Big( - \delta_{\vert k - (c-1)l \vert}^{k-(c-1)l} \delta_{\vert cl-k \vert}^{cl-k}\delta_{\vert k- (d-1)l\vert}^{k- (d-1)l} \delta_{\vert dl-k \vert}^{dl-k}- \nonumber \eeq
\beq - (\delta_{\vert k - (c+3)l \vert}^{k-(c+3)l} (\delta_{\vert  (c+4)l -k\vert}^{(c+4)l-k}\delta_{\vert (d+4)l-k \vert}^{(d+4)l-k}\delta_{\vert k- (d+3)l\vert}^{k-(d+3)l}  -  i\delta_{\vert k - (c-1)l \vert}^{k-(c-1)l} \delta_{\vert cl-k \vert}^{cl-k}\delta_{\vert (d+4)l-k \vert}^{(d+4)l-k} \delta_{\vert k- (d+3)l \vert}^{k-(d+3)l} + \nonumber \eeq
\beq +  i \delta_{\vert k - (c+3)l \vert}^{k-(c+3)l} \delta_{\vert (c+4)l  -k\vert}^{(c+4)l-k}\delta_{\vert dl-k \vert}^{dl-k} \delta_{\vert k-(d-1)l \vert}^{k-(d-1)l} ) \Big)+ \nonumber \eeq
\beq + \frac{A^{d/2}}{(2 \pi)^{d/2}} \sum_{k=1}^{16l} \sum_{j=1}^{16l}( igA^a_{\mu}T^a_{ij}  (\gamma^0 \gamma^0)_{ij} +  igA^a_{\mu}T^a_{cd}  (\gamma^0 \gamma^k)_{cd} ) (\theta_k \wedge \theta_j) e^{- \frac{A}{2} \vert \vec{x}_j - \vec{x}_k \vert^2} \times \nonumber \eeq
\beq \times \Big( \delta_{\vert k - cl \vert}^{k-cl} \delta_{\vert lc-k \vert}^{lc-k}\delta_{\vert ld-j \vert}^{ld-j}\delta_{\vert j-l(d-1)\vert}^{j-l(d-1)} + \delta_{\vert k - l(c+3) \vert}^{k-l(c+3)} \delta_{\vert l(c+4)-k \vert}^{l(c+4)-k}\delta_{\vert l(d+4)-j \vert}^{l(d+4)-j}\delta_{\vert j-l(d+3)\vert}^{j-l(d+3)} +  \nonumber \eeq
\beq +\delta_{\vert k - l(c-1) \vert}^{k-al} \delta_{\vert lc-k \vert}^{l(a+1)-k}\delta_{\vert l(d+4)-j \vert}^{l(d+4)-j}\delta_{\vert j-l(d+3)\vert}^{j-l(d+3)} -  \delta_{\vert k - l(c+3) \vert}^{k-l(c+3)} \delta_{\vert l(c+4)-k \vert}^{l(c+4)-k}\delta_{\vert ld-j \vert}^{ld-j}\delta_{\vert j-l(d-1)\vert}^{j-l(d-1)} \Big) \Big) \nonumber \eeq
Now, in order to write down the Lagrangian generator, we will do the following two steps:

1) Compare the above with \ref{J} and "read off" $J$

2) Substitute the value of $J$ into \ref{eqn:FermionGenerator} and this will be our definiton of Lagrangian generator specific for fermions.

Quick look at \ref{J} makes it obvious that the first step merely amounts to removing the "summation over $j$ and $k$" as well as the Grassmann products $\theta_k \wedge \theta_j$ from the expression for $\cal L$, and then identifying what is "left" with $\cal J$.  Thus, we obtain
\beq J= \frac{iA^{d/2}}{(2 \pi)^{d/2} \delta t} \sum_{c=1}^4    e^{- \frac{A}{2} \vert \vec{x}_j - \vec{x}_k \vert^2} \Big( - 2 i \delta_{\vert k - (c-1)l \vert}^{k-(c-1)l} \delta_{\vert cl-k \vert}^{cl - k}\delta_{\vert k- (c+3)l \vert}^{k-(c+3)l} \delta_{\vert (c+4)l-k \vert}^{(c+4)l-k} \Big) +\nonumber \eeq
\beq + \frac{i\alpha^{d/2}}{(2 \pi)^{d/2} \delta t} \sum_{c=1}^4   e^{- \frac{\alpha}{2} \vert \vec{x}_j - \vec{x}_k \vert^2}  \Big(\delta_{\vert k - (c-1)l \vert}^{k-(c-1)l} \delta_{\vert cl -k \vert}^{cl-k} \delta_{\vert (c+8)l-j \vert}^{(c+8)l-j} \delta_{\vert j- (c+7)l \vert}^{j-(c+7)l}+ \nonumber \eeq
\beq + \delta_{\vert k - (c+3)l \vert}^{k-(c+3)l} \delta_{\vert  (c+4)l -k \vert}^{(c+4)l-k}\delta_{\vert j - (c+7)l\vert}^{j- (c+7)l} \delta_{\vert (c+8)l-j \vert}^{(c+8)l-j}+  i\delta_{\vert k - (c-1)l \vert}^{k-(c-1)l} \delta_{\vert cl-k \vert}^{cl-k}\delta_{\vert (c+12)l-j \vert}^{(c+12)l-j} \delta_{\vert j- (c+11)l\vert}^{j-(c+11)l} - \nonumber \eeq
\beq - i \delta_{\vert k - (c+3)l \vert}^{k-(c+3)l} \delta_{\vert  (c+4)l -k \vert}^{(c+4)l-k}\delta_{\vert (c+8)l-j \vert}^{(c+8)l-j} \delta_{\vert j- (c+7)l\vert}^{j- (c+7)l}\Big) + \label{Final1}\eeq
\beq +  \sum_{c=1}^4 \sum_{d=1}^4 \Big( \frac{i(\gamma^0 \gamma^k)_{cd} \alpha^{(d+2)/2}}{(2 \pi)^{d/2}}  e^{- \frac{\alpha}{2} \vert \vec{x}_j - \vec{x}_k \vert^2}(x_j^i-x_k^i)  \Big(   \delta_{\vert k - (c-1)l \vert}^{k-(c-1)l} \delta_{\vert cl-  k \vert}^{cl- k}\delta_{\vert dl-j \vert}^{dl-j}  \delta_{\vert j- (d-1)l \vert}^{j-(d-1)l} + \nonumber \eeq
\beq +  \delta_{\vert k - (c+3)l \vert}^{k-(c+3)l} \delta_{\vert  (c+4)l -k \vert}^{(c+4)l-k}\delta_{\vert (d+4)l-j \vert}^{(d+4)l-j} \delta_{\vert j- (d+3)l \vert}^{j-(d+3)l}  +  i \delta_{\vert k - (c-1)l \vert}^{k-(c-1)l} \delta_{\vert cl-k\vert}^{cl-k}\delta_{\vert (d+4)l-j \vert}^{(d+4)l-j} \delta_{\vert j-(d+3)l \vert}^{j- (d+3)l}-  \nonumber \eeq
\beq - i \delta_{\vert k - (c+3)l \vert}^{k-(c+3)l} \delta_{\vert  (c+4)l - k\vert}^{(c+4)l-k}\delta_{\vert dl-j \vert}^{dl-j} \delta_{\vert j- (d-1)l\vert}^{j- (d-1)l} + \nonumber \eeq
 \beq +   \frac{i(\gamma^0 \gamma^k)_{cd} A^{(d+2)/2}}{(2 \pi)^{d/2}}  e^{- \frac{A}{2} \vert \vec{x}_j - \vec{x}_k \vert^2}(x_j^i-x_k^i)  \Big( - \delta_{\vert k - (c-1)l \vert}^{k-(c-1)l} \delta_{\vert cl-k \vert}^{cl-k}\delta_{\vert k- (d-1)l\vert}^{k- (d-1)l} \delta_{\vert dl-k \vert}^{dl-k}- \nonumber \eeq
\beq - (\delta_{\vert k - (c+3)l \vert}^{k-(c+3)l} (\delta_{\vert  (c+4)l -k\vert}^{(c+4)l-k}\delta_{\vert (d+4)l-k \vert}^{(d+4)l-k}\delta_{\vert k- (d+3)l\vert}^{k-(d+3)l}  -  i\delta_{\vert k - (c-1)l \vert}^{k-(c-1)l} \delta_{\vert cl-k \vert}^{cl-k}\delta_{\vert (d+4)l-k \vert}^{(d+4)l-k} \delta_{\vert k- (d+3)l \vert}^{k-(d+3)l} + \nonumber \eeq
\beq +  i \delta_{\vert k - (c+3)l \vert}^{k-(c+3)l} \delta_{\vert (c+4)l  -k\vert}^{(c+4)l-k}\delta_{\vert dl-k \vert}^{dl-k} \delta_{\vert k-(d-1)l \vert}^{k-(d-1)l} ) \Big)+ \nonumber \eeq
\beq + \frac{A^{d/2}}{(2 \pi)^{d/2}} ( igA^a_{\mu}T^a_{ij}  (\gamma^0 \gamma^0)_{ij} +  igA^a_{\mu}T^a_{cd}  (\gamma^0 \gamma^k)_{cd} ) (\theta_k \wedge \theta_j) e^{- \frac{A}{2} \vert \vec{x}_j - \vec{x}_k \vert^2} \times \nonumber \eeq
\beq \times \Big( \delta_{\vert k - cl \vert}^{k-cl} \delta_{\vert lc-k \vert}^{lc-k}\delta_{\vert ld-j \vert}^{ld-j}\delta_{\vert j-l(d-1)\vert}^{j-l(d-1)} + \delta_{\vert k - l(c+3) \vert}^{k-l(c+3)} \delta_{\vert l(c+4)-k \vert}^{l(c+4)-k}\delta_{\vert l(d+4)-j \vert}^{l(d+4)-j}\delta_{\vert j-l(d+3)\vert}^{j-l(d+3)} +  \nonumber \eeq
\beq +\delta_{\vert k - l(c-1) \vert}^{k-al} \delta_{\vert lc-k \vert}^{l(a+1)-k}\delta_{\vert l(d+4)-j \vert}^{l(d+4)-j}\delta_{\vert j-l(d+3)\vert}^{j-l(d+3)} -  \delta_{\vert k - l(c+3) \vert}^{k-l(c+3)} \delta_{\vert l(c+4)-k \vert}^{l(c+4)-k}\delta_{\vert ld-j \vert}^{ld-j}\delta_{\vert j-l(d-1)\vert}^{j-l(d-1)} \Big) \Big) \nonumber \eeq
Now, we will plug the result of \ref{Final1} into the equation  \ref{eqn:FermionGenerator}. Since the expression is quite long, we will make it easier for the reader by moving ${\cal J}$ term of \ref{Final1} towards the very bottom. Thus, our expression will take a form
\beq {\cal K} = ({\rm several \; terms})_1 (N_p^{-1}, N_p^{-2}, N_p^{-3}) + iN_p^{-1} \ln ({\rm several \; terms})_2 \nonumber \eeq
\beq J = ({\rm several \; terms} )_2 \eeq
Now, as we said earlier, all of the specific physics, including $\gamma$-matrices, is "hidden" inside $J$, while the terms outside of $J$ are generic to \emph{any} kind of Grassmann integration, and, therefore, lack $\gamma$-matrixes or any other "physical" information. Now, since $J$ will be under logarithm, one will find $\gamma$-matrixes under logarithm as well, while none of them will be present outside of the logarithm. Let us finally write that expression:

\beq {\cal K} (A^{\mu} (j_1), \phi (j_1), {\cal E} (j_1), {\cal E}' (j_1); A^{\mu} (j_2), \phi (j_2), {\cal E}^{\prime}; A^{\mu} (j_2), \phi (j_2), {\cal E} (j_2), {\cal E}' (j_2); A^{\mu} (j_3), \phi (j_3), {\cal E} (j_3), {\cal E}' (j_3)) =  \nonumber \eeq
\beq = - \pi N^{-1} \delta^{j_1-j_2}_{\vert j_1-j_2 \vert} (\delta_{a_{I (\mu_{I_1}, \mu_{I_2})}}^{j_1}+ \delta_{b_{I (\mu_{I_1}, \mu_{I_2})}}^{j_1}) {\cal E}' (j_2)  - \pi N^{-1} \delta^{j_1-j_2}_{\vert j_1-j_2 \vert} {\cal E}' (j_2)  + i \pi N^{-1} \delta_{\vert j_1 - j_2 \vert}^{j_1 - j_2} \delta^a_{a_k} \delta^b_{b_k}+ \nonumber \eeq
\beq +   i N_p^{-2} \ln (1- ({\cal E}' (j_1) - {\cal E} (j_1))^2) + i N^{-2} \ln (1 - ((\delta_{a_{I (\mu_{I_1}, \mu_{I_2})}}^{j_1} + \delta_{b_{I (\mu_{I_1}, \mu_{I_2})}}^{j_1}) {\cal E}' (j_1) - {\cal E} (j_1))^2)  + \nonumber \eeq
\beq + i N^{-2} \ln {\cal E}' (j_1)  + i N^{-2} \ln (1 - (\delta_{a_{I (\mu_{I_1}, \mu_{I_2})}}^{j_1} + \delta_{b_{I (\mu_{I_1}, \mu_{I_2})}}^{j_1}) {\cal E}'(j_1))  + i N^{-2}\ln (1- {\cal E} (j_1)) + N^{-3} \pi + 2 i N_p^{-3} \; \ln \delta^k_{n-1}  \nonumber \eeq
\beq + iN^{-1} \delta_{a_k}^{j_1} \delta_{b_k}^{j_2} \ln \Bigg[\frac{iA^{d/2}}{(2 \pi)^{d/2} \delta t} \sum_{c=1}^4    e^{- \frac{A}{2} \vert \vec{x}_j - \vec{x}_k \vert^2} \Big( - 2 i \delta_{\vert k - (c-1)l \vert}^{k-(c-1)l} \delta_{\vert cl-k \vert}^{cl - k}\delta_{\vert k- (c+3)l \vert}^{k-(c+3)l} \delta_{\vert (c+4)l-k \vert}^{(c+4)l-k} \Big) +\nonumber \eeq
\beq + \frac{i\alpha^{d/2}}{(2 \pi)^{d/2} \delta t} \sum_{c=1}^4   e^{- \frac{\alpha}{2} \vert \vec{x}_j - \vec{x}_k \vert^2} \Big(\delta_{\vert k - (c-1)l \vert}^{k-(c-1)l} \delta_{\vert cl -k \vert}^{cl-k} \delta_{\vert (c+8)l-j \vert}^{(c+8)l-j} \delta_{\vert j- (c+7)l \vert}^{j-(c+7)l}+ \nonumber \eeq
\beq + \delta_{\vert k - (c+3)l \vert}^{k-(c+3)l} \delta_{\vert  (c+4)l -k \vert}^{(c+4)l-k}\delta_{\vert j - (c+7)l\vert}^{j- (c+7)l} \delta_{\vert (c+8)l-j \vert}^{(c+8)l-j}+  i\delta_{\vert k - (c-1)l \vert}^{k-(c-1)l} \delta_{\vert cl-k \vert}^{cl-k}\delta_{\vert (c+12)l-j \vert}^{(c+12)l-j} \delta_{\vert j- (c+11)l\vert}^{j-(c+11)l} - \nonumber \eeq
\beq - i \delta_{\vert k - (c+3)l \vert}^{k-(c+3)l} \delta_{\vert  (c+4)l -k \vert}^{(c+4)l-k}\delta_{\vert (c+8)l-j \vert}^{(c+8)l-j} \delta_{\vert j- (c+7)l\vert}^{j- (c+7)l}\Big) + \label{Final1}\eeq
\beq +  \sum_{c=1}^4 \sum_{d=1}^4 \Big( \frac{i(\gamma^0 \gamma^k)_{cd} \alpha^{(d+2)/2}}{(2 \pi)^{d/2}} e^{- \frac{\alpha}{2} \vert \vec{x}_j - \vec{x}_k \vert^2}(x_j^i-x_k^i)  \Big(   \delta_{\vert k - (c-1)l \vert}^{k-(c-1)l} \delta_{\vert cl-  k \vert}^{cl- k}\delta_{\vert dl-j \vert}^{dl-j}  \delta_{\vert j- (d-1)l \vert}^{j-(d-1)l} + \nonumber \eeq
\beq +  \delta_{\vert k - (c+3)l \vert}^{k-(c+3)l} \delta_{\vert  (c+4)l -k \vert}^{(c+4)l-k}\delta_{\vert (d+4)l-j \vert}^{(d+4)l-j} \delta_{\vert j- (d+3)l \vert}^{j-(d+3)l}  +  i \delta_{\vert k - (c-1)l \vert}^{k-(c-1)l} \delta_{\vert cl-k\vert}^{cl-k}\delta_{\vert (d+4)l-j \vert}^{(d+4)l-j} \delta_{\vert j-(d+3)l \vert}^{j- (d+3)l}-  \nonumber \eeq
\beq - i \delta_{\vert k - (c+3)l \vert}^{k-(c+3)l} \delta_{\vert  (c+4)l - k\vert}^{(c+4)l-k}\delta_{\vert dl-j \vert}^{dl-j} \delta_{\vert j- (d-1)l\vert}^{j- (d-1)l} + \nonumber \eeq
 \beq +   \frac{i(\gamma^0 \gamma^k)_{cd} A^{(d+2)/2}}{(2 \pi)^{d/2}} e^{- \frac{A}{2} \vert \vec{x}_j - \vec{x}_k \vert^2}(x_j^i-x_k^i)  \Big( - \delta_{\vert k - (c-1)l \vert}^{k-(c-1)l} \delta_{\vert cl-k \vert}^{cl-k}\delta_{\vert k- (d-1)l\vert}^{k- (d-1)l} \delta_{\vert dl-k \vert}^{dl-k}- \nonumber \eeq
\beq - (\delta_{\vert k - (c+3)l \vert}^{k-(c+3)l} (\delta_{\vert  (c+4)l -k\vert}^{(c+4)l-k}\delta_{\vert (d+4)l-k \vert}^{(d+4)l-k}\delta_{\vert k- (d+3)l\vert}^{k-(d+3)l}  -  i\delta_{\vert k - (c-1)l \vert}^{k-(c-1)l} \delta_{\vert cl-k \vert}^{cl-k}\delta_{\vert (d+4)l-k \vert}^{(d+4)l-k} \delta_{\vert k- (d+3)l \vert}^{k-(d+3)l} + \nonumber \eeq
\beq +  i \delta_{\vert k - (c+3)l \vert}^{k-(c+3)l} \delta_{\vert (c+4)l  -k\vert}^{(c+4)l-k}\delta_{\vert dl-k \vert}^{dl-k} \delta_{\vert k-(d-1)l \vert}^{k-(d-1)l} ) \Big)+ \nonumber \eeq
\beq + \frac{A^{d/2}}{(2 \pi)^{d/2}} ( igA^a_{\mu}T^a_{ij}  (\gamma^0 \gamma^0)_{ij} +  igA^a_{\mu}T^a_{cd}  (\gamma^0 \gamma^k)_{cd} ) (\theta_k \wedge \theta_j) e^{- \frac{A}{2} \vert \vec{x}_j - \vec{x}_k \vert^2} \times \nonumber \eeq
\beq \times \Big( \delta_{\vert k - cl \vert}^{k-cl} \delta_{\vert lc-k \vert}^{lc-k}\delta_{\vert ld-j \vert}^{ld-j}\delta_{\vert j-l(d-1)\vert}^{j-l(d-1)} + \delta_{\vert k - l(c+3) \vert}^{k-l(c+3)} \delta_{\vert l(c+4)-k \vert}^{l(c+4)-k}\delta_{\vert l(d+4)-j \vert}^{l(d+4)-j}\delta_{\vert j-l(d+3)\vert}^{j-l(d+3)} +  \nonumber \eeq
\beq +\delta_{\vert k - l(c-1) \vert}^{k-al} \delta_{\vert lc-k \vert}^{l(a+1)-k}\delta_{\vert l(d+4)-j \vert}^{l(d+4)-j}\delta_{\vert j-l(d+3)\vert}^{j-l(d+3)} -  \delta_{\vert k - l(c+3) \vert}^{k-l(c+3)} \delta_{\vert l(c+4)-k \vert}^{l(c+4)-k}\delta_{\vert ld-j \vert}^{ld-j}\delta_{\vert j-l(d-1)\vert}^{j-l(d-1)} \Big) \Big) \Bigg] \nonumber \eeq  
The above expression "goes together" with the conditions that we have imposed in \ref{Conditions}, which, for convenience of the reader, will be restated here:
\beq  \ln 0 = N_{ln} \in \mathbb{R}\; ; \; \exp (z_1) = z_2 \Rightarrow^{def} \ln z_2 = \frac{z_1}{\vert z_1 \vert} \min (\vert z_1 \vert, N_{\ln}) \nonumber \eeq
\beq {\cal E} (a) \in \{0, 1 \} \; ; \; {\cal E}' (a) \in \{0, 1 \} \nonumber \eeq
\beq \frac{de_{I1}}{dt} = \frac{\omega e_{I2}}{\sqrt{e_{I1}^2 + e_{I2}^2}} \; ; \; \frac{de_{I_2}}{dt} = -\frac{\omega e_{I1}}{\sqrt{e_{I1}^2 + e_{I2}^2}}  \nonumber \eeq
\beq \nabla_s^{\alpha} \nabla_{s \alpha} \mu_{I_1} + m_I^2 \mu_{I_1} = e_{I_1} \delta^3 (\vec{x} - \vec{x}_I) e_{I_1}   \eeq
\beq \nabla_s^{\alpha} \nabla_{s \alpha} \mu_{I_2} + m_I^2 \mu_{I_2} = e_{I_1} \delta^3 (\vec{x} - \vec{x}_I) e_{I_2} \nonumber \eeq
\beq k (\mu_{I_1}, \mu_{I_2}) = -1 + \sum l T \Big( l - \epsilon < C \tan^{-1} \frac{\mu_{I_2}}{\mu_{I_1}} < l + \epsilon \Big) \nonumber \eeq
\beq T({\rm true})= 1 \; ; \; T({\rm false}) = 0 \nonumber \eeq
As one can see from the quick inspection of the above expression, there are no Grassmann numbers involved. In other words, our model of "fermions" is "bosonic"! There are few key differences between the ordinary bosonic expressions and the one we have produced:

a) In the "ordinary" bosonic case, the Lagrangian is real valued, whereas now it is complex valued

b) We have "very large" (but finite) terms in our expression, such as $\ln 0$, together with its "large but finite" definition

c) We have a lot of $\delta$-s in our expression

d) The "bosonic" field $J$ is restricted to the values of $0$ or $1$

As far as part $d$ is concerned, we could have replaced $J$ by some function over continuous parameter with values $0$ or $1$. The main reason we didn't do it is that it seems like it is "better" rather than "worse" to make smaller domain; but in principle we could have done it for aesthetic reasons. Also, again from aesthetic point of view, we could have used $\tan^{-1}$ with a "very large" coefficiet in place of all of the $\delta$-s; and, likewise, we could have identified $J$ with $\tan^{-1}$ of some continuous field, with similarly large coefficient. 

The issues that \emph{can not} be gotten rid of are $a$ and $b$. Both of these issues are problematic because they involve things that were "never observed" in the "real world" situations. In particular, complex-valued Lagrangian has never been used for bosons. Similarly, the "very large" constant hasn't been observed (after all, if it was, we would have been able to assign to it the exact value instead of saying it is "very large"). Nevertheless, since the nature of the research presented in this paper is speculative, introducing the above is approrpiate in this particular context.

\subsection*{3.2.8 Which time scales are smaller ones: Section 3.2 or Chapter 4?}

Since it is now a final section of Chapter 3, it is important to warn a reader not to confuse what has been done here with what we are about to do in Chapter 4. In both cases, we are using similar technique: we propose "very high frequency" processes in order to "explain" seemingly counterintuitive things that are being observed on a larger scales. However, the problems we are tackling are very different. In Chapter 3.2 we are focused on turning "quantum fermions" into "quantum bosons", while in Chapter 4 we are focused on turing "quantum bosons" into "classical ones". Thus, the question is: do these two chapters interfere in any way?

Now, the reason why Chapter 4 is exclusively focused on bosons is that we assume that, due to Sections 3.2.5 and 3.2.6,  we "already agree" that fermions are "bosons" as well. In principle, we \emph{could have} went another route. In particular, if we are to remove Section  3.2.5 while retaining sections 3.2.1--3.2.4, we would have had "true fermions" as a well defined mathematical objects. In this case, we would have redone 3.2.6 in a "fermionic" way, and then we would have written a "fermionic" part of Chapter 4 that accommodates the "fermionic" version of 3.2.6. But, as far as this paper is concerned, we have chosen the "rout" of "converting" fermions into bosons per Section 3.2.5. That is why it is only "logical" that Chapter 4 is focused exculsively on bosons. If Chapter 4 were to have fermionic part, the Lagrangian generator proposed in Section 3.2.6 would have been a "nuisace information" that never gets used. 

Now, since we have decided to take the rout of retaining Section 3.2.6 and skipping the fermionic part of Chapter 4, we are implicily saying that fermions are subject to \emph{two different} kinds of "processing". First, they are "bosonified" per Section 3.2.6, and then they are "thrown together with other bosons" into "further processing" of Chapter 4. \emph{But} if \emph{both} chapters have "short time scales" in them, can either of them continue to be viewed as "short", given that they are not necesserely "short" with respect to each other? This, of course, is closely related to the question which of these two time scales do we expect to be shorter?  In fact, in both chapters we introduced "rotating parameters". On the one hand, according to equation \ref{eqn:270} in Chapter 3.2, we have
\beq {\rm Chapter \; 3.2 :} \; \frac{de_{I1}}{dt} = \frac{\omega e_{I2}}{\sqrt{e_{I1}^2 + e_{I2}^2}} \; ; \; \frac{de_{I_2}}{dt} = -\frac{\omega e_{I1}}{\sqrt{e_{I1}^2 + e_{I2}^2}} \eeq
On the other hand, according to the equation \ref{eqn:459} in Chapter 4, we have 
\beq {\rm Chapter \; 4:} \frac{de_{j1}}{d t} = \frac{\nu e_{j2}}{\sqrt{e_{j1}^2+e_{j2}^2}} \; ; \; \frac{de_{j2}}{d t} = - \frac{\nu e_{j1}}{\sqrt{e_{j1}^2+e_{j2}^2}}  \eeq
In both cases we agree that the respective frequencies ($\omega$ in Chapter 3.2 and $\nu$ in Chapter 4) are "very large". But \emph{which} of these two frequencies is larger? 

 The quick answer to this question is that the time scales of the cyclical processes in chapter 4 are much shorter than the scales of Grassmann integration in Chapter 3.2 (and, therefore, $\nu \gg \omega$). After all, Chapter 4 assumes that everything is "bosonic". Now, according to Chapter 3.2, the fermions only "look bosonic" on "very small" scales. Thus, \emph{unless} we stick to \emph{much smaller} scales in Chapter 4, we would \emph{not} be able to assume that fermions have bosonic nature and, therefore, our argument would fail. Of course, it might be possible to design another "version" of Chapter 4, specifically for fermions; in this case, we would not need to make this assumption. But, as far as this paper is concerned, Chapter 4 is "geared" only to bosons for the reasons we have stated in previous paragraph. Thus, as long as Chapter 4 makes this assumption, its time scales should be "much smaller" than any and every "small" time scales mentioned in Chapter 3.2.

Therefore, \emph{during} the "very small" time period when we are still \emph{in the process} of "tackling" one single Grassmann parameter, such as $\theta_{57}$, \emph{millions} of repetitions of the process in Chapter 4 will occur. \emph{The fact that} time scale of Chapter 4 is "much smaller" than time scale of the Section 3.2 implies that "classical physics" in Chapter 4 will produce "bosonic field theory" of the present chapter (whatever "bosonic field theory" might be). \emph{As a result of this} we can "trust" that the Lagrangian generator of Section 3.2.6 will, in fact, produce what it is designed to produce. Only \emph{after that} we can go to \emph{even larger} scales and appeal  "smallness" of the scales of Section 3.2 thus ultimately explaining fermionic nature of fields involved.

To make long story short, on smaller scales everything looks "more natural" than on larger ones. We know that "quatum bosons" are more natural than "quantum fermions" and "classical physics" is even more natural than "quantum bosons". Thus, on a "very small scale" (Chapter 4) we have "classical physics", the on the "small scale" (Section 3.2) we have "bosonic quantum theory" and then on "ordinary scale" we have fermionic theory.  What this also tells us is that we have to "take seriously" the fact that the "small scales" we introduce are, in fact, "finite". After all, they can be "much larger" than other "small scales"!

\subsection*{4. "Classical" mechanism that "enforces" Lagrangians}

\subsection*{4.1 Algorithm}

As we said in introduction, the goal of Chapter 3 was to write down a "computer program" (that is, Lagrangian), while the goal of the Chapter 4 is to find a mechanism through which we are building the computer. Our mechanism will involve superluminItal signals but, at the same time, it will be "local". In other words, while the signals travel "very fast", their speed is still finite. But, since the universe is compact, they are able to circle the univers within a very short period of time; thus, producing the informaiton consistent with infinite speed of the signals. Our final goal is to write down a set of differential equations and claim that they "generate" the mathematical information that can be obtained from quantum mechanics. 

However, in light of the number of signals involved, the overal picture will be clearer if we talk in terms of "signal emitted", "signal absorbed", and so forth, \emph{without} "clouding things" with equations. Thus, we will first start off with the equation-free, diagram-like presentation and show why it will produce the desired result (sections 4.1 - 4.7) and \emph{after that} we will "go back" and "fill in the gaps" by postulating the equations that would "enforced" any of the alleged "emissions", "absorptions" and so forth that we were talking about (Sections 4.7 -- 4.12). Furthermore, let us devote the Section 4.1 to simply listing the algorithm without any motivation. But then, throughout parts 4.2--4.7 we will show in detail why this algorithm \emph{if followed} produces the desired results; then, as we just said, we will devote sections 4.8--4.12 to the question \emph{why} the algoritm is followed.

So, our algorithm is as follows: 

1) Every point $k$ has "charges" $q_k$ and $q'_k$, as described earlier. 

2) Every point $i$ has evolving internal degrees of freedom $\phi_k (t)$, $\overline{\phi}_k (t)$, $\vec{\overline{X}}_k$, $V_k^{\mu} (t)$, $\overline{V}_k^{\mu} (t)$, $\overline{q}_k (t)$, ${\cal L}_k (t)$, $S_k (t)$, and $\psi_k (t)$. The "overline", such as $\overline{\phi}$ shoud not be confused with a "vector sign" such as $\vec{x}$. 

It will be shown in Section 4.2 that,  due to parts 5-7 (see below), $\phi_k$ and $V_k^{\mu}$ approach constants at equilibrium. In fact, according to the discussion in Chapter 2, we "want" them to be constants. However, for the reasons to be discussed in Chapter 2, they "start out" as dynamical variables. 

3) There are evolving fields $\Phi (\vec{x}, t)$, $V^{\mu} (\vec{x}, t)$, $Q (\vec{x},t)$, $\vec{X} (\vec{x}, t)$, $L (\vec{x}, t)$, $S (\vec{x}, t)$, and $\Psi (\vec{x}, t)$ that propagate throughout continuum. In particular, they are well defined in the space \emph{between} the "lattice points"; they serve as a local mechanism of lattice points to communicate with each other. Note that $V^{\mu}$ is a four-index vector while $\vec{X}$ is a three-index one. This is okay, in light of the non-relativistic framework of our theory (the appearance of relativity is explained away by "coincidence" in the Lagrangians). 

4) Every particle emits a pulse with periodicity $2 \pi / \nu$ and the duration of each pulse is approximately equal to $2 \chi$. However, the overall phase of the emission of a pulse is individual to each particle.  It is assumed that $\nu$ is very large, so that each particle emits a very large number of pulses during a very short period of time. \emph{However} it is also assumed that the duration of each pulse, $\chi$, is very small. In particular, $\chi \ll N/ \nu$, where $N$ is a total number of particles. As a result, it is statistically expected that none of the pulses emitted by any of the particles in the universe overlap in time. 

5) If the particle $j$ emits a pulse during the time interval $t_1 \leq t \leq t_2$, it will have an effect on points $(t, \vec{x})$ that satisfy \beq \Big\vert \frac{\vec{x} - \vec{x}_j}{t-t_1} \Big\vert < c_s < \Big\vert \frac{\vec{x} - \vec{x}_j}{t-t_2} \Big\vert  \eeq
where $c_s$ is a speed of the superluminal signals. This "effect" is expressed in terms of the behavior of fields $S (\vec{x}, t)$, ${\cal L}(\vec{x}, t)$, $\omega(\vec{x}, t)$, $Q(\vec{x}, t)$, $Q'(\vec{x}, t)$, $\Psi(\vec{x}, t)$, $\Phi(\vec{x}, t)$, $V^{\mu}(\vec{x}, t)$, and $\vec{X}(\vec{x}, t)$. In particular, 
\beq (t, \vec{x}) {\rm \; as \; above} \Longrightarrow (S(\vec{x}_j, t) \approx S_i (t) \; ; \; {\cal L} (\vec{x}_j, t) \approx {\cal L}_i (t) \; , \; \omega (\vec{x}_j, t) \approx \omega (q_i, q'_i) \nonumber \eeq
\beq Q' (\vec{x}_j, t) \approx q_i' (t)) \; , \; \Psi (\vec{x}, t) \approx \psi_i (t)) , \Phi (\vec{x}, t) \approx \phi_j \label{eqn:337}\eeq
\beq V^{\mu} (\vec{x}, t) \approx V_j^{\mu} \; ; \; Q (\vec{x}, t) = q_j \; ; \; \vec{X} (\vec{x}, t) = \vec{x} - \vec{x}_j  \nonumber \eeq
It should be noticed that these fields are defined throughout the \emph{whole} space; but, outside of the above-specified region, their behavior does not give us any meaningful information. 

6) There is a propagating field $\mu_Q$ that is coupled to the $q$-charge of the emitting particle and is emitted duringt the pulse together with other fields. In the absence of the pulse, the value of $\vert \mu_q \vert$ is very small. On the other hand, in the presence of the pulse, the value of $\mu_Q$ is large. This provides an unambiguous way of "deciding" whether or not a given point in space is a subjected to the "pulse" at any given point in time. 

7) When a pulse passes by particle $i$, it "updates" the values of $\phi_i$ and $V_i^{\mu}$ according to 
\beq \phi_i \rightarrow \phi_i + (k_{\phi} \delta t) (\Phi (\vec{x}_i, t) - \phi_i) T (\vert Q (\vec{x}_i, t) - q_i \vert < \epsilon_q) T (\vert \vec{X} (\vec{x}_i, t) \vert < r_c) \label{eqn:338}\eeq
\beq V_i^{\mu} \rightarrow V_i^{\mu} + (k_V \delta t) (V^{\mu} (\vec{x}_i, t) - V_i^{\mu}) T (\vert Q (\vec{x}_i, t) - q_i \vert < \epsilon_q) T (\vert \vec{X} (\vec{x}_i, t) \vert <  r_c), \label{eqn:339}\eeq
where $\epsilon_q \ll 1$ is some small, but finite, constant. There are, however, two "exceptions" when this does \emph{not} happen:

(i) $\mu_Q$ is smaller than some threshold then the pulse is not present and, therefore, the particle is "not allowed" to do any updating

(ii) If the particle is in the process of \emph{emitting} a signal, then it should "know" that the external fields are due to its own emission. Therefore, it is supposed to refrain from "updating" anything, in order to avoid self-interaction (which would have produced unpredictable results). 

8) During the \emph{beginning} of the external pulse passing by particle $i$, the particle "updates" the value of ${\cal L}_i$, according to 
\beq {\cal L}_i \longrightarrow {\cal L}_i + (k_{\cal L} \delta t) {\cal K} (q_i, q'_i, \overline{q}_i, Q(\vec{x}_i, t), \label{eqn:340}\eeq
\beq \phi_i, \overline{\phi}_i (t), \Phi(\vec{x}_i, t),  V_i^{\mu} , \overline{V}_i^{\mu}, V^{\mu} (\vec{x}_i, t), \vec{\overline{X}}_i (t), \vec{X} (\vec{x}_i, t) ) \nonumber \eeq
where $r_c$ is a constant that is "much larger" than the radius of the clusters and "much smaller" than the distance between neighboring clusters of points. As was stated in section 2.2 $\cal K$ denotes the "Lagrangian generator". In the context of our algorithm (which admittedly is less intuitive then what we were looking at in Section 2.2), ${\cal K}$ is given by
\beq {\cal K} (q_i, q'_i, \overline{q}_i (t), Q(\vec{x}_i, t), \cdots) =  T(\vert q_i -  \overline{q}_i (t) \vert < \epsilon_q) T(\vert q_i -  Q (\vec{x}_i, t) \vert < \epsilon_q) {\cal K}_{ss} (\cdots) +  \nonumber \eeq
\beq  + T(\vert q_i - \overline{q}_i (t) \vert < \epsilon_q) T(\vert q'_i - Q (\vec{x}_i, t) \vert < \epsilon_q) {\cal K}_{st} (\cdots) + \eeq
\beq + T(\vert q'_i- \overline{q}_i (t) \vert < \epsilon_q) T(\vert q_i- Q (\vec{x}_i, t) \vert < \epsilon_q) {\cal K}_{ts} (\cdots) + \nonumber \eeq
\beq + T(\vert q'_i- \overline{q}_i (t) \vert < \epsilon_q) T(\vert q'_i-  Q (\vec{x}_i, t) \vert < \epsilon_q) {\cal K}_{tt} (\cdots) \nonumber \eeq
where $T$ is a "truth value" defined by 
\beq T ({\rm true})=1 \; ; \; T({\rm false}) =0, \eeq
${\cal K}_{ss}$, ${\cal K}_{st}$, ${\cal K}_{ts}$ and ${\cal K}_{tt}$ are, respectively, spacelike-spacelike, spacelike-timelike, timelike-spacelike and timelike-timelike Lagrangian generators. Together, they form a single Lagrangian gemerator. 

There are, however, three conditions that need to be satisfied. The first two conditions similarly to part 7, are necessary to identifiy the external signal and avoid self interaction:

(i) $\mu_Q$ is greater than some threashold

(ii) The particle is \emph{not} in the process of emitting any signal

The third condition, however, is needed in order to avoid time overlap between parts 8 and 9:

(iii) The above should be completted during the \emph{first half} of the pulse. Towards the second half of the pulse, the above process should be "frozen" so that part it would not react to what we are about to describe in part 9. 

9) \emph{If} the "external pulse" happens to satisfy 
\beq T( \vert Q(\vec{x}_i, t) - q_i \vert < \epsilon_q) + T(\vert Q (\vec{x}_i, t) - q'_i \vert < \epsilon_q) \geq 1 \eeq
\emph{then}, towards its \emph{second half}, the particle $i$ updates the values of $\overline{\phi}_i$ and  $\overline{V}_i^{\mu}$ according to 
\beq \overline{\phi}_i \rightarrow \Phi (\vec{x}_i, t) \; ; \; \overline{V}_i ^{\mu} \rightarrow V_i^{\mu} (\vec{x}_i, t) \; ; \; \overline{q}_i \rightarrow Q (\vec{x}_i, t) \label{eqn:344}\eeq
\emph{If}, on the other hand, $T(\vert Q (\vec{x}_i, t) - q_i \vert < \epsilon_q) + T(\vert Q (\vec{x}_i, t) - q'_i \vert < \epsilon_q) =0$, then the parameters $\overline{\phi}_i$, $\overline{V}_i^{\mu}$ and $\overline{q}_i$ will remain unchanged. 

Again, we have similar three conditions as in part 8: 

(i) $\mu_Q$ is greater than some threashold

(ii) The particle is \emph{not} in the process of emitting any signal

(iii) Lack of overlap:  we don't want the "updating" to interfere with part 8. Therefore, part 8 is limitted to the "first half" of the signal, while the "updating" of part 9 has to "wait" until the "second half". The mechanism for this will be provided in 4.10. 

10) During the time interval between pulses, ${\cal L}_i$ decays according to 
\beq \frac{d{\cal L}_i}{d t} = - \alpha_L {\cal L}_i \eeq
However, $\alpha_L$ is so small that its effect won't be felt on the scale of few pulses. It can only be felt on the scale of millions of them. 

11) When a singnal in a form of field $L$ passes a particle $k$, it causes it to "update" its internal degree of freedom $S_k$ according to 
\beq S_j \longrightarrow S_j +  (k_S \delta t) T(\vert \omega (q_j, q'_j) - \omega (\vec{x}, t) \vert < \epsilon_{\omega}) ({\cal L} (\vec{x}_i, t) +  k_S (S (\vec{x}_j , t) - S_j))  \label{eqn:346}\eeq
\beq \psi_j \longrightarrow \psi_j + (k_{1 \psi} \delta t) \Psi' (\vec{x}_k, t) e^{iS (\vec{x}_j, t))} T(\vert q_j - Q'(\vec{x}_j, t) \vert< \epsilon_q) +  \nonumber \eeq
\beq + (k_{2 \psi} \delta t) (\psi (\vec{x}_j , t) - \psi_j) T(\vert q_j - Q (\vec{x}_j, t) \vert < \epsilon_q) \eeq
where $\epsilon_q \ll 1$ is the same as in part 9, and $\epsilon_{\omega} \ll 1$ is some other small, but finite, constant. 

This, however, is conditional upon two of the following things: 

(i) $\mu_Q$ is greater than some threashold

(ii) The particle is \emph{not} in the process of emitting any signal

12) During the signal-free period of time, $S_k$ decays according to 
\beq \frac{dS_k}{dt} = - \alpha S_k \eeq
where $\alpha_S$ is a very small number. Similar to the situation with ${\cal L}_i$,  $\alpha_S$ is so small that its effect won't be felt on the scale of few pulses. It can only be felt on the scale of millions of them.

\subsection*{4.2 Equilibrium values of $\phi_i$ and $V_i^{\mu}$ (parts 5-7)}

Even though we have outlined our algorithm in the great detail, we have not explained the purpose of any of its ingredients which might have left the reader in confusion. The reason for this is that we kept in mind that, after "carrying out" that "prescription", the meaning will become clear once we get to final results. Therefore, we will devote the sections 4.2-4.7 to do just that.  In particular, we will treat $S$, $\cal L$ and $\psi_i$ as "arbitrary parameters"; and then we will show that their behavior "ends up" being consistent with action, Lagrangian, and probability amplitude, respectivily.  It is easy to see that the scheme we presented in the previous section violates time reversal; thus, the desired "equilibrium" can be conceivable as $t \rightarrow + \infty$ and \emph{not} $t \rightarrow - \infty$. The time reversal symmetry of quantum field theory is merely due to the "coincidence in Lagrangian", similarly to Lorentzian symmetry. 

Our present task is to show that the algorithm actually produces constraints discussed in Chapter 2, which are necessary in order to make sure that we produce the mathematical information consistent with quantum field theory. It is important to keep in mind that, just like with "classical physics" the initial conditions can be anything we like. Thus, we would like to \emph{violate} those constraints at $t=t_0$ and show that these constraints get \emph{produced} at $t \rightarrow \infty$ through the above-discussed algorithm. For the convenience of a reader we will devote each section to just one constraint. We will cover all of the constraints in Sections 4.2 to 4.7. 

The constraint we will discuss in this particular section has to do with enabling us to establish one-to-one correspondence between the "charge" $q$ and the "fields" $\phi_q \colon T \rightarrow \mathbb{C}$ and $V_q^{\mu} \colon T \rightarrow \mathbb{X}^4$. Technically, we define this correspondence based on 
\beq \Phi_q (P, t) = \frac{1}{\sharp (Q_q (P))} \sum_{p \in Q_q (P)} \phi (p, t) \; ; \; V_q^{\mu} (P, t) = \frac{1}{\sharp (Q_q (P))} \sum_{p \in Q_q (P)} V_q^{\mu} (p, t) \eeq
Now, in light of the fact that we would like to have a "local" theory, the averages can not play any physical role. Thus, in order for it to "look like" they do, we would like to be able to approximate these averages based on the internal degrees of freedom of each individual point:
\beq \phi (k, t) \approx \Phi_{q_k} (t) \; ; \; V^{\mu} (k, t) \approx V_{q_k} (t)^{\mu} (t) \eeq
These constraints are equivalent to 
\beq ((C(i) = C(j)) \wedge (q_i = q_j)) \Longrightarrow ((\phi(i, t) \approx \phi (j, t)) \wedge (V^{\mu} (i, t) \approx V^{\mu} (j, t))) \eeq
Like we said earlier, we can \emph{not} simply postulate those. After all, the "initial conditions" are supposed to be arbitrary which means that these constraints can be violated at $t=t_0$. Our task is to show that these constraints will be satisfied at the equilibrium point, if the algorithm is followed step by step. In light of the fact that the treatments of $\phi$ and $V^{\mu}$ by the "algorithm" are nearly identical, let us fucus on showing that the constraint is satisfied for $\phi$; we will then trust ourselves that we can "redo" it for $V^{\mu}$ by merely replacing $\phi$ by $V^{\mu}$ at appropriate places. 

In the prevous section we described a propagating field $\Phi (\vec{x}, t)$. We will distinguish it from $\Phi_q (P, t)$ we just mentioned mainly by listing variables (which are $(P, t)$ in the former case and $(\vec{x}, t)$ in the latter). Now, according to part 9, whenever the signal passes, $\phi_i$ is "updated" according to 
\beq \phi_i \rightarrow \phi_i + (k_{\phi} \delta t) (\Phi (\vec{x}_i, t) - \phi_i) T (\vert Q (\vec{x}_i, t) - q_i \vert < \epsilon_q) T (\vert \vec{X} (\vec{x}_i, t) \vert < r_c) \eeq
Now, according to part 4, if the source of a pulse was a particle "j", then the values of $\Phi (\vec{x}, t)$ and $Q(\vec{x},t)$ coincide with $\phi_j$ and $q_j$, respectively. If we now "apply" this to $\vec{x} = \vec{x}_i$, we obtain
\beq \Phi (\vec{x}_i, t) = \phi_j \; ; \; Q (\vec{x}_i, t) = q_j \; ; \; T (\vert \vec{X} (\vec{x}_i, t) \vert < r_c) = T (C(i) = C(j)). \eeq
By substitutting these parameters, we see that $\phi_i$ is being "updated" according to 
\beq \phi_i \rightarrow \phi_i + (k_{\phi} \delta t) (\phi_j - \phi_i) T (q_j = q_i) T (C(i)= C(j)) \eeq
Now, we recall that every single particle emits a pulse with periodicity $\delta t$. Therefore, during the interval between $t$ and $t+ \delta t$, the particle number $i$ will be subjected to pulses from \emph{every single} other particle. As a result, $\phi_i$ will change according to
\beq \phi_i (t+ \delta t) = \phi_i (t) + (k_{\phi} \delta t) \sum_j ((\phi_j (t) - \phi_i (t)) T (q_j = q_i) T (C(i)= C(j))) + O ((\delta t)^2)\eeq
We can now eliminate $T (\cdots)$ by instead putting "$\cdots$" as one of the restrictions of the sum. Thus, we get 
\beq \phi_i (t+ \delta t) = \phi_i (t) + (k_{\phi} \delta t) \sum_{j \in Q_{q_i} (C(i))} (\phi_j (t) - \phi_i (t)) \eeq
We now notice that inside the sum only the first term ($\phi_j$) is a function of the parameter we are summing over ($j$), while the second term ($\phi_i$) is treated as "constant". Therefore, we can pull $\phi_i$ out of the sum as long as we remember to multiply it by the number of terms of the sum, $\sharp (Q_{q_i} (C(i))))$. We can further combine it with the \emph{other} $\phi_i$ term on the right hand side. This gives us 
\beq \phi_i (t+ \delta t) = (1 - (k_{\phi} \delta t) \sharp (Q_{q_i} (C(i)))) \phi_i (t) + (k_{\phi} \delta t) \sum_{j \in Q_{q_i} (C(i))} \phi_j (t) \eeq
Now, our goal is to show that $\phi (i, t)$ approaches $\Phi_{q_i} (C(i), t)$ as $t \rightarrow \infty$. This is the same as saying that $\phi (i, t) - \Phi_{q_i} (C(i), t)$ approaches zero. Therefore, we will subtract $\Phi_{q_i} (C(i), t)$ from both sides of the above equation in order to see what we get. We will, furthermore, write out explicitly $\Phi_{q_i} (C(i), t)$ in the form of a "ratio" of a "sum" and the "number of elements": 
\beq \phi_i (t+ \delta t) - \frac{\sum_{j \in Q_{q_i} (C(i))} \phi_j (t)}{\sharp (Q_{q_i} (C(i)))}= (1 - (k_{\phi} \delta t) \sharp (Q_{q_i} (C(i)))) \Big( \phi_i (t) - \frac{\sum_{j \in Q_{q_i} (C(i))} \phi_j (t)}{\sharp (Q_{q_i} (C(i)))} \Big) \label{doublesum2}\eeq
Now, there is one suddlety here that needs to be taken care of. In order to explore the behavior of $\phi (i, t) - \Phi_{q_i} (C(i), t)$ we need to compare $\phi (i, t) - \Phi_{q_i} (C(i), t)$ with $\phi (i, t+ \delta t) - \Phi_{q_i} (C(i), t + \delta t)$. However, in the above equation we have evaluated $\phi (i, t+ \delta t) - \Phi_{q_i} (C(i), t)$. It turns out, however, that  $\Phi_{q_i} (C(i), t+ \delta t) =\Phi_{q_i} (C(i), t)$ which means that we \emph{did} evaluate $\phi (i, t+ \delta t) - \Phi_{q_i} (C(i), t+ \delta t)$ after all. Intuitively, each individual $\phi_i (t)$ is being "pulled" towards their average $\Phi_{q_i} (C(i), t)$ which is why the latter stays the same. Let us, however, show explicitly that this is, in fact, the case. In order to evaluate $\Phi_{q_i} (C(i), t + \delta t)$ we will take the expression for $\phi (i, t+ \delta t)$, replace $i$ with $k$, and then "take a sum" of all $k \in  Q_{q_i} (C(i))$:
\beq \sum_{k \in  Q_{q_i} (C(i))} \phi_k (t+ \delta t) = \sum_{k \in  Q_{q_i} (C(i))} \phi_k (t) + (k_{\phi} \delta t)  \sum_{k \in  Q_{q_i} (C(i))} \sum_{j \in Q_{q_k} (C(k))} (\phi_j (t) - \phi_k (t)) \label{doublesum} \eeq
Now I claim that the "double sum" on the right hand side is zero.  First of all, we know that if $k \in C(i)$, then $C(k) = C(i)$. Thus, we can rewrite the double sum as 
\beq \sum_{k \in  Q_{q_i} (C(i))} \sum_{j \in Q_{q_k} (C(k))} (\phi_j (t) - \phi_k (t))  = \sum_{k \in  Q_{q_i} (C(i))} \sum_{j \in Q_{q_k} (C(i))} (\phi_j (t) - \phi_k (t)) \eeq
It is now easy to see from antisymmetry the right hand side is zero. Therefore, the left hand side is zero as well: 
\beq \sum_{k \in  Q_{q_i} (C(i))} \sum_{j \in Q_{q_k} (C(k))} (\phi_j (t) - \phi_k (t))  = 0 \eeq
Going back to the equation \ref{doublesum} we now drop the double sum from the right hand side, and obtain
\beq \sum_{k \in  Q_{q_i} (C(i))} \phi_k (t+ \delta t) = \sum_{k \in  Q_{q_i} (C(i))} \phi_k (t) \eeq
We can now go back to the equation \ref{doublesum2} and replace the sum over $\phi_j (t)$ on the left hand side with the sum over $\phi_j (t + \delta t)$ (after all, we have just shown that these two sums are the same). At the same time, on the right hand side we can leave it in the form of the sum over $\phi_j (t)$ (again, we can do it, seeing that these two sums are the same). Thus, we obtain 
\beq \phi_i (t+ \delta t) - \frac{\sum_{j \in Q_{q_i} (C(i))} \phi_j (t + \delta t)}{\sharp (Q_{q_i} (C(i)))}= (1 - (k_{\phi} \delta t) \sharp (Q_{q_i} (C(i)))) \Big( \phi_i (t) - \frac{\sum_{j \in Q_{q_i} (C(i))} \phi_j (t)}{\sharp (Q_{q_i} (C(i)))} \Big) \eeq
If we now start off from $t=t_0$ and "iterate" the above procedure $n$ times, we obtain
\beq \phi_i (t_0+ n \delta t) - \frac{\sum_{j \in Q_{q_i} (C(i))} \phi_j (t_0 + n \delta t)}{\sharp (Q_{q_i} (C(i)))}= \eeq
\beq = (1 - (k_{\phi} \delta t) \sharp (Q_{q_i} (C(i))))^n \Big( \phi_i (t_0) - \frac{\sum_{j \in Q_{q_i} (C(i))} \phi_j (t_0)}{\sharp (Q_{q_i} (C(i)))} \Big) \nonumber \eeq
In the limit of $n \rightarrow \infty$ the right hand side of the above equation goes to zero. Therefore, if we move the second term on the left hand side to the right, we obtain  
\beq \lim_{n \rightarrow \infty} \phi_i (t_0 + n \delta t) =  \frac{\sum_{j \in Q_{q_i} (C(i))} \phi_j (t + \delta t)}{\sharp (Q_{q_i} (C(i)))} = \Phi_{q_i} (C(i)), \eeq
as desired. We now assume that at the times we are living at the sufficient time has been passed for the above limit to be nearly achieved. This is what is behind the fact that we "observe" the $\Phi_{q_i} (C(i))$-based phenomena. Furthermore, if in the above argumet we replace $\Phi$ with $V^{\mu}$, we will similarly show that $V_q^{\mu} (C(i))$- based phenomena holds as well at the equilibrium. In light of the similarlity of the argument it is not necessary to explicitly write down the procedure for $V^{\mu}$. 

\subsection*{4.3 Predicted behavior of ${\cal L}_k$ and the meaning of ${\cal K}_k$ (parts 4-10)}

So far we have demonstrated the "consistency" of $\phi$ and $V^{\mu}$, as defined based on $\phi_i$ and $V_i^{\mu}$ over same-charged particles. Let us now turn to Lagrangian density. In other words we would like to show that ${\cal L}_i$ approximates the desired value of Lagrangian density of these fields. There is an important difference though. While the values of $\phi_i$ and $V_i^{\mu}$ were supposed to be arbitrary, the value of ${\cal L}_i$ is supposed to be a pre-assigned \emph{function} of $\phi_k$ and $V_k^{\mu}$. At the same time, the "mechanistic" view of the universe demands that the initial conditions of  ${\cal L}_i$ at $t=t_0$ are arbitrary, which logically implies that our "desired" equation for ${\cal L}_i$ is most likely \emph{false}. 

Apart from that, parts 10-12 in our algorithm illustrate that ${\cal L}_i$ changes in time. At the same time, we would \emph{like} to believe that ${\cal L}_i$ represents quasi-local transition amplitude \emph{from} $(\phi_{q_i}, V_{q_i}^{\mu})$ \emph{to} $(\phi_{q_i'}, V_{q_i'}^{\mu})$. Now, both the former configuration and the latter is time-independant. The only "time dependance" that we have is a selection of "probability amplitudes" of each "charge" (see section 2) while, at the same time, \emph{we already know} what configuration is "behind" each charge. Thus, the "time dependance" of ${\cal L}_i$ per parts 10-12 seems to contradict what we \emph{believe} to be true about ${\cal L}_i$ based on our interpretation of quantum field theory. 

The way we answer both of the above objections is by claiming that evolution of $\cal L$ per parts 8 and 10 eventually produces an equilibrium point. At the equilibrium, the time dependance due to part 8 almost cancels out the one due to part 10. While the cancelation can't be exact (after all the timing of these two parts is not the same), the cancelation is nearly perfect on a "larger time scale". Thus,  ${\cal L}_i$ \emph{appears to be} time independant. Furthermore, the \emph{static} "equlibrium configuration" of ${\cal L}_i$ is a function of its \emph{dynamics}. This enables us to enforce the desired static picture by postulating appropriate time-dependant dynamics, which harmonizes our theory with the "mechanistic" world view. Of course, we had this goal in mind while postulating the dynamics in parts 8 and 10.

Let us now follow the steps 4-10 to see whether or not the desired value of ${\cal L}_i$ is, in fact, produced. Let us start off by looking at how ${\cal L}_i$ being updated during each pulse, per part 8, and after that discuss its decay "between the pulses" per part 10.
Now, according to part 4, each particle emits pulse with \emph{the same} periodicity $\delta t$. From this,  it is easy to see that if we "watch" the particles emit pulses "for a long time", the sequence in which these particles emit pulses will stay the same. We can, therefore, re-number the particles in such a way that first the pulse is being emitted by particle number $1$, then by particle number $2$, and so forth, all the way to particle number $N$; after which, again, the pulse is being emitted by particle number $1$. 

Let us now see how ${\cal L}_i$ is being "updated" when the pulse is being emitted by the particle number $n$. According to step 8, during the \emph{beginning} of a pulse, ${\cal L}_i$ is updated according to 
\beq {\cal L}_i \longrightarrow {\cal L}_i + (k_{\cal L} \delta t) {\cal K} (q_i, q'_i, \overline{q}_i, Q(\vec{x}_i, t), \label{eqn:366}\eeq
\beq \phi_i, \overline{\phi}_i (t), \Phi(\vec{x}_i, t),  V_i^{\mu} , \overline{V}_i^{\mu}, V^{\mu} (\vec{x}_i, t), \vec{\overline{X}}_i (t), \vec{X} (\vec{x}_i, t) ) \nonumber \eeq
We would now like to subsitute appropriate values for the "arguments" of $\cal K$ in the above equation. According to part 5, the values of $Q(\vec{x}_i, t)$, $\Phi (\vec{x}_i, t)$, $V^{\mu} (\vec{x}_i, t)$ and $\vec{X} (\vec{x}_i, t)$ are given by  
\beq Q(\vec{x}_i, t) \approx q_n \; ; \; \Phi (\vec{x}_i, t) \approx \phi_n \; ; \; V^{\mu} (\vec{x}_i, t) \approx V_n^{\mu} \; ; \; \vec{X} (\vec{x}_i, t) \approx \vec{x}_n - \vec{x}_i \label{eqn:367}\eeq
Now, according to part 9, the parameters $\overline{q}_i$, $\overline{\phi}_i$, $\overline{V}_i^{\mu}$ and $\vec{\overline{X}}_i$ "borrow" the values of $Q (\vec{x}_i, t)$, $\Phi (\vec{x}_i, t)$, $V_i^{\mu} (\vec{x}_i, t)$ and $\vec{X} (\vec{x}_i, t)$, respectively. However, this occurs only during the \emph{second} half of the pulse. At the same time, the "updating" of ${\cal L}_i$ is taking place during the \emph{first} half of the pulse, which means that $\overline{q}_i$, $\overline{\phi}_i$, $\overline{V}_i^{\mu}$ and $\vec{\overline{X}}_i$ have not "borrowed" $Q (\vec{x}_i, t)$, $\Phi (\vec{x}_i, t)$, $V_i^{\mu} (\vec{x}_i, t)$ and $\vec{X} (\vec{x}_i, t)$ \emph{yet}. However, they \emph{did} borrow $Q (\vec{x}_i, t)$, $\Phi (\vec{x}_i, t)$, $V_i^{\mu} (\vec{x}_i, t)$ and $\vec{X} (\vec{x}_i, t)$ from the \emph{previous} pulse. If we assume that $n \geq 2$, then the previous pulse was emitted by a particle number $n-1$. Thus, the values of $\overline{q}_i$, $\overline{\phi}_i$, $\overline{V}_i^{\mu}$ and $\vec{\overline{X}}_i$ can be obtained by "copying" the respective equations for $Q (\vec{x}_i, t)$, $\Phi (\vec{x}_i, t)$, $V_i^{\mu} (\vec{x}_i, t)$ and $\vec{X} (\vec{x}_i, t)$, while replacing $n$ with $n-1$:
\beq \overline{q}_i \approx q_{n-1} \; ; \; \overline{\phi}_i \approx \phi_{n-1} \; ; \; \overline{V}_i^{\mu} (\vec{x}_i, t) \approx V_{n-1}^{\mu} \; ; \; \vec{\overline{X}}_i \approx (\vec{x}_i, t) \approx \vec{x}_{n-1} - \vec{x}_i \eeq
By substitutting equations (\ref{eqn:344}) and (\ref{eqn:367}) (or, equivalently, (\ref{eqn:344}) and (\ref{eqn:337})) into the equation (\ref{eqn:366} or \ref{eqn:340}), we see that, up to the corresponding approximations, ${\cal L}_i$ is being updated according to 
\beq {\cal L}_i \longrightarrow {\cal L}_i + (k_{\cal L} \delta t) {\cal K} (q_i, q'_i, q_{n-1}, q_n, \phi_i, \phi_{n-1}, \phi_n,  V_i^{\mu} , V_{n-1}^{\mu}, V_n^{\mu} , \vec{x}_{n-1} - \vec{x}_i, \vec{x}_n - \vec{x}_i ) \label{eqn:369} \eeq
The above was true assuming that $n \geq 2$. On the other hand, $n=1$, then the "previous" particle was \emph{not} particle number $0$ (such particle does not exist!) but rather particle number $N$. Thus, in order to produce the "updating formula", we have to replace $n$ with $1$ and $n-1$ with $N$:
\beq {\cal L}_i \longrightarrow {\cal L}_i + (k_{\cal L} \delta t) {\cal K} (q_i, q'_i, q_N, q_1, \phi_i, \phi_N, \phi_1,  V_i^{\mu} , V_i^{\mu}, V_1^{\mu} , \vec{x}_N - \vec{x}_i, \vec{x}_1 - \vec{x}_i ) \label{eqn:370}\eeq
We can now combine the equations  (\ref{eqn:369}) and (\ref{eqn:370}) into a much simpler expression by defining ${\cal K}_{in}$ as
\beq {\cal K}_{in} =
\left\{
	\begin{array}{ll}
		{\cal K} (q_i, q'_i, q_{n-1}, q_n, \phi_i, \phi_{n-1}, \phi_n,  V_i^{\mu} , V_{n-1}^{\mu}, V_n^{\mu} , \vec{x}_{n-1} - \vec{x}_i, \vec{x}_n - \vec{x}_i ) \; , \;  n \geq 2 \\
		{\cal K} (q_i, q'_i, q_N, q_1, \phi_i, \phi_N, \phi_1,  V_i^{\mu} , V_i^{\mu}, V_1^{\mu} , \vec{x}_N - \vec{x}_i, \vec{x}_1 - \vec{x}_i )  \; , \;  n=1
	\end{array}
\right.
\eeq
which allows us to rewrite the "updating formula" as 
\beq {\cal L}_i \rightarrow {\cal L}_i + k_{\cal L} {\cal K}_{in}  \delta t \eeq
Now, since all of the particles emit pulse with \emph{the same} periodicity $\delta t$, it is clear that, within the time interval $\delta t$, the particle number $i$ will be subjected to the pulses comming from \emph{all} other particles. Thus, ${\cal L}_i$ will change according to  
\beq {\cal L}_i (t + \delta t) = {\cal L} (t) + (k_{\cal L} \delta t) \sum_{n=1}^N {\cal K}_{in} \eeq
Also, according to part 10, during the \emph{absence} of pulses, ${\cal L}_i$ decays according to 
\beq \frac{d {\cal L}_i}{d t} = - \alpha_{\cal L} {\cal L}_i \eeq
From these two conditions it is easy to see that, after the time interval $m \delta t$ (where $m$ is "very large" of the order of $1/ \delta t$), the evolution of ${\cal L}_i$ is given by
\beq {\cal L}_i (t_0 + m \delta t) \approx e^{- \alpha_{\cal L} m \delta t} {\cal L}_i (t_0) + \sum_{l=1}^m \Big(e^{- \alpha_{\cal L} l \delta t} \Big(k_{\cal L} \sum_{n=1}^N {\cal K}_{in} \Big) \Big) \eeq
In the above equation, the $l$-summands are identical, except for the factor of $e^{- \alpha_{\cal L} l \delta t}$. This is due to the fact that we have a periodic process and $l$ merely enumerates its "identical" repetitions. The factor $e^{- \alpha_{\cal L} l \delta t}$ tells us that the "older" repetitions decayed "more" than the "younger"; otherwise, their effects are identical. Therefore, we can pull the $n$-sum out of the $l$-sum, and obtain
\beq {\cal L} (t_0 + m \delta t) = e^{- \alpha_{\cal L} m \delta t} {\cal L} (t_0) + k_{\cal L} \Big( \sum_{l=1}^m e^{- \alpha_{\cal L} l \delta t} \Big)  \Big(\sum_{n=1}^N {\cal K}_{in} \Big). \eeq
We can now evaluate the $l$-sum by using the equation 
\beq a + \cdots + a^n = \frac{1-a^n}{a^{-1}-1} \eeq
where we identify the constant $a$ with $e^{- \alpha_{\cal L} l \delta t}$. This produces
\beq {\cal L} (t_0 + m \delta t) \approx e^{- \alpha_{\cal L} m \delta t} {\cal L} (t_0) + \frac{k_{\cal L}(1-e^{- \alpha_{\cal L} m \delta t})}{e^{\alpha_{\cal L} \delta t}-1} \sum_{n=1}^N {\cal K}_{in} \eeq
Now the "equilibrium" is achieved at the limit of $m \rightarrow \infty$, which is given by 
\beq \lim_{m \rightarrow \infty} {\cal L} (t_0 + m \delta t) = \frac{k_{\cal L}}{\alpha_{\cal L} \delta t} \sum_{n=1}^N {\cal K}_{in} \eeq
The above, in fact, is a constant. Furthermore, it happens to be proportional to the sum of the Lagrangian generators, as desired. The only problem is that, as will be amply evident from Section 2.2, we would \emph{like} to sum over \emph{three-point} Lagrangian generator,
\beq {\cal K}_{i, n_1, n_2} = {\cal K} (q_i, q'_i, q_{n_1}, q_{n_2}, \phi_i, \phi_{n_1}, \phi_{n_2},  V_i^{\mu}, V_{n_1}^{\mu}, V_{n_2}^{\mu}, \vec{x}_{n_1} - \vec{x}_i, \vec{x}_{n_2} - \vec{x}_i )  \eeq
where $n_1$ and $n_2$ are arbitrary. Intead, as a result of the "fixed" sequence in which the signals are emitted, we have \emph{unwanted} constraint $n_1 = n_2 -1$, which limits our sum to  
\beq {\cal K}_{in} = {\cal K}_{i,n,n-1} \eeq
What comes to our rescue is the fact that the sequence in which signals are emitted (or, equivalently, the numbering of the particles) is completely random with regards to their locations as well as their internal parameters. As a result, our "restricted" sum closely approximates the "non-restricted" sum, up to an overall coefficient. That coefficient is simply the ratio of terms of each sum. In light of the fact that $i$ is fixed (namely, it is the specific particle we are looking at), the "restricted sum" has only one parameter, $n$; thus, it has $N$ terms. On the other hand, non-restricted sum has two parameters, $n_1$ and $n_2$; thus, it has $N^2$ terms. Therefore, the relation between the two sums is given by
\beq \sum_{n=1}^N {\cal K}_{in} = \sum_{n=1}^N  {\cal K}_{i,n,n-1} \approx \frac{N}{N^2} \sum_{n_1, n_2}^N  {\cal K}_{i,n_1,n_2} = \frac{1}{N} \sum_{n_1, n_2}^N  {\cal K}_{i,n_1,n_2} \eeq
This allows us to re-express the expressions for ${\cal L}$ in terms of non-restricted sums:
\beq {\cal L} (t_0 + m \delta t) \approx e^{- \alpha_{\cal L} m \delta t} {\cal L} (t_0) + \frac{k_{\cal L}(1-e^{- \alpha_{\cal L} m \delta t})}{N(e^{\alpha_{\cal L} \delta t}-1)} \sum_{n=1}^N {\cal K}_{in} \eeq
\beq \lim_{m \rightarrow \infty} {\cal L} (t_0 + m \delta t) \approx \frac{k_{\cal L}}{N \alpha_{\cal L} \delta t} \sum_{n_1, n_2} {\cal K}_{i,n_1, n_2} \eeq
Thus, the equilibrium, as defined through limit over $m \rightarrow \infty$, is proportional to the desired "sum". Now, we would \emph{like} the right hand side to be defined by an integral, with the overall coefficient being $1$. Now, we will assume that the density of the clusters is $\rho_{\rm cl}$ and the density of the points is $\rho_{\rm pt}$ (which is equal to the product of $\rho_{\rm cl}$ with the average number of points per cluster). In light of the fact that the signals are being emitted by individual points, the "sum" of "something" is equal to the "integral" of "something multiplied by $\rho_{\rm pt}$. Therefore, in order for the overall coefficient of the integral to be $1$, we have to identify the overall coefficient of the sum, namely $k_{\cal L} / (N \alpha_{\cal L} \delta t)$, with $\rho_{\rm pt}^{-1}$. Thus, 
\beq \rho_{\rm pt} = \frac{N \alpha_{\cal L} \delta t}{k_{\cal L}} \Rightarrow \lim_{m \rightarrow \infty} {\cal L} (t_0 + m \delta t) \approx \sum_{n_1, n_2} {\cal K}_{i,n_1, n_2} \eeq
In reality, we probably have some physical considerations for assuming some "acceptable" range of values of $\rho_{\rm pt}$, $N$ $\alpha_{\cal L}$ and $\delta t$. After all, somehow we "know" that $\rho_{\rm pt}$  and $N$ are "very large", while $\alpha_{\cal L}$ and $\delta t$ are "very small". The only parameter that is "to be defined" is $k_{\cal L}$. We, therefore, rewrite the expression for $\rho_{\rm pt}$ in the form
\beq k_{\cal L} = \frac{N \alpha_{\cal L} \delta t}{\rho_{\rm pt}} \eeq
In light of the fact that $k_{\cal L}$ is merely a scaling parameter we use while "updating" ${\cal L}_i$, we do not have any apriori reasons to believe whether it is "large" or "small" apart from the desired outcome. Since the "desired outcome" is "accommodated for" by the above equation, this is the most reasonable way of defining $k_{\cal L}$. 

\subsection*{4.4 Global correlation of $\psi_j$ (parts 11-12)}

As we have said, the goal of the sections 4.2--4.7 is for us to illustrate that the required constraints are, in fact, satisfied at the equilibrium point. So far we have shown that this is true for the constraints that require the "consistency" of $\Phi_q (P)$ and $V_q^{\mu} (P)$. Let us now move on to the consistency of probability amplitudes $\psi_i$. As we said earlier, due to the fact that $\psi_i$, "taken" at a single particle $i$, refers to a probability amplitude of \emph{entire configuration} $\Phi_{q_i} \colon T \rightarrow \mathbb{C}$, is consistency requires a set of global "correlations" to be satisfied. These "correlations" take the form 
\beq q_i = q_j \Rightarrow \psi_i (t) =^? \psi_j (t). \eeq
As before, we do \emph{not} assume that these correlations are satisfied innitially. We \emph{do} however want to show that they will be satisfied "at the equilibrium". This will be the subject of this section. 

Now, in order to show that the "global correlation" arizes at the equilibrium, we can "watch" $\vert \psi_j - \psi_k \vert$, and demonstrate that its value is zero at the equilibrium \emph{as long as} $q_j = q_k$. We recall that both $\psi_j$ and $\psi_k$ are nearly constant during pulse-free time, and both are being "updated" whenever the pulse passes by. They "update" according to 
\beq \psi_j \longrightarrow \psi_j + (k_{1 \psi} \delta t) \Psi' (\vec{x}_j, t) e^{iS (\vec{x}_j, t))} T(\vert q_j - Q'(\vec{x}_j, t) \vert< \epsilon_q) + \nonumber \eeq
\beq + (k_{2 \psi} \delta t) k_{\psi} (\psi (\vec{x}_j , t) - \psi_j) T(\vert q_j -  Q (\vec{x}_j, t) \vert < \epsilon_q) \eeq
\beq \psi_k \longrightarrow \psi_k + (k_{1 \psi} \delta t) \Psi' (\vec{x}_k, t) e^{iS (\vec{x}_k, t))} T(\vert q_k - Q'(\vec{x}_k, t) \vert < \epsilon_q) + \nonumber \eeq
\beq + (k_{2 \psi} \delta t) (\psi (\vec{x}_k , t) - \psi_k) T(\vert q_k - Q (\vec{x}_k, t) \vert < \epsilon_q) \nonumber \eeq
Now, suppose that the source of the pulse is a particle number $i$. Then, from part 5, we know the following three equations:
\beq S(\vec{x}_j, t) \approx S(\vec{x}_k, t) \approx S_i \; ; \;  Q (\vec{x}_j, t) \approx Q (\vec{x}_k, t) \approx q'_i \; ; \; \Psi (\vec{x}_j, t) \approx \Psi (\vec{x}_k, t) \approx \psi_i \eeq
If we now substittute these equations into the prescription for "updating" $\phi_j$ and $\phi_k$, we obtain
\beq \psi_j \longrightarrow \psi_j + (k_{1 \psi} \delta t) S_i \psi_i T(q_j = q'_i) + (k _{2 \psi}\delta t)  (\psi_i - \psi_j) T(q_j =q_i) \eeq
\beq \psi_k \longrightarrow \psi_k + (k_{1 \psi} \delta t) S_i \psi_i T(q_j = q'_i) + (k _{2 \psi} \delta t) (\psi_i - \psi_k) T(q_k =q_i) \eeq
Now, the only situation in which we expect "global correlation" to arize is when $q_j = q_k$. Therefore, this is exactly what we will assume. This leads to the following simplifications: 
\beq q_j = q_k \Longrightarrow T(q_k = q_i) = T(q_j =q_i) = \delta_{q_i}^{q_j} \eeq
\beq q_j = q_k \Longrightarrow T(q_k = q'_i) = T(q_j =q'_i) = \delta_{q'_i}^{q_j} \eeq
Upon substittuting these into the "updating" prescriptions for $\psi_j$ and $\psi_k$, we obtain
\beq \psi_j \longrightarrow \psi_j + (k_{1 \psi} \delta t) S_i \psi_i \delta_{q_i'}^{q_j} + (k_{2 \psi}\delta t) \delta_{q_i}^{q_j} (\psi_i - \psi_j) \eeq
\beq \psi_k \longrightarrow \psi_k + (k_{1 \psi} \delta t) S_i \psi_i \delta_{q_i'}^{q_k} + (k_{2 \psi} \delta t) \delta_{q_i}^{q_k} (\psi_i - \psi_k) \eeq
As we mentioned earlier, in order to show that $\psi_k \leftrightarrow_{t \rightarrow \infty} \psi_j$ we will, instead, show that $\vert \psi_k - \psi_j \vert \rightarrow_{t \rightarrow \infty} 0$. In order to do that, we will subtract the above two equations from each other. This produces
\beq \psi_k - \psi_j \longrightarrow \psi_k - \psi_j - (k_{2 \psi} \delta t) \delta_{q_i}^{q_j} (\psi_k - \psi_j) \label{eqn:396}\eeq
While we are assuming that $q_j = q_k$, we make no such assumption regarding $q_i$. Thus, we have to consider two cases: either $q_i = q_j$ or $q_i \neq q_j$. In case $q_i \neq q_j$, no "updating" occurs; or, equivalently, the "updating" returns exactly what we have started from: 
\beq (q_i \neq q_j) \Longrightarrow (\psi_k - \psi_j \rightarrow \psi_k - \psi_j) \eeq
On the other hand, if $q_i = q_j$, then by replacing $\delta_{q_i}^{q_j}$ with $1$, and factoring out $\psi_k - \psi_j$ on the right hand side of equation (\ref{eqn:396}), we obtain
\beq (q_i = q_j) \Longrightarrow (\psi_k - \psi_j \rightarrow (1 - k_{2 \psi} \delta t) (\psi_k - \psi_j) )\eeq
Now, we recall that all particles, regardless of their charges or any other parameters, have internal oscillations with common period $\delta t$. According to part 4, the emission of a signal is determined to occur during a specific place on this "circle". Different particles reach this place at different times due to the fact that their oscillations have different overall phases. Based on this picture, it is easy to see that during the interval of length $\delta t$ every single particle has emitted its pulse. Whenever the pulse was emittedy by particle $i$ whose charge "happened" to coincide with $q_j$, the value $\psi_k - \psi_j$ was "multiplied by" $1-k_{2 \psi} \delta t$. Thus, during the interval $\delta t$ it was multiplied by an appropriate power of $1-k_{\psi} \delta t$; namely, 
\beq \psi_k (t + \delta t) - \psi_j (t+ \delta t) = (1 - k_{2 \psi} \delta t)^{\sharp \{i \vert q_i = q_j \}} (\psi_k (t) - \psi_j (t)). \eeq
Now, if we take two arbitrary points in time, $t_1$ and $t_2$, they are generally separated by a \emph{very large number} of the $\delta t$-intervals. We can, therefore, neglect the fraction of $\delta t$ and assume that $(t_2 - t_1)/ \delta t$ is integer. Therefore, up to a very good approximation, we obtain
\beq \psi_k (t_2) - \psi_j (t_2) \approx (1 - k_{2 \psi} \delta t)^{(\sharp \{i \vert q_i = q_j \})(t_2-t_1)/ \delta t}(\psi_k (t_1) - \psi_j (t_1)) \rightarrow_{t_2 \rightarrow \infty} 0. \eeq
Thus, we have successfully shown that, in the limit $t_2 \rightarrow \infty$, the parameters $\psi_j$ are globally-correlated, as desired. 

\subsection*{4.5 Global correlation of $S_j$ (parts 11-12)}

In the previous section we have shown that $\psi_j$ does, in fact, obey the desired correlations.  Let us now similarly show that $S_j$ also satisfies the desired correlations, namely,
\beq \omega (q_j, q'_j) = \omega (q_k, q'_k) \Rightarrow^? S_j = S_k. \eeq
 The techniques we will use are similar to the ones for $\psi$ except that there will be one simplification: two \emph{different} parameters, $q_k$ and $q'_k$ will be replaced by "the same" parameter $\omega (q_k, q'_k)$. Now, according to Part 11, $S_j$ and $S_k$ are updated according to 
\beq S_j \longrightarrow S_j +  (k_S \delta t) T(\vert \omega (q_j, q'_j) - \omega (\vec{x}_j, t) \vert < \epsilon_{\omega}) ({\cal L} (\vec{x}_i, t) +  k_S (S (\vec{x}_j , t) - S_j))  \eeq
\beq S_k \longrightarrow S_k +  (k_S \delta t) T(\vert \omega (q_k, q'_k) - \omega (\vec{x}_k, t) \vert < \epsilon_{\omega}) ({\cal L} (\vec{x}_i, t) +  k_S (S (\vec{x}_j , t) - S_j))  \eeq
Similarly to what we did in the previous section, we are going to isolate the emission of the signal by an arbitrary particle $i$. As this signal passes the particles $j$ and $k$, it produces relations
\beq {\cal L}(\vec{x}_j, t) \approx {\cal L}(\vec{x}_k, t) \approx {\cal L}_i (t) ) \; ; \;  S(\vec{x}_j, t) \approx S(\vec{x}_k, t) \approx S_i (t) \eeq
\beq  \omega (\vec{x}_j, t) = \omega (\vec{x}_k, t) \approx \omega (q_i, q'_i) \eeq
According to part 11, $S_j$ and $S_k$ are upated according to 
\beq S_j \rightarrow S_j + (k_S \delta t)  ({\cal L}_i + S_i - S_j) T(\omega (q_i, q'_i) =\omega (q_j, q'_j)) \eeq
\beq S_k \rightarrow S_k + (k_S \delta t)  ({\cal L}_i + S_i - S_k) T(\omega (q_i, q'_i) =  \omega (q_k, q'_k)) \eeq
Now, the only situation in which we expect global correlations to hold is when $\omega (q_k, q'_k) = \omega (q_j, q'_j)$. Therefore, we will assume just that. In this case, we can replace $T (\omega(q_i, q'_i) = \omega (q_k, q'_k))$ with $T (\omega(q_i, q'_i) = \omega (q_j, q'_j))$. If we now subtract the above two equation, while making this substitution, we obtain
\beq S_k - S_j \longrightarrow S_k - S_j - (k_S \delta t)  (S_k - S_j) T(\omega(q_j, q'_j) = \omega (q_i, q'_i)) \eeq
Now, while we assumed that $\omega (q_k, q'_k) = \omega (q_j, q'_j)$, we did \emph{not} make any assumptions regarding $\omega (q_i, q'_i)$. If, however, this equality is \emph{not} true, the signal produced by particle $i$ will be inconsequential: 
\beq (\omega (q_i, q'_i) \neq \omega (q_j, q'_j)) \Longrightarrow (S_k - S_j \rightarrow S_k - S_j) \eeq
On the other hand, if $\omega (q_i, q'_i)$ happened to coincide with $\omega (q_j, q'_j)$, then the updating will occur; the "difference" between the "updates" of $S_k$ and $S_j$ will be given by 
\beq (\omega (q_i, q'_i) = \omega (q_j, q'_j)) \Longrightarrow (S_k - S_j \rightarrow (1-k_S \delta t)(S_k - S_j)) \eeq
Now, since every particle emits signals with periodicity $\delta t$, we know that within this interval each of these signals will be emitted exactly once. As a result, it is easy to see that the incriment of $S_k - S_j$ within the interval $\delta t$ is given by
\beq S_k (t+ \delta t) - S_j (t+ \delta t) \approx (1-k_S \delta t)^{\sharp \{i \vert \omega (q_i, q'_i) = \omega (q_j, q'_j) \}} ( S_k (t) - S_j (t)) \eeq
If we now take a "finite" time interval, $t_2 - t_1$, we will get close enough approximation by assuming that the "very large" number $(t_2 - t_1)/ \delta t$ happens to be an integer. Thus we obtain
\beq S_k (t_2) - S_j (t_1) \approx (1-k_S \delta t)^{\sharp \{i \vert \omega (q_i, q'_i) = \omega (q_j, q'_j) \} (t_2 - t_1)/ \delta t} ( S_k (t) - S_j (t)) \rightarrow_{t_2 \rightarrow \infty} 0 \eeq
Thus, desired global correlation (at $t_2$)  is produced in the limit $t_2 \rightarrow \infty$

\subsection*{4.6 Time independence of $S_k$ and its value at the equilibrium (parts 11-12)}

As we have stated earlier, $S_k$ is supposed to "encode" the transition probabily from the state $\{i \vert q_i = q_k \}$ to the state $\{j \vert q_j = q'_k \}$. We know from the Schr\"odinger's picture of quantum field theory that the probability of transition between any \emph{two} states is time-independant (time dependance is only envoked in probability amplitudes of \emph{individual} states, which is described by $\psi_k$). There is, however, one problem. Apart from time independance, we have specific form in mind that we would like $S_k$ to take. Now, from "mechanistic" point of view, the initial conditions can be anything we like. This means, in particular, that initially $S_k$ might well be different from what we "want" it to be. Thus, we are forced to envoke time-dependance in order to "maneur" $S_k$ into its desired form. We can then be clever enough to make sure that this "time-dependant" dynamics has equilibrium point. Thus, "in our day and age" $S_k$ is time-independant simply because it has already reached that equilibrium. 

The dynamics of $S_k$ has already been described in parts 11-12. Therefore, our present task is to see whether or not it, in fact, leads to the euqilibrium and that its equilibrium value is, in fact, what we desire it to be. According to part 11, after the signal is received, $S_k$ is "updated" according to 
\beq S_k \rightarrow S_k + (k_S \delta t) ({\cal L}_i + S_i - S_k) T(\omega (q_i, q'_i) =  \omega (q_k, q'_k)) \eeq
Now, in the prevous section we have established that at "some" equilibrium (which is not to be confused with the equilibrium we are discussing now), a "global correlations" of the form 
\beq \omega (q_j, q'_j) = \omega (q_k, q'_k) \Rightarrow S_j = S_k \eeq
arize. In this section we are proposing a \emph{different kind} of equilibrium; namely, time-independance of $S_k$. If the equilibrium of this section were to occur "before" the equilibrium of the previous section, then the "uncorrelated picture" would have been "frozen in time". This would have prevented the "correlations" from \emph{ever} being produced, contradicting the conclusion of the previous section. Thus, the only other options we have is either to assume that the equilibrium of the previous section comes first, or else the two of them occur at the same time. Now, "simulteneous occurence" can be viewed as a special case of the statement that "equilibrium of the previous section comes first", with "time delay" being zero. Thus, we will simply assume that the equilibrium of previous section comes first, \emph{without worrying} whether time delay is zero or not. 

Now, lets consider a time interval when the equilibrium of the previous section has already been reached, while the equilibrium of the current section has not been reached yet. During this time interval, both $S_j$ and $S_k$ evolve in time and, yet, maintain their correlation. This is possible in light of the existance of "preferred time" $t$. Their "correlations" are specified as:
\beq \omega (q_j, q'_j) = \omega (q_k, q'_k) \Rightarrow S_j = S_k \eeq
 Now, we can rewrite the above statement in the following form: 
\beq {\rm Correlations} \Longleftrightarrow (S_i (t) - S_k (t)) T(\omega (q_i, q'_i) = \omega (q_k, q'_k)) = 0 \eeq
Now, if we look at the equation of "updating" of $S_k$, we can factor it in the following way: 
\beq S_k \rightarrow S_k + (k_S \delta t)  {\cal L}_i T(\omega (q_i, q'_i) +  (k_S \delta t)  (S_i - S_k) T(\omega (q_i, q'_i)=  \omega (q_k, q'_k)) \eeq
Thus, based on what we have just found, the "correlations" cause the second term on the right hand side to drop out. As a result, the "updating" equation becomes
\beq {\rm Correlations} \Longrightarrow (S_k \rightarrow S_k + (k_S \delta t) {\cal L}_i T(\omega (q_i, q'_i) =  \omega (q_k, q'_k))) \eeq
Now, apart from being "updated" whenever a pulse passes by, there is also a \emph{countinuous} evolution of $S$, which is independant of the presence or absence of the pulse. This \emph{continuous evolution} of $S_k$ is specified in part 12 as
\beq \frac{dS_k}{d t} = - \alpha S_k \eeq
Let us now combine the continuous evolution with discrete "updatings" to see what we will get. One difficulty we encounter is the fact that we can not present these two evolutions as sequential in time. After all,  the continuous evolution takes the entire $\delta t$ interval. Thus, in order for "discrete updating" to come either "before" or "after" the latter, it has to take place either at the very beginning of $\delta t$-interval or at the very end. However, we know that the timing of the discrete updating is dictated to us based on the timing of the emission of signal by one of the \emph{other} particles. Since the phases of different particles are uncorrelated, the "updating" will, most likely, occur "in the middle" of the $\delta t$-evolution; thus, preventing us from selecting the order of the two operations.

What comes to our rescue is the overall factor of $\delta t$ that multiplies the effects of both kinds of modifications. Thus, the order in which we "act" with these "operations" gives an effect of $O ((\delta t)^2)$. From our general knowledge of integral calculus, we know that the only thing we have to worry about is $O(\delta t)$, while $O((\delta t)^2$ can be safely thrown away. This allows us to simply "assign" the order to the two operations in whatever way we wish, while we are guaranteed to get the same answer. Let us, therefore, pretend that the particle "first" undergoes all of the relevent discrete updatings and "then" it "catches up" on the "continuous evolution" it "should have" been performing during the interval $\delta t$. Thus, we obtain the following equation:
\beq {\rm Correlations} \Longrightarrow S_k (t+ \delta t) = e^{- \alpha \delta t} \Big(S_k (t) + k_S \delta t \sum_{\omega (q_i, q'_i) = \omega (q_k, q'_k)} {\cal L}_i \Big) + O (\delta t)^2 \eeq
Here, the sum represents that, during the time interval $\delta t$, a particle number $k$ receives a signal from every other particle ($i$) exactly once. This is due to the fact that the period of signal-emission of \emph{all} particles is the same; namely, $\delta t$. Now, if we rewrite the above expression, by substitutting $t+ \delta t$ in the place of $t$, we obtain
\beq {\rm Correlations} \Longrightarrow S_k (t+ 2 \delta t) = e^{- \alpha \delta t} \Big(S_k (t + \delta t) + k_S \delta t \sum_{\omega (q_i, q'_i) = \omega (q_k, q'_k)} {\cal L}_i \Big) \eeq
We can now take the expression for $S(t+ 2 \delta t)$, and replace $S (t+ \delta t)$ on the right hand side with the corresponding expression in terms of $S(t)$ (which we have written earlier). With that substitution, the equation for $S(t + 2 \delta t)$ becomes 
\beq {\rm Correlations} \Longrightarrow S_k (t+ 2 \delta t) =  (k_S \delta t) e^{- \alpha \delta t} \sum_{\omega (q_i, q'_i) = \omega (q_k, q'_k)} {\cal L}_i + \nonumber \eeq
\beq + e^{- \alpha \delta t} \Big( e^{- \alpha \delta t} \Big(S_k (t) + (k_S \delta t) \sum_{\omega (q_i, q'_i) = \omega (q_k, q'_k)} {\cal L}_i \Big) \Big) = \eeq
\beq = (k_S \delta t) e^{- \alpha \delta t} \sum_{\omega (q_i, q'_i) = \omega (q_k, q'_k)} {\cal L}_i + (k_S \delta t) e^{- 2 \alpha \delta t} \sum_{\omega (q_i, q'_i) = \omega (q_k, q'_k)} {\cal L}_i + e^{- 2 \alpha \delta t} S_k (t) \nonumber \eeq
It is easy to guess (and the guess can be easily verified by induction) that $S_k (t+ n \delta t)$ is given by 
\beq {\rm Correlations} \Longrightarrow S_k (t+ n \delta t) = e^{-n \alpha \delta t} S_k (t) +( k_S \delta t) (e^{- \alpha \delta t} + \cdots + e^{-n \alpha \delta t}) \sum_{\omega (q_i, q'_i) = \omega (q_k, q'_k)} {\cal L}_i  \eeq
By expressing the geometric series in a form 
\beq a + \cdots + a^n = \frac{1-a^n}{a^{-1}-1} \eeq
and identifying $a$ with $e^{-t \alpha \delta t}$, the expression for $S_k (t+ n \delta t)$ becomes 
\beq {\rm Correlations} \Longrightarrow S_k (t+ n \delta t) = e^{-n \alpha \delta t} S_k (t) + \frac{(k_S \delta t) (1 - e^{- n \alpha \delta t})}{ e^{ \alpha \delta t} -1} \sum_{\omega (q_i, q'_i) = \omega (q_k, q'_k)} {\cal L}_i  \eeq
We  can now expand the denominator as 
\beq \frac{1}{e^{\alpha \delta t}-1} = \frac{1}{1 + \alpha \delta t  + \alpha^2 (\delta t)^2/2 -1 + O ((\delta t)^3)} = \frac{1}{\alpha \delta t (1 + \alpha \delta t/2 + O ((\delta t)^2))} = \nonumber \eeq
\beq = \frac{1- \alpha \delta t/2 + O ((\delta t)^2)}{\alpha \delta t} = \frac{1}{\alpha \delta t} - \frac{1}{2} + O (\delta t) \eeq
We can, therefore, rewrite the above expression as 
\beq S_k (t+ n \delta t) = e^{-n \alpha \delta t} S_k (t) + (k_S \delta t) (1- e^{- n \alpha \delta t}) \Big( \frac{1}{\alpha \delta t} - \frac{1}{2} + O (\delta t) \Big) \sum_{\omega (q_i, q'_i) = \omega (q_k, q'_k)} {\cal L}_i  \eeq
Now, in a limit of $n \rightarrow \infty$, we obtain
\beq \lim_{n \rightarrow \infty} S_k  (t_0 + n \delta t) = \frac{k_S}{\alpha} \sum_{\omega (q_i, q'_i) = \omega (q_k, q'_k)} {\cal L}_i  \eeq
The fact that right hand side is $n$-independant shows that the action approaches constant if we wait long enough, as desired. Furthermore, the "sum" on the right hand side implies that the "global Lagrangian" $S_k$ can, indeed, be represented as a "discretized integral" of the local one ${\cal L}_i$. Let us evaluate the coefficient in front of the integral. Now, there are two kinds of densities involved: density of individual points ($\rho_{\rm pt}$) and density of clusters of points ($\rho_{\rm cl}$). They are related by
\beq \frac{\rho_{\rm pt}}{\rho_{\rm cl}} \approx \frac{N_{\rm pt}}{N_{\rm cl}},\eeq
where $N_{\rm pt}$ and $N_{\rm cl}$ represents the number of points and clusters of points, respectivey, and the approximation sign is due to the random fluctuations of density in space. It should be noticed that the signals are being emitted by \emph{individual points} as opposed to "clusters as a whole". In other words, within the above sum, each cluster is represented \emph{several times}, while each point is represented \emph{exactly once}. Thus, in order to convert the sum into the integral, we need to know the density of \emph{points}. Now, if the volume of compactified space is given by $\cal V$, then the density of poitns is given by 
\beq \rho_{pt} \approx \frac{N_{\rm pt}}{\cal V} \eeq
Now, the only "points" we care about are the ones satisfying $\omega (q_i, q'_i) = \omega (q_k, q'_k)$. Since $\omega$ is one to one, this is equivalent to saying that $q_i =q_k$ and $q'_i = q'_k$ independantly hold. If the number of "allowed" values of "charges" is $M$, the probability of the above is $1/M^2$. This means that the density of the points "relevant for $\omega$" is given by
\beq \rho_{{\rm pt} \; \omega} = \frac{\rho_{\rm pt}}{M^2} = \frac{N_{\rm pt}}{M^2 \cal V} \eeq
Now, in order to convert the sum into the integral, we have to assume that ${\cal L}_k$ is some sort of discrete approximation to a continuous function. If we look at \emph{all} points, this can't possibly be the case. After all, if $q_i \neq q_j$ and/or $q'_i \neq q'_j$ then ${\cal L}_i$ and ${\cal L}_j$ represents two separate kinds of transitions; thus, we expect them to be completely different from each other. However, what we \emph{can} assume is this. Suppose we have $M^2$ \emph{separate} differential functions, ${\cal L}_{qq'}$, where $q$ and $q'$ are arbitrary elements of $\{1, \cdots, M\}$. We can now assume that 
\beq {\cal L}_k \approx {\cal L}_{q(k), q'(k)} (\vec{x}_k) \eeq
We can now use this assumption and "convert" the sum into the integral in the expression for $S_k$:
\beq \lim_{n \rightarrow \infty} S_k  (t_0 + n \delta t) \approx \frac{k_S \rho_{{\rm pt} \; \omega}}{\alpha} \int d^3 x\;  {\cal L}_{qq'} (\vec{x})  \eeq
Finally, we substitute the value of $\rho_{{\rm pt} \; \omega}$ and obtain 
\beq \lim_{n \rightarrow \infty} S_k  (t_0 + n \delta t) \approx \frac{k_S N_{\rm pt}}{\alpha M^2 \cal V} \int d^3 x\;  {\cal L}_{qq'} (\vec{x})  \eeq
We can now immediately find the appropriate value of $k_s$ that would set the overall coefficient in the right hand side of the above equation to $1$: 
\beq k_s = \frac{\alpha M^2 \cal V}{N_{\rm pt}} \Longrightarrow \lim_{n \rightarrow \infty} S_k  (t_0 + n \delta t) \approx \int d^3 x\;  {\cal L}_{qq'} (\vec{x}) \label{k_S}\eeq
The idea of setting $k_s$ to the value we want it to have can be easilly accomplished by "absorbing" an appropriate coefficient into $\cal L$. The latter can be done by similar absorption of that coefficient into the equations that generate $\cal L$. This, however, need not be explicitly done in this paper. We can simply assume that we have "started out" from the "correct" values of coefficients.  

Finally, it is important to make it clear that the "original" degrees of freedom are ${\cal L}_k$, and \emph{not} ${\cal L}_{qq'} (\vec{x})$. After all, the goal of this paper is to avoid anything and everything that involves "too many" degrees of freedom (this was the ultimate reason why we wanted to "encode" configuration space within ordinary one). Now, ${\cal L}_k$ is "attached to a point" (namely, point $k$). Thus, at a point number $5$ we have only \emph{one} parameter, namely, ${\cal L}_5$. On the other hand, ${\cal L}_{qq'} (\vec{x})$, being "continuous" itself, depends on a \emph{continuous} parameter $\vec{x}$.  Thus, \emph{at any given $\vec{x}$}, all $M^2$ "versions" of $\cal L$ are relevent. We can not "allow" this to happen! After all, "too many degrees of freedom" is precisely what we "didn'nt like" about configuration space. Thus, it would ruin the purpose of this whole paper if we were to resort to something we were trying to avoid. 

The way out of it is to make sure that ${\cal L}_{qq'} (\vec{x})$ is never mentioned in the "formal" dynamical equations that we are "postulating". It is "okay" to mention it in our "analysis" of the outcomes of these equations, as long as it is not present in the original equations, themselves. In fact, the mention of ${\cal L}_{qq'} (\vec{x})$ was merely an "informal" way of saying that something we \emph{do} care about (namely, ${\cal L}_k$) is a discrete approximation to \emph{something} continuous (and we merely denoted this "something" by ${\cal L}_{qq'} (\vec{x})$). Now, there is no fundamental reason to assume that ${\cal L}_k$ is continuous to begin with; after all,  when we perform path integral of, say, scalar field $\phi$ \emph{over a lattice} we treat the neighboring lattice points as independant degrees of freedom, thus assuming discontinuity. As long as the discontinuity is "allowed", ${\cal L}_{qq'} (\vec{x})$ has no role to play!

The reason we "looked at" quasi-continuous behavior (and, therefore, ${\cal L}_{qq'}$) is that the criteria of "valid" discrete Lagrangians is based on the "desired" behavior in a continuous limit. Thus, all we did was to check that we are, in fact, "allowed" to define ${\cal L}_k$ and $S_k$ the way we did. Once we have convinced ourselves that we were "allowed" to do it, we then drop the whole issue of continuity (and, therefore, ${\cal L}_{qq'} (\vec{x})$ altogether and simply assume the discretized dynamics we have postulated to begin with.

\subsection*{4.7 Evolution of $\psi_j$ under $S_k$ (part 11)}

In the previous section we have established that the "action", $S_i$, does, in fact, approximate an "integral" of the "Lagrangian", ${\cal L}_j$ over the spacelike plane (corresponding to the "constant" value of "preferred time", $t$). We have also shown that, while the "unwanted" time-dependance of $S_k$ might take place "at earlier times", it goes away at the equilibrium. We have further assumed that $S_k$ has already reached the equilibrium at the time we are living in. Together, this sums up by saying that $S_k$ has a "mathematical structure" of the "action". We will now proceed to show that the "state", $\psi_j$, evolves under that "action" in a desired manner. According to part 11, $\psi_j$ is "updated" according to 
\beq \psi_j \longrightarrow \psi_j + (k_{1 \psi} \delta t) \Psi (\vec{x}_k, t) e^{iS (\vec{x}_j, t))} T(\vert q_j - Q'(\vec{x}_j, t) \vert < \epsilon_q) + \nonumber \eeq
\beq + (k_{2 \psi} \delta t) (\Psi (\vec{x}_j , t) - \psi_j) T(\vert q_j - Q (\vec{x}_j, t) \vert < \epsilon_q) \eeq
Now, according to part 5, when particle $i$ emits a signal, it causes the values of the fields $\Psi$, $Q'$ and $S$ to approximate, respectively, $\psi_i$, $q'_i$ and $S_i$ throughout the space. This, in particular, will be true at the locations of particles $j$ and $k$:
\beq \Psi (\vec{x}_j, t) \approx \Psi (\vec{x}_k, t) \approx \psi_i (t) \; ; \; Q' (\vec{x}_j, t) \approx Q' (\vec{x}_k, t) \approx q'_i \; ; \; S (\vec{x}_j, t) \approx S_i \eeq
By substitutting these values into the equation of "updating" of $\psi_i$, we obtain
\beq \psi_j \longrightarrow \psi_j + (k_{1 \psi} \delta t) \psi_i e^{iS_i} T(q_j = q'_i) + (k_{ 2 \psi} \delta t) (\psi_i  - \psi_j) T(q_i = q_j)  (\label{eqn:438})\eeq
Now, in Section 4.4 (NOTE: THIS SECTION GOES BEFORE THE ONE ON HAMILTONIAN) we have shown that, at the equilibrium, $\psi_i$ obeys desired "global correlations" among same-charged particles: 
\beq {\rm Equilibrium} \Longrightarrow (\psi_i = \psi_j \; {\rm if} \; q_j = q_i) (\label{eqn:439})\eeq
It is easy to show, case by case, that the above statement is equivalent to  
\beq {\rm Equilibrium} \Longrightarrow ((\psi_i - \psi_j) T(q_i = q_j) = 0) \eeq
We immediately notice that $(\psi_i - \psi_j) T(q_i = q_j)$ coincides with one of the terms of the equation of the "updating" of $\psi_j$ (equation (\ref{eqn:396})). Since we have now shown this term is equal to zero, it gets "thrown away", and the equation of "updating" of $\psi_i$ becomes
\beq {\rm Equilibrium} \Longrightarrow (\psi_j \longrightarrow \psi_j + (k_{1 \psi} \delta t) \psi_i e^{iS_i} T(q_j = q'_i)) \eeq
Now, the above equation tells us what happends when the particle $j$ receives a signal from \emph{only one} other particle, namely, particle $i$. However, within the time interval $\delta t$, it receives the signals from \emph{all} other particles, not just one. Thus, we obtain
\beq {\rm Equilibrium} \Longrightarrow \psi_j (t+ \delta t) = \psi_j (t) +  (k_{1 \psi} \delta t) \sum_{i} e^{iS_i} \psi_i (t) T(q_j = q'_i))\eeq
We can, equivalently, get rid of the multiple of $T(q_j = q'_i)$ and, instead, say we are only summing over $i$ that satisfy $q_j = q'_i$:
\beq {\rm Equilibrium} \Longrightarrow \psi_j (t+ \delta t) = \psi_j (t) +  (k_{1 \psi} \delta t) \sum_{\{ i \vert q'_i=q_j \} } e^{iS_i} \psi_i (t)\eeq
We can now group the terms in the sum based on the "common value" of $q_i$. In other words, we will go over all values of $q \in \{1 \cdots M \}$ and, for each $q$, we will "go over" the particles satisfying $q_i =q$, while assuming no restrictions for $q'_i$: 
\beq {\rm Equilibrium} \Longrightarrow \psi_j (t+ \delta t) = \psi_j (t) +  (k_{1 \psi} \delta t) \sum_q \sum_{\{ i \vert q_i = q \; ; \; q'_i=q_j \} } e^{iS_i} \psi_i (t) \eeq
In order to reproduce the mathematical information of unitary evolutiom of $\psi$, we need to single out the degrees of freedom that are \emph{not} set to zero through correlations. For this end, we will define the "same-$q$ average value of $\psi$" and "same-$\omega$ average of $S$" as 
\beq \tilde{\psi} (q, t) = \frac{1}{\sharp Q_q (S)} \sum_{p \in Q_q (S)} \psi (p, t) \eeq
\beq \tilde{S} (q, q', t) = \frac{1}{\sharp \Omega_{\omega(q, q')}  (S)} \sum_{p \in \Omega_{\omega(q,q')} (S)} S (p, t) \eeq
In terms of these averages, the "equilibrium" is defined as 
\beq {\rm Equilibrium} \Longrightarrow ((\psi_k (t) = \tilde{\psi} (q_k, t)) \wedge (S_k (t) = \tilde{S} (q_k, q'_k, t)))  \eeq
By plugging this into the equation for $\psi_j (t+ \delta t)$, we obtain 
\beq {\rm Equilibrium} \Longrightarrow \psi_j (t+ \delta t) = \psi_j (t) +  (k_{1 \psi} \delta t) \sum_q \sum_{\{ i \vert q_i = q \; ; \; q'_i=q_j \} } e^{i \tilde{S} (q_i, q'_i)} \tilde{\psi} (q_i, t) \eeq
Since one of the "conditions" of the sum i the above equation is $q'_i = q_j$ we an replace $\tilde{S} (q_i, q'_i)$ with $\tilde{S} (q_i, q_j)$. Furthermore, because of the condition $q_i =q$, we can further replace $\tilde{S} (q_i, q_j)$ with $\tilde{S} (q, q_j)$. Thus, we obtain 
\beq {\rm Equilibrium} \Longrightarrow \psi_j (t+ \delta t) = \psi_j (t) +  (k_{1 \psi} \delta t) \sum_q \sum_{q_i = q \; ; \; q'_i=q_j} e^{i \tilde{S} (q, q_j)} \tilde{\psi} (q, t) \label{eqn:449} \eeq
We now notice that, in the new form,  the summands in our sum "happen" to be independant of $q_i$ and $q'_i$. Therefore, we can remove the sum over $q_i$ and $q'_i$ in favor of multiplication by the number of choices of $q_i$ and $q'_i$: 
\beq {\rm Equilibrium} \Longrightarrow \psi_j (t+ \delta t) = \psi_j (t) +  (k_{1 \psi} \delta t) \sum_q (\sharp \{ i \vert q_i = q \; ; \; q'_i=q_j \}) e^{i \tilde{S} (q, q_j)} \tilde{\psi} (q, t) \label{eqn:450} \eeq
Let us now verify that the equilibrium is "sustained" under the above equation. By an "equilibrium" we mean the "correlation" of $\psi_i$ among same-charged particles: 
\beq {\rm Equilibrium \; at \; t}\;\Longrightarrow (q_j= q_k  \Rightarrow \psi_j (t) = \psi_k (t)) \eeq
Now, we have just shown in equation (\ref{eqn:449}) that, if "equlibirium condition" holds at $t$, then $\psi_j (t+ \delta t)$ is a specified function of $\psi_j (t)$ and $q_j$ (that function further assumes the set of values of $\tilde{S}_k$). Therefore, if $q_j = q_k$ then the "inputs" associated with $j$-particle and $k$-particle are the same. After all, the $q$-inputs are the same by assumption, and $\psi$-inputs are the same based on "equilibrium condition" at $t$. This automatically implies that $\psi_j (t+ \delta t)$ and $\psi_k (t+ \delta t)$ are likewise the same: 
\beq {\rm Equilibrium \; at \; t} \Longrightarrow  (q_j = q_k \Rightarrow \psi_j (t+ \delta t) = \psi_k (t + \delta t)) \eeq
But, the right hand side is precisely the "equilibrium condition" \emph{at $t+ \delta t$}. Thus, we can rewrite it as 
\beq {\rm Equilibrium \; at \; t} \Longrightarrow {\rm Equilibrium \; at \; t+ \delta t} \eeq
As we poitned out earlier, the "global correlations" that define equilibrium are equivalent to a statement that any individual value of $\psi_i$ is equal to its average over a group of same-charged particles that include particle $i$. Thus, we obtain 
\beq (\forall i (\psi_i (t) = \tilde{\psi} (q_i, t))) \Rightarrow  (\forall i ( \psi_i (t+ \delta t) = \tilde{\psi} (q_i, t+ \delta t))) \eeq
We can now replace $\psi_j (t+ \delta t)$ on the left hand side of Equation (\ref{eqn:450}) with the above-mentioned "average" and obtain
\beq \tilde{\psi} (q_j, t+ \delta t) = \tilde{\psi} (q_j, t) +   (k_{1 \psi} \delta t) \sum_q (\sharp \{q_i = q \; ; \; q'_i=q_j \}) e^{i \tilde{S} (q, q_j)} \tilde{\psi} (q, t) \eeq  
Now, if the total number of available "partiles" is $N$, and the total number of "allowed" values of $q$ is $M$ then, statistically, one might expect 
\beq \sharp \{q_i = q \; ; \; q'_i=q_j \} \approx \frac{N}{M}. \eeq
By substitutting this into the equation (\ref{eqn:450}), we obtain
\beq \tilde{\psi} (q_j, t+ \delta t) \approx  \tilde{\psi} (q_j, t) +  \frac{N k_{1 \psi} \delta t}{M} \sum_q  e^{i \tilde{S} (q, q_j)} \tilde{\psi} (q, t) \eeq 
Now, in order to match the evolution equation, we would "like" the coefficient in front of the sum to be equal to $\delta t$. We can do it by setting $k_{1 \psi}$ to be 
\beq k_{1 \psi} = \frac{M}{N} \eeq
Now, since $N$ is the total number of particles while $M$ is the total number of allowed "charges", $N/M$ is an expected number of particles of each charge. This means that $N/M$ is "very large" while $k_{1 \psi} = M/N$ is "very small". This raises a question: why do we need "very small" $k_{1 \psi}$ if we already have "very small" $\delta t$? The answer to this question is that, during the interval $\delta t$, the particle is updated "very large" number of times (after all, it receives the signals from \emph{all} of the particles of the same charge). Therefore, in order for the \emph{overall} change throughout time interval $\delta t$ to be "small of the order 1", we would like the change produced by each signal to be "small of the order 2". That is why we have a coefficient $k_{1 \psi} \delta t$. 

It should be pointed out, however, that "small of the order 2" is not completely accurate. After all, the reason why $\delta t$ is "small" has nothing to do with the reason why $k_{\psi}$ is small; so it is entirely possible that $k_{\psi}$ is of the order of either $(\delta t)^2$ or $(\delta t)^{1/2}$, or something else entirely. What we do say, however, is that we would like our result to be of the order of $\delta t$. If the impact of each pulse was, itself, of the order of $\delta t$, then we would have gotten order of "small" $\delta t$ multiplied by "large" number of same-charged particles. We do not know whether the result would have been "small" or "large"; we \emph{do} know, however, that it would \emph{not} match $\delta t$ (simply because we know that the number of same-charged particles is \emph{not} of the order of $1$). Therefore, we need something to cancel the number of same-charged particles. The $k_{1 \psi}$ defined as above does, in fact, cancel them. Since the number of same-charged particles happened to be "large", $k_{1 \psi}$ is, accordingly, "small". 

\subsection*{4.8 Emission of a signal (parts 4-6)}

Let us now describe the mechanism of emission of signals. As we have stated in part 4, the pulses are emitted with periodicity $2 \pi / \nu$. This can be modelled by introducing two additional internal degrees of freedom, $e_{j1}$ and $e_{j2}$ (letter "$e$" stands for "emittor"). These evolve according to  
\beq \frac{de_{j1}}{d t} = \frac{\nu e_{j2}}{\sqrt{e_{j1}^2+e_{j2}^2}} \; ; \; \frac{de_{j2}}{d t} = - \frac{\nu e_{j1}}{\sqrt{e_{j1}^2+e_{j2}^2}} \label{eqn:459} \eeq
which, intuitively, looks like a "motion around the circle" on a so-called "$e_1e_2$-plane". The signal is being "emitted" during the time period when $e_{j2}/ \vert e_{j1} \vert < \chi$, where $\chi$ is some small constant. Thus, any given particle emits a signal with well-defined periodicity. The "randomness" is due to the fact that the "initial phases" of various particles are not "matched". The value of $\chi$ should be \emph{so small} that the "emission times" of various particles do not overlap, despite the fact that the number of particles is very large. Now, according to part 5, our aim is to make sure that during the "passage" of the "signal" the approximations 
\beq {\cal L} (\vec{x}, t) \approx {\cal L}_j \; ; \; \psi (\vec{x}, t) \approx \psi_j \; ; \; \Phi (\vec{x}, t) \approx \phi_j \eeq
\beq  V^{\mu} (\vec{x}, t) \approx V_j^{\mu} \; ; \; Q^{\mu} (\vec{x}, t) = q_j \; ; \; \vec{X} (\vec{x}, t) = \vec{x} - \vec{x}_j  \nonumber \eeq
are satisfied. Unfortunately, this can not be done directly. Instead, we will introduce a "messenger" fields, $\mu_{\cal L}$, $\mu_{\psi_1}$, $\mu_{\psi_2}$, $\mu_{\phi_1}$, $\mu_{\phi_2}$, $\mu_V^{\mu}$, $\mu_Q$ and $\mu_X$. These fields will then be, respectively, "converted" to $\Phi_1 = Re \; (\Phi)$, $\Phi_2 = Im \; (\Phi)$, $V^{\mu}$, $Q$ and $\vec{X}$ on a pointwise basis. Now, in order for these fields to carry the information about the internal degrees of freedom of "source point" $j$ (namely, $\phi_{j1} =\phi_1 (j) = Re (\phi (j))= Re (\phi_j)$ and $\phi_{j2} =\phi_2 (j) = Im (\phi (j))= Im (\phi_j)$), their "sources" at $j$ have to be coupled to the corresponding parameters. Furthermore, since we would like the "emission" \emph{only} to occur when $e_{j2}/ \vert e_{j1} \vert < \chi$, there should be a coupling to $T(e_{j2}/ \vert e_{j1} \vert < \chi)$. Thus, the "emission" equations are given by 
\beq \nabla^{\alpha}_{s} \nabla_{s; \alpha} \mu_{\cal L} + m_{\mu_{\cal L}}^2 \mu_{\cal L} + \zeta_{\mu_{\cal L}} \partial_0 \mu_{\cal L} = \sum_{j=1}^N {\cal L}_j \delta^3   (\vec{x}- \vec{x}_j)  T \Big(\frac{e_{j2}}{\vert e_{j1} \vert}   < \chi \Big) \nonumber\eeq
\beq \nabla^{\alpha}_s \nabla_{s; \alpha} \mu_{\psi_1}+ m_{\mu_{\psi}}^2 \mu_{\psi_1} + \zeta_{\mu_{\psi}} \partial_0 \mu_{\psi_1} = \sum_{j=1}^N \psi_1 (j) \delta^3   (\vec{x}- \vec{x}_j) T \Big(\frac{e_{j2}}{\vert e_{j1} \vert} < \chi \Big) \nonumber \eeq
\beq \nabla^{\alpha}_{s} \nabla_{s; \alpha} \mu_{\psi_2}+ m_{\mu_{\psi}}^2 \mu_{\psi_2} + \zeta_{\mu_{\psi}} \partial_0 \mu_{\psi_2}= \sum_{j=1}^N \psi_2 (j) \delta^3   (\vec{x}- \vec{x}_j)  T \Big(\frac{e_{j2}}{\vert e_{j1} \vert}   < \chi \Big)  \nonumber \eeq
\beq \nabla^{\alpha}_s \nabla_{s; \alpha} \mu_{\phi_1}+ m_{\mu_{\phi}}^2 \mu_{\phi_1} + \zeta_{\mu_{\phi}} \partial_0 \mu_{\phi_1}= \sum_{j=1}^N \phi_1 (j) \delta^3   (\vec{x}- \vec{x}_j) T \Big(\frac{e_{j2}}{\vert e_{j1} \vert} < \chi \Big) \label{eqn:461}\eeq
\beq \nabla^{\alpha}_s \nabla_{s; \alpha} \mu_{\phi_2}+ m_{\mu_{\phi}}^2 \mu_{\phi_2} + \zeta_{\mu_{\phi}} \partial_0 \mu_{\phi_2}= \sum_{j=1}^N \phi_2 (j) \delta^3   (\vec{x}- \vec{x}_j) T \Big(\frac{e_{j2}}{\vert e_{j1} \vert} < \chi \Big)  \nonumber \eeq
\beq \nabla^{\alpha}_{s} \nabla_{s; \alpha} \mu_V^{\mu}+ m_V^2 \mu_V^{\mu} + \zeta_V \partial_0 \mu_V^{\mu}= \sum_{j=1}^N V^{\mu} (j) \delta^3   (\vec{x}- \vec{x}_j)  T \Big(\frac{e_{j2}}{\vert e_{j1} \vert}   < \chi \Big) \nonumber \eeq
\beq \nabla^{\alpha}_{s} \nabla_{s; \alpha} \mu_Q + m_{\mu_q}^2 \mu_Q + \zeta_{\mu_q} \partial_0 \mu_Q= \sum_{j=1}^N q_j \delta^3   (\vec{x}- \vec{x}_j)  T \Big(\frac{e_{j2}}{\vert e_{j1} \vert}   < \chi \Big) \nonumber\eeq
\beq \nabla^{\alpha}_{s} \nabla_{s; \alpha} \mu_{Q'} + m_{\mu_q}^2 \mu_{Q'} + \zeta_{\mu_q} \partial_0 \mu_{Q'}= \sum_{j=1}^N q'_j \delta^3   (\vec{x}- \vec{x}_j)  T \Big(\frac{e_{j2}}{\vert e_{j1} \vert}   < \chi \Big) \nonumber\eeq
Here, the ''first time derivatives'' are introduced in order for the fields to attinuate so that they won't ''come back'' once they circle the universe. At the same time, the coefficients are small enough so that they can be approximated as non-attinuating on the small scales we are working on. The violation of relativity per first derivative is obvious, but we consider it acceptable since we violate relativity anyway for other purposes. The "summation" over $j$ is due to the fact that the fields are simultaneously coupled to \emph{all} of the particles \emph{at the same time}. After all, if we had separate set of fields for each of the particles, we would have "very large" number of fields propagating through space. This would cause the same "philosophical" problems as the large number of coordinates in a configuration space which we try to avoid. Thus, we claim that we have \emph{few} fields that are coupled to \emph{all} particles \emph{at the same time}.  The only reason their influences are distinguishable from each other is that they come in a form of "pulses" and the values of $Q$ and $\vec{X}$ measured during each given pulse allows us to specify the location and position in space of the particle that emitted it. This, in fact, is a prime reason for the parameters $Q$ and $\vec{X}$; this is also the main reason why we "want" the signals to come in a form of "pulses". 

Let us now find a way of "decoding" these signals and thus determining the parameters of the particles that emitted them. In order to avoid relativistic delays we will do the following trick. We will assume that the signals obey "superluminal" ($c_s$-based) relativity, while the trajectory of the particles obeys "ordinary" ($c_o$-based) relativity. That is, the signals propagate with velocity $c_s$, which corresponds to the metric $g_{s; \mu \nu}$; at the same time, the trajectories of the particles are derived (according to Chapter 5) based on \emph{another} metric, $g_{o; \mu \nu}$, which corresponds to the speed $c_o$. As a result, the $c_o$-based relativity would prevent the particles from moving "fast enough" for the $c_s$-based "relativistic delay" to be felt. Thus, 
\beq (\forall j (v_j \ll c_s))  \Rightarrow \Big( (\mu_{\phi_1}, \mu_{\phi_2}, \mu_V^{\mu}, \mu_Q)  \approx \sum_{j=1}^N \frac{(\phi_1 (j), \phi_2 (j), V_j^{\mu} , q_j)}{\vert \vec{x} - \vec{x}_j \vert} T \Big( \frac{e_{j2}}{\vert e_{j1} \vert}  < \chi \Big) \Big) \eeq
From the above equation, in combination with spherical symmetry around $\vec{x}_j$, it is easy to see that $\mu_Q$ obeys
\beq \vec{\nabla} \mu_Q \approx - \frac{q_j (\vec{x} - \vec{x}_j)}{\vert \vec{x} - \vec{x}_j \vert^2} \; ; \; \vert \vec{\nabla} \mu_Q \vert \approx \frac{q_j}{\vert \vec{x} - \vec{x}_j \vert} \eeq
This immediately implies that 
\beq \vert \vec{x} - \vec{x}_j \vert \approx \frac{\mu_q}{\vert \vec{\nabla} \mu_q \vert} \eeq
In order to find the direction of $\vec{x} - \vec{x}_j$, we should notice that it is supposed to be anti-parallel to $\vec{\nabla} \mu_Q$. Thus, we have to multiply its magnitude, given above, by $\vec{\nabla} \mu_Q / \vert \vec{\nabla} \mu_Q \vert$, and put minus sign in front. This immediately tells us that 
\beq \vec{x} - \vec{x}_j \approx  - \frac{\mu_q \vec{\nabla} \mu_q}{\vert \vec{\nabla} \mu_q \vert^2} \eeq
What we have just done \emph{could have} gotten us into trouble with sign. In particular, if instead of using $\mu_Q$ we were to use $\mu_{\phi_1}$ or $\mu_{\phi_2}$ the gradient \emph{could} have been pointing in the opposite direction from what we assumed if $\phi_{i1}$ or $\phi_{ij}$ were negative. If, however, we use $\mu_Q$ then we are "safe" since all of the particles have positive "charge". 
Now, the above expression allows us to \emph{replace} "non-local" expression $\vec{x} - \vec{x}_j$ with a "local" one, $\vec{X} (\vec{x}, t)$, where 
\beq \vec{X} = - \frac{\mu_q \vec{\nabla} \mu_q}{\vert \vec{\nabla} \mu_q \vert^2}. \eeq
It should be noticed that, while the expression for $\vec{x} - \vec{x}_j$ is an approximation, the expression for $\vec{X}$ is exact. This is due to the fact that $\vec{X}$ has no meaning apart from the definition we provide. Thus it is up to us how we define it, and we \emph{choose} the equality to be exact. On the other hand, $\vec{x} - \vec{x}_j$ has a meaning independant of out theory; thus we do not have such freedom. Let us now determine the values of ${\cal L}_j$, $\psi_{1j}$, $\psi_{2j}$, $\phi_{1j}$, $\phi_{2j}$, $\Vec{V}_j^{\mu}$, $q_j$ and $q'_j$. We know from the above dynamics that the corresponding $\mu$-fieds satisfy 
\beq \mu_{\psi_1} (\vec{x}, t) \approx \frac{\psi_{j1}}{\vert \vec{x} - \vec{x}_j \vert} \; ; \; \mu_{\psi_2} (\vec{x}, t) \approx \frac{\psi_{j2}}{\vert \vec{x} - \vec{x}_j \vert}  \; ; \; \mu_{\cal L} (\vec{x}, t) \approx \frac{{\cal L}_j}{\vert \vec{x} - \vec{x}_j \vert} \nonumber \eeq
\beq \mu_{\phi_1} (\vec{x}, t) \approx \frac{\phi_{j1}}{\vert \vec{x} - \vec{x}_j \vert} \; ; \; \mu_{\phi_2} (\vec{x}, t) \approx \frac{\phi_{j2}}{\vert \vec{x} - \vec{x}_j \vert}  \; ; \; \mu_V^{\mu} (\vec{x}, t) \approx \frac{V_j^{\mu}}{\vert \vec{x} - \vec{x}_j \vert} \eeq
\beq \mu_Q (\vec{x}, t) \approx \frac{q_j}{\vert \vec{x} - \vec{x}_j \vert} \; ; \;  \mu_{Q'} (\vec{x}, t) \approx \frac{q'_j}{\vert \vec{x} - \vec{x}_j \vert} \nonumber \eeq
 The above expression imediatey implies that the values of ${\cal L}_j$, $\psi_{1j}$, $\psi_{2j}$, $\phi_{1j}$, $\phi_{2j}$, $\Vec{V}_j^{\mu}$, $q_j$ and $q'_j$ are given by 
\beq \psi_{1j} \approx  \vert \vec{x} - \vec{x}_j \vert \mu_{\psi_1} (\vec{x}, t) \; ; \;  \psi_{2j} \approx  \vert \vec{x} - \vec{x}_j \vert \mu_{\psi_2} (\vec{x}, t) \; ; \; {\cal L}_j \approx  \vert \vec{x} - \vec{x}_j \vert \mu_{\cal L} (\vec{x}, t) \nonumber \eeq
\beq \phi_{1j} \approx  \vert \vec{x} - \vec{x}_j \vert \mu_{\phi_1} (\vec{x}, t) \; ; \;  \phi_{2j} \approx  \vert \vec{x} - \vec{x}_j \vert \mu_{\phi_2} (\vec{x}, t) \; ; \; V_j^{\mu} \approx  \vert \vec{x} - \vec{x}_j \vert \mu_V^{\mu} (\vec{x}, t) \eeq
\beq q_j \approx  \vert \vec{x} - \vec{x}_j \vert \mu_q (\vec{x}, t) \; ; \;  q'_j \approx  \vert \vec{x} - \vec{x}_j \vert \mu_{q'} (\vec{x}, t) \nonumber \eeq
Now we have already established that, \emph{during the passage of the pulse}, $\vec{x} - \vec{x}_j$ can be approximated by $\vec{X} (\vec{x}, t)$. Thus, we obtain 
\beq \psi_{1j} \approx  \vert \vec{X} (\vec{x},t) \vert \mu_{\psi_1} (\vec{x}, t) \; ; \;  \psi_{2j} \approx   \vert \vec{X} (\vec{x},t) \vert \mu_{\psi_2} (\vec{x}, t) \; ; \; {\cal L}_j \approx   \vert \vec{X} (\vec{x},t) \vert \mu_{\cal L} (\vec{x}, t) \nonumber \eeq
\beq \phi_{1j} \approx  \vert \vec{X} (\vec{x},t) \vert \mu_{\phi_1} (\vec{x}, t) \; ; \;  \phi_{2j} \approx   \vert \vec{X} (\vec{x},t) \vert \mu_{\phi_2} (\vec{x}, t) \; ; \; V_j^{\mu} \approx   \vert \vec{X} (\vec{x},t) \vert \mu_V^{\mu} (\vec{x}, t) \eeq
\beq q_j \approx   \vert \vec{X} (\vec{x},t) \vert \mu_Q (\vec{x}, t) \; ; \; q'_j \approx   \vert \vec{X} (\vec{x},t) \vert \mu_{Q'} (\vec{x}, t) \nonumber \eeq
Finally, we will turn the above "approximations" into the \emph{exact} expressions by \emph{defining} the fields $\cal L$, $\psi_1$, $\psi_2$, $\phi_1$, $\phi_2$, $\Vec{V}_j^{\mu}$, $Q$ and $Q'$ based on these expressions. Since none of these fields have been defined so far \emph{on a continuum}, we are "allowed" to define them however we want; and, in particular, we are "allowed" to demand that they \emph{exactly} coincide with our "approximate" expressions: 
\beq \psi_1 (\vec{x}, t) = \mu_{\psi_1} (\vec{x}, t) \vert \vec{X} (\vec{x}, t) \vert \; ; \; \psi_2 (\vec{x}, t) = \mu_{\psi_2} (\vec{x}, t) \vert \vec{X} (\vec{x}, t) \vert \nonumber \eeq
\beq  {\cal L} (\vec{x}, t) = \mu_{\cal L} (\vec{x}, t) \vert \vec{X} (\vec{x}, t) \vert \; ; \; V^{\mu} (\vec{x}, t) = \mu_V^{\mu} (\vec{x}, t) \vert \vec{X} (\vec{x}, t) \vert  \eeq
\beq \phi_1 (\vec{x}, t) = \mu_{\phi_1} (\vec{x}, t) \vert \vec{X} (\vec{x}, t) \vert \; ; \; \phi_2 (\vec{x}, t) = \mu_{\phi_2} (\vec{x}, t) \vert \vec{X} (\vec{x}, t) \vert \nonumber \eeq
\beq Q (\vec{x}, t) = \mu_Q (\vec{x}, t) \vert \vec{X} (\vec{x}, t) \vert \; ; \; Q' (\vec{x}, t) = \mu_{Q'} (\vec{x}, t) \vert \vec{X} (\vec{x}, t) \vert \nonumber \eeq
This allows us to make sure that the information about particle $j$ is \emph{locally available} at a point $\vec{x}$ which is "far away" from $j$. Now, if we place \emph{another} particle, $i$, at $\vec{x}$, then particle $i$ will "have access" to the information about particle $j$. This will allow us to "reproduce" the Lagrangian, despite the apparent "non-locality" which is inherent in discretizations. 

\subsection*{4.9 Reaction towards signal emitted by other particles within \emph{the same} cluster (part 7)}

Up till now we have outlined our algorithm and then illustrated why that algorithm produces the mathematical information we are looking for at the "equilibrium points". There is, however, a gap that needs to be filled: we can not simply postulate an algorithm. Rather, we have to postulate physical laws, \emph{in a form of differential equaitons} that would produce a desired algorithm. We will devote the sections 4.9 through 4.12 for this purpose. Now, since we have already satisfied ourselves that the algorithm, as defined in Section 4.1, produces the results of sections 4.2 through 4.9, our \emph{only} task is to "produce" the Section 4.1; the rest will follow through. Again, we will devote separate sections for each of the parts of the algorithm of seciton 4.1. 

This section will be devoted to part 7. Thus, our task is to come up with a differential equation that would produce the following change whenever a pulse passes particle number $i$: 
\beq \phi_i \rightarrow \phi_i + (k_{\phi} \delta t) (\Phi (\vec{x}_i, t) - \phi_i) T (\vert Q (\vec{x}_i, t) - q_i \vert < \epsilon_q) T (\vert \vec{X} (\vec{x}_i, t) \vert < r_c) \eeq
\beq V_i^{\mu} \rightarrow V_i^{\mu} + (k_V \delta t) (V^{\mu} (\vec{x}_i, t) - V_i^{\mu}) T (\vert Q (\vec{x}_i, t) -  q_i \vert < \epsilon_q) T (\vert \vec{X} (\vec{x}_i, t) \vert <  r_c) \eeq
The easiest way to enforce such dynamics is to postulate that \emph{during} the pulse, the time derivatives of $\phi_i$ and $V_i^{\mu}$ are, respectively, given by $\delta \phi_i / (2 \chi)$ and $\delta V_i^{\mu} / (2 \chi)$, where $2 \chi$ is the duration of a pulse, while $\delta \phi_i$ and $\delta V_i^{\mu}$ are "desired" incriments of $\phi_i$ and $V_i^{\mu}$ that are "to be accomplished" while the pulse is in progress; the respective values of $\delta \phi_i$ and $\delta V_i^{\mu}$ are to be read off from the equations above:
\beq \delta \phi_i = (k_{\phi} \delta t) (\Phi (\vec{x}_i, t) - \phi_i) T (\vert Q (\vec{x}_i, t) - q_i \vert < \epsilon_q) T (\vert \vec{X} (\vec{x}_i, t) \vert < r_c) \eeq
\beq \delta V_i^{\mu} = (k_V \delta t) (V^{\mu} (\vec{x}_i, t) - V_i^{\mu}) T (\vert Q (\vec{x}_i, t) - q_i \vert < \epsilon_q) T (\vert \vec{X} (\vec{x}_i, t) \vert <  r_c) \eeq
 At the same time, the time derivatives should be zero if the particle is not subjected to any pulse.  Therefore, the particle somehow needs to "know" whether it is subjected to the pulse or not, in order to "decide" whether it "wants" the time derivatives to be zero or the above pre-assigned constants. It can made this "decision" by "watching" the value of $\mu_Q (\vec{x}_i, t)$: if it is smaller than a threshold $\mu_c$ (where "c" stands for "close"; there will be another threshold, $\mu_f$ for "far away" signals for parts 11-12) then "there is no pulse"; otherwise, there is one. We can do that by selecting the value of $\mu_c$ to be much smaller than the "peak" value of $\mu_Q (\vec{x}_i, t)$ during each pulse and, at the same time, much larger than any values $\mu_Q (\vec{x}_i, t)$ might have without a pulse. Our next task is to avoid self-interaction of a particle. In other words we have to distinguish the pulse that has been produced by the particle itself from the pulses that are produce by external particles, and only "react" to the latter. This can be done by realizing that the particle is "supposed to" emit the pulse if and only if $e_{j2}/ \vert e_{j1} \vert > \chi$. Therefore, we identify an \emph{external} pulse based on a combination of "existence of a pulse" ($\mu_Q (\vec{x}_i, t) > \mu_c$) \emph{together with} the \emph{lack} of any internal pulses ($e_{j2}/ \vert e_{j1} \vert > \chi$). This is identified by a \emph{product} of two truth values, 
\beq T({\rm external \; pulse}) = T (\mu_Q (\vec{x}_i, t) > \mu_c) T \Big(\frac{e_{j2}}{\vert e_{j1} \vert}  > \chi \Big) \eeq
Therefore, we would like the time derivatives of $\phi_i$ and $V_i^{\mu}$ to be $\delta \phi_i / (2 \chi)$ and $\delta V_i^{\mu} / (2 \chi)$ if the above is $1$, and we would like them to be zero otherwise. This can be accomplished through 
\beq \frac{d \phi_i}{d t} = \frac{k_{\phi} \delta \phi}{2} T (\mu_Q (\vec{x}_i, t) > \mu_c) T \Big(\frac{e_{j2}}{\vert e_{j1} \vert}   > \chi \Big) \eeq
\beq \frac{d V_i^{\mu}}{d t} = \frac{\delta V_i^{\mu}}{2} T (\mu_Q (\vec{x}_i, t) > \mu_c) T \Big(\frac{e_{j2}}{\vert e_{j1} \vert}  > \chi \Big) \eeq
Now by substitutting the values of $\delta \phi_i$ and $\delta V_i^{\mu}$ given by equations (\ref{eqn:438}) and (\ref{eqn:439}), we obtain 
\beq \frac{d \phi_i}{d t} = \frac{k_{\phi} \delta t}{2} (\Phi (\vec{x}_i, t) - \phi_i) T (\vert Q (\vec{x}_i, t) -  q_i \vert < \epsilon_q) T (\vert \vec{X} (\vec{x}_i, t) \vert < r_c) \times \nonumber \eeq
\beq \times T (\mu_Q (\vec{x}_i, t) > \mu_c) T \Big(\frac{e_{j2}}{\vert e_{j1} \vert} > \chi \Big) \eeq
\beq \frac{d V_i^{\mu}}{d t} = \frac{k_V \delta t}{2} (\Phi (\vec{x}_i, t) - \phi_i) T (\vert Q (\vec{x}_i, t) -  q_i \vert < \epsilon_q) T (\vert \vec{X} (\vec{x}_i, t) \vert < r_c) \times \nonumber \eeq
\beq \times T (\mu_Q (\vec{x}_i, t) > \mu_c) T \Big(\frac{e_{j2}}{\vert e_{j1} \vert}  > \chi \Big) \eeq

\subsection*{4.10 Reaction towards signal emitted by sources in \emph{other} clusters (parts 8, 9 and 10}

Let us now discuss parts 8, 9 and 10 in our "scheme". As we have stated, we do not want these two processes to interfere in time. Thus, we would like part 8 to take place during the \emph{first half} of the time interval the pulse passes by, and we would like part 9 to take place during the \emph{second half} of that time interval. Therefore, we need a "clock" that would let us know just "how long" has the signal been passing by. We will identify that "clock" with an internal parameter $\xi_i$. We will introduce its dynamics in such a way that $\xi_i$ "resets to zero" during the time interval \emph{between} signals, and then increases with constant rate during the time a particle $i$ is subjected to new signal. In particular, we will postulate its dynamics to be 
\beq \frac{d\xi_i}{d t} = - k_{\xi} \xi + K_{\xi} T (\mu_Q (\vec{x}_i, t) > \mu_c) T \Big(\frac{e_{j2}}{\vert e_{j1} \vert}  > \chi \Big) \eeq
Thus, we have a quantitative definition of a statement "particle $i$ is subjected to a signal". Namely, that statement is true if and only if the value of $\mu_Q$-field at a location of particle $i$ exceeds the value $\mu_c$. Furthermore, we also have a definition of signal being "external". In particular, if $e_{j2}/\vert e_{j1} \vert \geq \chi$, then the particle $i$ "should be" emitting a signal which makes it internal. Otherwise, the signal is external. Therefore, the product of these two truth values is $1$ if there is an external signal (that is, it is true that it is a signal and it is also true it is external), and the product is zero if there is no external signal (that is, either there is no signal at all, or there is a signal, but it happens to be internal). The above equation, therefore, tells us that external signals serve as a "sources" of oscillatory process. 

Now, we assume that $k_{\xi} \ll K_{\xi}$. As a result, during the reception of an external signal the rate of change of $\xi$ is 
\beq \frac{d \xi_i}{d t} = - k_{\xi} \xi + K_{\xi} \times 1 \times 1 \approx K_{\xi} \eeq
which ammounts to $\xi_i$ increasing at "approximately constant" rate. The deviation due to $k_{\xi} \xi$-term is too small to be felt during the short duration of the pulse. Thus, it evolves according to 
\beq \xi_i \approx (\xi_i)_0 +  K_{\xi} t \eeq
On the other hand, during the time interval between the external pulses we have 
\beq \frac{d \xi_i}{d t} = - k_{\xi} \xi + K_{\xi} \times 0 \times 0 = - k_{\xi} \xi \eeq
if no pulses are present, or, during the internal pulse, we have 
\beq \frac{d \xi_i}{d t} = - k_{\xi} \xi + K_{\xi} \times 1 \times 0 = - k_{\xi} \xi \eeq
Even though $k_{\xi} \xi$ is "too small" to be felt during the duration of the "external pulse", it is still "large enough" for the time interval \emph{between the external pulses}. As long as $\chi$ is small enough, the time interval between the external pulses can be made arbitrarily many times larger than duration of each of the pulses. As a result, we can make sure that $k_{\xi} \xi$ is "large enough" to make sure that $\xi$ becomes close to zero upon the arrival of the next pulse, which amounts to "resetting a clock". In other words, this will assure us that 
\beq (\xi_i)_0 \approx 0, \eeq
which will imply that during the \emph{new} external pulse $\xi_i$ will evolve as 
\beq \xi_i \approx K_{\xi} t \eeq
where we agree to identify $t=0$ with the "beginning" of the external pulse. In other words, our "clock" works well. 

Finally, we need to agree on the exact time at which the "first part" of the signal ends and the "second part" begins.  From the dynamics of $(e_{j1}, e_{j2})$ it is easy to see that these parameters are "rotating" with frequency $\nu$. Since the emission of a signal by particle $j$ takes place when $e_{j2} / \vert e_{j1} \vert < \chi$, the duration of the emission of the signal is $2 tan^{-1} \chi \approx 2 \chi$. In light of the "superluminal" $c_s$-based relativity, the signal emitted by outside source will also be perceived to have the same duration. Therefore, we can define the "first part" and "second part" of the pulse in terms of $\xi_i < \chi$ and $\xi_i > \chi$, respectively. Due to the very small value of $\chi$ (small enough to make sure that duration of each pulse is far shorter than the inverse of the number of its sources in the entire universe!), the difference between "$\chi$" and "$\tan \chi$" is immaterial. Nevertheless, for the sake of mathematical rigour, it is important to be clear whether we are being "exact" or "making an approximation". In answering this question, we can remind ourselves that our choice of splitting into "halfs" was arbitrary. We could have split it into "first third" and "remaining two thirds" or visa versa. The only thing that is important is that we make sure to set up dynamics in a way consistent with how we chose to split it. In light of this, we will split it into the "first $\chi$" and "remaining $(2\tan \chi) - \chi$. Of course, both of these happen to \emph{very closely} approximate $\chi$; but that is not so important. 

Now that we have defined the "first part" and "second part" of the pulse, we are actually ready to do part 8 and 9 (which is what this section is supposed to be all about!) Now, part 8 takes place during the "first half" of the pulse (that is, $\xi < \chi$), and it is supposed to involve "updating" of ${\cal L}_i$, which, for the sake of compactness, we can write as 
\beq {\cal L}_i \rightarrow {\cal L}_i + {\cal K}_i ,\eeq
where ${\cal K}_i$ is a "Lagrangian generator" (or, in other words, a"desired" incriment of Lagrangian $\cal L$), which is given by
\beq {\cal K} (q_i, q'_i, \overline{q}_i (t), Q(\vec{x}_i, t), \cdots) =  T(\vert q_i -  \overline{q}_i (t) \vert < \epsilon_q) T(\vert q_i- Q (\vec{x}_i, t) \vert < \epsilon_q) {\cal K}_{ss} (\cdots) +  \nonumber \eeq
\beq  + T(\vert q_i- \overline{q}_i (t) \vert< \epsilon_q) T(\vert q'_i- Q (\vec{x}_i, t) \vert<\epsilon_q) {\cal K}_{st} (\cdots) + \eeq \beq + T(\vert q'_i- \overline{q}_i (t) \vert<\epsilon_q) T(\vert q_i -  Q (\vec{x}_i, t) \vert < \epsilon_q) {\cal K}_{ts} (\cdots) + \nonumber \eeq
\beq + T(\vert q'_i- \overline{q}_i (t) \vert< \epsilon_q) T(\vert q'_i- Q (\vec{x}_i, t) \vert<\epsilon_q) {\cal K}_{tt} (\cdots) \nonumber \eeq
Since we do \emph{not} want ${\cal L}_i$ to evolve during the second part of the external pulse, we have to multiply its derivative by $T(\xi_i < \chi)$. However, if we leave it at that, then ${\cal L}_i$ will evolve during the time interval \emph{between} the pulses. After all, after sufficient time has passed for $\xi$ to "decay", it will be close to zero and, therefore, will automatically meet this condition. The evolution of $\cal L$ might be non-trivial: after all, all of the "fields" which are "supposed to" decode the information about the particle-source, continue to be \emph{formally} defined between the pulses as well. Their behavior would be chaotic and, therefore, "destroy" whatever the results we are trying to achieve. In order to avoid this, we have to put another condition: $\cal L$ can not evolve \emph{unless} there is an external pulse. This can be accomplished by another factor, $T(\mu (\vec{x}, t) > \mu_c) T(e_{j2}/ \vert e_{j1} \vert > \chi)$, which defines the presence of "external pulse". After all, we have stated earlier that "$\mu (\vec{x}, t) > \mu_c$" is a definition of "being subjected to a pulse" and $e_{j2}/ \vert e_{j1} \vert > \chi$ is a sign that the pulse is "external". Thus, the dynamics for $\cal L$ is given by  
\beq \frac{d{\cal L}_i}{d t} = - \alpha_L {\cal L}_i + \frac{{\cal K} (\cdots)}{\chi} T(\xi_i < \chi) T(\mu (\vec{x}, t) > \mu_c) T \Big(\frac{e_{j2}}{\vert e_{j1} \vert}  > \chi \Big) , \label{eqn:489}\eeq
where the division by $\chi$ is meant to "cancel" the length of the time interval, $\chi$ upon the time integration. The $- \alpha_L {\cal L}_i$ term has been directly copied from part 10 in order for ${\cal L}_i$ to slowly attinuate \emph{between} the pulses. It is so small that it is not felt on a scale of few pulses. But, as discussed earlier, this would allow us to ensure that ${\cal L}_i$ is approximately constant in time, instead of linearly increasing due to the arrival of new pulses. 

Let us now do part 9. In other words, we would like to do $\overline{\phi}_i \rightarrow \Phi (\vec{x}_i, t)$ and $\overline{V}_i^{\mu} \rightarrow V_i^{\mu} (\vec{x}_i, t)$. This can be accomplished by setting a dynamics of a form 
\beq \frac{d \overline{\phi}_i}{d t} = \frac{K_{\phi}}{\chi} (\Phi (\vec{x}_i, t) - \overline{\phi}_i) \times (\cdots) \nonumber \eeq
\beq \frac{d \overline{V}_i^{\mu}}{dt} = \frac{K_V}{\chi} (V^{\mu} (\vec{x}_i, t) - \overline{V}_i^{\mu}) \times (\cdots) \eeq
\beq \frac{d \overline{q}_i}{d t} = \frac{K_q}{\chi} (Q (\vec{x}_i, t) - \overline{q}_i) \times (\cdots) \nonumber \eeq
Here, $K_{\phi}$ and $K_V$ are assumed to be "very large", and by $(\cdots)$ we denote some $T$-s. Let us now see what they are. Since we would like "h" to take place during the "second half" of the time interval, we replace $T(\xi_i < \chi)$ with $T(\xi_i > \chi)$. On the first glance, it might seem that we no longer need $T(\mu (\vec{x}, t)) > \mu_c)$ since during the time free of pulses we have $0 \approx \xi_i < \chi$. There is, however, a problem with this argument. In particualr, it will take some time for $\xi_i$ to decay; thus it would continue to be "large" right after the pulse is gone. Now, the "justification" for throwing away $T(\mu (\vec{x}, t)) > \mu_c)$ might be based on the fact that the time it takes for $\xi_i$ to decay is much smaller than the time interval between the pulses (which has to be true anyway, in order for our theory not to produce unwanted overlaps between different pulses). However, it is important to realize that all of the "evolution" that we propose takes place \emph{during} a pulse. Therefore, we have to compare the "unwanted" time to the duration of a pulse \emph{itself}. This comparison leads to just the opposite conclusion. In light of the fact that $k_{\xi} \ll K_{\xi}$, it would take much longer for $\xi_i$ to decay then it took for it to be "bumped". Therefore, the "mess" that would be created during its decay will \emph{dominate} whatever "desired" contribution we have produced during the pulse. In order to avoid this, we \emph{have} to retain $T(\mu (\vec{x}, t)) > \mu_c)$. Finally, according to part 9 the "updating" only happens \emph{if} $T(Q (\vec{x}_i, t) = q_i) + T(Q (\vec{x}_i, t) = q'_i) \geq 1$. Since that sum is an integer, the only other option is for this sum to be zero. In this case, we can "block" the updating by simply multiplying relevent time derivatives by $T(Q (\vec{x}_i, t) = q_i) + T(Q (\vec{x}_i, t) = q'_i)$. Finally, as always, we need to have coefficients $T ( e_{j2}/ \vert e_{j1} \vert > \chi )$ in order to avoid self-interaction of each particle.Thus, our dynamics is 
\beq \frac{d \overline{\phi}_i}{d t} = \frac{K_{\phi}}{\chi} (\Phi (\vec{x}_i, t) - \overline{\phi}_i) T(\xi_i > \chi) T(\mu (\vec{x}, t) > \mu_c) T \Big(\frac{e_{j2}}{\vert e_{j1} \vert}  > \chi \Big)  \times \nonumber \eeq
\beq \times (T(\vert Q (\vec{x}_i, t) - q_i \vert < \epsilon_q) + T(\vert Q (\vec{x}_i, t) - q'_i \vert< \epsilon_q)) \nonumber \eeq
\beq \frac{d \overline{V}_i^{\mu}}{dt} = \frac{K_V}{\chi} (V^{\mu} (\vec{x}_i, t) - \overline{V}_i^{\mu}) T(\xi_i > \chi) T(\mu (\vec{x}, t) > \mu_c) T \Big(\frac{e_{j2}}{\vert e_{j1} \vert} > \chi \Big)  \times \nonumber \eeq
\beq \times  (T(\vert Q (\vec{x}_i, t) - q_i \vert< \epsilon_q) + T(\vert Q (\vec{x}_i, t) - q'_i \vert< \epsilon_q)) \eeq
\beq \frac{d \overline{q}_i}{dt} = \frac{K_V}{\chi} (Q (\vec{x}_i, t) - \overline{q}_i) T(\xi_i > \chi) T(\mu (\vec{x}, t) > \mu_c) T \Big(\frac{e_{j2}}{\vert e_{j1} \vert}  > \chi \Big) \times \nonumber \eeq
\beq \times  (T(\vert Q (\vec{x}_i, t) - q_i \vert< \epsilon_q) + T(\vert Q (\vec{x}_i, t) - q'_i \vert< \epsilon_q)) \nonumber \eeq
The equation for $d \overline{q}_i /d t$ can be rewritten in the following form:
\beq \frac{d \overline{q}_i}{dt} = \frac{K_V}{\chi}  T(\xi_i > \chi) T(\mu (\vec{x}, t) > \mu_c) T \Big(\frac{e_{j2}}{\vert e_{j1} \vert} > \chi \Big)\times \eeq
\beq \times  ((q_i - \overline{q}_i) T(\vert Q (\vec{x}_i, t) - q_i \vert < \epsilon_q) +(q_i' - \overline{q}_i) T(\vert Q (\vec{x}_i, t) - q'_i \vert  < \epsilon_q)) \nonumber \eeq
It is important to emphasize similarities and differences between parts 8 and 9. While part 8 described linear evoluton, part 9 describes an "exponential decay" of $\overline{\phi}_i - \Phi$ and $\overline{V}_i^{\mu} - V^{\mu}$. As a result, the "linear evoution" of part 8 required an \emph{exact} value of a coefficient. On the other hand, the "exponential decay" of part 9 simply requires the coefficients to be "very large". 

\subsection*{4.11 Mechanism of "integrating" $\cal L$ to get $S$ (parts 11-12)}

So far we have described a way in which a \emph{quazi-local} lagrangian density, $\cal L$ is "generated" through the \emph{local} information ${\cal L}_i$. Let us now proceed to generating a \emph{global} action, $S$, again through the "local" parameter $S_i$. We would like $S_i$ to be a "sum" of ${\cal L}_j$ over \emph{all} $j$ that obey $\omega (q_j, q'_j) = \omega (q_i, q'_i)$. As was shown earlier, this has been accomplished in parts 11 and 12 of the algorithm. Our task is to come up with "mechanism" in which these two parts are "enforced".  As we recall, according to part 11, in the reaction to the "pulse" $S_j$ gets "updated" according to
\beq S_j \longrightarrow S_j +  (\delta t) T(\vert \omega (q_j, q'_j) -  \omega (\vec{x}, t) \vert < \epsilon_{\omega}) ({\cal L} (\vec{x}_i, t) +  k_S (S (\vec{x}_j , t) - S_j)),  \eeq
and, in the absence of the pulse, it evolves as
\beq \frac{dS_k}{dt} = - \alpha S_k \eeq
 There is a complication which we have not encountered before. In particular, in order to make sure that the delays due to the "finite" speed of "superluminal" signals to stay "small", we have to assume that the universe is compact. Now, the compactness of the universe implies that the signals will keep comming back and, therefore, each particle will be affected multiple times by a pulse. 

This has not been an issue before. After all, as far as the previous sections are concerned, in order for the particle to react to the pulse the condition $\mu (\vec{x}, t) > \mu_c$ needed to be satisfied. Now, $\mu_c$ has been chosen to be so large, that this condition can't possibly be satisfied by a pulse emitted by one of the "far away" particles. As long as the universe is \emph{not} spherical (that is, it takes different time to circle it for signals traveling in different directions), the signal will remain weak even after it returns to the point where it was emitted. It would take extremely long time for resonance to "accidentally" occur; by that time the signal will die off due to its non-zero mass. 

In our present situation, however, we are unable to retain a "large" constant $\mu_c$. After all, we would like $S_i$ to approximate the sum of \emph{all} ${\cal L}_j$ satisfying $\omega (q_j, q'_j) = \omega (q_i, q'_i)$, \emph{regardless of distance}. Thus, we replace $\mu_c$ by \emph{much smaller} constant $\mu_f$, so that even a signal emitted by "far away" points is "large enough" to "update" $S_i$. \emph{But} if we do that, then the signal will \emph{continue to be} large enough to update $S_j$ as it keeps circling the universe. After all, circling it several times will not make it "diffuse" any more than passing through the universe once. As a result, $S_i$ will keep getting "updated" over and over agian, every time the signal passes a particle $i$. 

On the first glance, it is possible to argue that, on average, the effects of all pulses get multiplied "roughly" by the same thing (even if the number is very large). This number is finite: due to the "very small mass" the signal will eventually die off. The longer it takes for the signal to die off, the larger this finite number would be; making this number "large" will ultimately make its relative fluctuations "small". However, it is also possible that global topology of the universe will imply that the signal emitted at different places will circle the universe different number of times. As a result, one of the two points inside the laboratory will have "much larger" contribution to the Lagrangian than the other point within that same laboratory, due to some anamoly taking place on astronomical scale. Again, it is possible that such effects average out to something undetectabe, but it is not safe to bet on this assertion without further investigation.

As far as this paper is concered, we will take a "safer rout" by postulating extra dynamics "by hand" that would prevent a particle from "listening" to the signal it has "already heard before". We will attach a "clock" $\zeta$ to each particle that measures the amount of time the particle has been subjected to a signal. This clock will evolve according to 
\beq \frac{d\zeta_i}{d t} = - k_{\zeta} \zeta_i + K_{\zeta} T (\mu_Q (\vec{x}_i, t) > \mu_f) T \Big(\frac{e_{j2}}{\vert e_{j1} \vert}  > \chi \Big) \eeq
The signal can have an effect on evolution of $S_i$ \emph{only if} $\zeta < \chi$. Now, the "speed of the clock" $K_{\zeta}$ is chosen to be so high, that the clock reaches $\chi$ within a \emph{small fraction} of the time it takes a signal to pass. Since the duration of a signal "happens to be" equal to $2 \chi$, this can be accomplished by 
\beq K_{\zeta} \gg 1. \eeq
As a result, a particle "can't even" finish listening to the "first signal", let alone "its repetition" after it circles the universe. Now, due to the $-k_{\zeta} \zeta_i$-term, the clock eventually "resets itself". The time needed for the clock to "reset" is \emph{much smaller} than the time interval between signals, $2 \pi/ (N \nu)$ (where $N$ is the number of particles):
\beq k_{\zeta} \gg N \nu \eeq
 As a result, the particle \emph{will} "hear" the \emph{next} signal. \emph{At the same time}, however, the "resetting time" is \emph{much larger} than the time it takes for a signal to "die off":
\beq k_{\zeta} \ll m_{\mu} \eeq 
As a result, the particle "wont hear" the signal as it keeps returning back after circling the universe several times, simply because it happens "too soon" and the clock "hasn't finished resetting yet". By the time that the clock is "reset", the signal had already died off. 

Let us now write down the dynamical equation for $S_i$; or, in other words, an expression for $dS_i/ d t$. As before, we will use $T$-factors in order to "enforce" the constraints. In order to "block" the evolution when the "clock" $\zeta_i$ exceedes the value of $\chi$, we need to put in a factor of $T(\zeta_i < \chi)$. Now, \emph{after} long enough time passes for $\zeta_i$ to "reset" to values less than $\chi$, the signal will still technically be non-zero, albeit very small (due to the attenuation from its mass term). Therefore, in order to prevent that signal from having any effect, we need to put in $T(\mu (\vec{x}, t) > \mu_f)$. Here, $\mu_f$ is selected to be so small that the signal circling the universe once will exceed it:
\beq \mu_f \ll \frac{1}{L^{1/3}} \eeq
where $L$ is the expected length of the geodesic circling the universe. At the same time, $\mu_f$ is large enough to prevent the "old" signals from "making it" once the new signal is emitted:
\beq \mu_f \gg \Big(\frac{N \nu}{c_s} \Big)^{1/3}, \eeq
where $N$ is the number of particles and $2 \pi / \nu$ is a periodicity of pulses emitted by each particle. Finally, just like we did before, we also insert the factor of $T(e_{j2}/ \vert e_{j1} \vert$ in order to avoid self-interaction. Thus, we have a total of three different $T$-s
\beq T(\zeta_i < \chi) T(\mu (\vec{x}, t) > \mu_f) T \Big(\frac{e_{j2}}{\vert e_{j1} \vert}  > \chi \Big) \nonumber \eeq
 As before, as long as all three $T$-s are equal to unity, we would like $dS_i/ d t$ to be some afore-given constant. This "constant" should be equal to the desired incriment of $S$ (will call it $\delta  S$) divided by the duration during which the particle "responds" to the pulse. That duration is equal to the time it takes for clock $\zeta_i$ to reach $\chi$; this time is $\chi / K_{\zeta}$. Thus, the value of $dS_i / d t$ is $K_{\zeta} \delta S / \chi$. In addition, as was explained in detail in Section 4.6, we need an attinuation term, $- \alpha_S S_i$ (see part 12) that will allow $S_i$ to "decay" in the absence of external pulses. This term will make sure that $S_i$ will eventually approach some equilibrium as opposed to indefinitely increasing. Thus, we have 
\beq \frac{dS_i}{d t} = - \alpha_S S_i + \frac{K_{\zeta} \delta S}{\chi} T(\zeta_i < \chi) T(\mu (\vec{x}, t) > \mu_f) T \Big(\frac{e_{j2}}{\vert e_{j1} \vert}  > \chi \Big) , \eeq
where $\delta S$ is a desired incriment of $S$ during the passage of a signal. Now, by looking at the equation of "updating" of $S_i$ (\ref{eqn:346}), we can "read off" $\delta S$ to be 
\beq \delta S = (k_S \delta t) T(\vert \omega (q_j, q'_j) -\omega (\vec{x}, t) \vert < \epsilon_{\omega}) ({\cal L} (\vec{x}_i, t) +  k_S (S (\vec{x}_j , t) - S_j)),  \eeq
By substitutting this into the expression for $d S_i / d t$, we get 
\beq \frac{dS_i}{d t} = - \alpha_S S_i + \frac{K_S K_{\zeta} \delta t}{\chi} T(\vert \omega (q_j, q'_j) -  \omega (\vec{x}, t) \vert < \epsilon_{\omega}) ({\cal L} (\vec{x}_i, t) +  k_S (S (\vec{x}_j , t) - S_j))   \times \nonumber \eeq
\beq \times T(\zeta_i < \chi) T(\mu (\vec{x}, t) > \mu_f) T \Big(\frac{e_{j2}}{\vert e_{j1} \vert}  > \chi \Big) , \eeq
Finally, we substitute $k_S$ from equation \ref{k_S}, namely, 
\beq k_S = \frac{\alpha M^2 {\cal V}}{N_{\rm pt}}, \eeq
where $\alpha$ is a coefficient of the "decay" equation for $S$ in the absence of pulses, $M$ is the number of allowed "charges", $\cal V$ is the volume of the universe, and $N_{\rm pt}$ is the number of points in the universe. Substituting this, produces our final expression for the dynamics of $S_i$:
\beq \frac{dS_i}{d t} = - \alpha_S S_i + \frac{\alpha M^2 {\cal V} K_{\zeta} \delta t}{N_{\rm pt} \chi}  ({\cal L} (\vec{x}_i, t) +  k_S (S (\vec{x}_j , t) - S_j))   \times \eeq
\beq \times T(\vert \omega (q_j, q'_j) - \omega (\vec{x}, t) \vert < \epsilon_{\omega}) T(\zeta_i < \chi) T(\mu (\vec{x}, t) > \mu_f) T \Big(\frac{e_{j2}}{\vert e_{j1} \vert}  > \chi \Big) . \nonumber \eeq

\subsection*{4.12 Mechanism of "evolution" of $\psi$ under $S$ (part 11)}

We have now finished describing the way in which the "transition amplitude" $S_i$ is produced. The next logical step (which happens to be much simpler) is to show how the values of $S_i$ "guide" the evolution of probability amplitudes of the quantum states in Schr\"odinger's representation. Such probability amplitudes are "encoded" in terms of $\psi_i$. We have already shown in Chapter 4.7 that, if $\psi_i$ behaves according to part 11, then on a "larger scale" its evolution will, in fact, approximate the one dictated by $S_i$. Our only task, therefore, is to come up with a dynamics that would "generate" whatever part 11 prescribes. Now, according to part 11, $\psi_j$ gets "updated" in response to the pulse according to
\beq \psi_j \longrightarrow \psi_j + (k_{1 \psi} \delta t) \Psi' (\vec{x}_k, t) e^{iS (\vec{x}_j, t))} T(\vert q_j - Q'(\vec{x}_j, t) \vert< \epsilon_q) +  \nonumber \eeq
\beq + (k_{2 \psi} \delta t) (\psi (\vec{x}_j , t) - \psi_j) T(\vert q_j - Q (\vec{x}_j, t) \vert < \epsilon_q). \eeq
Now, we notice that $\psi_k$ is subject to "global correlations", just like $S_k$ is. Thus, we face the same dilemma as we did in case of $S_k$. Namely, we would like signals emitted by "far away" sources to have influence; but, at the same time, we do \emph{not} want the signals to have any influence after they have circled the universe. We address this problem in the identical way to what we did for $S_k$. Namely, we will use the "clock" $\zeta_i$ which, according to the previous chapter, evolves according to 
\beq \frac{d\zeta_i}{d t} = - k_{\zeta} \eta + K_{\zeta} T (\mu_Q (\vec{x}_i, t) > \mu_f) T \Big(\frac{e_{j2}}{\vert e_{j1} \vert}  > \chi \Big). \eeq
The reception of a signal is "blocked" whenever we have $\zeta_i > \chi$. We have chosen $K_{\zeta}$ in such a way that this becomes the case after only a small portion of the duration of the pulse. As a result, a particle receives only a very beginning of the pulse while blocks out the rest. Now, due to the smallness of $k_{\zeta}$, it takes a long time for $\zeta_i$ to "recover" (that is, become smaller than $\chi$) even \emph{after} the pulse was gone. As a result, when pulse circles the universe and "comes back", the particle $i$ is unable to receive it. At the same time, $\zeta_i$ "recovers" long \emph{before} the \emph{next} pulse is emitted, so that the particle $i$ \emph{will} be able to receive the latter. But, at the same time, the \emph{old} pulse will "die off" by that time due to its mass. The relationships between coefficients that produce such effect were outlined in Section 4.11. 

Now, in order for $\psi_i$ to "stop evolving" when $\zeta_i > \chi$, we have to insert a coefficient of $T(\zeta_i< \chi)$ in the equation for $d \psi_i / d t$. In order for $\psi_i$ not to react to random fluctuations that are too weak to be classified as a "pulse" we need to insert $T (\mu (\vec{x}, t) > \mu_f)$, where, as explained in Chapter 4.11, $\mu_f$ is used instead of $\mu_c$ in order to have long range effects. Finally, in order to avoid self-interaction, the coefficient $T(e_{j2}/ \vert e_{j1} \vert > \chi)$ is inserted. Thus, the $T$-part of the equation takes the form 
\beq T(\zeta_i < \chi) T(\mu (\vec{x}, t) > \mu_f) T \Big(\frac{e_{j2}}{\vert e_{j1} \vert}  > \chi \Big) \nonumber \eeq
This part is being multiplied by the "constant" value of $d \psi_i / d t$ we would "like" it to have when the above is unity. This value is equal to the desired incriment of $\psi$, which we will call $\delta \psi$, divided by relevent time interval, $\chi /K_S$. Thus, we have 
\beq \frac{d\psi_i}{d t} = \frac{K_S \delta \psi_i}{\chi} T(\zeta_i < \chi) T(\mu (\vec{x}, t) > \mu_f) T \Big(\frac{e_{j2}}{\vert e_{j1} \vert}  > \chi \Big)  \eeq
Now, we will look at the "updating" equation for $\psi_i$ to "read off" $\delta \psi_i$:
\beq \delta \psi_i = (k_{1 \psi} \delta t) \Psi' (\vec{x}_k, t) e^{iS (\vec{x}_j, t))} T(\vert q_j - Q'(\vec{x}_j, t) \vert < \epsilon_q) + \nonumber \eeq
\beq +  (k_{2 \psi} \delta t) (\psi (\vec{x}_j , t) - \psi_j) T(\vert q_j - Q (\vec{x}_j, t) \vert < \epsilon_q). \eeq
By substitutting this into the equation or $d \psi_i / d t$, we obtain 
\beq \frac{d\psi_i}{d t} = \frac{K_S k_{1 \psi} \delta t}{\chi} \Psi' (\vec{x}_k, t) e^{iS (\vec{x}_j, t))} T(\vert q_j - Q'(\vec{x}_j, t) \vert < \epsilon_q) +  \eeq
\beq + \frac{K_S k_{2 \psi} \delta t}{\chi} \psi (\vec{x}_j , t) - \psi_j) T(\vert q_j - Q (\vec{x}_j, t) \vert< \epsilon_q)) T(\zeta_i < \chi) T(\mu (\vec{x}, t) > \mu_f) T \Big(\frac{e_{j2}}{\vert e_{j1} \vert}  > \chi \Big) \nonumber \eeq
It is important to point out that much of what we did here is analogous to what we did for $S_i$ in Section 4.11. At the same time there is one difference. In case of $S_i$ we had an extra term that implies attenuation in the absence of signals. In case of $\psi_i$ we do not. As one could see from Section 4.6, the dynamics in the presence of attenuation term leads to an equilibrium point; that is, $S_i$ will "end up" being "nearly constan" if we wait "long enough". At the same time, it is also clear that in the \emph{absence} of attenuation term, $\psi_i$ will \emph{never} be a constant. 

Intuitively, in case of $S_i$ the attenuation in the absence of signals will "cancel out" the increments in their presence, which leads to nearly-constant overall behavior. In case of $\psi_i$, there is "nothing there" to cancel out the incriments. It can be also understood from the perspective of attinuation constant going to zero. In case of $S_i$, the smaller attenuation constant is, the longer we have to "wait" for an equilibrium. Thus, in case of $\psi_i$, the attenuation constant is zero; thus we have to "wait" infinite amount of time for the equilibrium of $\psi_i$ to occur, which is why it hasn't occured "yet". Obviously the fact that $S_i$ does not evolve while $\psi_i$ does is what we physically expect. This is why we have attenuation term for the former but not for the latter. 

\subsection*{5. Gravity and measurement}

\subsection*{5.1 Gravity: a "classical" field or "quantum mechanical"?}

Let us now include gravity in the theory. The first question that should be asked is how are we planning to deal with the issue of its quantization.  From what we have seen before, when we quantize non-gravitational fields, we have to "attach" these fields \emph{to} the point particles and leave them undefined elsewhere. The key reason for this is that we would like to "embed" the "configuration space" inside the "ordinary space". Since the former is much larger than the latter, we have to \emph{first} find a way of reducing the former. We do that through discretization. However, we can not repeat this for gravity. After all we are assuming the presence of \emph{some} metric \emph{between the points} the moment we attempt to write down the wave equation for the signals that these points exchange. 

At first one might argue that we might first attach the metric to the points and then find some way of "extrapolating it" into the space between them (in fact, this is what we will do with non-gravitational fields in Section 5.6). But upon closer look we see that this would not work either. The very act of extrapolating the information about the point to the space between points requires apriori knowledge of a metric. In particular we have to assume that nearby points are "more important" than "far away" ones. Thus, we need metric in order to determine which points are "nearby" and which ones are "far away". Yet, we \emph{don't have} the metric \emph{until} we do our extrapolation trick; this makes the theory circular. 

Of course, the "circular argument" we ran into is logically parallel to the circular arguments people face when they attempt any other way of quantizing gravity, independently of the specific theory we are proposing. The general issue is this: in order to do "quantization", we need a geometrical background; yet, if gravity is being quantized, we do not have the "assumed" geometrical background. In the specific case of our theory, the "geometrical background" is needed in order to describe the way particles exchange signals (see Section 4.8, and, specifically, equation (\ref{eqn:461})). Outside of our framework, of course, this specific issue does not arize. But still, geometrical background is needed for \emph{something}, whatever that "something" might be. Hence, the problems we are facing and the problems others are facing are logically parallel to each other. 

In this paper we borrow the view expressed by Dyson in \cite{Dyson}; namely, we claim that gravity is purely classical field and does not have "quantum" nature to begin with. At the same time, \emph{other} fields, such as spin $0$, spin $1$ and spin $1/2$ \emph{are} "quantum mechanical" and they propagate in a background of classical gravitational field. Now, when these fields "collapse" due to quantum measurement, the "classical" outcome of the collapse, $T_{\mu \nu}$, will serve a role of a "source" of gravitational field which, in turn, obeys the modified Einstein's equation, 
\beq G_{\mu \nu} + \epsilon_{\rm nc} \partial^{\rho} \partial_{\rho} g_{\mu \nu} = T_{\mu \nu}\eeq
where the Einstein tensor $G_{\mu \nu}$ is defined as 
\beq G_{\mu \nu} =^{\rm def} R_{\mu \nu} - \frac{1}{2} Rg_{\mu \nu} \eeq
and $\epsilon_{\rm nc}$ is a small, but finite, constant ("nc" stands for "non covariant"). The presence of $\partial$-s instead of $\nabla$-s in $\epsilon_{\rm nc} \partial^{\rho} \partial_{\rho} g_{\mu\nu}$ term breaks Lorentz covariance and, therefore, implies that Einstein's equation continues to have the solution for non-conserved sources \emph{despite} Bianchi identities. This is crucial since the "classical" source $T_{\mu \nu}$ might fail to be conserved as a result of the "collapse" during measurement. If we insist that gravity is classical, the presence of quantum mechanical collapse is crucial in order to avoid the gravitational interaction between "parallel universes". Thus, we will attempt to come up with a theory that addresses the questions of collapse and gravity simultaneously: the "measurment outcome" will be used as a "classical source" of gravity and gravity will be used as a "classical background" of both the unitary evolution \emph{and} the collapse mechanism.

Before we proceed, let us slow down and ask ourselves the following question: In light of the fact that in this paper \emph{everything} is claimed to be "classical", what exactly do we mean when we are saying "gravity is classical while other fields are quantum mechanical"? The logical way to answer this question is to cross out the word "quantum mechanical" and replace it with the "classical" model that we have proposed "to this effect". One thing we have done in this paper is that we have replaced "continuous" spin $0$ and spin $1$ fields with corresponding "internal degrees of freedom" that we attach to the point particles. We have to ask ourselves though: is discretization logically linked to quantization, or is it something we simply \emph{happened} to introduce for mathematical convenience? If the answer is "yes" then we can simply "write down" the word "classical discretization" in place of the crossed-out word "quantization"; otherwise, we still have to "cross out" the word "quantization" but we will have to find something else "classical" to replace it with. 

Fortunately for us, the answer to the question is "yes, there is a clear logical link between the discretization and quantization". After all,  if something is "attached to the particles", we can proceed to postulate "global correlations" among "same charged" particles, thus "splitting" the picture into "$M$ parts". Each of the $M$ "parts" corresponds to a given "common" value of "charge" of the particles that "make up" this particular "part" of the picture; and \emph{at the same time} that "part" also corresponds to one of the "configurations" we are attaching "probability amplitude" to. For example, the common value of \emph{classical} degree of freedom $\psi_i$ among the particles whose charge obeys $q_i = q$ corresponds to the "probability amplitude" of a wavefunction being $\Psi (C) = \Psi_q (C)$. On the other hand, if a field is defined in the space between the particles, we no longer have such way of "splitting it up" into different "parts"; thus, we are forced to say that the \emph{one} field that we have is "all there is"; in other words, we have one single "classical" reality. 

Thus, we have established that, indeed, discretization and quantization \emph{are} logically linked to each other, as desired. In light of this, we propose to "translate" the statement "gravity is classical while everything else is quantum mechanical" into a statement "gravity is defined in the spacetime continuum while everything else is attached to the points that we use to discretize the space with". As far as our "new" statement is concerned, it is understood that in both cases the fields we are dealing with are viewed as classical. This, in fact, answers the question we have asked a bit earlier; namely, can we use the metric while writing down the equation of signals being exchanged between particles? The answer to this question is "yes". After all, gravity is a "classical field" so, "by definition", it is defined throughout the entire spacetime continuum and, of course, we can freely use it. 

\subsection*{5.2 Measurement outcome: classical or quantum mechanical?}

In the previous section we agreed to stick to the view expressed by Dyson in \cite{Dyson} that gravity is purely classical. Furthermore, we have described the way to "convert" that statement into the famework of our model. Namely, "quantum mechanical" quantities are represented by internal degrees of freedom of lattice points, while gravity, being "classical", fills the entire space.   However, the question regarding the nature of sources of gravitational field is a bit less obvious. We have stated that the source of gravitational field is the outcome of collapse of the wave function.  Now, if the wave function is "quantum mechanical" while the outcome of its collapse is "classical", does it mean that the former is "attached to the particles" while the latter is "spread out throughout the space"? If so, exactly how does "spreading out" gets accomplished during the time of the "collapse"? Or, if such is not the case, does it mean that outcome of the collapse is also "quantum mechanical"? If so, can it be used as a source of "classical" gravity?

In this paper, we pick the latter option. We view the outcome of collapse of wave function as "quantum mechanical" entity. In spirit, we will borrow the ideas of GRW model due to Gherardi, Remini, Weber and Pearle (see \cite{GRW1}, \cite{GRW2}, \cite{Pearle1}, \cite{Pearle2}); but, of course, we will have to "redo" it in order for it for it to be incorporated into our framework. The core idea of these models is that the wave function is subjected to a series of "hits" which, essentially, amount to "multiplication" of a wave function by "gaussians" around "random" points. These Gaussians are very wide and the frequency of "hits" is quite rare. Thus, the effect on few-particle systems is negligeble. At the same time, in case of multiparticle entanglement, any given particle experiences an indirect influence from the "hits" that affect any other particle it is "entangled" with. Thus, their accumulated effect is no longer negligible, which results in a "localization" of a given particle.

The implication of this model is that the outcome of the "collapse" is "quantum mechanical". After all, the appearance of the "collapse" is part of the natural evolution of a wave function that occurs with or without entanglement; and, unlike the Bohmian case, there are no "beables" or any other "classical" entities existing separately from the wave function. Therefore, the "translation" of this framework into out model implies that the outcome of the "collapse" of the wave function is identified with the internal degrees of freedom $\psi_i$ of each particle and, therefore, is \emph{not} defined in the space between them. Similarly, the "source" of gravitational field, $T_{i; \mu \nu}$ is also defined \emph{at} each given particle $i$. The specific way of obtaining $T_{i; \mu \nu}$ is, in spirit, similar to the way we obtained ${\cal L}_i$ and it will be explicitly worked out in \ref{U}  

There are, of course, some ''classical'' issues that need to be addressed. In particular, if the lattice points produce, either directly or indirecty, a continuum gravitational field, it appears that we would obtain unwanted ''black holes'' around each of the lattice points. Furthermore, even outside these ''black holes'' the behavior of gravitational field would be reflecting distances to nearest lattice points as opposed to larger-scale behavior of the ''collapsed'' fields on the lattice. These issues, however, are relatively easy to deal with by ''cutting off'' the ''very large'' values of relevent fields (thus avoiding black holes) and further introducing dispersion in order to ''get rid'' of small scale variations of ''smoothed out'' fields. The mathematical model to this effect will be presented in Section 5.6. 

\subsection*{5.3 Extra machinery which preserves general relativistic covariance}

Throughout this paper, we are assuming that lattice particles are stationary. From the point of view of lattice field theory, the demand that particles are stationary with respect to preferred frame is needed anyway, independently of the above issues. After all, the setup of quantum field theory assumes the background of stationary lattice. At the same time, due to the Lorentz covariance of the Lagrangians, the "preferred frame" defined by that lattice is "hidden" and we still obtain Lorentz covariant results. Now, if we were to allow the lattice particles to move, we would have several different "lattices" superimposed upon each other; each "lattice" consisting of particles moving with some common velocity. But, according to quantum field theory, one single, stationary, lattice can accomodate the non-zero momenta. Thus, the presence of moving lattices will lead to reduncancies: instead of having a "particle moving with momentum $\vec{p}$", we will now have "particle moving with momentum $\vec{p}$ \emph{and} living in the lattice which moves with velocity $\vec{v}$". The values of $\vec{p}$ and $\vec{v}$ will be unrelated to each other, creating unwanted extra degrees of freedom. 

As we have said, the natural way of avoiding these difficulties is to demand that all of the particles are stationary in preferred frame. This will permanently identify $\vec{v}$ with $\vec{0}$ in that frame and, therefore, leave the only non-trivial parameter to be $\vec{p}$, as desired. As we have repeatedly stated, we openly violate special relativistic covariance and trust our experience with quatum field theory that all of our predictions will be covariant. This, however, is a bit more tricky when it comes to general relativistic covariance. After all, as a result of violation of the metric, the "distance" between the particles will vary as well. This will imply the violation of untraviolet cutoff and, accordingly, the variation of "renormalized" mass, charge and so forth, provided that we assume that their corresponding "bare" parameters are constant. In order to address this issue, we have to "hide" some of the particles whenever they are "overcrowded" in order to maintain their density the same. 

In order to be able to "hide" a particle number $i$, we will introduce an internal degree of freedom, $l_i (t)$. As one can easilly see, if this internal parameter is zero, then the particle $i$ will have no effect on $T_{\mu \nu} (\vec{x}, t)$, which makes it "invisible" in that respect. We will further specify the dynamics of $l_i (t)$ in such a way that, while it evolves continuously, it is either very close to $0$ or very close to $1$ "most of the time" (except for very brief "transition periods); thus, any given particle is either "completely visible" or "completely invisible".

However, if we are "serious" about particle $i$ being "invisible", we do not want it to emit any of the signals we discussed in Chapter 4, either. As we recall, the signal is "emitted" when 
\beq \frac{e_{j2}}{\vert e_{j1} \vert} < \chi \eeq
where $e_{j1}$ and $e_{j2}$ evolve according to 
\beq \frac{de_{j1}}{d t} = \frac{\nu e_{j2}}{\sqrt{e_{j1}^2+e_{j2}^2}} \; ; \; \frac{de_{j2}}{d t} = - \frac{\nu e_{j1}}{\sqrt{e_{j1}^2+e_{j2}^2}} \eeq
Therefore, in order to prevent the signal from being emitted, we can simply modify the dynamics of $e_{j1}$ and $e_{j2}$ in such a way that the region $e_{j2}/ \vert e_{j1} \vert$ is never reached. This can be done by inserting the multiples of $l_j$ inside the equations for $de_{j1}/d t$ and $de_{j2}/ d t$. This, however, might have unwanted "side effect": if the particle "turns invisible" at the time when $e_{j2}/ \vert e_{j1} \vert$ holds (which is statistically highly unlikely but still possible) it will be emitting "infinitely long" singal; which is just the opposite from what we wanted! In order to prevent this from happening, we have to make sure that $e_{j1}$ and $e_{j2}$ "keep moving" if they are in that specific region, and "stop" once they exit. This amounts to adding another term to the derivatives, which is proportional to $T(e_{j2}/ \vert e_{j1} \vert < \chi)$. Since that term is only "needed" for "invisible" particles, we have to multiply it by an extra factor $1- l_j$. Thus, we propose our "modified" dynamics for $e_{j1}$ and $e_{j2}$ to be
\beq \frac{de_{j1}}{d t} = \frac{\nu e_{j2}}{\sqrt{e_{j1}^2+e_{j2}^2}} \Big(l_i + (1-l_i) T \Big( \frac{e_{j2}}{\vert e_{j1} \vert} < \chi \Big) \Big)\eeq
\beq \frac{de_{j2}}{d t} = - \frac{\nu e_{j1}}{\sqrt{e_{j1}^2+e_{j2}^2}} \Big(l_i + (1-l_i) T \Big( \frac{e_{j2}}{\vert e_{j1} \vert} < \chi \Big) \Big)\eeq
Let us now turn to the dynamics of $l_i (t)$. As we mentioned earlier, we would like to "hide" some particles when the metric changes in such a way that they are "overcrowded". Thus, the particle needs to have some way of "finding out" the average spacing of the surrounding particles and "decide" whether it "wants" to hide or not. Since it only "looks" at the particles that are \emph{not} "hidden", it would not feel a need to "hide" if some of the \emph{other} particles have already "hid themselves", instead. This will ensure that the average density of "unhid" particles and, therefore, ultraviolet cutoff, stays constant. 

In principle, we could have introduced new kinds of "pulses" (similar to the ones of Section 2.3) that would allow particle to "count" the particles in surrounding areas. We recall, however, that the main reason we were using "pulses" was the fact that we wanted to isolate the effect of specific "charge"; thus we wanted to separate it in time from all other "charges". In our present situation, however, we do \emph{not} want to work out the spacing between points for any specific charge; rather, we want to do it for all of the charges "at the same time". After all, the spacing between lattice points corresponds to ultraviolet cutoff. Thus, separately evaluating "particle spacing" for each charge would amount to have a separate "ultraviolet cutoff" for each specific field configuration; this, of course, is absurd. On the other hand, in order to have a single \emph{universal} cutoff for the \emph{entire} path integral is to operate with a \emph{total} densities of \emph{all} particles. This means that we no longer have to separate their influences in time; and, therefore, we no longer need "pulses", either. 

While the above would make equations look a lot simpler and a lot more natural, this comes with the price. In particular, the singular nature of self-interaction of any given particle might falsely make that particle "believe" that there is "infinitely large" density of surrounding particles.  As a result, every single particle will "hide" in order to reduce the alleged "infinite density", no matter how small the actual density of surrounding particles might be! In the scenario when we assumed the presence of pulses, we have "switched off" self-interaction by adding extra coefficient $T (e_{j2}/ \vert e_{j1} \vert > \chi)$ on every equation that governs the internal evolution of the particle "in response" to "external" signals. In order for this to "work", however, we have to make sure that the \emph{emission} of a pulse is limitted to $e_{j2}/ \vert e_{j1} \vert < \chi$; but this is precisely what we now claim we are \emph{not} doing!

We, therefore, propose another way of avoiding self-interaction of the particles. In particular, we "split" the particles on the ones with even $q'$ and odd $q'$. Likewise, there are two separate "fields", $\rho_e$ and $\rho_o$, which "compute" the "density" of the particles with "even" and "odd" $q'$, respectively: 
\beq \nabla_s^{\alpha} \nabla_{s; \alpha} \rho_e + m^2_{\rho} \rho_e + \zeta \partial_0 \rho_e = G_{\psi} \sum_{j=1}^N l_j (1 + (-1)^{q'_j}) \delta^3 (\vec{x} - \vec{x}_j) \label{eqn:520}\eeq
\beq \nabla_s^{\alpha} \nabla_{s; \alpha} \rho_o + m^2_{\rho} \rho_o + \zeta \partial_0 \rho_o= G_{\psi} \sum_{j=1}^N l_j (1 - (-1)^{q'_j}) \delta^3 (\vec{x} - \vec{x}_j) \label{eqn:521}\eeq
Now, we demand that the particle with even $q'$ is only sensitive to $\rho_o (\vec{x}, t)$ while it \emph{ignores} $\rho_e (\vec{x}, t)$; on the other hand, the particle with odd $q'$ is only sensitive to $\rho_e (\vec{x}, t)$, while ignoring $\rho_o (\vec{x}, t)$. This assures that both groups of particles will "ignore" the signular influence they exert on themselves. 

Now, if an ''odd-indexed'' particle ''found out'' that $\rho_e$ is too high, it should do ''something'' in order to ''equilize the situation''. If that particle decides to ''lower'' its own ''visibility'', this would lower $\rho_o$ rather than $\rho_e$. So, instead, it should send a signal \emph{to} the even-valued particles ''informing'' them that their density is ''too high'' which would lead one of the \emph{even-valued} particles lowering its visibility. This can be done by the particle emitting a field $\rho'_e$ (as contrasted to $\rho_e$). While $\rho_e$ is emitted by even-indexed particles, $\rho'_e$ is emitted by \emph{odd}-indexed particles \emph{in reaction to} $\rho_e$. Likewise, $\rho_o$ is emitted by odd-indexed partiles, while $\rho'_o$ is emitted by \emph{even}-indexed ones \emph{in reaction to} $\rho_o$:
\beq \nabla_s^{\alpha} \nabla_{s; \alpha} \rho'_e + m^2_{\rho} \rho'_e + \zeta' \partial_0 \rho'_e = G_{\psi} \sum_{j=1}^N l_j \rho_e (\vec{x}_j) (1 - (-1)^{q'_j}) \delta^3 (\vec{x} - \vec{x}_j) \label{eqn:520}\eeq
\beq \nabla_s^{\alpha} \nabla_{s; \alpha} \rho_o + m^2_{\rho} \rho_o + \zeta' \partial_0 \rho_o'= G_{\psi} \sum_{j=1}^N l_j \rho_o (\vec{x}_j) (1 + (-1)^{q'_j}) \delta^3 (\vec{x} - \vec{x}_j) \label{eqn:521}\eeq
Thus, even-indexed particles ''learn'' about ''their own'' density by ''listening'' to odd-indexed ones that ''tell'' them the ''outsider perspective'' of what their density is. Likewise, the odd-indexed particles learn about their own density from the ''outsider perspective'' of even-indexed particles. Finally, the particle $i$ will ''indirretly learn'' about the density of ''its own kind'', $\rho_i (t)$, according to  
\beq \rho_i (t) = (-1)^{q'_i} \rho'_o (\vec{x}_i, t) + (-1)^{q'_i+1} \rho'_{\psi e} (\vec{x}_i, t) \label{eqn:522}\eeq
Now, the reason we have used $q'$ instead of $q$ in the above "trick" is that quantum states are associated with $q$. Thus, using $q$ would imply a "splitting" of quantum states into two categories which might potentially lead to some unwanted effects. On the other hand, a particle with any given value of $q'$ can potentially have any "available" value of $q$, whether even or odd; thus, these effects are avoided. It should also be pointed out that what we have done is completely unrelated to the intended purpose of $q'$. Thus, it would have been "more logical" if we were to introduce $q''$. But, for the sake of making the theory as concise as possible, we chose to use $q'$; in principle there is nothing "wrong" with the same parameter playing two different roles. The only important part is that $q'$ is separate from $q$. 

Now, once the particle "has computted" the density in a form of $\rho'_i (t)$, the next step is for it to become either "visible" or "invisible" in order to "equilize the situation". We will introduce a "very large" coefficient $K_l$ and claim that the "mechanism" for the particle becomming "invisible" (or, in other words, a mechanism of "sending" $l_i$ to $0$) is 
\beq \frac{d l_i}{dt} = - K_l l_i, \eeq
while the mechanism of the particle becoming "visible" (or, in other words, the mechanism of "sending" $l_i$ to $1$) is 
\beq \frac{d l_i}{dt} = K_l (1-l_i) \eeq
Now, in order to have a "mathematical criteria" of deciding whether the density is "too high" or not, we will introduce the constants $\rho_1$ and $\rho_2$. If $\rho_i > \rho_2$, then the particle $i$ will "decide" that the density is "too high" and therefore "become invisible". On the other hand, if the particle sees that $\rho_i < \rho_1$, then it will "decide" that the density is "not high enogh" and therefore "become visible" (unless, of course, it was "visible" to begin with). This can be enforced through
\beq \frac{d l_i}{dt} = K_l (1-l_i) T (\rho'_i < \rho_1) - K_l l_i  T (\rho'_i > \rho_2) \eeq
We notice that if $\rho_1 > \rho_2$ then the particle will "jump back and forth" between "visible" and "invisible"; we don't want that. Instead, we claim that $\rho_1 < \rho_2$ which will imply that $l_i$ is unambiguously "sent" to $1$ if $\rho' < \rho_1$  and it is unambigously "sent" to $0$ for $\rho' > \rho_2$; on the other hand, in the region $\rho_1 < \rho' < \rho_2$ it is left the way it is.  This implies that density of "visible" particles is free to fluctuate between $\rho_1$ and $\rho_2$ but is prevented from going "too far" outside that region \emph{regardless of behavior of the metric}. This means, of course, that ultraviolet cutoff will fluctuate accordingly; but we can still expect for its effects to "average out" to the situation where the cutoff is fixed. This requires lack of any obvious correlation between the cutoff and gravity. Indeed, the correlation is absent: the "smallness" of $g_{\mu \nu}$ makes things "look" more dense and, at the same time, encourages more particles to "hide" which, together, leads to ambiguous effect. 

\subsection*{5.4 Review of "standard" GRW collapse model}

So far, we have established that we would like to \emph{first} attach to the particles an internal degree of freedom $T_{i; \mu \nu}$; \emph{after that} we would like to "smear it out" in a form of $T_{\mu \nu} (\vec{x}, t)$, and, finally, we would like to use $T_{\mu \nu} (\vec{x}, t)$ as a source of gravitational field. The last two steps assure us that we can "afford" specifying $T_{i; \mu \nu}$ at a point, without any unwanted singularities.  Our next goal is to go ahead and actually define $T_{i; \mu \nu}$.  However, we recall from our previous discussion that we would like the source of gravitational field to be the \emph{outcome} of the collapse of wave function. The next logical step, therefore, is to discuss the collapse mechanism we would "like" to have.  We propose to borrow the spirit, but not the letter, from GRW approach. Therefore, we will devote this section towards the "standard" version of this approach (free from gravity), and then in the next section we will describe our own, gravity-based, version. 

Let us therefore describe GRW model for the case of $N$-particle non-relativistic quantum mechanics in flat space. They start out from the function $\psi (\vec{x}_1, \cdots, \vec{x}_N)$. A random particle (say, particle number $a_i$, corresponding to coordinate $\vec{x}_{a_i}$), is selected at a random time $t_i$. After that, a random point in that three-fold, $\vec{x}_{a_i} = \vec{x}_{a_i 0}$, is selected according to the "biased coin" that will be specified shortly. Once that point is selected, the "mini-collapse" will occur (which Ghirardi, Rimini, Weber and Pearl called a "hit"). That "hit" involves multiplication of the wave function by a Gaussian around $\vec{x}_{a_i} = \vec{x}_{a_i 0}$:
\beq \psi (\vec{x}_1, \cdots \vec{x}_N) \longrightarrow N (\phi; a_i; \vec{x}_{a_i 0}) e^{\frac{\alpha}{2} \vert \vec{x}_{a_i} - \vec{x}_{a_i 0} \vert^2} \eeq
where the constant $\alpha$ is very small, and $N (\phi; a_i; \vec{x}_{a_i 0})$ is selected in such a way that the resulting state is properly normalized. Now, we specify our "biased coin" in such a way that the probability of selection of $\vec{x}_{a_i} = \vec{x}_{a_i 0}$ is \emph{inversely} proportional to $N (\phi; a_i; \vec{x}_{a_i 0})$:
\beq q (\phi; a_i; \vec{x}_{a_i 0}) = \beta N^{-2} (\phi; a_i; \vec{x}_{a_i 0}) \eeq
where $\beta$ is some very small constant. Since $\alpha$ is small, the effect of each mini-collapse is negligible. Since $\beta$ is small, these collapses are very rare. Thus, in case of non-entangled system, their overall effect is not noticeable. But, in case of the entanglement of several particles, each particle experience the indirect effect of the "mini-collapse" of any other particle within that entangled system. Thus, the effects multiply. As a result, a given particle stays "semi-localized" on the macroscopic scales. At the same time, the effects of entanglement, however large they might be, are still finite. Thus, a particle hitting the screen gets localized on the macroscopic scales but \emph{not} on the scales of atoms of the screen  

In light of the fact that one of the "philosophical" purposes of our paper is to "get rid" of the configuration space, the above model is unacceptable for the simple fact that it is \emph{based on} the latter. We will, therefore, propose our own model in the next section that fits our purposes. However, it turns out that the argument that a given model leads to Born's rule is independent of the specifics of the model. Thus, a carbon copy of the argument "in favor" of GRW model can be also used "in favor" of our own model that we will propose in the next section. For this reason, we will attempt to present that argument in a way that emphasizes the common principle between the two models and de-emphasizes the specifics described in the equation above. 

One good argument to this effect was produced by Pearl in \cite{Pearle1} and \cite{Pearle2}. Pearl has been using the analogy between the above-described random process and "gambler's ruin" game. Suppose we have player $1$ and player $2$ which, at the beginning, have $n_1$ dollars and $n_2$ dollars, respectively. Throghout the game, a coin is being tossed. If it lands on one side, then player $1$ pays one dollar to player $2$; if it lays on the other side then player $2$ pays one dollar to player $1$. The player that is left without money loses. The idea presented by Pearl is that quantum states play the role of such "players". The "money" is the value of $\vert \psi \vert^2$ of the states in question, and the "winning" state is the state the wave function collapsed \emph{into}. Thus, a collapse of a wave function can be produced through a sequence of "small" steps, which can be viewed as "mini-collapses", involving exchanges of probability amplitude between various states.

The general concept, which is shared by "standard" GRW model \emph{as well as} our paper is that we need to introduce some "extra dynamics" \emph{in addition to} unitary evolution. We can think of "unitary evolution" as analogous to earning money by going to work or losing money by going to the store and buying things. On the other hand, we can think of "additional dynamics" as "gambling" that we do \emph{on top of} these activities.  The amount of money exchanged "through gambling" is significantly \emph{smaller} than the amount of money exchanged by going to work and/or store. Thus, these "mini-collapses" are not noticeable on their own. At the same time, as stated earlier, in case there is an entanglement,their effects multiply.  

In light of the "gambling model" just presented, the crucial aspect of arriving towards Born's rule is finding out the probability of a given player "winning" in the above described game. Suppose we have two players; the player $1$ has $n_1$ dollars and the player $2$ has $n_2$ dollars. Then, the probability that player $1$ wins is $f(n_1, n_2)$. At the same time, if after \emph{next} step the player $1$ has one more dollar, the probability he wins will be $f(n_1+1, n_2-1)$; if, on the other hand, he will have one \emph{less} dollar, the probability that he wins will be $f(n_1-1, n_2+1)$. Since each case will happen with probability $1/2$, the probability of player $1$ winning is the average of these two quantities. But, at the same time, we also said that probability of player $1$ winning is $f(n_1, n_2)$. Therefore,
\beq f(n_1, n_2) = \frac{f(n_1-1, n_2+1) + f(n_1+1, n_2 -1)}{2} \eeq
Now, if we first multiply both sides by $2$, and then subtract $f(n_1, n_2)+f(n_1-1, n_2+1)$ from both sides, we obtain
\beq f(n_1, n_2) - f(n_1 -1, n_2 +1) = f(n_1 +1, n_2 -1) - f(n_1, n_2)\eeq
This means that $f(n_1 +1, n_2-1)- f(n_1, n_2)$ is the same number for all $n_1$ and $n_2$, provided that $n= n_1+n_2$ is fixed. Now, if the total number of coins is $n$, then we know that $f(n, 0)= 1$ and $f(0,n) =0$. This implise that 
\beq f(n_1+1, n_2-1) - f(n_1, n_2) = \frac{1}{n} \eeq
Again, by using $f(0,n) =0$, this implies that 
\beq f(n_1, n_2) = \frac{n_1}{n} = \frac{n_1}{n_1 +n_2} \eeq
This means that the probability that a given player wins is proportional to the money that he posesses. Thus, in order to reproduce Born's rule, $\vert \psi \vert^2$ (as opposed to, say, $\vert \psi \vert$) should be used as "money"; in other words, $\vert \psi \vert^2$ should be "raised" or "lowered" by "one step".

There is, however, an important difference between that game and quantum mechanics. As far as that game is concerned, it ends once one of the players "loses". In case of quantum mechanics, the "physics" continues to proceed even after wave function has collapsed. This means that the state that "lost" has an apportunity to "recover". Of course, since the amount of money (or $\vert \psi \vert^2$) can't be negative, a state can't "participate" when $\vert \psi \vert^2$ is closer to zero than the "length" of "one step". But we have to remember that \emph{apart from that game} we also have an ordinary, collapse-free, unitary evolution. So, once the value of $\vert \psi \vert^2$ of a given state increases \emph{slightly} due to that evolution (which is analogous to a "losing" player "earning" money by going to work, independanty of the "game"), the state can continue to "participate" in the game, and this time "win".  

Of course, this is consistent with what we observe: after the wave function "collapses" it then "spreads out" again, and it can then "collapse" somewhere else afterwords. At the same time, however, we are faced with the question: what is so "special" about collapse, if it is simply one of the "numbers" between $0$ and $1$ (which "happens" to actually approximate $0$ or $1$), while the game continues regardless? In order to answer this question, we have to notice that the typical "outcome" of the "measurement" is \emph{not} a "very common" state. For example, the state of "electron hitting a screen" is considerably less common than the state when all particles are stable. From this point of view, the "players" in this game are limitted to rare states such as "electron hitting the screen". The fact that they are "rare" means that the amount of "money" the players typically "earn" is very small. Thus, once a given player lost the money in the game, they would be away for a "long time" until they "earn" again. 

At the same time, however, the theory of measurement also applies to seemingly "more common" states as well. For example, it should explain why the wave function in a configuration space that describes  macroscopical objects does not "spread out". This part of the phenomena can be addressed by noticing that, in multidimensional configuration space, any given state is "very rare". Even if we do \emph{not} have any electron incident onto our system, it is very "unusual" to have the specific destribution of defects that it has. Now, in a situation free of "gambling", the "unusual" set of defects gets "transferred" into another "equally-unusual" set that can be predicted from unitary evolution. If, however, the "gambling" does occur, and one of these "unusual" states loses, it would be highly unlikely for the unitary evolution to produce that exact state afresh.

From a more geometrical point of view, the "complexity" of quantum system implies that the wave function does not "spread out" evenly but, instead, it takes a form of a very complicated shape that reflects the "complexity" of the system at hand. The fact that the probability of "going" from one set of "fingerprints" to another is very small implies that in some of the dimensions the wave function "spreads" very little, if at all. This assention is, in fact, implicitly made in Bohmian mechanics since it is assumed that during the measurements the wave function splits into branches (each branch corresponding to measuring outcome), that will \emph{never} overlap again. Clearly, this argument will fall through if braches "spread out" equally in all directions. So the clear assumption here is that they spread out only in \emph{few} directions; thus, the directions in which different braches were spreading are unlikely to match. 

The claim regarding lack of future overlap, however, has been disputted by Tony Leggett (see \cite{Leggett1} and \cite{Leggett2}). In the context of Bohm, I agree with his criticism. After all, in order for different measurement outcomes not to be "confused" in the far future, we have to assume that the braches will \emph{never} overlap, \emph{no matter how long we wait}. Regardless of the rest of the statement, the clausure "no matter how long we make" is very risky, by default. In the GRW context, however, we no longer have to make such clausure. In fact, we admit that \emph{after some time passes} the state that "lost" might eventually "earn" the money again. \emph{But} if we take some other state that never had money to begin with, it might \emph{also} earn something "once time passes". Thus, after a given state "lost all its money" it became "indistinguishable" from the states that never had any money on the first place. This is the ultimate definition of collapse of wave function. 

We \emph{do} however need to make a claim that "some time needs to pass" before the state that "lost the game" has "money" again. After all, this is what makes the "collapse" event significant. This is a "weaker" claim than the one made in Bohmian context: instead of saying something will never happen we say it will happen "after sufficient time passes". But, in both cases, we appeal to the fact that the unitary evolution in configuration space does \emph{not} cause a wave function to spread evenly. Therefore, in case of one particle configuration the "spreading" is instantaneous but, in case of multiparticle one we need to "wait" until the wavefunction "decides" to spread in the direction we "want" it to. It should be emphasized that at this point we are \emph{not} introducing anything "extra" that will "keep" the wave function from spreading. Rather, we are making a \emph{mathematical conjecture} regarding the behavior of solution of wave equation in multidimensional configuration space. 

The above argument, however, falls through once we introduce the "gambling". As we said earlier, the reason we had to "wait" for quantum state to recover was related to the specifics of the solution of wave equation; namely, the fact that wave function fails to "spread evenly". On the other hand, if the states are allowed to "gamble", they can "get back" money \emph{apart from} the unitary evolution, which means that they don't have to "wait". This means that we have to prohibit the "poor people" from "gambling"; this way, they will have to wait to "first" earn their "salary" from the wave function, and only "then" be re-admitted into the "game". For example, we can say that only the states that satisfy $\vert \psi \vert^2 > n_{\rm loss}$ are allowed to participate.  Thus, the state satisfying $\vert \psi \vert^2 = n_{\rm loss} + (\delta n)/2$ will "play". If it happens to lose, it will then satisfy $\vert \psi \vert^2 = n_{\rm loss} + (\delta t)/2$. Because it is now below $n_{\rm loss}$ threshold, it will have to "wait" until it gets above the threshold again through "earning" money; and only \emph{after that} it will be allowed to continue to gamble. 

There is, however, a problem with this argument. Namely, if the steps are "very small" (which we have to assme in order to get continuum spectrum of probabilities), then it would be "very quick" for a state in question to "earn" the small portion of "money" to be back in the game again. We will therefore propose to replace "lower cutoff" with something more sophisticated. First, we replace money with something more continuous, such as sand. Then we claim that if the respective amounts of sand that players have is $n_1$ and $n_2$, then the probability that player $1$ "wins" next time is $q_1 (n_1, n_2)$, while the probability that player $2$ "wins" is $q_2 (n_1, n_2)$. In the event that player $1$ wins, the player $2$ "gives" him the amount of sand equal to $\epsilon g_1 (n_1, n_2)$, where $\epsilon$ is some very small constant. In case of player $2$ wins, the player $1$ gives him the amount of sand equal to $\epsilon g_2 (n_1, n_2)$. One obvious idea that comes to mind is to manipulate $g_1$, $g_2$, $q_1$ and $q_2$ in such a way that they "push" the "losing" player "downwards". At the same time, these functions are continuous, so a losing player will have to travel "very far" upwards in order to stop being "pushed down".

 Let us first describe a general situation with arbitrary well behaved $g_1$, $g_2$, $q_1$ and $q_2$ and then see what specific forms they should take in order to help us. From this setup, one can easilly see that the "probability of final win" of player $1$, $p_1 (n_1, n_2)$, obeys 
\beq p_1 (n_1, n_2) = q_1 (n_1, n_2) p_1 (n_1 + \epsilon g_1 (n_1, n_2), n_2 - \epsilon g_1 (n_1, n_2)) + \nonumber \eeq
\beq + q_2 (n_1, n_2) p_1 (n_1 - \epsilon g_2 (n_1, n_2), n_2 + \epsilon g_2 (n_1, n_2)) \label{eqn:532}\eeq
If we now look at the first order in $\epsilon$, we can rewrite the above in differential form as
\beq g_1 q_1 \Big( \frac{\partial p_1}{\partial n_1} - \frac{\partial p_1}{\partial n_2} \Big) + g_2 q_2  \Big( \frac{\partial p_1}{\partial n_2} - \frac{\partial p_1}{\partial n_1} \Big)= 0 \eeq
We can now combine the above expression in the following form
\beq (g_1 q_1 - g_2 q_2) \Big( \frac{\partial p_1}{\partial n_1} - \frac{\partial p_1}{\partial n_2} \Big) = 0 \label{eqn:534} \eeq
Now, our intuition tells us that $\partial p_1 / \partial n_1$ is positive and, therefore, $\partial p_1 / \partial n_2$ is negative. Therefore, \emph{if} $g_1 q_1 \neq g_2 q_2$, both of the above will be zero. In order to see what is going on, let us take an extreme case: suppose $g_1$ is always zero, while $g_2$ is positive definite. In this case, whenever the first player "wins", nothing will happen; but whenever he "loses" he will give more and more sand. As a result, player $1$ is guaranteed to lose \emph{all} of his sand at the end. This of course implies that $p_1 (n_1, n_2)=0$ and $p_2 (n_1, n_2)=1$, making their derivatives zero. We can, therefore, generalize it to saying that player $1$ is guaranteed to win whenever $g_1 q_1 > g_2 q_2$ and he is guaranteed to lose whenever $g_1 q_1 < g_2 q_2$, thus explaining why partial derivatives are zero in these cases. On the other hand, if we assume that 
\beq g_1 q_1 = g_2 q_2 \label{eqn:535}\eeq
we will no longer be forced to sent derivatives to zero (after all, the $g_1 q_1 - g_2 q_2$ will be zero and, therefore, will "explain" the zero on the right hand side of (\ref{eqn:534})). However, setting $g_1 q_1 - g_2 q_2$ to zero will produce an argument that \emph{second} derivatives are zero. In order to see it, we will "fix" the value of $n=n_1+n_2$ and define $f_n (x)$, $g_{1n} (x)$, $g_{2n} (x)$, $q_{1n} (x)$, $q_{2n} (x)$according to 
\beq f_n (x) = p_1 (x, n-x) \; ; \; g_{1n} (x) = g_1 (x, n-x) \; ; \; g_{2n} (x) = g_2 (x, n-x) \nonumber \eeq
\beq q_{1n} (x) = q_1 (x, n-x) \; ; \; q_{2n} (x) = q_2 (x, n-x) \eeq
In this case, the equation (\ref{eqn:532}) becomes 
\beq (q_{1n} g_{1n} + q_{2n} g_{2n}) f_n''= 0\eeq
Since now we have the \emph{sum}, $q_{1n} g_{1n} + q_{2n} g_{2n}$, instead of a difference, its value is no longer zero, which implies $f_n''=0$. This implies linearity, thus reproducing our desired result. 

Now, from what we have found, the player $1$ is guaranteed to lose if we have $q_1 g_1 < g_2 g_2$ \emph{throughout the entire domain}. Let us now be a bit more creative and say that $q_1 g_1 < g_2 g_2$ for $n_1 < n_{\rm crit}$, $q_1 g_1 > g_2 g_2$ if $n_2 < n_{\rm crit}$, and $q_1 g_1 = g_2 g_2$ everywhere else. This can be accomplished by defining them as
\beq g_1 (n_1, n_2) =
\left\{
	\begin{array}{ll}
		 n_1 /n_{crit} & \mbox{if } 0 \leq n_1 <n_{\rm crit} \\
		1 & \mbox{if } n_{\rm crit}< n_1 
	\end{array}
\right.
\eeq

\beq g_2 (n_1, n_2) =
\left\{
	\begin{array}{ll}
		 n_2 /n_{crit} & \mbox{if } 0 \leq n_2 <n_{\rm crit} \\
		1 & \mbox{if } n_{\rm crit}< n_2 
	\end{array}
\right.
\eeq
Now, if $K$ is a very large number, then it would take "much quicker" for a player $1$ go lower the amount of money to $n_{\rm crit}/K$ than it would take for him to "recover". This is due to the fact that the dynamics is irreversible in the region $n_{\rm crit}/K < n < n_{\rm crit}$. The time it takes to travel from, say, $n_1=n/2$ to $n_1=n_{\rm crit}$ is $T_1$. The time it takes to travel from $n_1=n_{\rm crit}$ to $n_1 = n_{\rm crit}/K$ is $T_2$, and the time it takes to travel from $n_1 = n_{\rm crit}/K$ back to $n_1 = n_{\rm crit}$ is $T_3$. Thus, the time it takes "to lose" is $T_1 + T_2$, while the time it takes to "recover" from having "lost" is $T_1 + T_3$. Now, if $K$ is sufficiently large, then $T_1 \ll T_2 \ll T_3$. This implies that $T_1 + T_2 \ll T_1 + T_3$. This is what allows us to insist that the time it takes for wave function to "collapse" is very small, and, at the same time, also claim that it takes a long time to "recover" from the consequences of the collapse. 

Let us now return to the GRW model described in the beginning of this section. As was stated earlier, the "gambling event" involves the multiplication of the wave function by the Gaussian, 
\beq \psi (\vec{x}_1, \cdots \vec{x}_N) \longrightarrow N (\phi; a_i; \vec{x}_{a_i 0}) e^{\frac{\alpha}{2} \vert \vec{x}_{a_i} - \vec{x}_{a_i 0} \vert^2} \psi (\vec{x}_1, \cdots \vec{x}_N), \eeq
and the probability of that event is given by 
\beq q (\phi; a_i; \vec{x}_{a_i 0}) = \beta N^{-2} (\phi; a_i; \vec{x}_{a_i 0}) \eeq
This means that the "probability" of winning or losing is "weighted average" of the corresponding "probabilities" for the case of each of the collapses: 
\beq p (\psi) = \sum_{a_i} \int d^3 x_{a_i} \beta N^{-2} (\phi; a_i; \vec{x}_{a_i 0}) p \Big(N (\phi; a_i; \vec{x}_{a_i 0}) e^{\frac{\alpha}{2} \vert \vec{x}_{a_i} - \vec{x}_{a_i 0} \vert^2} \Big) \eeq
Now, if we use $p (\psi) = \vert \psi \vert^2$ on the right hand side, we can easily see that $p (\psi) = \vert \psi \vert^2$ holds true on the left hand side, as well. Of course, since we have "infinitely dimensional" functional space instead of "one dimensional" induction, this is not a very good argument. Nevertheless, for the purposes of this paper, let us agree to accept its conclusion, and leave the detailed mathematical exploration of this issue for some other paper. 

Now, as one can easilly see, GRW model does not match the game. For one thing, a given point-particle state can win or lose different amounts, depending on how far it happened to be from the center of Gaussian. Also, we have multiplication instead of addition; and we can only convert it into addition if we assume that $\alpha \vert \vec{x} \vert^2 \ll 1$, which is probably \emph{not} the case, given a very large size of the universe.  Nevertheless, the "spirit" of the game is reproduced through the statement that the weighted average, as described above, matches the value of $\vert \psi \vert^2$. Vaguely speaking, "weighted average" corresponds to our statement that $g_1 q_1 - g_2 q_2 =0$ (even though right now we have uncountably many, instead of just two, values of $g$ and $q$). 

From this, very vague, point of view, the situation is similar to the one where both $g_1$ and $g_2$ become "very small" when one of the two players is losing. This is evident from the fact that if we convert "multiplication" into "addition", the incriment that we are "adding" will be proportional to the quantity we are attempting to multiply. Thus, if the quantity in question is nearly zero, the incriments will be very small, as well. Thus, the "claim" of that specific GRW model is that the "losing" player exchange such a small amounts of money (or "sand") that it would take him very long time to recover. At the same time, the analogue of our assertion that $g_1 q_1 - g_2 q_2 <0$ does \emph{not} hold. On the contrary, since the "weighted averages" hold \emph{at all times}, it means that exact equality, $g_1 q_1 - g_2 q_2 =0$, holds at all times, as well. 

This immediately raizes a question. The difference between $g_1 q_1$ and $g_2 q_2$ was putting an element of irreversibility into the picture: as soon as a player has less than $n_{\rm crit}$ amount of sand, it takes him far less time to lose more send than it would take for him to get it back. We needed this irreversibility since the collapse of the wave function is irreversible, as well (in particular, it takes a very small time for collapse to happen and very large time for its consequences to be "reversed"). Once $g_1 q_1$ and $g_2 q_2$ are equal, the irreversibility disappears. So, if it takes a "long time" to "recover" from "poverty", it would likewise take a long time to "get" there; this seems to contradict our experience telling us that the collapse of wave function is quick.

This question might be answered if we remind ourselves that the collapse-free unitary evolution (or, in other words, "earning" of the money) takes the form of a superposition of "simpler" larger-scale behavior and "more complicated" small-scale fluctuations. If the former is significantly smaller than the latter, the "rich" players can proceed "without much interference". After all, he ignores large fluctuations because they are "too slow" and he also ignores small fluctuations because they are "too small". On the other hand, a "poor" player does not regard "small fluctuations" as "too small". Thus, while he still ignores large fluctuations, he \emph{no longer} ignores the small ones. The latter keep "messing up" his game results and prevent him from becomming rich.

Let us assume that the "expected fluctuations" of the "income" is $n_{\rm fluc}$. Now, in light of the fact that the states that have very little money are winning/losing in a small "baby steps", it would take considerably longer to "lose" to the level of $n_{\rm fluc}/K$ than it would take to "lose" to the level of $n_{\rm fluc}$, where $K$ is some very large number. At the same time, this does \emph{not} affect the "small fluctuations"; after all, the latter are parts of unitary evolution and, therefore, are linear. Thus, a state can take a "long time" $T_1$ to "lose" to the level of $n_{\rm fluc}$ through the "game", and then take a "small" time $\delta t$ to go from $n_{\rm fluc}$ to $n_{\rm fluc}/K$ via "small fluctuations". At the same time, the "reverse" of this process won't work. Sure enough, these same small fluctuations will allow the state to successfully "recover" from $n_{\rm fluc}/K$ back to $n_{\rm fluc}$ within time $\delta t$; \emph{but} after that they will bring it \emph{right back} to $n_{\rm fluc}$ to $n_{\rm fluc}/K$, thus effectively \emph{preventing it} from recovering from $n_{\rm fluc}$ back to "large" $n$ within time $T_1$. 

This argument, of course, is based on assumption that the time scale of small-fluctuations "component" of the unitary evolution is significantly smaller than the time scale of "large fluctuation" one. Thus, if a state "starts out" at $p \gg p_{\rm fluc}$, then we can neglect the small fluctuations and assume that the time for unitary evolution to have any significant effect is $(\delta T)_{\rm large}$. Now, in case of the presence of entanglement, $T_1$ decreases due to the fact that any particle indirecty experience the effects of "mini-collapses" of all other particles. Thus, we can assume that $(\delta T)_{\rm large} \gg T_1$ despite the fact that we previously said $T_1$ was "large". Thus, in term, will allow us to neglect unitary evolution on the scale of $T_1$ just like we did when we were deriving our probabilities of collapses.  On the other hand, once the player arrived at the scales of $n_{\rm fluc}$, the scales of relevent fluctuations is $(\delta t)_{\rm small}$, which happens to be \emph{much smaller} than $T_1$ \emph{despite} the fact that $T_1$ is lowered through entanglement. This effectively prevents a player from participating in the game without interference and, therefore, he remains in the "losing" state. 

The other assumption that was made is the one of preferred basis which we "choose" to identify as "players" in our game. One thing that would \emph{not} have worked is if we were to identify players with eigenstates of the "perturbed" Hamiltonian we are working with. After all, the crucial element in our argument is a statement that the "small fluctuations" shift money around "between players"; this will no longer be true for exact eigenstates. Indeed, GRW-type theories \emph{do} have a preferred basis. For example, according to "standard" GRW model, the "game" involves multiplying wave function by Gaussians. If we assume that Gaussians are "in space coordinates" this will single out space as "preferred" to momentum, or any other basis for that matter. Since points in space are \emph{not} eigenstates of Hamiltonians involved, our argument works. 

In case of the paper at hand we will \emph{not} use Gaussians and, in fact, our "game" will be completely different from the one proposed in other GRW-type models. But we will still have "preferred basis" as well. In our case, it will consist of "classical" wave functions. Now, from the example of one dimensional harmonic oscillator, a $\delta$-function can be represented as an infinite series of states generated by raising and lowering operators. Since we assume, as always, that our states are living in a finite lattice, we can view it as multidimensional harmonic oscillator in a "rotated" basis. The "classical" wave function will then be a "point" in that multidimensional space, and the corresponding quantum state will be a $\delta$-function around that point. Again, we can represent that state as an infinite sum of the states generated by creation and annihilation operators. Clearly, this sum will \emph{not} coincide with any of the eigenstates of the Hamiltonian; thus, our "small fluctuations" argument still works. 

What we have been saying can be summarised by a statement that "quantum mechanics works for 'losing' states but fails to work for 'winning' ones". After all, we claimed that in case of $n$ being large, the scale to which quantum mechanics is sensitive to is \emph{much larger} than the scale of our "game", which allows the latter to "go uninterrupted", while in the case of $n$ being small, the opposite is true. Likewise, we also stated that in the case of $n$ being small  it would take "long time" for unitary evolution to "bring it back" to large values; we based this statement strictly on the \emph{purely mathematical} conjecture regarding \emph{unitary evolution}, without referring to anything external, such as mini-collapses. On the other hand, in case of $n$ being large, the collapse happens \emph{despite} the unitary evolution and it \emph{is} caused by something external (namely, the game we are describing).

This immediately brings up a quesiton: why do we see quantum mechanics "working" for large-probability states? This question can be answered by pointing out that it does \emph{not} work for entangled states: after all, the wavefunctions of tables and chairs don't "spread out". Now, the entanglement is precisely what we have appealed for when we made an assertion that $(\delta T)_{\rm large} \gg T_1$; after all, a particle in entangled state is being affected by the mini-collapses of all of the other particles it is being entangled with, which "divides" the value of $T_1$ by the number of entangled particles. Now, in case of lack of entanglement, or even in case of entanglement of \emph{few} particles (such as individual atoms or individual molecules), the value of $T_1$ is considerably larger and $(\delta T)_{\rm large} \ll T_1$ holds. This makes the "winning" players totally analogous to the "losing" ones (for which $(\delta t)_{\rm small} \ll T_1$ holds). Thus, \emph{in case of lack of entanglement}, \emph{neither} winning \emph{nor} losing players can participate in the game and \emph{both} are subject to unitary evolution. This means that "absence of entanglement" has the same effect as "low value of $n$": in both cases the unitary evolution is restored. 

\subsection*{5.5 Our version of GRW model: preliminary overview}

In the previous section we have described a generic "game" that reproduces the desired probability amplitudes. We have also mentioned the specific "game" that was used in GRW model, which involves "multiplication by Gaussians". Unfortunately, that specific game is not suitable for our purposes. After all, the Gaussians involved are living in configuration space, while the purpose of our paper is to get rid of the latter. However, it is evident that the argument in favor of Born's rule was independant of the specifics of the game. The only key element of the argument is that there has to be "some" way in which states "slightly increase" or "slightly decrease" their probability amplitude "apart from" unitary evolution. In the case of GRW model, that slight increase or slight decrease was attributed to Gaussians. But it doesn't have to be! In this paper we propose, instead, that the source of sight increase or slight decrease has to do with nature's "reducing" the mismatch with Einstein's equation due to the extra $\partial^{\rho} \partial_{\rho} g_{\mu \nu}$. This will allow us to "kill two birds with one stone": we will be able to accommodate better gravity and \emph{at the same time} we no longer need Gaussians for "flat space quantum mechanics"- purposes, either. 

We propose the following approach. As we have stated, the gravitational field evolves according to \emph{modified} Einstein's equation, with extra term $\epsilon_{\rm nc} \partial^{\rho} \partial_{\rho} g_{\mu \nu}$. This extra term allows the non-conserved energy-momentum sources $T_{\mu \nu}$ to produce $G_{\mu \nu}$ in a mathematically consistent way. But, at the same time, the difference between $G_{\mu \nu}$ and $T_{\mu \nu}$ can serve as a measure of non-conservation. Now, in the discrete case, the collapse of the wave function is no longer the only reason for non-conservation of $T_{\mu \nu}$. Quite independantly from this, any discrete theory is guaranteed to violate any "continuum-based" rule, including momentum conservation, with absolute certainty. Now, we propose to define the degree of non-conservation and then systematically "modify" the probability amplitudes of various states in such a way that we lower it as much as possible while, at the same time, keeping our "modifications" small consistenly with GRW outlook. This will essentially means that we introduce a \emph{new} non-conservation (namely the collapse of wave function) with a specific purpose of \emph{cancelling out} the non-conservation that was there due to discreteness!

Now, we would like the specific way in which we "set to zero" the probability amplitudes of "unwanted states" to be consistent with the rules of "gambler's ruin" . In particular, we would like to satisfy the multi-state generalization of the condition $g_1 q_1=g_2q_2$ described in equation (\ref{eqn:535}) of Section 5.4. The "cartoon" of our "game"  is the following. We "look" at each lattice particle, one by one (that is, in every cluster we separately look at each of the particles). While considering a given particle $i$, we will "look at" $\psi (\{j \vert q_j = q_i \})$ and $A^{\mu} (\{j \vert q_j = q_i \})$ (as well as other \emph{bosonic} fields that might be involved). We then compute the value $T_{i; \mu \nu}$ according to the prescription of Sections 5.7 and 5.8 (as shown in Section 5.8, we can exploit charges $q$ and $q'$ in order to get both spacelike \emph{and} timelike components of $T_{i; \mu \nu}$), and then compare it with $G_{\mu \nu}$. Then, based on this comparison, we lower the probability amplitude of the \emph{entire set} of particles $\{j \vert q_j = q_i \}$. In other words, we lower $\psi_j$ for \emph{all} of the said particles $j$ \emph{regardless} of their distance to particle $i$, as long as their "charges" match. The amount by which we lower it is "larger" if the "mismatch" at the particle $i$ is larger. On the other hand, that amount by which we lower $\psi (q_i)$ is smaller if $\psi (q_i)$ happened to be smaller to begin with. After all, if the probability amplitude of a given state is "very small", it is "not important enough" for us to "fix". 

Just like in Pearl's case (see \cite{Pearle1} and \cite{Pearle2}), the values of our "steps" are quite small. The probability amplitudes of "unwanted states" are successfully "set to zero" only if this step is being repeated a very large number of times. Since the probability amplitude $\psi (\phi, A^{\mu})$ refers to $\phi$ and $A^{\mu}$ \emph{over entire space} (as opposed to any specified region), any given "one step decrease" affects the entire universe as well (namely, it affects \emph{all} particles $j$ satisfying $q_j=q_i$). At the same time, the \emph{cause} of that event is located in some specific region (in particular, the region surrounding particle $i$). After all, it is highly unlikely that  the value of $\vert T_{j; kl} - G_{kl} (\vec{x}_j)\vert$ will exceed $\epsilon_{\rm Ein}$ everywhere "at the same time". We would expect it to happen \emph{first} at the region where that difference is "the largest". Now, that specific region will, most likely, be \emph{outside} of our laboratory (and outside our solar system for that matter). Therefore, we will not be able to predict the nature of GRW-type "hit" by measuring the behavior of gravitational field and we we view the sequence of "hits" to be random, just like it is random in GRW case. This randomness is what allows us to recover Born's rule without any unwanted statistical correlations. 

At the same time, \emph{despite} the lack of correlation with gravity inside our laboratory, the unwanted gravitational effects are systematically gotten rid of. On the one hand, \emph{most} of the "collapse events" we observe in our laboratory are "intended" to "correct" the unwanted gravitational effects "far away" and, therefore, "look random" to us. On the other hand, \emph{if} there is some unwanted gravitational effect "in our laboratory", there will be "additional" collapse events intended to correct this, "too". This means that the "presence" of unwanted gravitational effects is \emph{sufficient} but \emph{not} necessary for the wave function to "collapse". The impact of gravity mismatch \emph{inside the laboratory} implies \emph{additional} criteria of "collapse of wave function", independant of entanglement. In particular, any two states with "macroscopically mismatching" $T_{\mu \nu} (\vec{x}_i)$ will not be able to "simultaneously" match the same $G_{\mu \nu} (\vec{x})$; thus, the probability amplitude of one or both of them will be "bound" to be sent to zero.  

It is interesting to note that both the presence of entanglement and the "large" value of $T_{\mu \nu}$ is linked to the large value of particles in the laboratory. At the same time, the \emph{ways} in which these two things produce collapse of wave function are completely different. It is also conceivable that the number of particles needed to create entanglement is vastly different from the number of particles needed to create "macroscopically significant" $T_{\mu \nu}$. If such is the case, one would expect that the ultimate mechanism of wave function collapse will be identified with whatever requires the "smallest" number of particles. Now, the experimental observed correlation between "measurement" and "temperature" can only be explained by entanglement. Thus, it stands to reason that the entanglement requires much smaller number of particles than one would need to create $T_{\mu \nu}$ capable of producing macroscopically detectable graviational field. At the same time, however, \emph{even if} there was no such thing as entanglement, the "gravity" of "quantum mechanical" things will \emph{continue} to be "too weak" since now the "second mechanism" would "switch on" that would "collapse" the superposition of any "gravitationally distinguishable" states. 

\subsection*{5.6 Sources of gravitational field: can singularity be avoided}

Now, just like with other GRW-type theories, the outcome of collapse is \emph{not} the exact $\delta$-function; we still have a superposition of several different states. But, at the same time, these states are \emph{not} macroscopically distinguishable. In light of the fact that gravity can only be observed on macroscopic level, we have two possible options of defining its source. On the one hand, we can \emph{first} take a "weighted average" of $\phi$-s and $A^{\mu}$-s associated with different states and then define a \emph{single} $T_{\mu \nu}$ based on that weighed average. Or, on the other hand, we can separately evaluate $T_{\mu \nu}$-s for each given $\phi$ and $A^{\mu}$ and \emph{then} take the weighted average of the results. 

In the previous version of the paper I sloppilly assumed that since these two options are macroscoplically indistinguishable it doesn't matter which to pick and i picked the last one. Unfortunately, I now realize it was a very bad choice, which, in fact, is the main reason I am now putting forth the new version of the paper. While the \emph{spacelike} components of $T_{\mu \nu}$ would, indeed, look similar, the time components would not. The ''smallness'' of time interval $\delta t$ implies that $T_{\mu \nu}$ can be ''very large'' if we choose to make a ''jump'' across the so-called ''classically indistinguishable'' region within the time interval $\delta t$. At the same time, the fact that $T_{\mu \nu}$ includes the \emph{square} of the time derivative (rather than time derivative itself), it can only be made ''large'' in the positive direction but not in the negative one. Thus, we would not be able to ''cancel out'' the ''very large'' contributions to $T_{\mu \nu}$ which would produce unwanted result. On the other hand, if we average out the fields, themselves, then their first derivatives can be both positive or negative. Thus, ''on average'' the time derivatives of the fields will have reasonable values and we will identify their squares with $T_{\mu \nu}$. 

Now, since we intend to use the values of $\phi$ and $A^{\mu}$ as sources of gravitational field, we would like them to be ''smeared out'' throughout the continuum. This can be accomplished by using $\phi_k (t)$ and $A_k^{\mu} (t)$ as "sources" of the $\phi (\vec{x}, t)$ and $A^{\mu} (\vec{x}, t)$, respectively. However, if we do it naively, we would encounter several problems. First of all, these fields will be singular in the vicinity of each of the lattice points, which would create unwanted black holes. Secondly, away from the ''black hole'' regions, the gradients of these fields would be pointing out towards the nearest lattice points, as opposed to the large-scale gradient of corresponding ''discretized'' fields. The first feature can be dealt with by ''cutting off'' the values of the fields according to 
\beq \phi_{\rm cut} (x^{\mu}) = \frac{\phi (x^{\mu}) \min (\vert \phi (x^{\mu}) \vert, \phi_{\rm max})}{\vert \phi (x^{\mu}) \vert} \eeq
\beq A^0_{\rm cut} (x^{\mu}) = \frac{A^0 (x^{\mu}) \min (\vert A^0 (x^{\mu}) \vert, A_{\rm max})}{\vert A^0 (x^{\mu}) \vert} \eeq
\beq \vec{A}_{\rm cut} (x^{\mu}) = \frac{\vec{A} (x^{\mu}) \min (\vert \vec{A} (x^{\mu}) \vert, A_{\rm max})}{\vert \vec{A} (x^{\mu}) \vert} \eeq
In order to address the second issue, we need to introduce a ''dispersion'' of these fields so that they would be ''fuzzy enough'' to reflect the behavior of the corresponding ''discretized'' fields on a scale seleral magnitudes larger than the distance between lattice points. This can be done by appealing to ''preferred frame'' and introducing first time derivatives into the propagation term:
\beq \nabla_s^{\alpha} \nabla_{s \alpha} \phi  - \zeta_{\phi} \partial_0 \phi = \sum_{k=1}^N \frac{l_k (t) \psi_k (t)}{K_{\psi} (t)} \phi_k (t) \sqrt{- g_{\mu \nu}} \delta^3 (\vec{x} - \vec{x}_k (t)) \label{56phi} \eeq
\beq \nabla_s^{\alpha} \nabla_{s \alpha} A^{\mu} - \zeta_A \partial_0 A^{\mu} = \sum_{k=1}^N \frac{l_k (t) \psi_k (t)}{K_{\psi} (t)} A_k^{\mu} (t) \sqrt{- g_{\mu \nu}} \delta^3 (\vec{x} - \vec{x}_k (t)) \label{56A}\eeq
where $K_{\psi}$ is a normalization constant for $\psi$, whose dynamics is defined in Section 5.9. Finally,, the source of gravitational field is identified with a \emph{continuum}-based versions of $T_{\mu \nu}$:
\beq T_{\mu \nu} = \frac{\delta {\cal L}_{\rm non-grav} (\vec{x}, t)}{\delta g^{\mu \nu} (\vec{x}, t)} \eeq
where
\beq {\cal L}_{\rm non-grav} (\vec{x}, t) = g^{\mu \nu} (\nabla_{\mu} \phi^*_{\rm cut} + g_{\mu \rho} A^{\rho} \phi^*_{\rm cut}) (\nabla_{\nu} \phi_{\rm cut} + g_{\nu \rho} A^{\rho} \phi_{\rm cut})  +  \nonumber \eeq
\beq + (\nabla_{\mu} A_{\rm cut}^{\nu} - g^{\rho \nu} g_{\sigma \mu} \nabla_{\rho} A_{\rm cut} ^{\sigma}) (g^{\rho \mu} g_{\sigma \nu} \nabla_{\rho} A_{\rm cut}^{\sigma} -  \nabla_{\nu} A_{\rm cut}^{\mu} \eeq
and the gravitational field evolves according to the modified Einstein's equation mentioned earlier,
\beq G_{\mu \nu} + \epsilon_{\rm nc} \partial^{\rho} \partial_{\rho} g_{\mu \nu} = T_{\mu \nu} \eeq
where the purpose of $\epsilon_{\rm nc}$-term is to break general relativistic covariance so that the equation continues to have solutions for non-conserved $T_{\mu \nu}$. This corresponds to the action
\beq S_{\rm class \; grav} = \int d^3 x \sqrt{-g} (R + {\cal L}_{\rm non-grav}) + \epsilon_{\rm nc} \int d^3 x \partial^{\rho} g^{\mu \nu} \partial_{\rho} g_{\mu \nu} \eeq
where the absence of $\sqrt{-g}$ in the last term corresponds to the violation of general relativistic covariance which, in turn, assures the existence of solutions for non-conserved $T_{\mu \nu}$. The dynamics of $g_{\mu \nu}$ is given by \emph{classical} equation
\beq \frac{\delta S_{\rm class \; grav}}{\delta g_{\mu \nu}} = 0 \eeq
which is devoid of path integral since we are viewing gravity as purely classical and avoiding its quantization. 

\subsection*{5.7. Do fermions produce gravity?}

Based on the Sections 5.3 and 5.6, we would like to use $T_{\mu \nu}$ as a source of gravitationanl field. Therefore, Section 5.8 will be devoted to definition of $T_{\mu \nu}$.  Before we proceed towards this goal, it is important to agree on the "sources" of gravity. Namely, do they include fermions? Since $T_{\mu \nu}$ is used as a "classical" source, it should have ontological meaning. Therefore, if we are to insist that fermionic fields are sources of gravity, it would "force" us to view Grassmann numbers as ontologically meaningful quantities per section 3.2, \emph{as opposed to} thinking of them as mere formalities as is commonly done in standard QFT. Logically, this implies that the results we obtain will be the function of the specific way in which we define Grassmann numbers. At the same time, our original goal was to produce "the same" information as QFT, not "more information". As long as we produce "the same" information, we are safe. But, as soon as we produce "more information" we put ourselves under the risk of being "wrong". So the question is: are we willing to take this risk? 

The first thing to notice is that \emph{even if we did} take Section 3.2 seriously, the "fermionic field" itself has assigned afore-given values, that are not subject to variation. This is evident from the "measure" we impose on Grassmann variables in equation \ref{eqn:94}, 
\beq d_i (\xi) \vert_{\xi = \theta_j} = \delta_i^j \theta_i \eeq
The $\delta_i^j$ in the "measure" implies that fermionic field has a certain value with absolute certainty and has probability zero of any other value. In particular, in section 3.2.2 we have defined Grassmann numbers in terms of "sets". According to Definition 2, 
\beq \theta_n (A) = \left\{
	\begin{array}{ll}
		1 \; , \; A = \{n \}\\
		0 \; , \; A \neq \{n \}
	\end{array}
\right.
\eeq
We also recall that $\theta_n$ was treated as \emph{constant}. Thus, in order for the "fermionic field" $\eta (x)$ to change values, it has to take the form 
\beq \eta (k) = \sum f_k \theta_k \eeq
in which case the "fermionic field" would have been identified with $f_n$. \emph{But} the above-given definition of $\delta_i^j$ implies that there is a "probability $1$" to have  
\beq f_k (j) = \delta_k^j \eeq
and "probability $0$" to have anything else. This "forces" any potentially meaningful information regarding fermionic field to be "constant". This can be seen from the fact that in Section 3.2.5 we identified the elements of the "sets" given in Definition 2 with lattice points. This implies that Definition 2 can be interpretted as one to one correspondence between the specific value of Grassmannian field and a point in space. Now, if $\nu (k)$ were to fluctuate arbitrarily, we could have had $\eta (k) = \theta_l$ where point $l$ can, in principle, be "far away" from $k$. The $\delta_k^l$ in the definition of the measure "prohibits" us from doing it. On the one hand, this gets rid of unphysical non localities. On the other hand, however, it implies that the fermionic field at $k$ is restricted to \emph{one} value:
\beq \eta (k) = \theta_k \eeq
This, in turns, makes it impossible for "variation of fermionic field" to produce gravity, for the simple fact that there is no such thing as "variation of fermionic field"! The only possible thing we can do is to claim that the "constant" fermionic field, defined by the above equation, produces "constant" gravity. Furthermore, in light of the fact that its derivative is zero, it is likely that the "constant gravity" will be sent to zero as well. 

There \emph{is}, however, a \emph{different} source of gravity that the formal reading of section 3.2 implies. In particular, during the process of "taking the integral" over each particular "time slice", we are "going through" all possible subsets of that slice. These subsets are identified with "evaluation fields" defined according to 
\beq ev_A (a) = 
\left\{
	\begin{array}{ll}
		1\; , \; a \in A \\
		0 \; , \; A \not\in A 
	\end{array}
\right.
\eeq
Therefore, \emph{unlike, and completely opposite to} fermionic field $\eta$, the above "evaluation fields" \emph{do} change, and they change \emph{a lot}: within a very small time interval $\delta t$, they take very large ($n$) values, which constitutes \emph{all} values avaiable! Thus, they shift from one value to the other within a time interval $\delta t/n$. At the same time, the "shift" between any two values is "of the order $1$" so to speak. In the context of Section 3.2, these discontinuities were reflected in the Lagrangians: the latter had "very large" terms, such as $\ln 0$ in Section 3.2.5. 

Strictly speaking, $\ln 0$ was still viewed as finite via Equation \ref{eqn:ln0}. After all, since everything is discrete, a finite quantity can play a role of infinity as long as it is significantly larger than the appropriate power of the inverse of the discreteness scale. Furthermore, in section 5.6 it was demonstrated that it is possible to prevent "very large" energy momentum from creating black hole singularities. Thus, in principle, it is "possible" to produce a non-singular gravitational theory that uses the Lagrangians of section 3.2.5 as one of the "sources". However, it is still true that the gravitational effects produced by these sources are "much larger" than the ones produced by bosons. Now, according to Section 3.2.4 and 3.2.5, the "order" in which we "go" from one set to the other is independant of any other physical processes, such as the behavior of bosonic fields. This means that the physics-independant gravity will dominate over physics-dependant one. While this is "mathematically conceivable" it clearly contradicts our experience.  

Apart from what we have just described, there is also another issue: Lagrangians proposed in Section 3.2.5 had "imaginary part" in order to "set to zero" whatever we needed to "get rid of" during any given "step" of "Grassmann integration". In the gravity-free context this is fine. After all, we have demonstrated that we reproduce the results of Feynmann path integral which, in term, implies that the effects of the afore-said "complex potentials" disapper on large enough scales. However, if we are to allow these complex-valued quantities to gravitate, we would have to deal with complex valued metric. We will no longer be able to say that complex valued metric "disappears on the larger scale" for the simple reason that we will no longer have a definition of "larger scale", since we "don't understand" the "complex valued" geometry to begin with. Of couse, it is possible to try and make sense of it (for example, complex coordinates are often used to describe world sheet of the string), but this is not something we are willing to deal with in this work. 

In light of the above difficulties, we propose to claim that bosons are the only quantities that emit gravitational field. This does \emph{not} imply that we are denying anything said in Chapter 3. After all, it is mathematically conceivable to claim that something is "gravitationally neutral", just like it is conceivable to claim that something is "electrically neutral". Violation of equivalence principle lines up with the spirit of our paper. For one thing, we had already violated Lorentz covariance. Furthermore, it can be argued that our definition of Grassann numbers forces us to view each of the fermionic fields separately, thus violating $SU(2)$, $SU(3)$ or any other symmetries between fermions that might arize.  In all such cases, we pointed out that the symmetries were restored in the lab. This implies that, according to our philosophy, we are free to add things we are violating as long as they, too, are restored in the lab. We will use this as a justification why we violate equivalence principle by saying that fermions don't gravitate. After all, the equivalence principle will be restored "in the lab" through the gravity of the bosons that fermions emit. The boson-produced effects are manifested in mass renormalization which (according to equation 7.29 of Peskin and Schroeder) is given by 
\beq \delta m = \frac{3 \alpha}{4 \pi} m_0 \ln \Big( \frac{\Lambda^2}{m^2} \Big) \eeq
The only "fermionic" contribution to gravity comes from $m_0$, which is of the order of $(\ln \Lambda)^{-1}$. Thus, by claiming that fermions don't gravitate we imply that equivalence principle is violated in the order of $(\ln \Lambda)^{-1}$. In light of the fact that the latter is very small, our claim can't be experimentally falsified.

Let us now ask a different question: if we can attribute gravitational field of fermions to the gravity of the bosons they emit, can we use the similar argument to discard gravity of one boson in favor of gravity of the other? For example, if we have charged scalar field, can we say that it does not gravitate, and only the spin $1$ field that it is coupled to does? The key to the answer is the theory of measurement. According to the framework layed out in Sections 5.1-5.3, the \emph{outcome} of the "collapse of wave function" produces gravity. Thus, if we are "measuring" only one bosonic field, it will be logical to assume that this field is the only one gravitating; or if we measure all of them, then all of them should gravitate. In principle, it is conceivable to say that we are only "measuring" one of the bosonic fields indirectly through the other one. For the purposes of this paper we will simply "assume" that we measure all of the bosonic fields, and we will leave the exploration of alternative situations for future research. 

In fermionic case, however, we \emph{know} that nothing is being measured. After all, as was said earlier, the fact that we have $\delta_i^j$ in our "measure" implies that fermionic field has afore-given values. This leaves no room for "measuring" it, since we already know what it \emph{is}. As far as "evaulation fields" are concerned, such as $ev_A$, they \emph{do} alter (and, in fact, alter a lot); but they alter according to the afore-given prescription outlined in Sections 3.2.4 and 3.2.5. This, again, means that we can not "measure" them, since we already \emph{know} what their behavior \emph{is}. Incidentally, the point of view that "we can not measure fermions" is also shared by the Struyve and Westmann (see \cite{Struyve}), although the context of their work is Bohmian. Likewise, the idea that the mass of electron is accounted for by photons \emph{alone} has been brought up by other physicists, as was mentioned in the Feynmann lectures. In his chapter on "electromagnetic mass" he mentioned that there are two points of view: some people say that only part of the electron's mass is "electromagnetic" while others claim that all of it is. 

\subsection*{5.8 Generation of "encoding" of $T_{\mu \nu}$ inside lattice "particles"}

In the past two sections we have cleared some controversies surrounding $T_{\mu \nu}$. In particular, in Section 5.6 we have argued that it should be produced through the differentiating of ''smeared out'' fields. Then, in Section 5.7, we have argued that only bosons produce gravity, which implies that only "bosonic fields" are to be taken into account in definition of $T_{\mu \nu}$. We are now ready to define the "encoding" of energy momentum tensor $T_{i \mu \nu}$. 

We know from field theory that $T_{\mu \nu}$ has the same differential structure as $\cal L$ does, except for the fact that the former has not been contracted while the latter has been, along with possible sign differences. This seems to suggest that the procedure of generating $T_{\mu \nu}$ should be closely parallel to the one of generating $\cal L$ in Chapters 3 and 4. In this section we will explicitly modify the key results of these two sections for $T_{\mu \nu}$.  We will essentially be redoing the same thing we have done for ${\cal L}_i$, except for extra $(\mu \nu)$ indexes that "come along for the ride". As we recall from part 8 of the algorithm in Chapter 4.1, ${\cal L}_i$ is being "updated" according to 
\beq {\cal L}_i \rightarrow {\cal L}_i + {\cal K}_i \eeq
whenever an external pulse is received. In the case of $T_{i; \mu \nu}$ the role of "Lagrangian generator" $\cal K$ is being played by a "stress tensor generator" $U_{\mu \nu}$. Thus, we rewrite the above "updating" formulae as 
\beq T_{i ; \mu \nu} \rightarrow T_{i ; \mu \nu} + U_{i ; \mu \nu} \eeq
Now, again by "copying" what we did for $\cal K$ and replacing $\cal K$ with $U_{\mu \nu}$ we can immediately propose that $U_{i; \mu \nu}$ is given by 
\beq U_{i; \mu \nu} =  U_{\mu \nu} (q_i, q'_i, \overline{q}_i (t_i), Q(\vec{x}_i, t), \cdots) \eeq
where $U_{\mu \nu} (\cdots)$ takes the form 
\beq U_{\mu \nu} (q_i, q'_i, \overline{q}_i (t_i), Q(\vec{x}_i, t), \cdots) =  T(\vert q_i- \overline{q}_i (t_i) \vert< \epsilon_q) T(\vert q_i - Q (\vec{x}_i, t) \vert < \epsilon_q) U_{ss; \mu \nu} (\cdots) +  \nonumber \eeq
\beq  + T(\vert q_i- \overline{q}_i (t_i) \vert< \epsilon_q) T(\vert q'_i- Q (\vec{x}_i, t) \vert< \epsilon_q) U_{st; \mu \nu} (\cdots) + \label{U}\eeq
\beq+ T(\vert q'_i- \overline{q}_i (t_i) \vert< \epsilon_q) T( \vert q_i- Q (\vec{x}_i, t) \vert < \epsilon_q) U_{ts; \mu \nu} (\cdots) + \nonumber \eeq
\beq + T(\vert q'_i- \overline{q}_i (t_i) \vert< \epsilon_q) T(\vert q'_i - Q (\vec{x}_i, t) \vert< \epsilon_q) U_{tt; \mu \nu} (\cdots) \nonumber \eeq
which, again, is completely analogous to the form $\cal K$ takes. In principle, we can come up with an explicit expression for $U_{\mu \nu}$ by writing $T_{\mu \nu}$ in differential form and then replacing derivatives with corresponding integral expressions we had written in Sections 3.1.2-3.1.4; after that, we can combine them under the same integral and, finally, identify the expression under the integral with $U_{\mu \nu}$. This, of course, is the exact procedure we have followed in order to derive the expression for $\cal K$. However, the need to explicitly carry out these steps can be avoided if we recall that $T_{\mu \nu}$ is simply a derivative of $\cal L$ with respect to the metric. The only thing we have to be careful about is to make sure that either all the vectors are written in the lower indexes and then use $g^{\mu \nu}$ to raise them whenever necessary. Likewise, $e^{- \frac{\alpha}{2} \vec{r}^2}$ should be replaced with $e^{+ \frac{\alpha}{2} g^{ij} r_i r_j}$ (the minus sign was replaced with plus due to the $(+ \;  - \; - \; -)$ metric convention). Thus, we postulate
\beq U_{\mu \nu} (\cdots) = \frac{\delta {\cal K} (\cdots)}{\delta g^{\mu \nu}} \eeq
Apart from saving us time, this expression also has a benefit of elegance of the theory. After all, we no longer have to separately "postulate" $\cal K$ and $U_{\mu \nu}$. We \emph{only} have to postulate $\cal K$ and then apply a "general principle" of how $U_{\mu \nu}$ is "defined" based on $\cal K$. Finally, since we now have defined $T_{i; \mu \nu}$ we can postulate a dynamics that "enforces" the desired "updating" equation. This will be a carbon copy of the equation (\ref{eqn:489}) for ${\cal L}_i$, with appropriate replacements: 
\beq \frac{dT_{i; \mu \nu}}{d t} = - \alpha_T T_{i; \mu \nu} + \frac{U_{\mu \nu} (\cdots)}{\chi} T(\xi_i < \chi) T(\mu (\vec{x}, t) > \mu_c) T \Big(\frac{e_{j2}}{\vert e_{j1} \vert}  < \chi \Big) , \eeq
As a general principle, we would like to allow the constants coupled to different things to be different; thus we are using $\alpha_T$ instead of $\alpha_{\cal L}$. But, in light of the fact that $T_{\mu \nu}$ is closely linked to ${\cal L}_{\mu \nu}$, it is justifiable to argue that $\alpha_L = \alpha_T$. Since we don't have strong arguments either in favor or against it, we will leave it up to the philosophy of the reader.

\subsection*{5.9 Generation of the "encoding" of the "normalization constant"}

Let us now discuss the issue of normalization. As we recall, the fields $\phi_i (t)$ and $A_i^{\mu} (t)$ is "smeared into" $\phi (\vec{x}, t)$ and $A^{\mu} (\vec{x}, t)$ according to equation 
\beq \nabla_s^{\alpha} \nabla_{s \alpha} \phi  - \zeta_{\phi} \partial_0 \phi = \sum_{k=1}^N \frac{l_k (t) \psi_k (t)}{K_{\psi} (t)} \phi_k (t) \sqrt{- g_{\mu \nu}} \delta^3 (\vec{x} - \vec{x}_k (t)) \eeq
\beq \nabla_s^{\alpha} \nabla_{s \alpha} A^{\mu} - \zeta_A \partial_0 A^{\mu} = \sum_{k=1}^N \frac{l_k (t) \psi_k (t)}{K_{\psi} (t)} A_k^{\mu} (t) \sqrt{- g_{\mu \nu}} \delta^3 (\vec{x} - \vec{x}_k (t)) \eeq
Now, the coefficient $K_{i \psi}$ in denomenator serves the purpose of normalization of $\psi_i$. The main reason that coefficient is necessary is that the "collapse mechanism" reduces the probability amplitudes of "losing" states, but \emph{fails to} increase the probability amplitude of "winning" ones. Thus, the normalization of $\psi_i$ decreases in time. This is compensated by dividing it by $K_{i \psi}$ which decreases in time "at the same rate". 

Now, from "quantum mechanical" perspetive, that coefficient should \emph{only} be a function of time, but \emph{not} of space. Furthermore, it should be the same for \emph{all} charges. After all, the normalization constant is an "integral" of $\vert \psi \vert^2$ over \emph{all} "states". Thus, it can \emph{not} be specific for any particular state. In our language, it can not be specific for any particular "charge". Now space-independance, together with charge-independance implies that it shoud be independant of $i$. The reason we write $K_{i \psi}$ as opposed to simply $K_{\psi}$ is the same as the reason why ${\cal L}_i$ and $S_i$ were $t$-dependant, despite the fact that we don't want them to be. In both cases the underlying issue is that we would like to have a "superluminally-local mechanism" that enforces our "desired" phenomena. In order to have any hope of making things "exact" we want our signals to move with infinite speed. But, since our intuition demands locality, we insist that the speed of signals is finite, albeit superluminal. As a result, we might have some unwanted variations of things we "want" to be constant due to the delays of propatation of these signals; this is what forces us to speak of $i$-dependance of $K_{i \psi}$, $t$-dependance of ${\cal L}_i$ and $S_i$, and so forth. 

Let us now go ahead and describe the mechanism through which the "normalization coefficient" is "generated". Since we would "like" to reduce $i$-dependance as much as possible, we do \emph{not} want to introduce pulse-based dynamics for $K_{\psi e}$ (after all, the ultimate purpose of pulse-based dyncamics is to \emph{create} charge-dependance, which is just the opposite to what we want to do). Therefore, we will borrow the equations we were using for $\rho_i$, while replacing $\rho_i$ with $K_{\psi i}$, along with other necessary replacements. As we recall from the case with $\rho$, the \emph{lack} of "splitting into pulses" creates the concern about signularity due to self-interaction of any given particle. We avoided such singularities by introducing \emph{two} fields, $\rho_e$ and $\rho_o$. The particles with even $q'$ were \emph{emitting} $\rho_e$ but were only \emph{reacting to} $\rho_e$; and, particles with odd $q'$ were \emph{emitting} $\rho_o$ while reacting to $\rho_e$. We will do similar trick right now, while replacing $\rho_e$ and $\rho_o$ with $K_{\psi e}$ and $K_{\psi o}$, respectively. We now "copy" the equations (\ref{eqn:520}) and (\ref{eqn:521}) with the above-mentioned replacements, and obtain
\beq \nabla_s^{\alpha} \nabla_{s; \alpha} K_{ \psi e} + m^2_K K_{\psi e} + \zeta_K \partial_0 K_{\psi e} = J_{\psi} \sum_{j=1}^N (1 + (-1)^{q'_j}) \vert \psi_i \vert^2 l_i (t) \delta^3 (\vec{x} - \vec{x}_j) \eeq
\beq \nabla_s^{\alpha} \nabla_{s; \alpha} K_{ \psi o} + m^2_K K_{\psi o} + \zeta_K \partial_0 K_{\psi o} = J_{\psi} \sum_{j=1}^N (1 - (-1)^{q'_j}) \vert \psi_i \vert^2 l_i (t) \delta^3 (\vec{x} - \vec{x}_j) \eeq
Again, similartly to what we have done for $\rho_i$, we formally define $K_{\psi i}$ in such a way that it only "takes into account" $\rho_o$ if $i$ is even and, on the other hand, it only "takes into account" $\rho_e$ if $i$ is odd. Again, we merely copy the equation (\ref{eqn:522}) for $\rho_i$ with appropriate substitutions: 
\beq K_{\psi;i} (t) = (-1)^{q'_i} K_{o\psi} (\vec{x}_i, t) + (-1)^{q'_i+1} K_{e\psi} (\vec{x}_i, t) \eeq
This completes discussion of the coefficients next to $\psi_i$ in equation (\ref{56phi}) and (\ref{56A}). Just to summarize, $\psi_i$ has been divided by $K_{i \psi}$ for the purposes of normalization, and it has been multiplied by $l_i (t)$ (which, in most cases, eitehr "very close to zero" or "very close to $1$") for the purposes of "hiding" the unwanted particles. Thus, in many cases, we are using 
\beq \frac{l_i (t)\psi_i (t)}{K_{i \psi} (t)} \nonumber \eeq
as a single "item". At the same time, in some situations there are good reasons \emph{not} to use either or both of these coefficients. We will be making sure to explain in any situation that might come up, the reasons why we chose to either iinclude them or not. 

\subsection*{5.10 The detailed mechanism of gravity-based "gambler's ruin"}

In light of the fact that we have normalization, we no longer have to worry about "increasing" the probability amplitude for "winning" states. We only have to worry about "decreasing" it for "losing" ones. Let us now work out the mechanism in which the probability amplitude is "decreased".  As we have stated earlier we want to look at each particle, one by one. Each time, we are to look at the value of $T_{j \mu \nu}$, compare it to $G_{\mu \nu} (\vec{x}_j)$, quantify their "mismatch", and "decide" how much to lower the probability amplitude of  a state $\{j \vert q_j = q_i \}$. In order to enforce it "mechanically", we propose that  particle $j$ sends  a "lowering signal" whenever its internal timer "clicks". That signal will be received by the particles $i$ satisfying $q_i = q_j$, in response to which the value of $\psi_i$ will be "lowered" by the value specified in that signal. 

Now, the easiest way for the particle $i$ to "know" the value of $q_j$ is to "time" the emission of the charge-information and the emission of lower-signal in such a way that the two are simultaneous. We have already seen in equation (\ref{eqn:459}) of chapter 4.8 that the information about "charges" $q_i$ and $q'_i$, is being emitted according to a "rotating" timer $(\chi_{i1}, \chi_{i2})$ attached to the particle $i$ (namely, the signal is emitted whenever $e_{j2}/ \vert e_{j1} \vert < \chi$). We would like, therefore, to "link" the "lowering signal" to that same "timer". Now, we recall from equations (\ref{eqn:461}) that the information about "charges" $q$ and $q'$ is carried through the signals $\mu_Q$ and $\mu_{Q'}$, which propagate according to
\beq \nabla^{\alpha}_{s} \nabla_{s; \alpha} \mu_Q + m_{\mu_q}^2 \mu_Q + \zeta_{\mu_q} \partial_0 \mu_Q= \sum_{j=1}^N q_j \delta^3   (\vec{x}- \vec{x}_j)  T \Big(\frac{e_{j2}}{\vert e_{j1} \vert}   < \chi \Big) \nonumber\eeq
\beq \nabla^{\alpha}_{s} \nabla_{s; \alpha} \mu_{Q'} + m_{\mu_q}^2 \mu_{Q'}  + \zeta_{\mu_q'} \partial_0 \mu_Q= \sum_{j=1}^N q'_j \delta^3   (\vec{x}- \vec{x}_j)  T \Big(\frac{e_{j2}}{\vert e_{j1} \vert}   < \chi \Big) \nonumber\eeq
The fields $\mu_Q$ and $\mu_{Q'}$ are then "converted" into the fields $Q$ and $Q'$, which are given by 
\beq Q (\vec{x}, t) = \mu_Q (\vec{x}, t) \vert \vec{X} (\vec{x}, t) \vert \; ; \; Q' (\vec{x}, t) = \mu_{Q'} (\vec{x}, t) \vert \vec{X} (\vec{x}, t) \vert \eeq
where the field $\vec{X}$ is given by 
\beq \vec{X} = - \frac{\mu_q \vec{\nabla} \mu_q}{\vert \vec{\nabla} \mu_q \vert^2} \eeq
The field $\vec{X}$ is claimed to approximate $\vec{x}_i - \vec{x}_j$, while the field $Q$ approximates $q_i$. Now, in order to "attach" the "lowering signal" to this, we have to "add" some other field that carries the information necessary to decide just "how much" recepient particle "should" lower its internal degree of freedom $\psi_j$. We can describe the "lowering signal" by introducing the field $\mu_{\rm err}$ (where "err" stands for "error" as in "error between $G_{\mu \nu}$ and $T_{\mu \nu}$"), which propagates according to 
\beq \nabla_s^{\alpha} \nabla_{s; \alpha} \mu_{\rm err} + m^2_{\mu_{\rm err}} \mu_{\rm err}  + \zeta_{\mu_{\rm err}} \partial_0 \mu_Q = \sum_{j=1}^N \Big( \delta^3 (\vec{x} - \vec{x}_j) \times \nonumber \eeq 
\beq \times g \Big(\sum_{\mu \nu} (\vert \psi_j \vert^2 ;  G_{\mu \nu} (\vec{x}_j, t) - T_{j; \mu \nu} (t))^2  \Big)  T \Big(\frac{e_{j2}}{\vert e_{j1} \vert}   < \chi \Big) \Big) \eeq
As one can immediately notice, we went out of the way in trying to break Lorentz covariance so that the plus-signs are used everywhere. The argument for doing so is the same as the one we had for the conversion coefficient between $T^{\prime}_{\mu \nu}$ and $T_{\mu \nu}$, which included similar sum. In both cases, we are trying to avoid the situation when the components of tensor quantity in question are large and, yet, the Lorentzian contraction is small. And, in both cases, we resort to breaking Lorentz covariance in such a way that we are able to judge whether or not the components of tensor quantity are large in our "preferred frame", but not in other frames.

The function $g (\eta, \xi)$ is supposed go to zero whenever either $\eta \rightarrow 0$ or $\xi \rightarrow 0$. Thus, if $\vert \psi_i \vert^2$ is "small", the "lowering signal" will be weak. After all, as long as the state is "inprobable", it might be "okay" for it to "violate general relativity" and we don't have a need to "get rid of it". On the other hand, if the difference between $G_{\mu \nu}$ and $T_{\mu \nu}$ is small then, again, the lowering signal is weak, even if $\vert \psi_i \vert^2$ is large. This time we simply say that the state falls within the "tolerance" of our theory; thus, it should be "allowed" to have large probability. In this paper, we are deliberately leaving $g$ unspecified beyond these general considerations. This will allow us to explore in future research different implications of different specifications of $g$ and thus make more informed decision of what $g$ is best to select. 

Since we can now easilly identify the charge of the particle that has emitted the field based on the value of $Q (\vec{x}, t)$, a particle $j$ can "look" at a vicinity of \emph{its own location} and "decide" what the value of $q_i$ is:
\beq \vec{x} = \vec{x}_j \Rightarrow q_i \approx Q (\vec{x}_j) \eeq
If a particle $j$ "sees" that $Q (\vec{x}_j, t)$ approximates $q_j$ (up to some afore-given "tolerance" value), it will "decide" to lower $q_j$; otherwise, it will "ignore" the signal. 
\beq \frac{d \psi_i}{dt} =  - \frac{K_{\psi; i} (t) \beta \mu_{\rm err} (\vec{x}_i, t) \vert \vec{X} (\vec{x}_i, t) \vert \psi_i (t)}{2 \vert \psi_i \vert^2} T (\vert \mu_{\rm err} (\vec{x}_i, t) \vert > \mu_{\rm err; 0}) T (\vert Q(\vec{x}_i, t) -  q_i  \vert < \epsilon_{qe})  \label{eqn:566}\eeq 
The parameter $\epsilon_{qe}$ refers to the tolerance in which we can rely on $Q (\vec{x}_i, t)$ as a measure of charge $q_j$ of the emitting particle. We call it $\epsilon_{qe}$ instead of $\epsilon_q$ because $\epsilon_q$ has already been used for a completely different purpose. The "e" in $\epsilon_{qe}$ stands for "Einstein". Now, seeing that "charges" are integer, we would like to have strict equality between $q_i$ and $q_j$. But, at the same time, the estimate $Q(\vec{x},t)$ can not be exact because  there might be some "small fluctuations" of $\mu_e$ and $\mu_E$ as a "fallout" from some other emissions; that is why instead of demanding exact equality, we demand approximation, with tolerance $\epsilon_{qe}$.

The ratio $\mu/ \vert \vec{X} \vert$ is meant to approximate the "correction" to $\psi_i$ that the "emitting" particle $j$ "wants" to make. The only way it can communicate it is through the signal $\mu_{\rm err}$. But, since $\mu_{\rm err}$ is weakened with distance, it should be balanced out by multiplication by the distance to source. The only "local" way of knowing that distance is through $\vert \vec{X} \vert$, which is why we multiply the former by the latter. Now, if no signals was emitted, then there might be some small fluctuations that would produce some "misleading" information about $\vert \vec{X} \vert$. If $\vert \vec{X} \vert$ "happens" to be "very large", the value of $\mu_{\rm err} \vert \vec{X} \vert$ will be large as well and produce unwanted large effect. We avoided this situation by introducing a multiple $T (\vert \mu_{err} \vert > \mu_{\rm err; 0}$. The constant $\mu_{\rm 0}$ is chosen in such a way that $\mu_{\rm err}$ produced by "small fluctuations" will fail to exceed it; thus, any such unwanted effect will be killed off by multiplication to by zero. The factor $\psi/ \vert \psi \vert^2$ comes from our attempt to make sure that $d \vert \psi_i \vert^2 /dt$ is constant; this can be achieved if $d \psi_i/ dt$ is proportional to $\psi/ \vert \psi \vert^2$. The fact that $d \vert \psi_i \vert^2/dt$ \emph{as opposed to} $d \psi/dt$ is constant impies that $\vert \psi_i \vert^2$, as oppose to $\psi$, is used as "money" in gambler's ruin, thus producing Born's rule. 

The constant $K_{\psi; i}$ is a normalization constant. It is needed in order to make sure that all the "wins" and all the "loses" balance each other out. This is a multi-player generalization of $q_1 g_1 = q_2 g_2$ (equation \ref{eqn:535}). Strictly speaking, in terms of $\psi_i$, all of the particles are "losing". But, since we define define wins and losses in terms of $\psi_i/K_{\psi i}$. This would mean that, while all particles "lose", the ones that "lose less" are the ones that win. Now, it is possible to stick to this definition of "winning" and "losing" and, at the same time, skip $K_{\psi i}$ in the dynamics itself. If we did that, however, it would imply that, as time progresses, the scales of $\psi_i$ decrease and, \emph{compared to that scale}, the scales of $d \psi_i/d t$ will increase. Thus, the "collapse" will become stronger and stronger as time goes on. The only way of accounting for \emph{lack} of such a phenomena is $K_{\psi i}$ in a numerator. Thus, as scales of $\psi_i$ decrease, $K_{\psi i}$ decreases as well. Thus, \emph{as long as $K_{\psi i}$ is in numerator} we have smaller and smaller "steps" as time goes on; this means that on the scale of decreasing $\vert \psi_i \vert^2$ the length of steps is still the same. Thus, the "strength" of "collapse" does not change in time. 

Another thing one might notice is that $dK_{\psi i}/dt \neq 0$ might imply that we might want to have $d (K_{\psi i} \vert \psi \vert^2)/dt =0$ which \emph{contradicts} $d \vert \psi \vert^2 /dt =0$. It turns out that we are able to avoid such conflict due to our "pulse by pulse" picture. At the time when $d \psi_i / dt \neq 0$, particle $j$, with charge $q_j = q_i$, is the \emph{only one} that emits a pulse \emph{at that exact time}. From this we know that, if $q_k \neq q_i$, then $d \psi_k /dt =0$. Now, since the value of $K_{\psi i}$ represents the sum of $\vert \psi_k \vert^2$ over \emph{all} $k$ (the only purpose of index $i$ is to "encode" that "global" quantity "locally", similarly to our encoding the "global" action $S$ locally as $S_i$), the value of $d K_{\psi i}/ dt$ is neglegebly small. 

Now, as we recall, the "pulses" the particle $j$ emits are "timed" based on the "rotating" $(\chi_{j1}, \chi_{j2})$. The reason the pulses don't overlap is that the phases of these "rotations" are different for different $j$. At the same time, the frequency, $\nu$, is the same. Thus, within a given time interval $\nu^{-1}$ \emph{all} particles emitted a pulse. Therefore, the average of incriment of $K_{\psi i}$ within that time interval is \emph{not} small. What happens is that $\psi_i$ reacts "strongly" to pulses emitted by "same charged" particles, and "completely ignores" all the other pulses. On the other hand, $K_{\psi i}$ reacts "weakly" to \emph{all} the pulses. As a result, during the "same charged" pulses, $\psi_i$ "leaps ahead" and $K_{\psi i}$ "falls behind"; then, during the rest of the time interval, $K_{\psi i}$ "catches up". \emph{But}, as far as the dynamics of $\psi_i$ is concerned, the \emph{only} thing relevent is what happens when same-charged pulses are emitted. \emph{That} is why we safely assuming that $dK_{\psi i}/dt$ is negligibly small, without worrying about "catching up" part. At the same time, the \emph{overall} value of $K_{\psi i}$ (as opposed to its derivative) \emph{is} affected by the "catching ups" in the past. That is why we \emph{do} have to include it as one of the "constant" factors that, despite being treated as "constant", is distinct from unity.  

In the above expression, we did \emph{not} include the coefficient $l_i (t)$ for the following two reasons. First of all, skipping $l_i (t)$  does \emph{not} mean "making a particle visible". After all, the process we are describing does \emph{not} involve the particle sending out any signals. On the contrary, it involves \emph{lowering} $\psi_i$ thus comming up with \emph{additional reason} for the particle to be "invisible". Thus, if $l_i$ is \emph{always} close to zero, the outcome is identical whether we include it or not. If we include it, then $\psi_i$ will "stay large" \emph{but} particle will "still" be "invisible" because of $l_i$ being "small". If, on the other hand, we skip $l_i$ then the particle will be invisible because $\psi_i$ is small \emph{and also} because $l_i$ is small. Whether there are two reasons for particle to be invisible, or only one, the outcome is the same. 

The second, and far more important, reason for skipping $l_i (t)$ is that we do not want the "unwanted" state to "reimerge" if/when $l_i (t)$ \emph{stops} being nearly-zero. Consider the following scenario. First, $g_{\mu \nu}$ increased. As a result, some particles needed to "hide". After that, an entanglement occurs for some other, \emph{unrelated}, reason. As a result of this entanglement, the "gambler's ruin" is played; the outcome is that $\phi_1 (\vec{x})$ "wins" while $\phi_2 (\vec{x})$ "loses". Finally, $g_{\mu \nu}$ decreases, so the hidden particles become "visible" again. Now, $\phi_1 (\vec{x})$ and $\phi_2 (\vec{x})$ correspond to "charges" $q_1$ and $q_2$. Thus, in order for $\phi_2 (\vec{x})$ to lose, we have to insist that 
\beq q_i = q_2 \Rightarrow \phi_i \approx 0 \eeq
\emph{But} if we we assume that the hidden particles were exempt from participating in "gambler's ruin" (by including $l_j (t)$ in the equation), then the above will \emph{not} be true for them. While $g_{\mu \nu}$ is still "large", and they continue to be hidden, this will not create any problem: even though $\psi_j$ is "large" we can not "see it" due to $l_j$ being small. But, the moment $g_{\mu \nu}$ decreases "back to normal", they reimerge. Thus, we will be able to "see" the superposition of two different measurement outcomes \emph{long after} one of them was destroyed "through measurement"! 

The obvious way to avoid it, of course, is to make sure that \emph{all} particles participate in "gambler's ruin" and, therefore, $\psi_j$ becomes "small" \emph{even among the invisible particles}. The lowering of $g_{\mu \nu}$ only changes $l_i (t)$ from $0$ back to $1$, but it does \emph{not} affect $\psi_i (t)$. Thus, $\psi_i (t)$ will continue to stay small \emph{as a result of gambler's ruin} and, therefore, no "trouble" will occur. This, however, requires $\psi_i (t)$ to actually participate in "gambler's ruin" while the particle $i$ is in "invisible state". The "participation in 'gambler's ruin' " is simply the change of $\psi_i$ per equation (\ref{eqn:566}). Thus, we want the right hand side of (\ref{eqn:566}) to be non-zero \emph{even when} the particle is invisible (per $l_i (t)$ being near-zero). This amounts to \emph{avoiding} the coefficient $l_i (t)$ on the right hand side.  

What we have just said is related to a more general issue. As we recall from Chapter 2, we expect $\psi_i$ to be "globally correlated" among same-charged particles. Yet, what we were saying right now implies that, if $l_i (t)$ were included, then we would have $\psi_i$ "large" and $\psi_j$ small \emph{despite} $q_i = q_j$ as long as $j$ is "visible" while $i$ is not. What this tells us is that we are in danger of violating the "global correlations" that we were insisting upon throughout this paper. Now, the source of "global correlations" was the fact that the reception of signals was \emph{always} coupled to $q_i$ \emph{alone}. By introducing any additional coupling, whether it be a coupling to $l_i (t)$ or anything else, we are at risk of "destroying" the "global correlation" we worked so diligently to preserve. 

The obvious cure for this problem is to make sure that whenever we describe particles \emph{response} to external signal, we should \emph{not} include any quantities other than $q$; which, by implication, means we should not inclue $l_i (t)$ either. On the other hand, when we describe particle's \emph{emission} of a signal, it is okay to include quantities unrelated to $q$: after all, we have already been using $q'$ instead of $q$ and this did not cause any problems. Thus, we can safely include $l_i (t)$ "as well". This implies that the visibility is a barrier \emph{only} for emission and \emph{not} for absorption. As long as the "emission barrier" is enforced, our goals of "blocking" particle's influence are met. 

\subsection*{6. Conclusion}

In this paper we have proposed a way of discretizing quantum field theory and, after that, proposing a classical model that "generates" the mathematical information we would "like" to obtain. Upon cursory read of this paper, one question comes to mind. Namely, we have no evidence that the specific "classical mechanism" we are proposing takes place. The obvious answer to this kind of criticism is a proposal of a research direction that would lead to some specific predictions. In fact,  we can think of one such proposal. In particular, Chapter 5 of this paper implies that gravity can be modeled as entirely classical theory. If that is the case, every single problem should, in principle, be solvable in terms of classical Einstein's equation. This should include the "unanswered" problems such as information paradox.

At the same time, however, this comes with the price. The set of differential equations involved will \emph{not} just be Einstein's equation, but also a detailed description of its sources, per Chapter 4. Cursory reading of Chapter 4 shows that it might be unreasonable to hope to solve these equations analytically. Nevertheless, again by reading Chapter 4, one might notice that at first we had a specific "pattern" in mind, and only after that we have written down the equations that fit that pattern. As a result, it might be possible to analyze our "classical" system in a "qualitative" way and attempt to see what it has to say regarding things like information paradox.

The other "interesting" question to explore involves superluminal singlas. We had assumed that they circle the universe within a very short period of time in order to produce a mathematical informaiton consistent with "globally defined" probability amplitudes. But what if such is not the case? What if they \emph{only} pass through our galaxy but not the entire universe within a short time? In this case it is logical that the results of quantum field theory break down on cosmological scales. But what will happen instead? The fact that we have an explicit "classical" theory allows us to explore this question.

\end{document}